\pdfoutput=1 
\documentclass[twocolumn,tighten]{aastex701}

\hypersetup{
    colorlinks=true,
    linkcolor=red,
    filecolor=magenta,      
    urlcolor=cyan,
    citecolor=blue,
    }
    
\usepackage[all]{hypcap}

\shortauthors{Gregg et al.}

\begin{document}

\title{The Calibration of Short Wavelength Polycyclic Aromatic Hydrocarbon Emission as Star Formation Rate Indicators with JWST}

\correspondingauthor{Benjamin Gregg}
\email{bagregg@astro.umass.edu}

\author[0000-0003-4910-8939]{Benjamin Gregg}
\affiliation{Department of Astronomy, University of Massachusetts, 710 North Pleasant Street, Amherst, MA 01003, USA}
\email{bagregg@astro.umass.edu}

\author[0000-0002-5189-8004]{Daniela Calzetti}
\affiliation{Department of Astronomy, University of Massachusetts, 710 North Pleasant Street, Amherst, MA 01003, USA}
\email{calzetti@astro.umass.edu}

\author[0000-0002-8192-8091]{Angela Adamo}
\affiliation{Department of Astronomy, The Oskar Klein Centre, Stockholm University, AlbaNova, SE-10691 Stockholm, Sweden}
\email{angela.adamo@astro.su.se}

\author[0000-0002-8222-8986]{Alex Pedrini}
\affiliation{Department of Astronomy, The Oskar Klein Centre, Stockholm University, AlbaNova, SE-10691 Stockholm, Sweden}
\email{alex.pedrini@astro.su.se}

\author[0000-0002-1000-6081]{Sean T. Linden}
\affiliation{Steward Observatory, University of Arizona, 933 N. Cherry Avenue, Tucson, AZ 85719, USA}
\email{seanlinden@arizona.edu}

\author[0009-0008-4009-3391]{Varun Bajaj}
\affiliation{Space Telescope Science Institute, 3700 San Martin Drive, Baltimore, MD 21218, USA}
\email{vbajaj@stsci.edu}

\author[0000-0002-2918-7417]{Jenna E. Ryon}
\affiliation{Space Telescope Science Institute, 3700 San Martin Drive, Baltimore, MD 21218, USA}
\email{ryon@stsci.edu}

\author[0000-0001-8068-0891]{Arjan Bik}
\affiliation{Department of Astronomy, The Oskar Klein Centre, Stockholm University, AlbaNova, SE-10691 Stockholm, Sweden}
\email{arjan.bik@astro.su.se}

\author[0009-0003-6182-8928]{Giacomo Bortolini}
\affiliation{Department of Astronomy, The Oskar Klein Centre, Stockholm University, AlbaNova, SE-10691 Stockholm, Sweden}
\email{giacomo.bortolini@astro.su.se}

\author[0000-0001-6464-3257]{Matteo Correnti}
\affiliation{INAF Osservatorio Astronomico di Roma, Via Frascati 33, 00078, Monteporzio Catone, Rome, Italy}
\affiliation{ASI-Space Science Data Center, Via del Politecnico, I-00133, Rome, Italy}
\email{matteo.correnti@inaf.it}

\author[0000-0002-0846-936X]{Bruce T. Draine}
\affiliation{Department of Astrophysical Sciences, Princeton University, 4 Ivy Lane, Princeton, NJ 08544, USA}
\email{draine@astro.princeton.edu}

\author[0000-0002-1723-6330]{Bruce G. Elmegreen}
\affiliation{Katonah, NY 10536, USA}
\email{belmegreen@gmail.com}

\author[0000-0002-2199-0977]{Helena Faustino Vieira}
\affiliation{Department of Astronomy, The Oskar Klein Centre, Stockholm University, AlbaNova, SE-10691 Stockholm, Sweden}
\email{helena.faustinovieira@astro.su.se}

\author[0000-0001-8608-0408]{John S. Gallagher}
\affiliation{Department of Astronomy, University of Wisconsin-Madison, 475 N. Charter Street, Madison, WI 53706, USA}
\email{jsg@astro.wisc.edu}

\author[0000-0002-3247-5321]{Kathryn Grasha}
\affiliation{Research School of Astronomy and Astrophysics, Australian National University, Canberra, ACT 2611, Australia}   
\affiliation{ARC Centre of Excellence for All Sky Astrophysics in 3 Dimensions (ASTRO 3D), Australia}   
\email{kathryn.grasha@anu.edu.au}

\author[0000-0001-8348-2671]{Kelsey E. Johnson}
\affiliation{Department of Astronomy, University of Virginia, Charlottesville, VA 22904, USA}
\email{kej7a@virginia.edu}

\author[0000-0001-8490-6632]{Thomas S.-Y. Lai}
\affiliation{IPAC, California Institute of Technology, 1200 E. California Blvd., Pasadena, CA 91125, USA}
\email{shaoyu@ipac.caltech.edu}

\author[0000-0003-1427-2456]{Matteo Messa}
\affiliation{INAF - Osservatorio di Astrofisica e Scienza dello Spazio di Bologna, Via Gobetti 93/3, I-40129 Bologna, Italy}
\email{matteo.messa@inaf.it}

\author[0000-0002-3005-1349]{G\"{o}ran \"{O}stlin}
\affiliation{Department of Astronomy, The Oskar Klein Centre, Stockholm University, AlbaNova, SE-10691 Stockholm, Sweden}
\email{ostlin@astro.su.se}

\author[0000-0002-0806-168X]{Linda J. Smith}
\affiliation{Space Telescope Science Institute, 3700 San Martin Drive, Baltimore, MD 21218, USA}
\email{lsmith@stsci.edu}

\author[0000-0002-0986-4759]{Monica Tosi}
\affiliation{INAF - Osservatorio di Astrofisica e Scienza dello Spazio di Bologna, Via Gobetti 93/3, I-40129 Bologna, Italy}
\email{monica.tosi@inaf.it}

\received{July 3 2025}
\revised{Oct. 23 2025}
\accepted{Nov. 4 2025}
\submitjournal{ApJ}

\begin{abstract}
We use JWST/NIRCam and MIRI imaging acquired by the Feedback in Emerging extrAgalactic Star clusTers (FEAST) program along with archival HST imaging to map ionized gas (Pa$\alpha$, Br$\alpha$, and H$\alpha$) and Polycyclic Aromatic Hydrocarbon (PAH) emission (3.3 and 7.7~$\mu$m) across a sample of four nearby galaxies (NGC~5194, 5236, 628, and 4449). These maps are utilized to calibrate the PAH features as star formation rate (SFR) indicators in 40~pc size regions around massive emerging young star clusters (eYSCs). We find a tight, sub-linear (power-law exponent, $\alpha{\,}{\sim}{\,}0.8$) relation between the PAH luminosities (3.3 and 7.7~$\mu$m) and SFR (extinction corrected Pa$\alpha$) in near solar metallicity environments. PAH destruction in more intense ionizing environments and/or variations in the age of our sources may drive the deviation from a linear relation. In the metal-poor environment of NGC~4449 (${\sim}$1/3~Z$_{\odot}$), we see substantial deficits in the PAH feature strengths at fixed SFR and significantly higher scatter in the PAH-SFR relations. We determine that the 3.3/7.7~$\mu$m PAH luminosity ratio increases towards lower metallicity environments. This is interpreted as a result of a shift in the size distribution towards smaller PAHs at lower metallicities, possibly due to inhibited grain growth. Focusing on the regions in NGC~4449, we observe a decreasing 3.3/7.7~$\mu$m ratio towards higher SFR, which could indicate that small PAHs are preferentially destroyed relative to larger PAHs in significantly sub-solar metallicity conditions. We estimate that ${\sim}$2/3 of the PAH emission in typical local star-forming galaxies is excited by older stars and unrelated to recent ($<$10~Myr) star formation.
\end{abstract} 


\hypertarget{1}{\section{Introduction}}

Polycyclic Aromatic Hydrocarbons (PAHs) are dust molecules that are generally an abundant and highly influential component of the interstellar medium (ISM) in actively star-forming galaxies. They are considered to be a type of nanoparticle-sized dust ``grain", although their physical properties are much different from the orders of magnitude larger grains that comprise the majority of the dust mass in galaxies \citep[e.g.][]{2021ApJ...917....3D,2023ApJ...948...55H}. PAH molecules are composed of carbon (C) and hydrogen (H) atoms that are predominantly bonded in an aromatic ring structure, normally with 20$-$1000 C atoms total \citep{2008ARA&A..46..289T}. In the ISM, PAHs absorb ultraviolet (UV) and optical photons, typically in the energy range of ${\sim}$3$-$9 eV \citep{2021ApJ...917....3D}, which excites various vibrational modes of the C-H and C-C bonds and the energy is then re-emitted in the infrared (IR). This process has been associated with the broad near-infrared (NIR) and mid-infrared (MIR) emission features that are widely observed in star-forming galaxies \citep[e.g.][]{1984A&A...137L...5L,1985ApJ...290L..25A,1989ApJS...71..733A,2008ARA&A..46..289T,2020NatAs...4..339L}. These emission features are very luminous in normal star-forming galaxies, corresponding to roughly 10$-$20$\%$ of the total IR luminosity \citep{2000ApJ...532L..21H,2007ApJ...656..770S,2008ARA&A..46..289T,2020NatAs...4..339L}.

The high intensity of UV radiation in the regions around massive, newly formed stars or star clusters results in the efficient heating of PAH molecules and bright, localized emission. As PAHs are some of the smallest dust particles in the ISM, they are much more susceptible to dissociation/destruction by the high-energy ionizing UV radiation from newly formed stars, since the heat capacity of dust grains is proportional to their volume \citep{1989ApJ...345..230G,2003ApJ...584..316L,2021ApJ...917....3D}. The survival of PAHs typically requires shielding from this ionizing radiation by intervening material, i.e., hydrogen gas \citep{2004ApJS..154..253H,2007ApJ...660..346P,2008MNRAS.389..629B,2009ApJ...699.1125R}, and thus in star-forming regions, PAH emission is widely thought to emanate from the photodissociation regions (PDRs) that encompass HII regions, rather than within them \citep[e.g.][]{2009ApJ...699.1125R}. This has been generally confirmed with James Webb Space Telescope (JWST) observations of the Orion Nebula in the Milky Way \citep[e.g.][]{2024A&A...685A..73H}. 

As a result of the tight spatial connection between PAH emission and star-forming regions in galaxies, PAHs have a long history of being used as a tracer of the star formation rate (SFR) \citep[e.g.][]{2004ApJS..154..253H,2004ApJ...613..986P,2006ApJ...652..283B,2007ApJ...666..870C,2007ApJ...657..810D,2007ApJ...656..770S,2009ApJ...703.1672K,2016ApJ...818...60S}. PAH emission is a less direct SFR tracer compared to the far-ultraviolet (FUV) continuum that traces massive, short-lived O-type stars, and to the nebular line emission from the gas that those stars ionize \citep[e.g. hydrogen recombination lines;][]{2013seg..book..419C}; however, the latter are inherently tracers of the component of SFR that is unobscured and require a correction for dust attenuation, which can be highly uncertain when limited to only UV and optical wavelengths. On the other hand, dust (including PAH) emission can inherently trace the dust-obscured SFR. The bolometric IR emission (${\sim}$5$-$1000~$\mu$m) depends sensitively on the temperature of the dust, which increases on average in galaxies that are more strongly star-forming \citep[e.g.][]{1986ApJ...311L..33H}, and has been used and calibrated extensively as a tracer of the obscured SFR \citep[e.g.][]{1998ARA&A..36..189K,2012ARA&A..50..531K,2013seg..book..419C, 2023A&A...678A.129B}. The comparatively shorter wavelength of the PAH emission features, and the fact that they result from single-photon heating, can be useful, particularly for studying dust-obscured star formation at high redshift. The most luminous PAH emission feature at about 7.7~$\mu$m has been widely investigated, calibrated, and used as an SFR indicator out to high redshift \citep[e.g.][]{2011A&A...533A.119E,2024ApJ...970...61R}. However, there are a few significant complications that previous studies have identified. 

The abundance of PAHs in the ISM, as well as their emission properties, can significantly depend on the local ISM environment (metallicity, radiation field, etc.) and the star formation history of the galaxy. A substantial deficit in the 7.7~$\mu$m PAH luminosity for low metallicity galaxies has been observed by many studies \citep[e.g.][]{2005ApJ...628L..29E,2007ApJ...666..870C,2007ApJ...663..866D,2007ApJ...656..770S,2008ApJ...682..336G,2012ApJ...744...20S,2014MNRAS.445..899C,2017ApJ...837..157S,2022ApJ...928..120G}, typically interpreted as a result of a decreased PAH abundance. The decreased abundance of PAHs at low metallicity has widely been attributed to PAH destruction by the harder (or higher average energy per photon) radiation field \citep{2006A&A...446..877M,2008ApJ...682..336G,2010ApJ...712..164H}. Yet, some studies have found evidence for a smaller PAH size distribution in metal-poor environments, disfavoring PAH destruction as the main mechanism for the observed deficit, and argue instead that inhibited PAH grain growth at low metallicity is the dominant driver of the trend, given that metals can act as catalysts for the formation and growth of PAHs \citep[e.g.][]{2012ApJ...744...20S,2024ApJ...974...20W,2025ApJ...991L..56L,2025ApJS..280....4Z}. The harsh radiation field in the vicinity of an active galactic nucleus (AGN) has a strong effect on the observed emission from PAHs, and evidence of destruction is clear in these regions \citep[e.g.][]{2022Univ....8..356S,2023ApJ...957L..26L}. Spatially resolved studies show that PAH emission can be excited by relatively cool stars \citep[e.g.][]{2001AAS...199.9713L} and that a significant fraction of the 7.7~$\mu$m emission is associated with the diffuse ISM \citep{2008MNRAS.389..629B,2013ApJ...762...79C,2014ApJ...784..130C,2014ApJ...797..129L}, implying that a large component of the PAH emission may be excited by older stellar populations that do not trace recent ($<$10 Myr) star formation. All of these issues complicate the use of the PAH emission features as SFR indicators.

Despite these complications, to the extent that PAH emission can trace SFR, it may be of key importance to surveys of high redshift galaxies, particularly in the era of JWST. In fact, in some distant systems, it may be the only feasible way to trace the dust-obscured star formation. Star formation is highly dust-obscured by nature, and over half of the photons that trace recent star formation in galaxies at z$\leq$5 are obscured and absorbed by dust and re-radiated in the IR \citep{2018ApJ...862...77C,2020ApJ...902..112B}. The shortest wavelength PAH emission feature at 3.3~$\mu$m can be observed with JWST/MIRI/MRS spectroscopy out to a redshift of at least z${\sim}4.5$ and could be used to push dust-obscured measures of SFR well past Cosmic Noon \citep[e.g.][]{2020ApJ...905...55L}. An additional strength of the 3.3~$\mu$m PAH feature as a tracer of SFR is that it is ${\sim}$2.5 times less sensitive to dust extinction compared to the hydrogen recombination line Pa$\alpha$ (1.87~$\mu$m) and a factor of 3$-$10 times brighter than Br$\alpha$ (4.05~$\mu$m) in normal star-forming galaxies \citep{2018A&A...617A.130I}, allowing it to be more easily detected at high redshift. The 3.3~$\mu$m feature has been detected at high redshift with Spitzer/IRS spectroscopy for ultraluminous infrared galaxies (ULIRGs) at z${\sim}$2 \citep{2009ApJ...703..270S} and for a strongly lensed galaxy at z${\sim}$3 \citep{2009ApJ...698.1273S}, and with JWST/MIRI/MRS spectroscopy for a lensed galaxy at z${\sim}$4.2 \citep{2023Natur.618..708S}. Yet, to understand and interpret the emission feature at high redshift, we must first develop a detailed understanding at high resolution in local systems and across diverse environments. In the past, this feature was difficult to study observationally in detail in extragalactic objects due to a lack of coverage by the spectroscopic and imaging instruments on Spitzer for targets in the local Universe. JWST is now providing new insights into the nature of the 3.3~$\mu$m feature \citep[see][]{2023ApJ...944L..12C,2023ApJ...957L..26L,2023ApJ...944L...7S,2024A&A...685A..75C,2024ApJ...971..115G,2024A&A...685A..73H,2025ApJS..280....4Z}.

The various NIR/MIR PAH emission features are known to trace different PAH species. The 3.3~$\mu$m feature is emitted primarily by the smallest, neutral PAHs \citep[e.g.][]{2020MNRAS.494..642M,2021ApJ...917....3D} and contributes ${\sim}0.1\%$ of the total IR luminosity and ${\sim}$1.5$-$3$\%$ of the total PAH power in nearby star-forming galaxies \citep{2020ApJ...905...55L}. The 7.7~$\mu$m feature originates from relatively larger, ionized (positively charged) PAHs \citep[e.g.][]{2020MNRAS.494..642M,2021ApJ...917....3D} and carries ${\sim}40\%$ of the total PAH power \citep{2007ApJ...656..770S}. The relative strengths or ratios of the different PAH features can probe the properties of PAH molecules \citep[e.g.][]{2008ARA&A..46..289T,2012ApJ...744...20S,2020MNRAS.494..642M}. In particular, the ratio of the 3.3 to 7.7~$\mu$m features traces both the size distribution and ionization such that an increased ratio traces a smaller and/or less ionized PAH population \citep[e.g.][]{2021ApJ...917....3D}. In this way, PAH emission can be a powerful probe of the ISM in galaxies, both in terms of its physical properties and its distribution \citep[see][]{2023ApJ...944L...8S,2024A&A...685A..78S}.

In this study, we explore and calibrate two short wavelength PAH emission features (3.3 and 7.7~$\mu$m) as SFR indicators on the fundamental scale of individual newly formed star clusters and HII regions across diverse nearby galaxy environments. We utilize Cycle 1 JWST/NIRCam and MIRI imaging from the FEAST (Feedback in Emerging extrAgalactic Star clusTers) survey (ID 1783, PI: A. Adamo) in combination with previously obtained Hubble Space Telescope (HST) imaging to map PAH (3.3 and 7.7~$\mu$m) and ionized gas emission (Pa$\alpha$, Br$\alpha$, and H$\alpha$) across four local galaxies (NGC 5194, 5236, 628, and 4449) at the angular resolution of ${\sim}{\,}$0.07$-$0.27$''$. From these emission line maps, candidate emerging young star clusters (eYSCs) are selected as tightly spatially connected, compact peaks in PAH and ionized gas emission. PAH and ionized gas properties are measured in 40 pc size regions around the eYSCs, and the relationship between the properties is evaluated. The eYSC stage represents an early phase ($\lesssim \,$5$-$7 Myr) of star cluster evolution where sources are still embedded in their natal gas and dust, before clearing their surroundings and emerging as optically visible young star clusters (YSCs). As such, eYSCs represent the best sources to study PAH emission and its relationship to star formation on the scale of individual star clusters in nearby galaxies. In the regions around these sources, we also investigate the 3.3 to 7.7~$\mu$m PAH luminosity ratios and how they vary as a function of the local ISM environment and physical properties. This study is a direct extension of the work by \cite{2024ApJ...971..115G}, which included only one galaxy target, NGC 628, and only the 3.3~$\mu$m feature. 

This paper is structured in the following way. In Section \hyperlink{2}{2}, we present our sample of galaxy targets and their characteristics, the new JWST/NIRCam and MIRI imaging data, and the basic data reduction. In Section \hyperlink{3}{3}, we describe the analysis of our data, including the continuum subtractions, eYSC selection, aperture photometry, and derivation of PAH, ionized gas, and physical properties. The results of our work are given in Section \hyperlink{4}{4}. We discuss our results and their implications in the context of former studies in Section \hyperlink{5}{5}. In Section \hyperlink{6}{6}, our major conclusions are highlighted. In Appendix \hyperlink{A}{A}, we test the continuum subtraction of MIRI/F770W and discuss its effect on our results. In Appendix \hyperlink{B}{B}, we test if overlapping photometric measurements have an impact on our results. 

\hypertarget{2}{\section{Sample and Data}}

Our sample consists of four galaxies: NGC 5194 (M51), NGC 5236 (M83), NGC 628 (M74), and NGC 4449 (Caldwell 21). These galaxies are selected from the parent sample of the Local Volume Legacy (LVL) survey \citep{2009ApJ...703..517D} and are well-studied with coverage from many space- and ground-based telescopes across the electromagnetic spectrum. They represent some of the diversity of star-forming environments found in the nearby ($<$10 Mpc) Universe. Three are massive, well-defined spiral galaxies (NGC 5194, 5236, 628), while one is a Magellanic-type, irregular dwarf galaxy (NGC 4449). NGC 5194 exhibits strong spiral arms and high SFRs throughout the disk, likely induced by its interaction with the companion NGC 5195 \citep{2010MNRAS.403..625D}. NGC 5236 hosts a strong nuclear starburst and a prominent central bar. NGC 628 is slightly lower in mass and metallicity compared with the other spirals and exhibits no evidence of a central bar or elevated nuclear star formation. The irregular dwarf NGC 4449 has significantly lower mass and metallicity and is an interacting galaxy, hosting a central starburst. The galaxy sample and important physical properties are listed in Table \hyperlink{t1}{1}.

In this study, we use both newly obtained and archival JWST data, as well as archival HST data. The new JWST data for each of the four galaxies in our sample were obtained in Cycle 1 as part of the JWST--FEAST program (ID 1783, PI: A. Adamo). We utilize both JWST/NIRCam and MIRI imaging from this program for each galaxy with various filters including the NIRCam/F150W, F187N, F200W, F300M, F335M, F405N, and F444W, and the MIRI/F560W and F770W.

These JWST data are reduced in the following way. We obtain stage two calibrated data products from the Mikulski Archive for Space Telescopes (MAST), produced with (or more recent than) the calibration pipeline 1.12.5 using the calibration reference data context number 1174 for NIRCam, or the pipeline 1.11.4 using context number 1141 for MIRI. We extract catalogs containing point spread function (PSF) fit positions and fluxes from the stage two products using the Python package \texttt{one$\_$pass$\_$fitting}\footnote{\url{https://github.com/Vb2341/One-Pass-Fitting}}, with PSF models created by WebbPSF \citep{2014SPIE.9143E..3XP}. The JWST images and corresponding catalogs cannot be consistently aligned to Gaia \citep{2016A&A...595A...1G,2023A&A...674A...1G} due to the difference in wavelength and sensitivity.  Therefore, a laddered approach is used. We first align the archival HST/ACS (or WFC3) F814W image to Gaia \citep[see][]{2017wfc..rept...19B} and then extract a catalog from F814W on the Gaia astrometric frame.  We then provide this reference F814W catalog and the PSF-fitted NIRCam/F200W catalogs as custom, user-supplied catalogs in stage three of the JWST calibration pipeline. The resulting aligned F200W catalogs are combined and used as a reference for the remaining NIRCam and MIRI data, yielding an overall astrometric precision of ${\sim}$15 milliarcseconds or less. Aligned images are then combined into a single mosaic for each filter using the stage three pipeline, projected onto the same pixel grid with a scale of 0.04 $''$/pixel for NIRCam and 0.08 $''$/pixel for MIRI.  We convert the mosaics from units of MJy/sr to Jy/pixel. A more detailed discussion of the entire data reduction process is presented in A. Adamo et al. (in prep.). Also, see \cite{2024ApJ...971...32P} for a short description of our method of removing gradients in the background of the final MIRI mosaics. 

In addition to the JWST--FEAST data, we utilize two other JWST imaging bands that have been observed for one of our targets, NGC 628, which are publicly available on MAST, obtained as part of the PHANGS--JWST program (ID 2107; PI: J. C. Lee) and presented in \cite{2023ApJ...944L..17L}. We acquire the stage two NIRCam/F300M and MIRI/F1000W data products from this program for NGC 628 and run them through our full data reduction process. These observations align very well within the larger footprint of the FEAST mosaics, but only cover a fraction (${\sim}$2/3) of the field of view (FOV). The inclusion of these data ensures uniform coverage of the JWST imaging filters across our sample, since in NGC 628 alone, the FEAST program obtained the NIRCam/F277W rather than the F300M. The F1000W allows us to investigate a variety of continuum subtraction techniques for the 7.7~$\mu$m PAH emission in NGC 628. 

\begin{figure}
\centering
\includegraphics[width=0.47\textwidth]{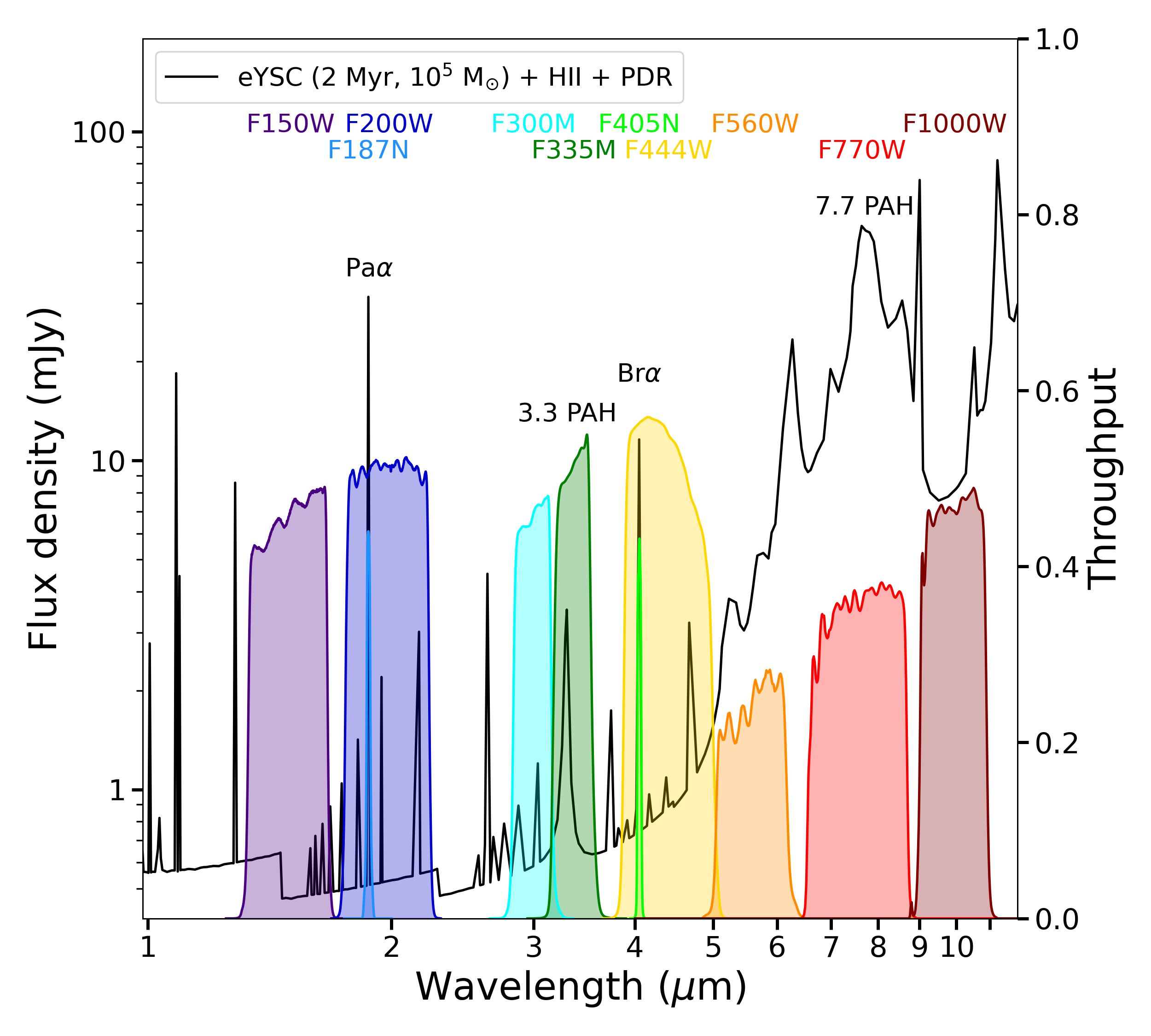}
\caption{Spectrum of a 2 Myr old, 10$^{5}$ M$_{\odot}$ eYSC powering an HII region and PDR from the MAPPINGS III derived models of \cite{2008ApJS..176..438G} (black line). The JWST/NIRCam and MIRI filter throughputs (colored curves) are overlaid on the model to outline the spectral regions sampled by our data. The F187N and F405N target the Pa$\alpha$ and Br$\alpha$ hydrogen recombination lines, respectively. The F335M and F770W target the 3.3 and 7.7~$\mu$m PAH emission features, respectively. All other filters target continuum emission. Note that the MIRI/F1000W filter is only available in one of our targets, NGC 628. }
\label{fig:f1}
\end{figure}

In Figure \ref{fig:f1}, we show the total system throughput curves for all JWST/NIRCam and MIRI imaging filters investigated in this study, on top of a model spectrum, corresponding to a 2 Myr old, 10$^{5}$ M$_{\odot}$ eYSC that powers both an HII region and the surrounding PDR, from the study by \cite{2008ApJS..176..438G}. The model assumes the Starburst99 models \citep{1999ApJS..123....3L} as input stellar spectra, dust models that consist of graphites, silicates, and PAHs, a one-dimensional dynamical evolution model of HII regions, and the MAPPINGS III photoionization code \citep{2004PhDT.......183G} to generate the spectral energy distributions (SEDs). This model spectrum may roughly represent the star-forming regions around eYSCs that are investigated in this study. Therefore, Figure \ref{fig:f1} provides a rough outline of the regions of the SED sampled by the JWST filters.

We also utilize archival HST imaging data for each galaxy for the specific goal of tracing the H$\alpha$ emission. We obtain the HST/ACS/F555W, F658N, and F814W for NGC 5194 (Program: 10452; PI: S. Beckwith), the WFC3/F547M, F555W, F657N, and F814W for NGC 5236 (Program: 11360; PI: R. O$'$Connell), the ACS/F555W, F658N, and F814W for NGC 628 (Programs: 9796, 10402; PIs: J. Miller, R. Chandar), and the ACS/F550M, F658N, and F814W for NGC 4449 (Programs: 10522, 10585; PIs: D. Calzetti, A. Aloisi). For these HST images, we apply standard data reduction steps and match the astrometric frame and sampling with the JWST/NIRCam data.

For this study, we work at a few different angular resolutions: the native resolution, NIRCam/F444W, MIRI/F770W, and F1000W. Unless otherwise stated, the default working angular resolution is F770W, which is used for the majority of our analysis/results. To achieve this common resolution, we match the PSFs of each JWST/HST imaging band to the lowest resolution PSF, which in this case is the F770W with a full-width-at-half-maximum (FWHM) of 0.269$''$ \citep{2023PASP..135d8001R}. To do this, we first create effective PSFs \citep[ePSFs; see][]{2000PASP..112.1360A}  for each filter. These are based on the PSF models of \cite{2006acs..rept....1A} for HST/ACS, \cite{2016wfc..rept...12A} for HST/WFC3 \citep[also see][]{2022wfc..rept....5A}\footnote{\url{https://www.stsci.edu/~jayander/HST1PASS/LIB/PSFs/STDPSFs/}}, and WebbPSF \citep{2014SPIE.9143E..3XP} for JWST. We place grids of these PSF models in blank copies of the individual frames for each filter and then drizzle the frames together using the same parameters as our science images. PSFs are then extracted from the drizzled frames and combined to construct the ePSF for each filter. Using these ePSFs, we then create convolution kernels to convert the PSF of each filter to the PSF of F770W via the method described in \cite{2011PASP..123.1218A} and implemented in the \texttt{make$\_$jwst$\_$kernels}\footnote{\url{https://github.com/thomaswilliamsastro/jwst_beam_matching}} code. The reduced science images are convolved with these kernels to generate a dataset with all imaging filters matched to the PSF of F770W. PSF matching is important for this study for the accurate subtraction of point sources (e.g. stars) in the derived emission line maps and to allow for a direct comparison between the various emission lines, particularly in crowded regions. However, we also make use of the native resolution images for the selection and cleaning of source catalogs from the derived emission line maps.

\begin{center}
\begin{table*}
\centering
\hypertarget{t1}{\caption{Galaxy Sample and Properties}}\label{tab1:properties}
\begin{tabular}{ l l c c c c c c c c c }
\hline
\hline
\rule{0pt}{3ex} 
$\,\,\,$Galaxy & $\,\,\,\,\,$D & $i$ & z & SFR & log(M$_{*}$) & 12+log(O/H) & Gradient & R$_{25}$ & s & A$_{corr}$\\
\rule{0pt}{3ex} 
& [Mpc] & [$^{\circ}$] & & [M$_{\odot}$ yr$^{-1}$] & [M$_{\odot}$] & [dex] & [dex R$_{25}^{-1}$] & [$''$] & [pc/$''$] & [pc$^{2}$] \\
\rule{0pt}{3ex} 
$\,\,\,\,\,\,\,\,$(1) & $\,\,\,\,$(2) & (3) & (4) & (5) & (6) & (7) & (8) & (9) & (10) & (11) \\[1mm]
\hline 
\rule{0pt}{3ex}
\hspace{-2mm} NGC 5194 & 7.50$\,^{\mbox{\textit{a}}}$ & 22.0$\,^{\mbox{\textit{f}}}$ & 0.001534 & 6.88$\,^{\mbox{\textit{j}}}$ & 10.38$\,^{\mbox{\textit{j}}}$ & 8.75$\,^{\mbox{\textit{k}}}$ & $-0.270$$\,^{\mbox{\textit{k}}}$ & 336.6$\,^{\mbox{\textit{k}}}$ & 36.4 & 1355.3 \\[1mm]
NGC 5236 & 4.66$\,^{\mbox{\textit{b}}}$ & 24.0$\,^{\mbox{\textit{g}}}$ & 0.001711 & 4.17$\,^{\mbox{\textit{g}}}$ & 10.53$\,^{\mbox{\textit{g}}}$ & 8.95$\,^{\mbox{\textit{l}}}$ & $-0.240$$\,^{\mbox{\textit{l}}}$ & 386.4$\,^{\mbox{\textit{l}}}$ & 22.6 & 1375.6 \\[1mm]
NGC 628 & 9.84$\,^{\mbox{\textit{c}}}$$^{,}$$^{\mbox{\textit{d}}}$ & 8.9$\,^{\mbox{\textit{h}}}$ & 0.002190 & 3.67$\,^{\mbox{\textit{j}}}$ & 10.04$\,^{\mbox{\textit{j}}}$ & 8.71$\,^{\mbox{\textit{k}}}$ & $-0.400$$\,^{\mbox{\textit{k}}}$ & 315.0$\,^{\mbox{\textit{k}}}$ & 47.7 & 1272.0 \\[1mm]
NGC 4449 & 4.01$\,^{\mbox{\textit{e}}}$ & 68.0$\,^{\mbox{\textit{i}}}$ & 0.000690 & 0.94$\,^{\mbox{\textit{j}}}$ & 9.04$\,^{\mbox{\textit{j}}}$ & 8.26$\,^{\mbox{\textit{m}}}$ & $-0.207$$\,^{\mbox{\textit{m}}}$ & 184.2$\,^{\mbox{\textit{m}}}$ & 19.4 & 3354.6 \\[1mm]
\hline
\end{tabular}
\begin{flushleft} 
\rule{0pt}{3ex}
\currtabletypesize{\sc Note}--- \\
\rule{0pt}{4ex}
\hspace{-2mm} Columns: 1) Galaxy name, 2) distance, 3) inclination angle, 4) Heliocentric redshift as listed in the NASA/IPAC Extragalactic Database (NED), 5) total SFR, 6) total stellar mass (M$_{*}$), 7) nebular oxygen abundance or 12+log(O/H) at the galaxy center, 8) radial oxygen abundance gradient, 9) radius in the B band equal to 25 mag (R$_{25}$), 10) physical size (pc) to angular size ($''$) scale factor given the assumed distance, and 11) the inclination corrected physical area of a 20 pc radius aperture. \\[0.4mm]
References: $^{\mbox{\textit{a}}}$\citep{2023A&A...678A..44C}, $^{\mbox{\textit{b}}}$\citep[][as listed in NED as the mean of Cepheids and TRGB measurements]{2013AJ....146...86T}, $^{\mbox{\textit{c}}}$\citep{2009AJ....138..332J}, $^{\mbox{\textit{d}}}$\citep{2021MNRAS.501.3621A}, $^{\mbox{\textit{e}}}$\citep{2018ApJS..235...23S}, $^{\mbox{\textit{f}}}$\citep{2014ApJ...784....4C}, $^{\mbox{\textit{g}}}$\citep{2021ApJS..257...43L}, $^{\mbox{\textit{h}}}$\citep{2020ApJ...897..122L}, $^{\mbox{\textit{i}}}$\citep{2005ApJ...634..281H}, $^{\mbox{\textit{j}}}$\citep{2015AJ....149...51C}, $^{\mbox{\textit{k}}}$\citep{2020ApJ...893...96B}, $^{\mbox{\textit{l}}}$\citep[][O3N2]{2016ApJ...830...64B,2004MNRAS.348L..59P}, $^{\mbox{\textit{m}}}$\citep{2015MNRAS.450.3254P}\\
\end{flushleft}
\end{table*}
\end{center}
\vspace{-3.8mm} 

\hypertarget{3}{\section{Analysis}}
\hypertarget{3.1}{\subsection{Continuum subtractions}}

We produce emission line maps across each galaxy for the hydrogen recombination lines Pa$\alpha$, Br$\alpha$, and H$\alpha$, and for the PAH emission features at 3.3 and 7.7~$\mu$m by continuum subtracting the reduced JWST and HST images. Our continuum subtraction methods are based on those outlined in the study by \cite{2024ApJ...971..115G}. The method utilizes both a shorter and a longer wavelength filter to derive the continuum in the emission line filter, which is then subtracted. The continuum in the NIRCam/F335M filter is estimated at each pixel in the image by linearly interpolating the SED between the F300M and F444W filters at the location of F335M. This results in a continuum image at F335M, which is subtracted from the original F335M image to derive the 3.3~$\mu$m PAH emission feature map for each galaxy. Here, we use the F300M as the short wavelength filter rather than the F277W, unlike \cite{2024ApJ...971..115G}, since we expect it to provide a slightly more accurate continuum subtraction as discussed in that study, particularly in regions with a large stellar contribution like the central bulge. To create the Pa$\alpha$ emission line maps, the F150W and F200W are used to remove the continuum at F187N. For Br$\alpha$, we use the F300M and F444W to remove the continuum at F405N. 

The F200W and F444W filters are contaminated by the Pa$\alpha$ and Br$\alpha$ lines, respectively (see Figure \ref{fig:f1}). To correct for this contamination, we implement iterative subtraction techniques \citep[see][]{2021ApJ...909..121M} as described in detail in \cite{2024ApJ...971..115G}. We find that 3 iterations are more than sufficient to remove this contamination, which is relatively small for the F200W and F444W filters, typically no more than a few percent in star-forming regions. The corrected F200W and F444W images are used to produce the final emission line maps. 

Similar to the study by \cite{2024ApJ...971..115G}, we test a variety of scaling factors to apply to the derived F335M continuum images to optimize the subtraction of the F335M. The continuum scaling factor can help account for nonlinearity in the SED between the interpolated continuum tracing filters. We visually inspect/test a range of scaling factors between 1.0 and 1.2. We determine that scaling up the continuum before subtraction results in reduced stellar residuals in the derived 3.3~$\mu$m PAH emission feature maps, which we find useful for the selection/cleaning of source catalogs from these maps. In this study, we utilize a continuum scaling factor of 1.06 for F335M solely for the purpose of source selection and catalog cleaning. There are a number of indications from recent studies \citep[e.g.][]{2024ApJ...971..115G,2024ApJ...971...32P} and from newly obtained JWST/NIRSpec spectroscopy (ID 3503, PI: A. Adamo) that suggest the NIRCam/F444W filter may be contaminated by various bright emission features in star-forming regions; in particular at ${\sim}$4.7~$\mu$m, possibly from deuterated hydrocarbons \citep[e.g.][]{2004ApJ...604..252P,2025ApJ...984L..42D}. Scaling up the F335M continuum (from F300M and F444W) before subtraction may result in a slight underestimate of the 3.3~$\mu$m PAH feature strength in these regions. As a result, we use the unscaled F335M continuum to derive the 3.3~$\mu$m PAH emission feature maps used for all photometric measurements made in this study. 

Our final 3.3~$\mu$m PAH emission feature maps receive a contribution from both the aliphatic 3.4~$\mu$m feature and the 3.47~$\mu$m plateau feature, however, these features are much weaker than the bright, aromatic 3.3~$\mu$m feature and represent typically ${\sim}$10$-$30$\%$ of the total power of the 3~$\mu$m complex in the presence of star formation \citep[e.g.][]{2012A&A...541A..10Y}. Spectroscopy (e.g. with JWST/NIRSpec) is required to separate these additional features and their relative contribution to the F335M within our sample of star-forming regions, as well as to determine the best method for subtracting the F335M continuum. This work is underway in our team using Cycle 2 JWST/NIRSpec/MOS spectroscopy (ID 3503, PI: A. Adamo) from 1--5 $\mu$m for ${\sim}$100 eYSCs in NGC 628 (see H. Faustino Vieira et al. in prep.). 

We derive the 7.7~$\mu$m PAH emission feature maps simply by subtracting the MIRI/F560W image from the F770W. Our final 7.7~$\mu$m PAH emission feature maps receive some contribution from the relatively fainter 8.6~$\mu$m PAH feature. In Appendix \hyperlink{A}{A}, we discuss a variety of continuum subtraction techniques for the F770W, including the linear interpolation between F560W and F1000W, as well as scaling up the F560W image prior to the subtraction. In short, we determine that for the sources investigated in this study in NGC 628, the continuum in F770W estimated from an interpolation between F560W and F1000W can be accurately approximated using only the F560W when it is scaled up by a factor of 1.6. However, the F560W receives contributions from various emission features, including the wing of the bright 6.2~$\mu$m PAH feature. The F1000W may receive significant contamination (PAH emission, silicate absorption at about 10~$\mu$m) as well. Given this contamination, the best technique is unclear, and scaling up the F560W further could result in an over-subtraction when it is used to remove the F770W continuum. As a result, we elect to use the unscaled F560W for the subtraction of F770W for the results presented in this study. Spectroscopy (e.g. with MIRI/MRS) is needed to properly understand the optimal subtraction method for the continuum in F770W. New studies are starting to develop our understanding \citep[e.g.][]{2025A&A...698A..86C,2025ApJ...983...79D}; however, our sources are different and need to be studied in detail with spectroscopy. We find that the assumed continuum subtraction technique for F770W has an insignificant impact on our results (see Appendix \hyperlink{A}{A}).

To obtain the H$\alpha$ emission line maps, we use a combination of the HST/ACS/F555W and F814W filters to remove the continuum in F658N for both NGC 5194 and NGC 628. For NGC 4449, we use the HST/ACS/F550M, F658N, and F814W. For NGC 5236, we produce two versions: one using the HST/WFC3/F547M, F657N, and F814W that covers the majority of the galaxy, and one using the F555W as the short wavelength continuum tracing filter that covers the central regions where the F547M is unavailable. The ACS and WFC3 F555W filters are contaminated by the [O III] 0.5008~$\mu$m emission line. However, the filter responses are very broad, and in metal-rich systems like NGC 5194, 5236, and 628, the oxygen line is relatively weak \citep{2015ApJ...806...16B}. We expect the impact of the [O III] line on the interpolated stellar continuum for the F657N/F658N to be $\lesssim \,$2$\%$ \citep{2024ApJ...971..118C} in NGC 5194 and NGC 628, and in the fraction of regions in NGC 5236 that use the F555W. At low metallicity where the line is expected to be bright (e.g. NGC 4449), we use the ACS/F550M as the short wavelength continuum tracing filter, which excludes the [O III] line. Therefore, contamination by the [O III] line will not affect the analysis presented in this study. 

\begin{figure*}
\centering
\includegraphics[width=0.98\textwidth]{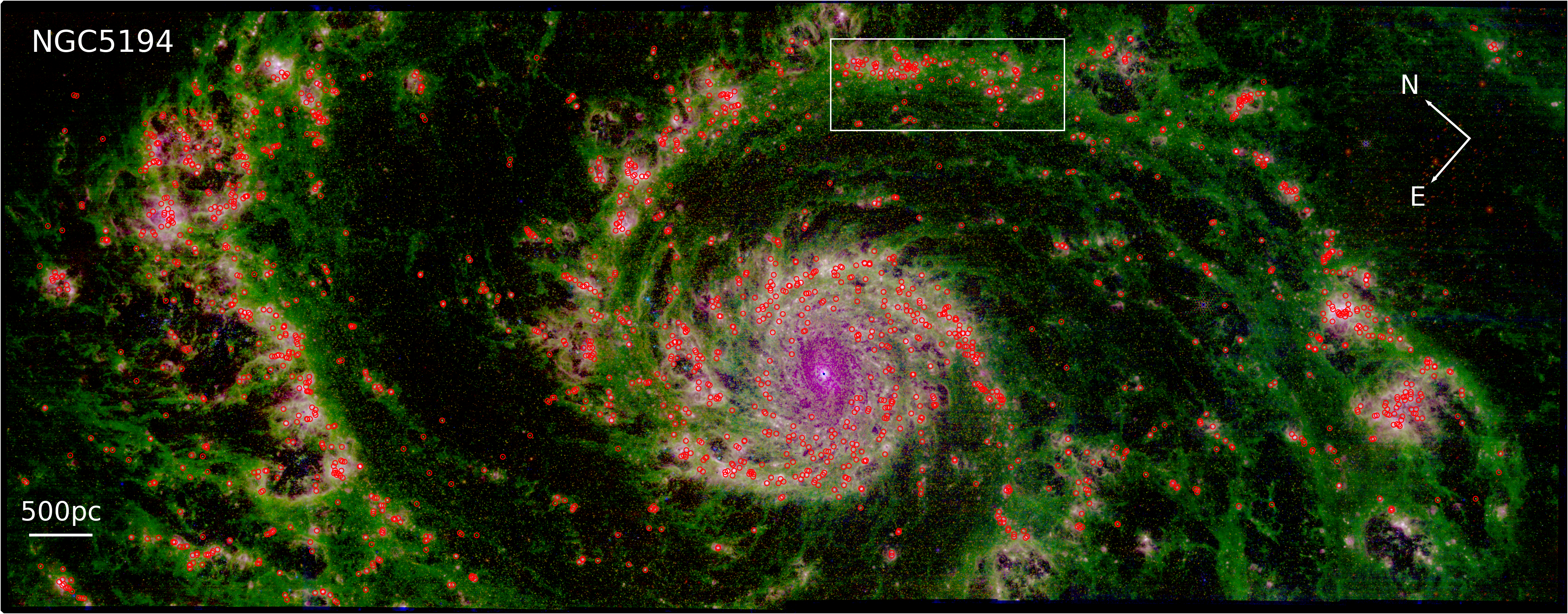}
\includegraphics[width=0.98\textwidth]{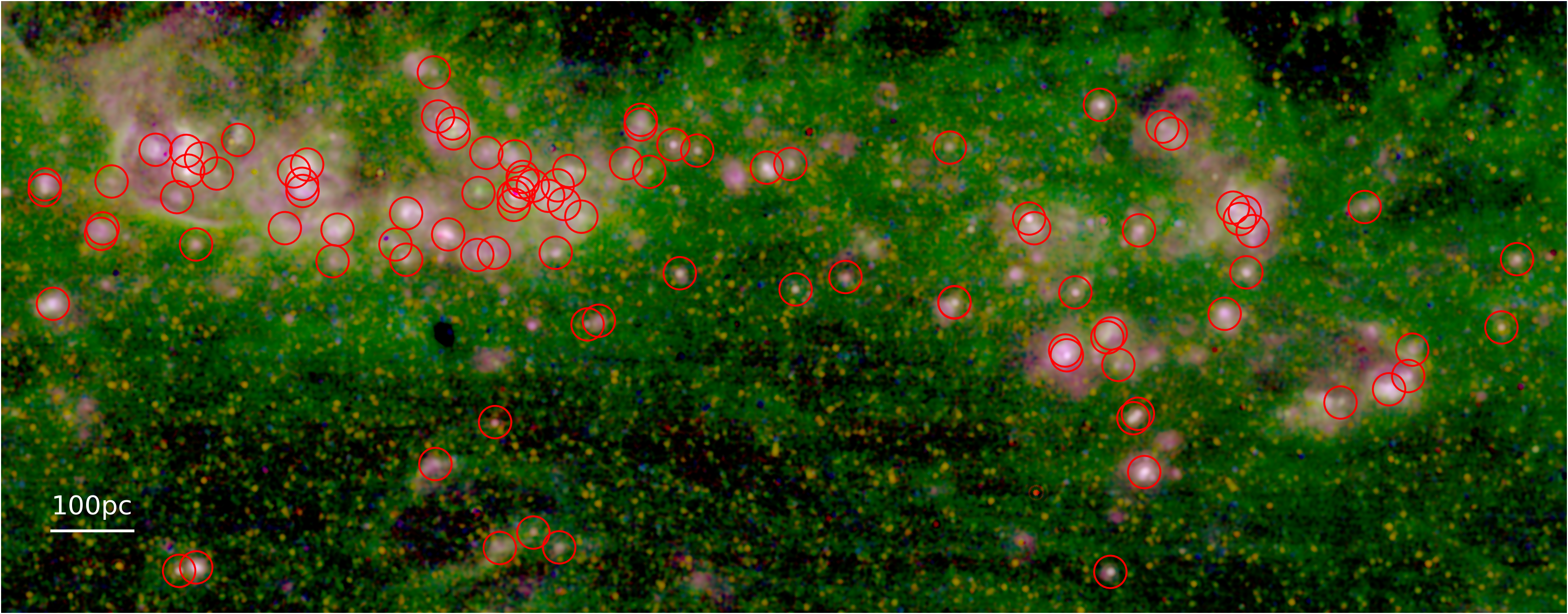}
\caption{Top panel: A three-color composite image showing the full Br$\alpha$ (red channel), 3.3~$\mu$m PAH (green channel), and Pa$\alpha$ (blue channel) emission maps (at the F444W resolution) for NGC 5194 (M51). Overlaid on top of the image are the 20 pc ($\sim$0.55$''$) radius apertures (red circles) used to measure the photometry of the regions around each identified eYSC--I source (cospatial Pa$\alpha$, Br$\alpha$, and 3.3~$\mu$m PAH emission peaks). The white bar shows the 500 pc scale. Bottom panel: A zoom-in corresponding to the star-forming spiral arm region shown by the white box in the top panel. }
\label{fig:f2}
\end{figure*}

Figure \ref{fig:f2} shows a three-color composite image of the final Pa$\alpha$, Br$\alpha$, and 3.3~$\mu$m PAH emission feature maps for NGC 5194 covering the full NIRCam FOV (top panel) and a zoom-in on a representative star-forming spiral arm region (bottom panel). The two hydrogen lines Pa$\alpha$ and Br$\alpha$ trace gas that is ionized by massive, young stars in HII regions, and the emission closely follows the spiral arm structure of the galaxy where the vast majority of star formation takes place. The 3.3~$\mu$m PAH emission is also strong in star-forming regions and traces the PDRs that surround HII regions. Diffuse PAH emission can also be seen far outside the boundaries of massive star-forming regions, possibly arising as a result of the heating of PAHs by older stellar populations. 

\hypertarget{3.2}{\subsection{The selection of emerging young star clusters}}

The specific details of how eYSCs are selected in the FEAST galaxies are reported in A. Adamo et al. (in prep.). Here, we provide a summary of the relevant steps. 

New catalogs of candidate eYSCs are selected from our NIRCam emission line maps for each of the galaxies in our sample. The average size of young, massive star clusters is ${\sim}$3 pc \citep[e.g.][]{2017ApJ...841...92R,2021MNRAS.508.5935B}, while the native resolution of our NIRCam maps of ionized gas and PAH emission is ${\sim}{\,}$0.07--0.15$''$, corresponding to 3--7 pc in the furthest target, NGC 628, or smaller scales in the others. Therefore, individual clusters are generally marginally resolved in the maps for each galaxy. 

The eYSCs are selected as bright, compact peaks in the Pa$\alpha$, Br$\alpha$, and 3.3~$\mu$m PAH emission feature maps, using the Python library for Source Extraction and Photometry \citep[\texttt{SEP};][]{1996A&AS..117..393B,2016JOSS....1...58B}. We independently extract the peaks in each of the three maps for each galaxy. Each of the resulting catalogs is then visually inspected on all three emission line maps and cleaned to remove contaminants, such as residual stellar point sources, sources on the edge of the mosaics, hot pixels, etc. We perform the extraction and cleaning of the catalogs on the native resolution emission line maps to maximize the number of sources and obtain precise source locations. Also, we find it easier to identify and remove contaminants at the native resolution.

We measure aperture photometry on all the NIRCam images of each galaxy for each catalog of emission line peaks using the FEAST pipeline, detailed in A. Adamo et al. (in prep.). We use circular apertures of radius 0.16, 0.20, 0.16, and 0.20$''$ for NGC 5194, 5236, 628, and 4449, respectively, where the larger apertures correspond to the closer galaxies. The measurements are corrected for the local background via an annulus with an inner radius of r$_{in}$ and an outer radius of r$_{out}$, where r$_{in}$=[0.24, 0.28, 0.24, 0.28]$''$ and r$_{out}$=[0.32, 0.36, 0.32, 0.36]$''$ for NGC 5194, 5236, 628, and 4449, respectively. Concentration index (CI) based aperture corrections are also applied. We use these measurements to place brightness cuts (magnitude error $\leq$ 0.3) on the catalogs of compact emission peaks. We require a magnitude error $\leq$ 0.3 in both the F187N and F200W for the Pa$\alpha$ selected catalogs. The magnitude error is $\leq$ 0.3 in F405N and F444W for the Br$\alpha$ selected catalogs, or $\leq$ 0.3 in F335M and F444W for the 3.3~$\mu$m PAH peaks. An error cut on both continuum and emission line tracing filters can help ensure that the sources are not spurious. The remaining sources are detected above these thresholds for the local background subtracted photometry and are thus ``compact". These catalogs give a fairly complete census of the bright, compact ionized gas and PAH emission peaks across the regions covered by our maps. We select three distinct classes of sources from these catalogs, defined as eYSC--I, eYSC--II, and PAH compact.

The eYSC--I sources are selected from the cleaned and cut catalogs as bright Pa$\alpha$ peaks with both Br$\alpha$ and 3.3~$\mu$m PAH emission peaks within ($\leq$) 0.16$''$ (about the FWHM of the PSF of F444W). This results in 1980, 1127, 816, and 258 eYSC--I sources for NGC 5194, 5236, 628, and 4449, respectively. Small variations in the assumed matching radius (i.e. 0.16$''$) do not significantly change these numbers \citep[see][]{2025arXiv250508874K}. For these sources, there is a tight spatial connection between the compact PAH and ionized gas emission, corresponding to a scale smaller than about 3--7 pc, depending on the target. These sources likely correspond to newly formed massive star clusters that remain embedded in their natal gas and dust, and thus are expected to be the youngest.

The eYSC--II or ``hydrogen recombination line compact" sources are selected as Pa$\alpha$ peaks with a Br$\alpha$ peak within 0.16$''$, but no corresponding bright, compact 3.3~$\mu$m PAH emission peak. PAH compact sources are selected as 3.3~$\mu$m PAH emission peaks with no corresponding Pa$\alpha$ or Br$\alpha$ peak within 0.16$''$. The latter two classes of sources are expected to be slightly older and more emerged (eYSC--II), or much older and/or less massive (PAH compact) than eYSC--I. See the studies by \cite{2024ApJ...971..115G}, \cite{2025arXiv250508874K}, and S. T. Linden et al. (in prep.) for a more detailed discussion of these different sources. The source catalogs for each FEAST target galaxy will be released publicly on the FEAST webpage\footnote{\url{https://feast-survey.github.io/}}, along with other relevant data products.

For the remainder of this study, we utilize only the catalogs of eYSC--I sources for each galaxy. The regions of PAH and ionized gas emission around eYSC--I sources make ideal targets for investigating the heating of PAH grains on small scales in galaxies and the relationship between PAH emission and star formation. Given the close proximity of these regions to the local newly formed massive star cluster, we can be confident that the PAH heating is dominated by photons that are related to new star formation ($<$10 Myr). We can also properly account for the diffuse heating component from older stellar populations by estimating and removing the local PAH background, as discussed in the next section. In addition, the selection of these sources as tightly spatially connected peaks in both PAH and ionized gas emission minimizes any effects due to differences in the relative geometries. 

\hypertarget{3.3}{\subsection{Photometry}}

Aperture photometry is measured on each of the five emission line maps, Pa$\alpha$, Br$\alpha$, H$\alpha$, 3.3~$\mu$m PAH, and 7.7~$\mu$m PAH (all at the MIRI/F770W resolution), for each galaxy for the corresponding catalog of eYSC--I sources. First, for each source, 200 pixel by 200 pixel (or ${\sim}$8$''$ by 8$''$ in the case of NIRCam images) cutouts of each image are created, centered on the location of the source. We then measure the photometry using the Python package \texttt{photutils} as the sum in circular apertures of 20 pc in radius at each source location, corresponding to about 0.55, 0.89, 0.42, and 1.03$''$ for NGC 5194, 5236, 628, and 4449, respectively, given the assumed distances. 

The choice of this aperture size is physically motivated, selected to be about the size within which we expect the local young ionizing source to dominate the PAH heating in our brightest source. For instance, based on our calculations presented in the next section, the brightest source across our eYSC--I samples, located in NGC 5194, has an SFR$\, {\sim}\, $0.016 M$_{\odot}$ yr$^{-1}$, corresponding to an ionizing photon rate Q$\,{\sim}\, 2.16\times 10^{51}$ s$^{-1}$ for the calibration of \cite{2013seg..book..419C}. The Str\"{o}mgren radius of this source is R$_{s}\, {\sim}\, 17$ pc, assuming a recombination coefficient $\alpha _{B} \, {\sim}\, 3.5\times 10^{-13}$ cm$^3$ s$^{-1}$ for Case B recombination from \cite{1995MNRAS.272...41S} and a density n$\, {\sim}\, 100$ cm$^{-3}$ and temperature T$\, {\sim}\, 7000$ K. We expect that the bright, localized PAH emission in the PDRs around HII regions (roughly traced by the Str\"{o}mgren radius) represents the majority of the PAH emission that is excited by the nearby young ionizing source (i.e. eYSC). Our 20 pc radius apertures are slightly larger than the Str\"{o}mgren radius of the largest HII region in our sample, and therefore, should be sufficiently large to capture most of the relevant emission, while remaining small enough to avoid being significantly impacted by the heating from non-local sources. 

The local background around each source is determined as the iteratively 3$\sigma$ clipped mode within an annulus of equal area that encompasses the aperture, i.e. with an inner radius equal to the aperture radius and an outer radius equal to the aperture radius multiplied by $\sqrt{2}$. The total background in the aperture is estimated by multiplying the local background by the total number of pixels in the aperture, which is then subtracted from the measurements. We estimate the image uncertainty for each emission line map by measuring the iteratively 3$\sigma$ clipped standard deviation within a blank sky region. Uncertainties in our photometric measurements are then derived as the emission line map uncertainties multiplied by the square root of the number of pixels in the aperture. We require that each source be detected with greater than 3 times the photometric uncertainties in all five emission lines for our local background subtracted measurements. This results in 1744, 971, 420, and 223 eYSC--I sources remaining in NGC 5194, 5236, 628, and 4449, respectively. The ${\sim}$20$\%$ decrease in the total number of sources compared with the original eYSC--I catalogs is mostly due to the imperfect overlap of the MIRI and HST mosaics with the NIRCam in terms of FOV. In the case of NGC 5236, we measure the photometry on the H$\alpha$ emission line map derived with the WFC3/F547M as the short wavelength continuum tracing filter when available (${\sim}$65$\%$ of the sources), while for the remainder of the sources in NGC 5236, we make measurements on the H$\alpha$ map derived with F555W.

In Figure \ref{fig:f2}, we show the 20 pc radius apertures (red circles) used to measure the photometry for the final eYSC--I catalog of NGC 5194, overlaid on the three-color composite image of the Pa$\alpha$, Br$\alpha$, and 3.3~$\mu$m PAH emission feature maps. The tight spatial connection between these sources and the spiral arms and regions of massive star formation, or HII regions, is made apparent in the image. The sources sample the bright white knots of emission, corresponding to cospatial peaks in both ionized gas and PAH emission. Also apparent is that in dense regions of star formation, the measurements may overlap; a consequence of the selection on the native resolution maps and the large apertures used to capture the majority of PAH and ionized gas emission excited by the local young ionizing source. 

The data set used for these measurements is matched to the PSF of F770W, the lowest angular resolution filter, via convolution, and as a result, we are sampling the same physical scales across wavelengths. Our aperture size is generally much larger than the FWHM of the PSF of F770W; in the furthest target, NGC 628, it is about 3 times the FWHM. Additionally, in this study, we are only concerned with the regions of line emission around our sample of eYSC--I sources, rather than the properties of the local, newly formed star cluster. Therefore, a correction for our fixed apertures is unnecessary. Furthermore, the overlap of the measurements in dense regions is generally not a concern and does not affect our main results, as tested explicitly in Appendix \hyperlink{B}{B}.

The HST/ACS/F658N and WFC3/F657N filters are contaminated by the [NII] emission lines at 6548 and 6583 \AA, and thus so are our H$\alpha$ emission line maps. The average ratio between the two [NII] lines and H$\alpha$ is 0.6, 0.53, 0.4, and 0.23 for NGC 5194, 5236, 628, and 4449, respectively \citep{2008ApJS..178..247K}. We use these ratios to derive a basic correction to the measured H$\alpha$ flux densities that accounts for the [NII] contamination in our maps. The oxygen abundance gradients in our galaxies are sufficiently small (Table~\hyperlink{t1}{1}) that variations in the [NII]/H$\alpha$ ratio due to the gradients are small relative to the [NII]/H$\alpha$ measurement uncertainty ($\sim$6\%--10\%), which we propagate in all our measurements. We also apply a minor correction to the measured flux densities to account for the transmission through the filters at the location of the redshifted emission lines, assuming the redshifts for each galaxy as listed in Table \hyperlink{t1}{1}. 

\hypertarget{3.4}{\subsection{Physical properties}} 

From the measured emission line flux densities of each source, we derive the emission line luminosity using $L\,$(erg$\,$s$^{-1})\,=\,$($3\times 10^{-5}\, f_{\nu}\,\lambda^{-2}) \, 4 \pi d^{2} \, BW$, where $f_{\nu}$ is the flux density in Jy, $\lambda$ and $BW$ are the pivot wavelength and bandwidth of the relevant filter in \AA, and $d$ is the distance to the galaxy in cm. For the NIRCam filters, we use the pivot wavelengths (1.874, 3.365, and 4.055~$\mu$m) and bandwidths (240, 3470, and 460 \AA) for F187N, F335M, and F405N, respectively, as given in the NIRCam documentation\footnote{\url{https://jwst-docs.stsci.edu/jwst-near-infrared-camera/nircam-instrumentation/nircam-filters}}. For MIRI, we use the pivot wavelength (7.639~$\mu$m) and bandwidth (19500 \AA) for F770W as listed in the MIRI documentation\footnote{\url{https://jwst-docs.stsci.edu/jwst-mid-infrared-instrument/miri-instrumentation/miri-filters-and-dispersers}}. For the HST/ACS/WFC/F658N and HST/WFC3/UVIS/F657N, we use the pivot wavelengths (0.6584 and 0.6567~$\mu$m, respectively) listed in the FITS headers and bandwidths (87.487 and 121 \AA) given in the relevant documentation\footnote{\url{https://etc.stsci.edu/etcstatic/users_guide/appendix_b_acs.html}}$^{,}$\footnote{\url{https://hst-docs.stsci.edu/wfc3ihb/chapter-6-uvis-imaging-with-wfc3/6-5-uvis-spectral-elements}}. 

Monte Carlo calculations are performed to estimate the uncertainties in the emission line luminosities, assuming that the flux density errors are normally distributed with a standard deviation given by their photometric uncertainties. In addition, we assume a 10\% error on the distance to the galaxy to account for its uncertainty along with some additional contribution from unquantified sources of uncertainty (the continuum subtractions, flux calibrations, etc.). We simulate 10$^{4}$ random draws for each measurement and then derive the uncertainty as the standard deviation of the resulting distribution of emission line luminosity. 

We calculate the magnitude of dust attenuation for each source by measuring the color excess E(B$-$V) from the observed H$\alpha$/Pa$\alpha$ luminosity ratio. The Python code \texttt{PyNeb}\footnote{\url{https://pypi.org/project/PyNeb/}} \citep{2015A&A...573A..42L} is used to calculate the intrinsic H$\alpha$/Pa$\alpha$ luminosity ratio, assuming Case B recombination from \cite{1995MNRAS.272...41S} and typical HII region density n$_{e}{\sim}\, 100$ cm$^{-3}$ and temperature T$_{e}{\sim}\, 7000$ K for near solar metallicity. This gives an intrinsic $L_{H\alpha}$/$L_{Pa\alpha}\sim 7.88$. We assume a standard foreground dust geometry and the relation given by
\begin{displaymath}
L(\lambda)_{obs}=L(\lambda)_{int} 10^{[-0.4 \, E(B-V) \, k(\lambda)]}\,, 
\end{displaymath}
where $L(\lambda)_{obs}$ and $L(\lambda)_{int}$ are the observed and intrinsic luminosities \citep[e.g.][]{2021ApJ...913...37C}. The extinction curve $k(\lambda)$ determined by \cite{2021ApJ...916...33G} for the Milky Way is considered. The color excess is determined as 
\begin{displaymath}
E(B-V) = \frac{log\left( \frac{(L_{H\alpha}/L_{Pa\alpha})_{obs}}{(L_{H\alpha}/L_{Pa\alpha})_{int}}\right) }{-0.4 \, [k(H\alpha)-k(Pa\alpha)]}\, ,
\end{displaymath}
where $k(H\alpha){\sim}2.467$ and $k(Pa\alpha){\sim}0.364$ for the extinction curve given above. In this study, we chose to use the observed H$\alpha$/Pa$\alpha$ luminosity ratio rather than the Pa$\alpha$/Br$\alpha$ for deriving E(B$-$V) given the results of the studies by \cite{2024ApJ...971..115G} and \cite{2024ApJ...971...32P}, which suggest that using the Pa$\alpha$/Br$\alpha$ ratio determined via NIRCam imaging may result in a significant underestimate of the E(B$-$V). 

The color excess values are utilized to correct the emission line luminosities for the effects of dust attenuation. Using the relation for foreground dust given above and the measured E(B$-$V) for each source, we derive the intrinsic or extinction corrected luminosity ($L_{corr}$) for the Pa$\alpha$, Br$\alpha$, H$\alpha$, and 3.3~$\mu$m PAH emission features. We assume that the dust attenuation, and thus the value of E(B$-$V), is the same for the hydrogen recombination lines and the PAH features. Given the selection methods for our eYSC--I sources, we expect this assumption to be reasonable, although there may remain some differences in the distribution and relative geometries of the ionized gas and dust/PAHs. We do not correct the 7.7~$\mu$m PAH emission feature for dust attenuation as the effect is relatively negligible and the assumed empirical extinction curve does not extend to high enough wavelengths to allow for the correction.

SFRs are estimated from the extinction corrected Pa$\alpha$ luminosities. The standard H$\alpha$ calibration is SFR$_{H\alpha}$ (M$_{\odot}$ yr$^{-1}$) = 5.5$\times$10$^{-42}$ L$_{H\alpha}$ (erg s$^{-1}$) \citep{2013seg..book..419C}, considering the stellar IMF of \cite{2001MNRAS.322..231K}. Assuming this calibration and the same intrinsic H$\alpha$/Pa$\alpha$ luminosity ratio as above, we derive SFRs using SFR$_{Pa\alpha,corr}$ (M$_{\odot}$ yr$^{-1}$) = 4.33$\times$10$^{-41}$ L$_{Pa\alpha,corr}$ (erg s$^{-1}$).

UV photons are well-known to leak far outside the boundaries of HII regions. Between about 30$\%$ to 50$\%$ of the ionizing photons emitted by the local young star clusters are expected to leak out of the HII regions without ionizing the local gas \cite[e.g.][]{1997MNRAS.291..827O,2022A&A...659A..26B} and instead power the diffuse ionized gas (DIG). These photons will be missed by our 20 pc radius apertures. Therefore, the physical properties that we derive from the measured emission line luminosities may be affected; in particular, the SFRs, which rely on the intrinsic ionizing photon rates that may be underestimated. In the most extreme circumstance where half of the ionizing photons leak out of the HII regions and are missed by our apertures, the intrinsic SFRs would be a factor of ${\sim}$2 higher than our estimates from the extinction corrected Pa$\alpha$ luminosities. A similar effect to leakage can be expected in the presence of direct absorption of ionizing photons by dust in our HII regions. While this mechanism is difficult to characterize in extragalactic regions because of its degeneracy with other effects, it has been shown to absorb about 15\% of the ionizing photons in the Milky Way, increasing to about 50\% for the brightest HII regions \citep{Binder+2018, McCallum+2025}.

The equivalent width (EW) of nebular emission lines like Pa$\alpha$ is a direct tracer of the age of the local young star cluster and corresponding HII region \citep[e.g.][]{1981Ap&SS..80..267D,1996ApJS..107..661S,2021ApJ...909..121M}. We derive the EW of the Pa$\alpha$ emission line as the ratio of the line flux to the flux density of the corresponding continuum. We use a formulation based on the one listed in \cite{2021ApJ...909..121M} for other recombination lines, given by
\begin{displaymath}
EW_{Pa\alpha} (\textrm{\AA}) = \left(\frac{f_{Pa\alpha,L}}{f_{Pa\alpha,C}} \right) BW_{F187N} \, ,
\end{displaymath}
where $f_{Pa\alpha,L}$ is the measured Pa$\alpha$ flux density (Jy), $f_{Pa\alpha,C}$ is the flux density (Jy) of the continuum under Pa$\alpha$, and $BW_{F187N}$ is the bandwidth (\AA) of the F187N filter as listed above. For this calculation, we make measurements on the native resolution Pa$\alpha$ emission line maps and the corresponding continuum images (derived by the method presented in Section \hyperlink{3.1}{3.1}) using the same apertures and annuli, described in Section \hyperlink{3.3}{3.3}. Both the Pa$\alpha$ emission line and continuum measurements are local background subtracted. In this study, we find it important that the native resolution images are used for the EW calculation. If we use the dataset that is matched to the PSF of F770W, we find that a significant fraction of the sources are determined to have negative Pa$\alpha$ EW. This is likely an effect of confusion in the smoothed (PSF--matched) images, due to source crowding, which makes the determination of the continuum level highly uncertain.
Conversely, the native resolution images enable us to more reliably estimate the local continuum level and the EW. An aperture correction is unnecessary for this calculation, given that the PSFs of the native resolution Pa$\alpha$ emission line and continuum images are the same, with an FWHM many times smaller ($>$12) than our apertures. However, we note that in dense regions there may be some differences between the emission probed by our apertures for the different resolution (convolved and native) maps that could affect the measured value of the Pa$\alpha$ EW with respect to other derived physical properties. We assume that dust attenuation affects the emission line and the stellar continuum similarly, such that the EW is not affected by attenuation. 

The galactocentric radius ($''$) of each source is calculated as the distance between the location of the source and the galaxy center. We utilize the galactocentric radius in combination with the central nebular oxygen abundance and the radial oxygen abundance gradient to estimate the nebular oxygen abundance for each source, assuming the values for each galaxy derived by the studies listed in Table \hyperlink{t1}{1}. The nebular oxygen abundance or 12+log(O/H) is used as a tracer of the gas-phase metallicity, and we use the terms interchangeably. Note that solar metallicity on this scale corresponds to 12+log(O/H)$\,{\sim}\,$8.7 \citep{2009ARA&A..47..481A}. For these calculations, we assume the same galaxy center locations as the studies of the nebular abundance trends. 

The surface densities corresponding to our aperture measurements are calculated from the inclination corrected physical area of the aperture, or A$_{corr}$ (Table~\hyperlink{t1}{1}). We assume A$_{corr}\, $(pc$^{2}$)$\, =\, \pi r^{2} / $cos$(i)$, where $r$ is the physical radius of the apertures (20 pc) and $i$ is the inclination angle of the galaxy as listed in Table \hyperlink{t1}{1}. Emission line luminosity surface density is then calculated as $\Sigma_{L}\,$(erg$\,$s$^{-1}$$\,$pc$^{-2}$)$\, =\, L / $A$_{corr}$.

\begin{figure*}
\centering
\includegraphics[width=0.46\textwidth]{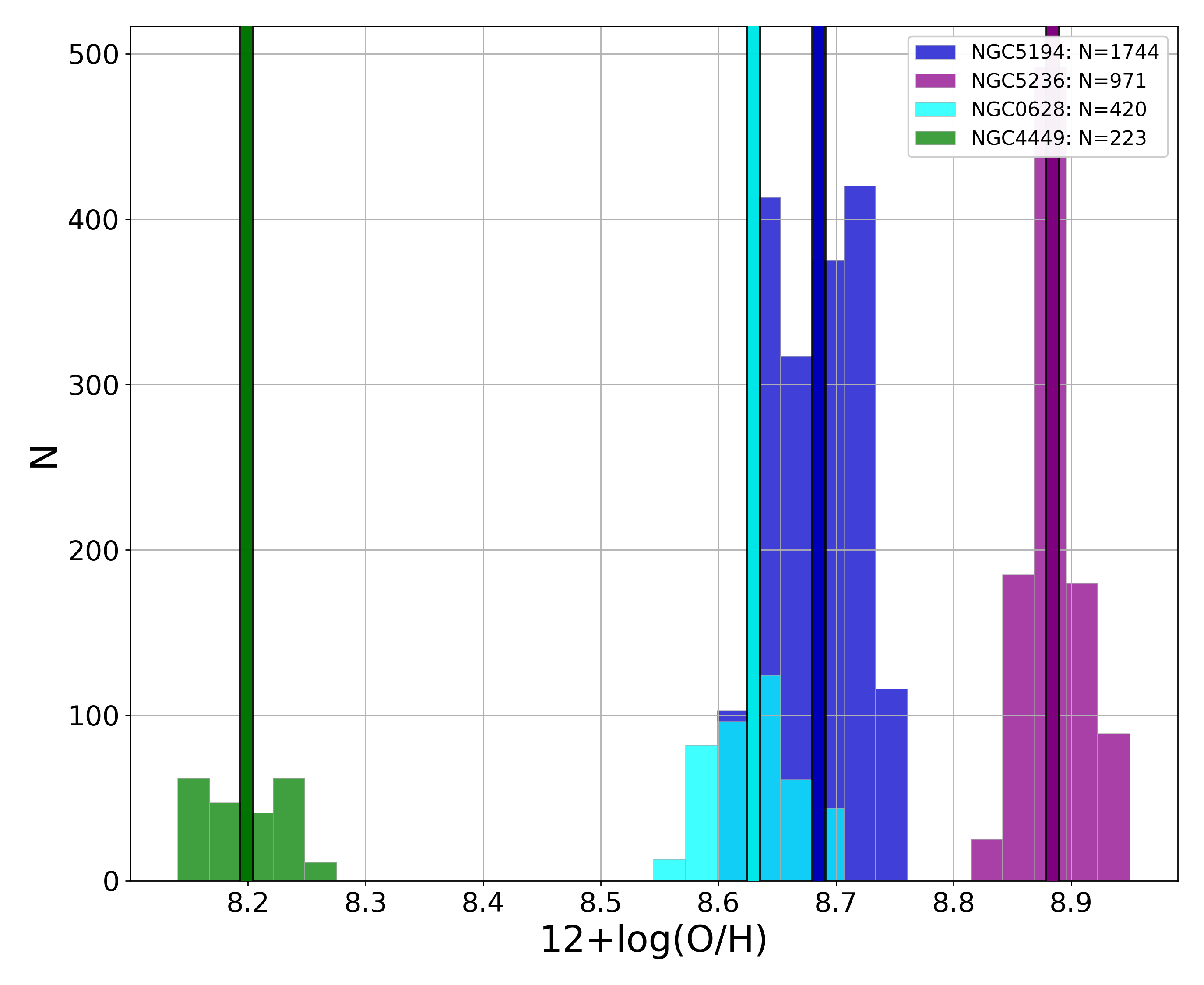}
\includegraphics[width=0.46\textwidth]{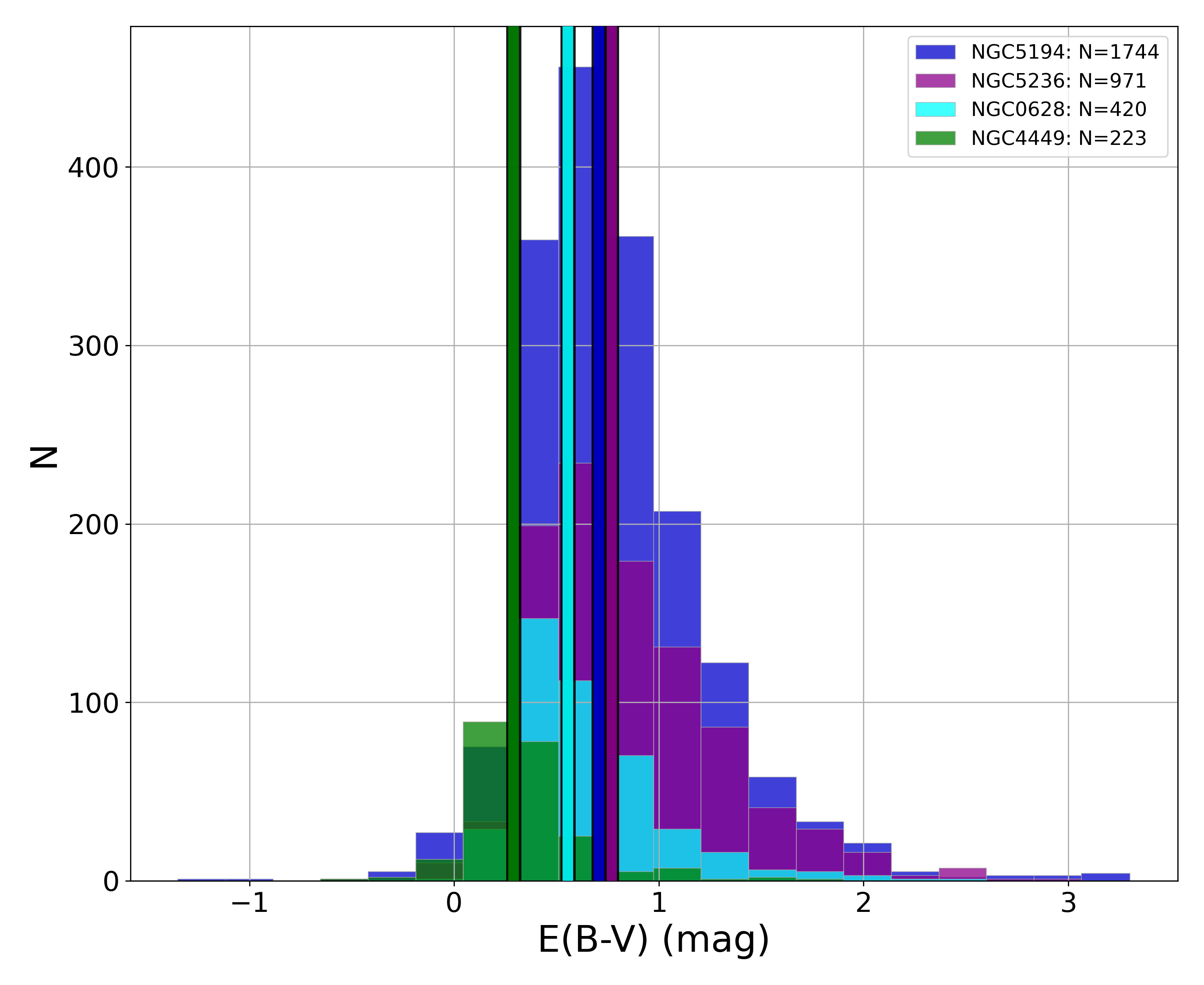}
\includegraphics[width=0.46\textwidth]{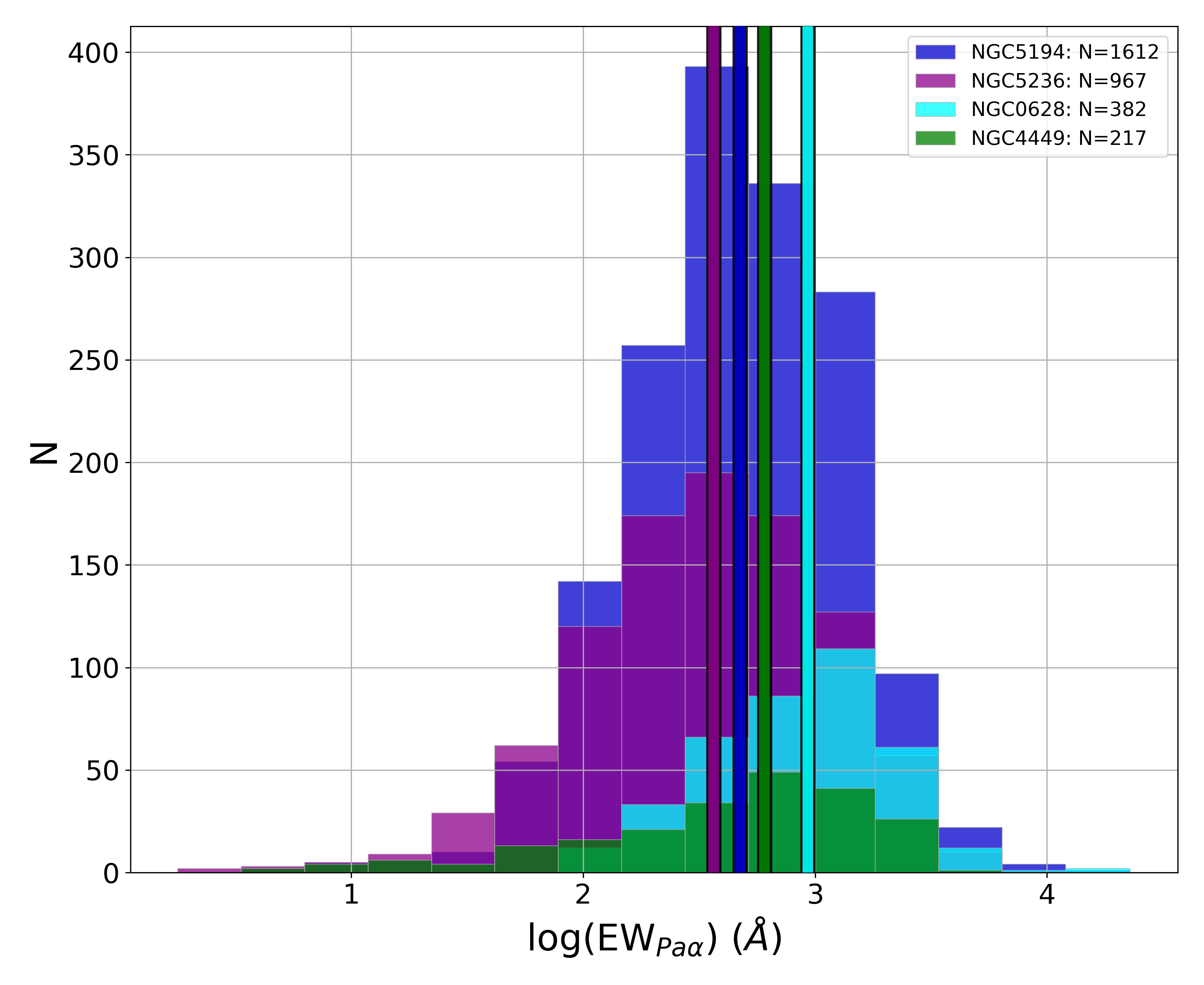}
\includegraphics[width=0.46\textwidth]{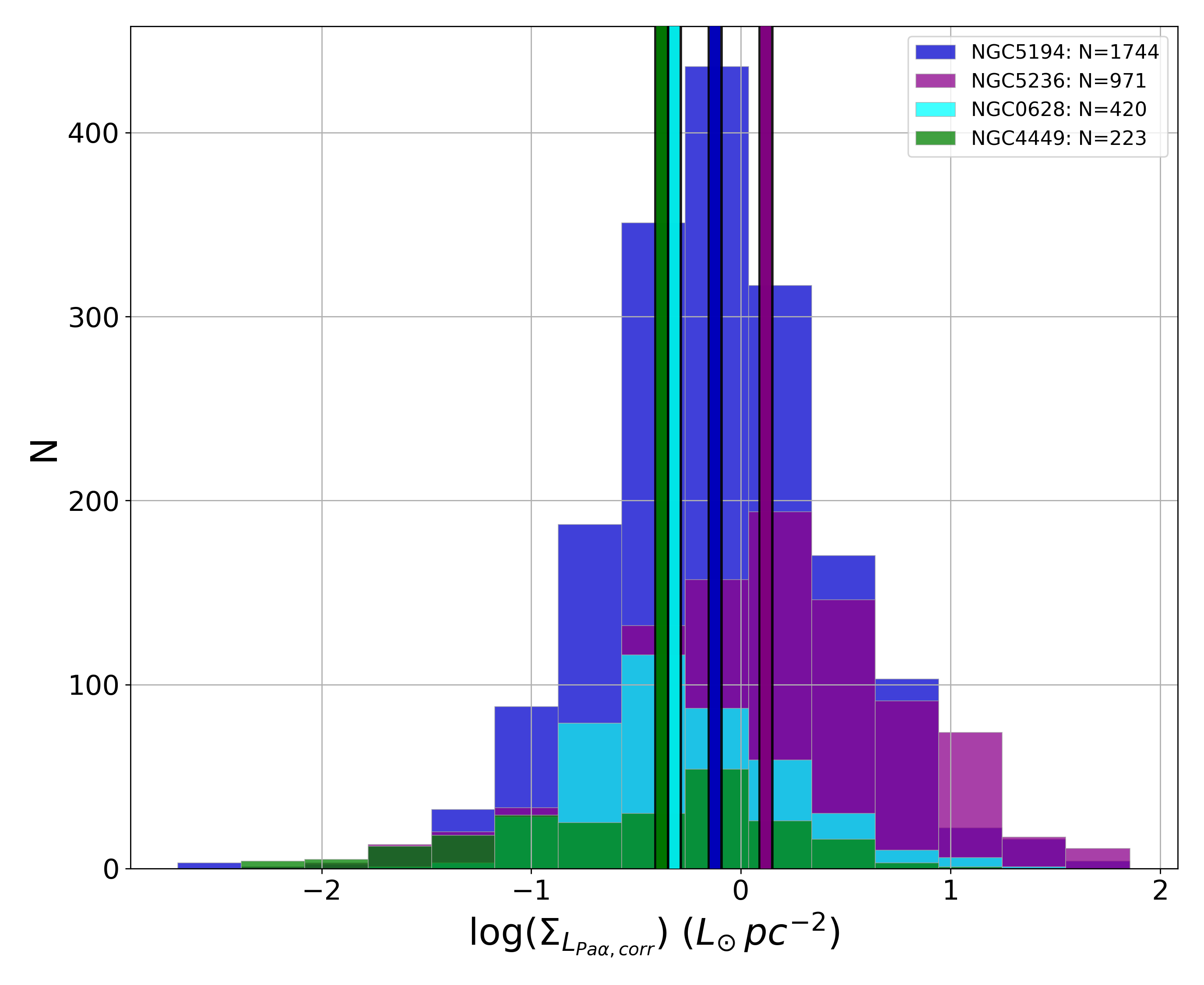}
\caption{Histograms showing the distribution of the nebular oxygen abundance or 12+log(O/H) (top left), the color excess or E(B$-$V) (top right), the equivalent width (EW) of Pa$\alpha$ (bottom left), and the extinction corrected Pa$\alpha$ luminosity surface density (bottom right) for our measurements around eYSC--I sources in each of the four galaxy targets, shown by the various colors. The bins are matched across the targets.  The oxygen abundance is derived from the observed radial gradient in combination with the galactocentric radius of each source. E(B$-$V) is derived from the observed H$\alpha$/Pa$\alpha$ luminosity ratio. The vertical colored lines show the median values for each galaxy. The lower numbers of sources in the bottom left panel result from the fact that the Pa$\alpha$ EW cannot be reliably estimated for all of our sources.}
\label{fig:f3}
\end{figure*}

Figure \ref{fig:f3} shows the distributions of various physical properties (nebular oxygen abundance, E(B$-$V), Pa$\alpha$ EW, and extinction corrected Pa$\alpha$ luminosity surface density) measured around eYSC--I sources in the four galaxies. The sources in NGC 5236 reside in the highest metallicity environments, with greater than solar metallicity, while in NGC 4449, the metallicities are much lower, significantly sub-solar. The metallicities are similar for NGC 5194 and NGC 628, at about solar, however, NGC 628 exhibits slightly lower values on average. We find that the sources in NGC 5236 and NGC 5194 have the highest values of dust attenuation on average, with a maximum E(B$-$V)$\,{\sim}\,$3 mag, while in NGC 4449, the attenuation is the lowest on average. Our sources are young ($<$10 Myr) by selection and exhibit high values of Pa$\alpha$ EW. We find the Pa$\alpha$ EW distributions to be similar across the targets, reaching the highest values on average in NGC 628, possibly a result of detection limits. The surface densities of ionized gas luminosity (and therefore SFR) are the highest on average for our sources within NGC 5236.    

\hypertarget{4}{\section{Results}}

\begin{figure*}
\centering
\includegraphics[width=0.49\textwidth]{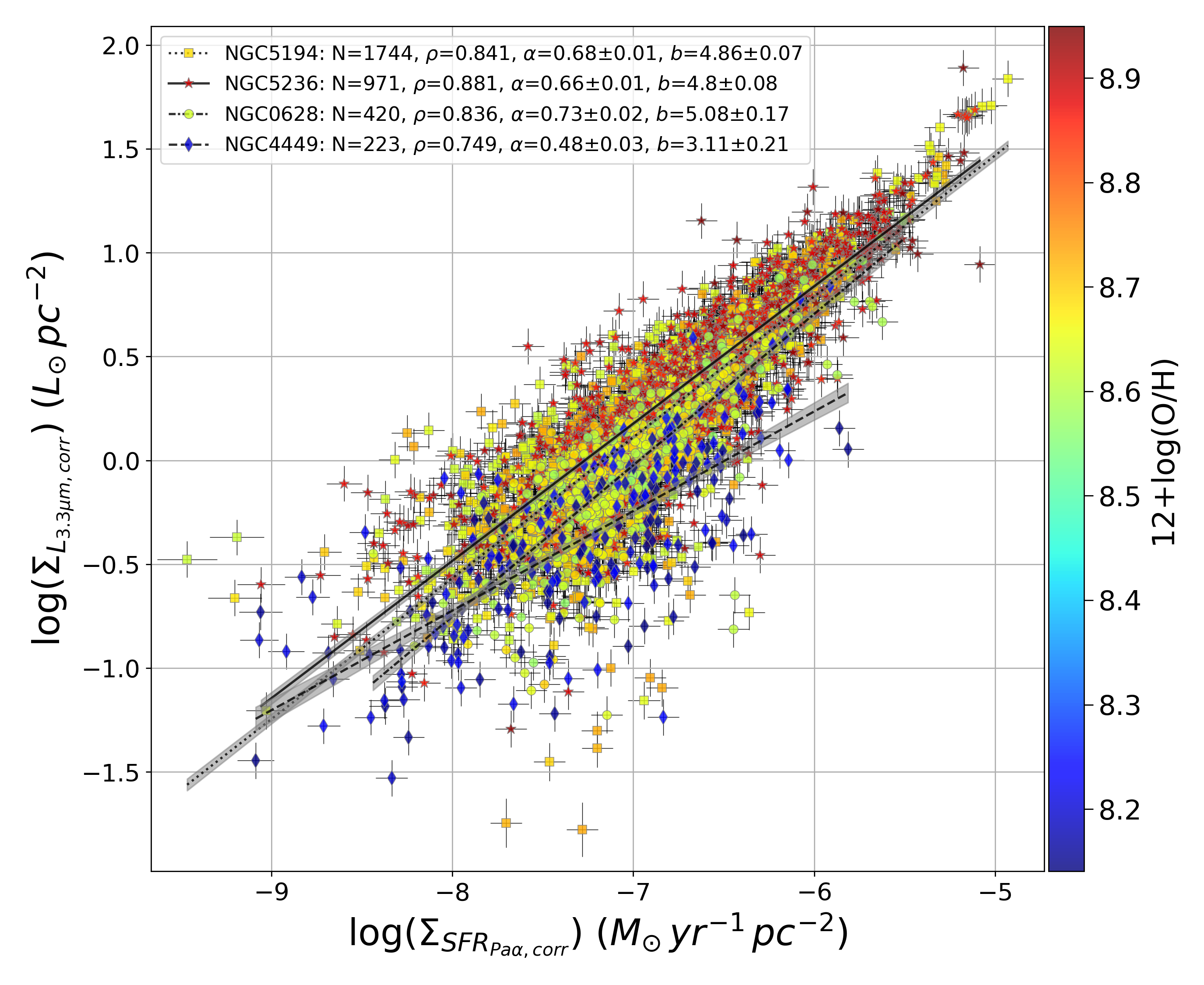}
\includegraphics[width=0.49\textwidth]{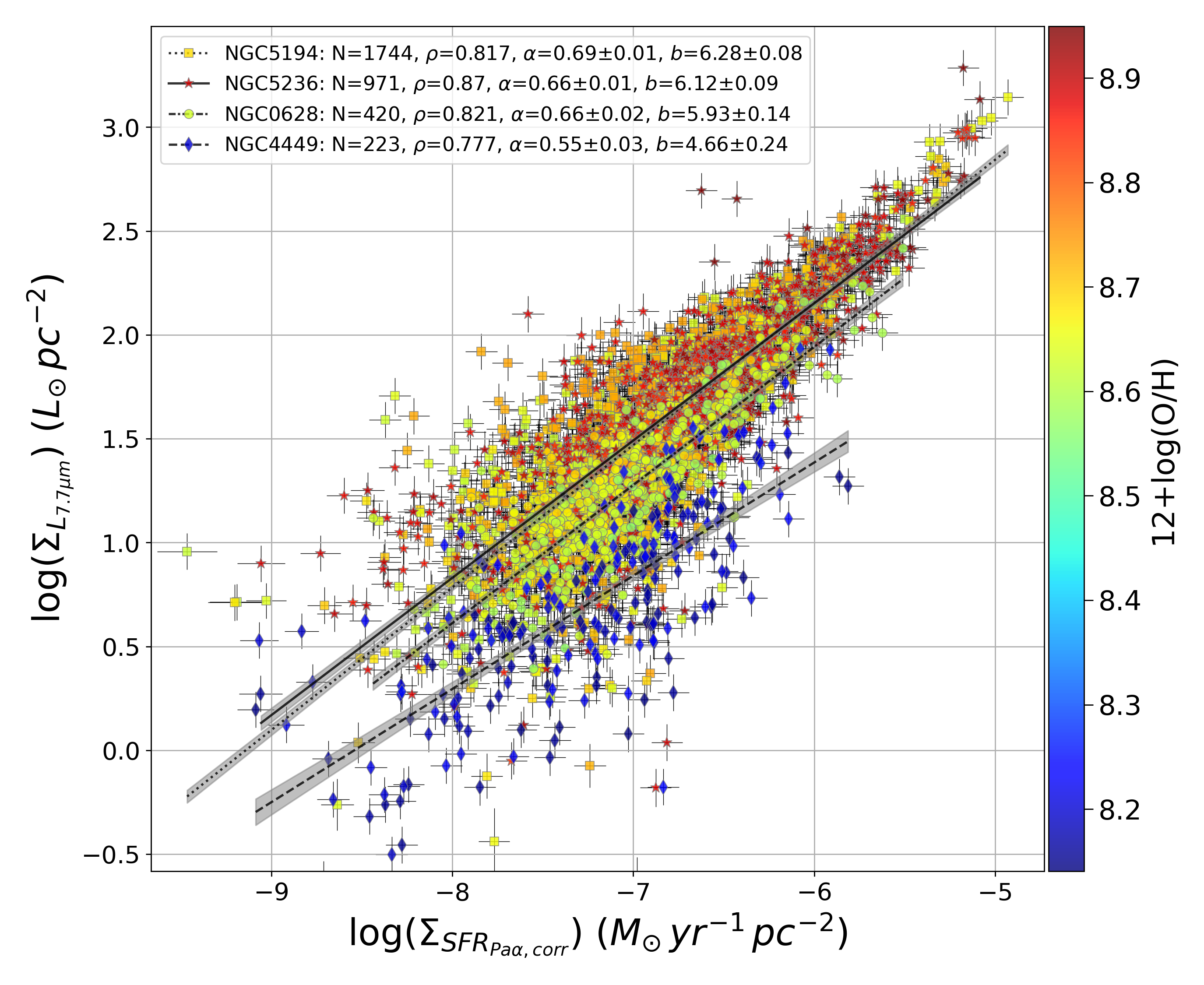}
\caption{Left panel: The extinction corrected 3.3~$\mu$m PAH luminosity surface density as a function of SFR surface density (corrected Pa$\alpha$). The data points show the measurements around eYSC--I sources for each galaxy and are color-coded by the nebular oxygen abundance (gas-phase metallicity), derived from the observed radial gradients. The various black lines show the best-fit relations for each galaxy determined from a Bayesian linear regression using the \texttt{LINMIX} package, while the shaded regions show the 1$\sigma$ confidence intervals. The figure caption gives the total number of sources (N), the Spearman correlation coefficient ($\rho$), and the values of the best-fit slope ($\alpha$) and y-intercept ($b$) and their 1$\sigma$ uncertainties determined from the Bayesian regression for each galaxy. Right panel: Same as the left, but instead for the 7.7~$\mu$m PAH luminosity surface density.}
\label{fig:f4}
\end{figure*}

In Figure \ref{fig:f4}, we show the relationship between the surface densities of PAH luminosity (extinction corrected 3.3~$\mu$m, left panel; 7.7~$\mu$m, right panel) and SFR derived from corrected Pa$\alpha$ for our measurements around eYSC--I sources in each of the four galaxies. The data points are color-coded by the gas-phase metallicity as traced by the nebular oxygen abundance or 12+log(O/H), derived from the observed radial abundance gradients. The best-fit relations are determined for each galaxy separately using the Bayesian linear regression algorithm implemented in the \texttt{LINMIX}\footnote{\url{https://github.com/jmeyers314/linmix}} Python code. The code uses the linear mixture model developed by \cite{2007ApJ...665.1489K} and can robustly fit data with uncertainties on two variables. The best-fits are shown by the various black lines in Figure \ref{fig:f4}, corresponding to the mean best-fit slope and y-intercept of the traces. The gray-shaded regions show the 1$\sigma$ confidence intervals given by the standard deviation of the best-fit parameters. 

We find a relatively tight relation between the PAH luminosities (3.3 and 7.7~$\mu$m) and SFR, especially in the three high metallicity spirals (NGC 5194, 5236, and 628), with a Spearman correlation coefficient ($\rho$) greater than 0.7 in all cases. However, it is quite clear in Figure \ref{fig:f4} that the scatter in the data is much greater towards low luminosities. It is also apparent that the best-fit relations do not fit the data well at high luminosity, particularly for the two galaxies NGC 5194 and NGC 5236. This is likely a consequence of the high scatter and high number of sources at low luminosity driving the determination of the slope. These results can likely be explained by a fraction of our sources being significantly affected by stochastic sampling of the stellar IMF, which can contribute to the substantial scatter at low luminosity (see the discussion in the next section). 

To mitigate the effects of stochastic sampling of the stellar IMF, we perform an additional cut on the luminosity of our sources. We calculate the H$\alpha$ luminosity expected for a 4 Myr old star cluster with a stellar mass of 3000 M$_{\odot}$ \citep{2011ApJ...741L..26F}, assuming the Starburst99 stellar population synthesis models \citep{1999ApJS..123....3L} with a metal mass fraction Z=0.02 and the Padova AGB stellar evolutionary tracks. This gives log(L$_{H\alpha}\,/$ erg s$^{-1}$)$\,=37.55$ or log(SFR$\,/$ M$_{\odot}$ yr$^{-1}$)$\,=-3.71$ for the calibration from \cite{2013seg..book..419C}. We take this value as the lowest limit to use for the remainder of our analysis. 

\begin{figure*}
\centering
\includegraphics[width=0.49\textwidth]{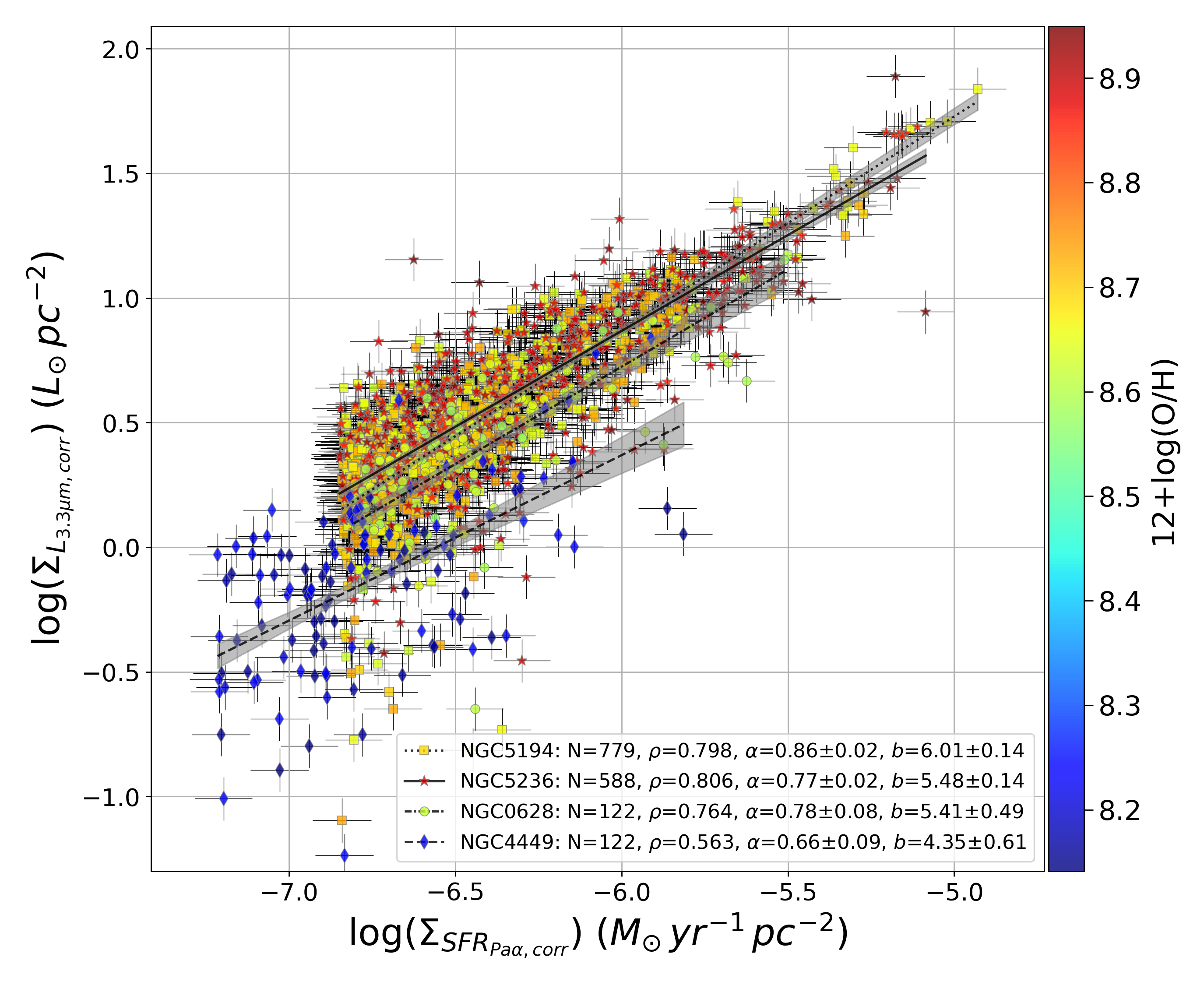}
\includegraphics[width=0.49\textwidth]{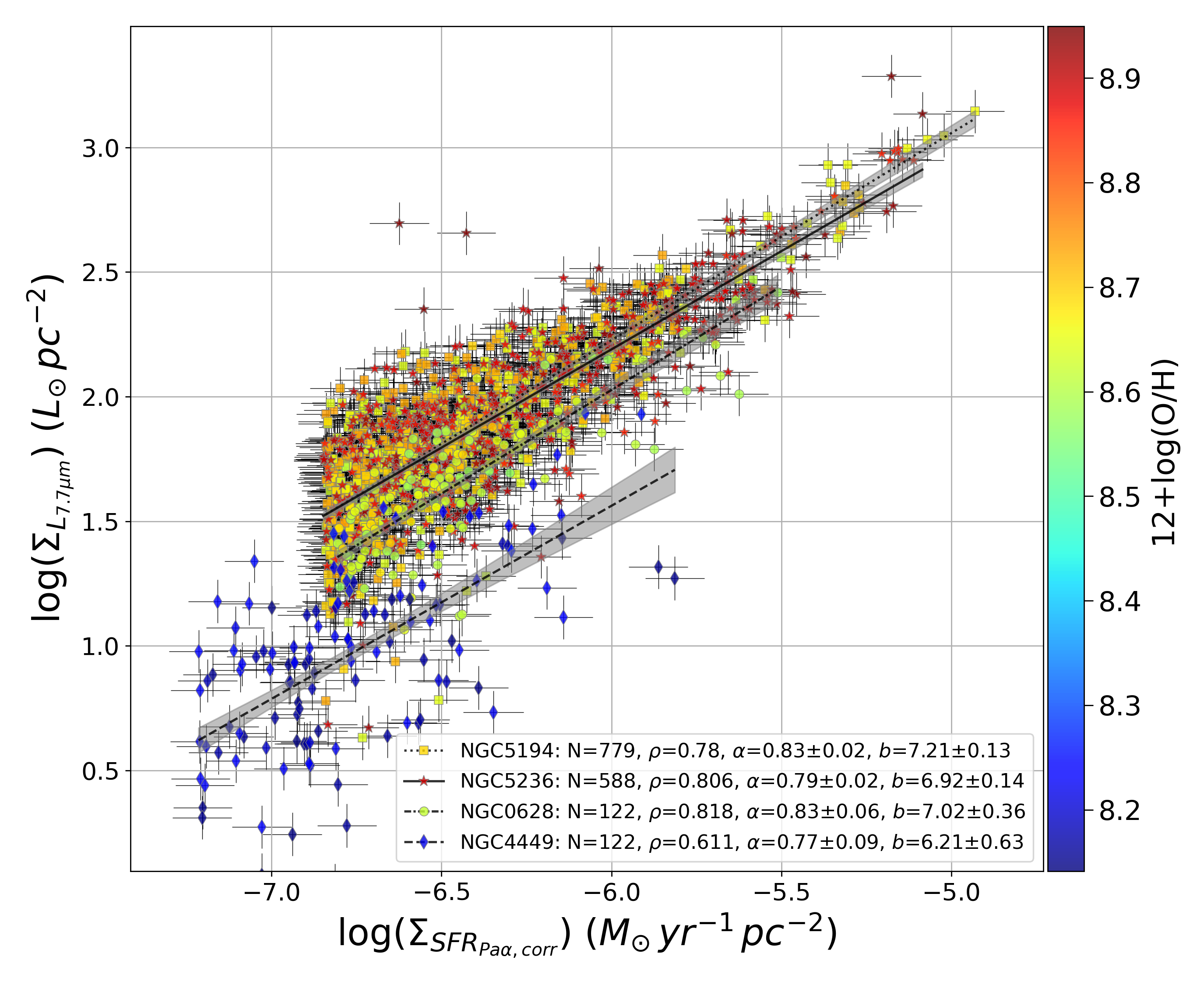}
\includegraphics[width=0.49\textwidth]{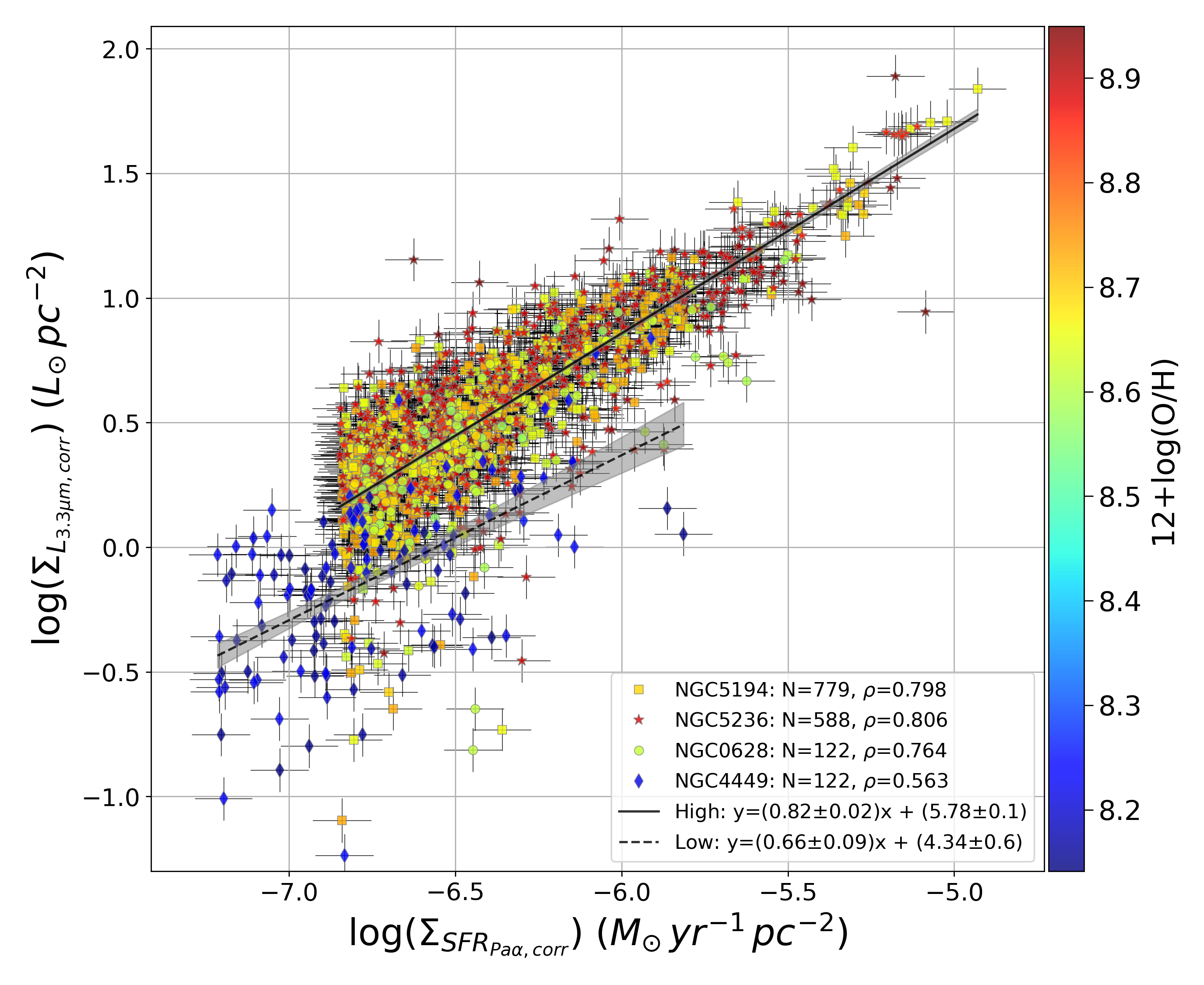}
\includegraphics[width=0.49\textwidth]{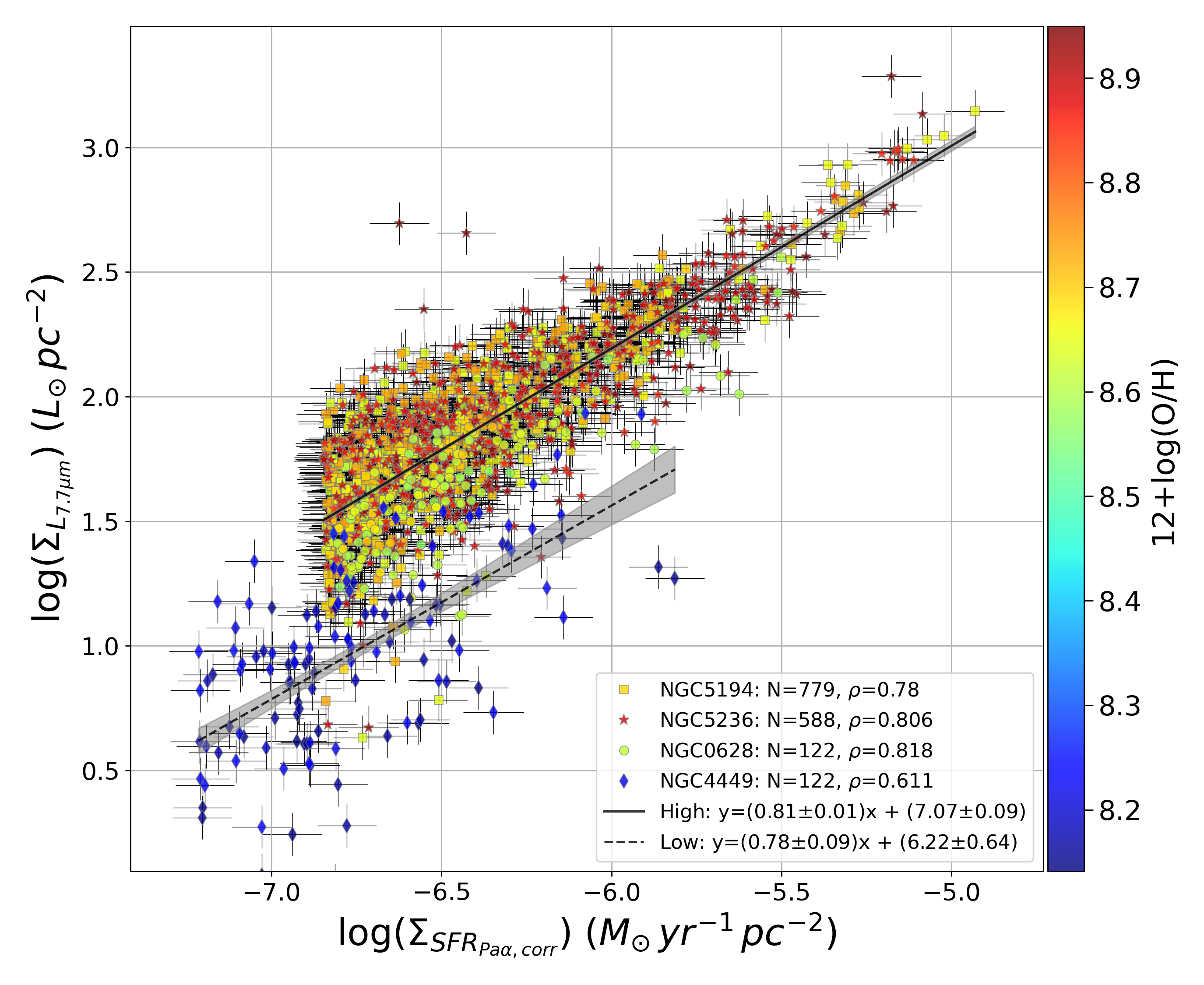}
\caption{The extinction corrected 3.3~$\mu$m PAH (left panels) or 7.7~$\mu$m PAH (right panels) luminosity surface density as a function of SFR surface density with an additional luminosity cut. The cut has been applied to limit the effects from stochastic sampling of the stellar IMF. Regions with H$\alpha$ luminosity below that expected from a 4 Myr old cluster with a stellar mass of 3000 M$_{\odot}$ are removed, based on Starburst99 models with Z=0.02 and the Padova AGB evolutionary tracks. The best-fit relations (black lines) are shown either for each galaxy separately (top panels) or in 2 bins (high/low) of metallicity (bottom panels). The three high metallicity spirals (NGC 5194, 5236, and 628) constitute the high bin, while the lower metallicity dwarf (NGC 4449) constitutes the low bin. Note that the luminosity cutoff is the same for all galaxies, however, the high inclination angle (I=68$^{\circ}$) of NGC 4449 makes it appear lower in the space of inclination corrected surface density. See Figure \ref{fig:f4} for a more complete description.}
\label{fig:f5}
\end{figure*}

Figure \ref{fig:f5} shows the extinction corrected 3.3~$\mu$m PAH (left panels) or 7.7~$\mu$m PAH (right panels) luminosity surface density as a function of SFR surface density as in Figure \ref{fig:f4}, however with the additional luminosity cut applied to mitigate the effects of stochastic IMF sampling at the low-mass end. It is important to note that the luminosity cutoff is the same for each of the galaxies; however, the much higher inclination angle (I=68$^{\circ}$) of NGC 4449 makes the cutoff appear lower for this galaxy in the space of inclination corrected surface density. The best-fit relations are shown separately for each galaxy in the top panels. We find that the best-fit relations provide a much better fit to the data when the sources affected by stochastic IMF sampling are removed. For all four galaxies, we determine that the relations are sub-linear, defined as having a power-law exponent $\alpha$ (or the slope of the best-fit linear regression in log-log space) less than one. The three high metallicities spirals are found to have best-fit relations that are fairly consistent, although NGC 628 exhibits a slight shift down by ${\sim}$0.1 dex. 

The bottom panels of Figure \ref{fig:f5} are the same as the top, however, the best-fit relations are shown for two bins of metallicity. The sources in the three high metallicity spirals (NGC 5194, 5236, and 628) are binned together and make up the high metallicity bin, while the sources in the low metallicity dwarf NGC 4449 comprise the low bin. Table \hyperlink{t2}{2} lists the results for the PAH-SFR relations (both 3.3 and 7.7~$\mu$m) for each galaxy and metallicity bin in the case where sources affected by stochastic sampling of the stellar IMF have been removed. For the high metallicity bin, we observe a tight relation between the PAH luminosities (3.3 and 7.7~$\mu$m) and SFR, with correlation coefficients $\rho \, {\sim}\, 0.8$ and best-fit slopes $\alpha \, {\sim}\, 0.8$. We find that at low metallicity, there is a significant deficit in the PAH emission at fixed luminosity for both the 3.3 and 7.7~$\mu$m features. At log($\Sigma_{SFR}\,/$ M$_{\odot}$ yr$^{-1}$ pc$^{-2}$)$\,=-6.5$, the PAH deficit at the metallicity of NGC 4449 is about 0.4 and 0.6 dex for the 3.3 and 7.7~$\mu$m features, respectively. We also find that at low metallicity, the scatter in the relations is much higher. The typical scatter, derived as the mean orthogonal distance between the data and the best-fit relation (d$_{orth}$), is determined to be 0.11 and 0.19 dex (or 0.12 and 0.18 dex) for the high and low metallicity bins, respectively, for the 3.3~$\mu$m PAH (or 7.7~$\mu$m PAH) feature. We determine that for the 7.7~$\mu$m PAH feature, the best-fit slope is fairly consistent between the two bins of metallicity, corresponding to about 0.81 and 0.78 for the high and low metallicity bins, respectively. However, for the 3.3~$\mu$m PAH, the low metallicity bin exhibits a slightly lower slope (${\sim}$2$\sigma$), corresponding to about 0.66, compared with the value at high metallicity at about 0.82. The 3.3 and 7.7~$\mu$m PAH features are found to be fairly consistent in terms of both the best-fit slope and scatter for the high metallicity bin.

The best-fit relations shown in the bottom panels of Figure \ref{fig:f5} for the high metallicity bin correspond to new SFR calibrations for the 3.3 and 7.7~$\mu$m PAH emission features. The new calibrations are given by:
\begin{eqnarray}
\label{eq:1}
    \text{log}\Big(\frac{\Sigma_{SFR}}{M_\odot \, yr^{-1} \, pc^{-2}}\Big) = (1.22 \pm & 0.02) \times \text{log}\Big(\frac{\Sigma_{L_{3.3\mu m}}}{L_{\odot} \, pc^{-2}}\Big) \nonumber \\
    & - (7.04 \pm 0.12)
\end{eqnarray} 
for the 3.3~$\mu$m feature, and,
\begin{eqnarray}
\label{eq:2}
    \text{log}\Big(\frac{\Sigma_{SFR}}{M_\odot \, yr^{-1} \, pc^{-2}}\Big) = (1.23 \pm & 0.02) \times \text{log}\Big(\frac{\Sigma_{L_{7.7\mu m}}}{L_{\odot} \, pc^{-2}}\Big) \nonumber \\
    & - (8.70 \pm 0.12)
\end{eqnarray}
for the 7.7~$\mu$m feature. These calibrations are for the JWST filters and treat the PAH features in the same way as the hydrogen recombination lines, i.e., the continuum subtracted flux densities are multiplied by the filter bandwidths. If we instead use the ``in-band" assumption for deriving the PAH luminosities, i.e., multiply the continuum subtracted flux densities by the pivot wavelength of the filters rather than the bandwidth, we obtain the following calibrations:
\begin{eqnarray}
\label{eq:3}
    \text{log}\Big(\frac{\Sigma_{SFR}}{M_\odot \, yr^{-1} \, pc^{-2}}\Big) = (1.22 \pm & 0.02) \times \text{log}\Big(\frac{\Sigma_{L_{3.3\mu m, band}}}{L_{\odot} \, pc^{-2}}\Big) \nonumber \\
    & - (8.25 \pm 0.14)
\end{eqnarray}
for the 3.3~$\mu$m feature, and,
\begin{eqnarray}
\label{eq:4}
    \text{log}\Big(\frac{\Sigma_{SFR}}{M_\odot \, yr^{-1} \, pc^{-2}}\Big) = (1.23 \pm & 0.02) \times \text{log}\Big(\frac{\Sigma_{L_{7.7\mu m, band}}}{L_{\odot} \, pc^{-2}}\Big) \nonumber \\
    & - (9.43 \pm 0.13)
\end{eqnarray}
for the 7.7~$\mu$m feature. The latter calibrations treat the PAH features more like broadband continuum emission, a common assumption by many previous studies, making comparisons with those studies easier. 

Expressing the variables in terms of luminosity rather than luminosity surface density, we obtain the following calibrations:
\begin{eqnarray}
\label{eq:5}
    \text{log}\Big(\frac{SFR}{M_\odot \, yr^{-1}}\Big) = (1.21 \pm & 0.02) \times \text{log}\Big(\frac{L_{3.3\mu m}}{L_{\odot}}\Big) \nonumber \\
    & - (7.71 \pm 0.06)
\end{eqnarray}
\begin{eqnarray}
\label{eq:6}
    \text{log}\Big(\frac{SFR}{M_\odot \, yr^{-1}}\Big) = (1.23 \pm & 0.02) \times \text{log}\Big(\frac{L_{7.7\mu m}}{L_{\odot}}\Big) \nonumber \\
    & - (9.40 \pm 0.06)
\end{eqnarray}
\begin{eqnarray}
\label{eq:7}
    \text{log}\Big(\frac{SFR}{M_\odot \, yr^{-1}}\Big) = (1.21 \pm & 0.02) \times \text{log}\Big(\frac{L_{3.3\mu m, band}}{L_{\odot}}\Big) \nonumber \\
    & - (8.91 \pm 0.07)
\end{eqnarray}
\begin{eqnarray}
\label{eq:8}
    \text{log}\Big(\frac{SFR}{M_\odot \, yr^{-1}}\Big) = (1.23 \pm & 0.02) \times \text{log}\Big(\frac{L_{7.7\mu m, band}}{L_{\odot}}\Big) \nonumber \\
    & - (10.12 \pm 0.06).
\end{eqnarray}

The calibrations given above correspond to ${\sim}$40 pc scales for tightly spatially connected, compact peaks in both ionized gas and PAH emission (eYSC--I) at near solar metallicity. The regime these calibrations apply to remains unclear, however, we expect they may describe the relations in the youngest, embedded phases of star formation for individual (or at most a few) HII regions/PDRs at high metallicity relatively well, with little contribution from PAH heating by older ($>$10 Myr) stellar populations. The calibration coefficients are likely affected by the leakage of UV photons, although possibly only the y-intercepts as the leakage of photons is not expected to depend strongly on luminosity. In the extreme case where half of the ionizing photons leak out of the HII regions, we would expect a factor of two higher SFR, corresponding to an increase in the value of the y-intercept by ${\sim}$0.3 relative to the values in the equations above. In the case of direct absorption of ionizing photons by dust, there is a potential dependence on luminosity \citep{Krumholz+2009, Draine+2011, Calzetti+2025}, which would mainly contribute to exacerbating the non-linearity of the above calibrations.

We estimate that the total observed Pa$\alpha$ flux density above 3$\sigma$ for the FEAST NIRCam mosaics is 0.269, 0.597, 0.143, and 0.119 Jy for NGC 5194, 5236, 628, and 4449, respectively. For NGC 5194, where the coverage of the final eYSC--I catalog corresponds to almost the entirety of the FEAST NIRCam mosaics, the total observed Pa$\alpha$ flux density for the eYSC--I sources is 0.040 Jy before (0.033 Jy after) the cut on luminosity that mitigates the effects of stochastic IMF sampling, estimated as the sum of the measurements for the source catalogs with no overlap. Therefore, the fraction of the total observed Pa$\alpha$ flux density in the NGC 5194 mosaics contributed by the eYSC--I sources is roughly 0.15 before (0.12 after) the luminosity cut. When including corrections for the eYSC--I sources below the luminosity cut and for the emission from both eYSC--II and line--emitting optical YSCs \citep{Calzetti+2025, 2025arXiv250508874K}, this fraction increases to a little over 50\%. This is in agreement with the findings that only about 50\% of the ionizing photons are associated with HII regions in disks, while the remaining 50\% leaks out of the regions and diffuse in the galaxy \citep[e.g.,][]{Reynolds+1990, Ferguson+1996, Hoopes+2003, Oey+2007, Zhang+2017}. Our estimated fractions are likely to be uncertain due to incompleteness in the eYSC--I source catalogs, local background subtraction, contamination in the mosaics, etc.

\begin{figure}
\centering
\includegraphics[width=0.47\textwidth]{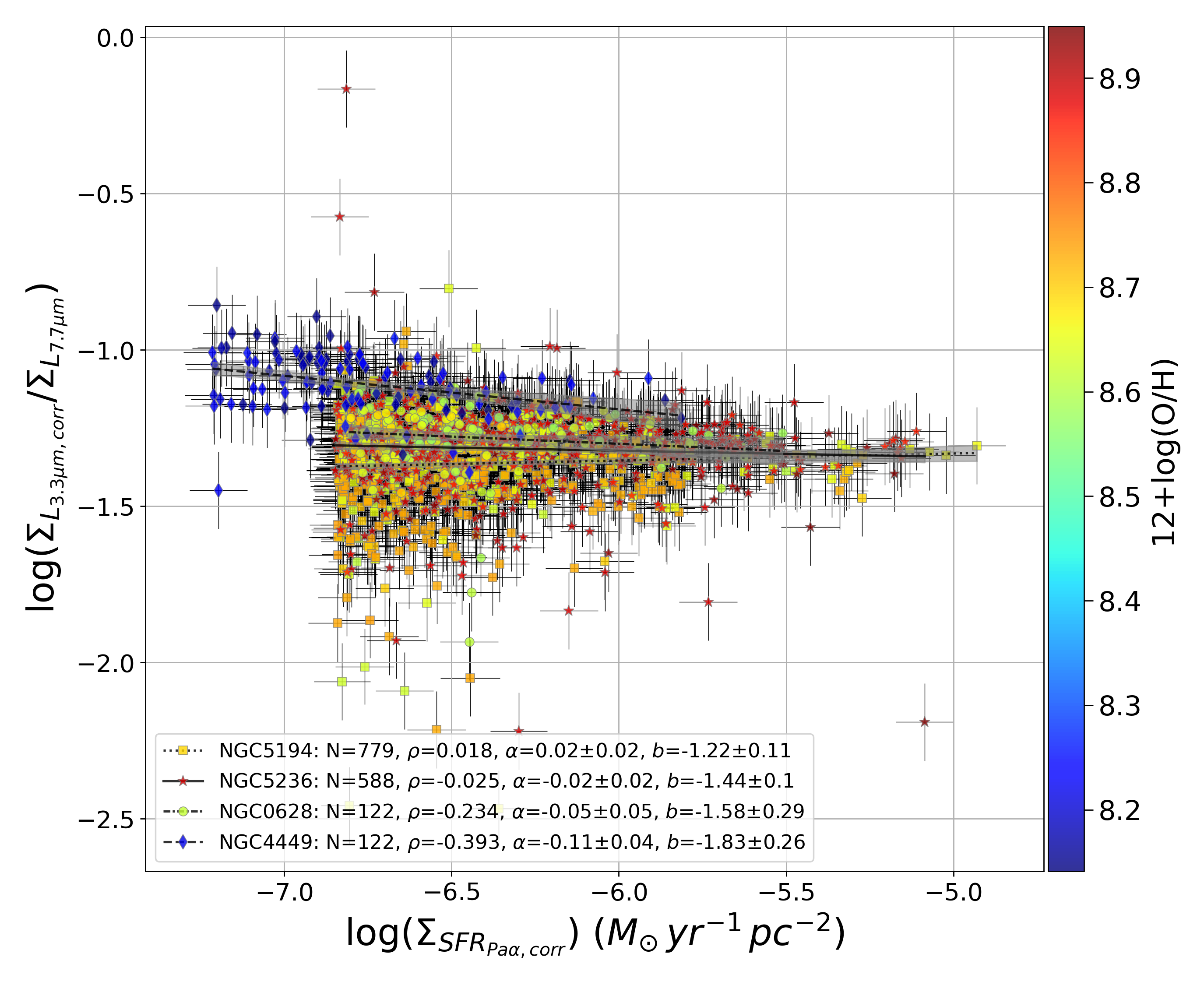}
\caption{The ratio of extinction corrected 3.3~$\mu$m PAH to 7.7~$\mu$m PAH luminosity surface density as a function of SFR surface density. See Figure \ref{fig:f4} for a more complete description.}
\label{fig:f6}
\end{figure}

In Figure \ref{fig:f6}, we show the ratio of extinction corrected 3.3~$\mu$m PAH to 7.7~$\mu$m PAH luminosity surface density (or the 3.3/7.7~$\mu$m PAH ratio) as a function of SFR surface density for our measurements of each of the galaxies sources that are above the stochastic sampling limit, along with the best-fit relations derived for each galaxy. We determine that the three high metallicity spirals show relatively flat relations between the 3.3/7.7~$\mu$m PAH ratio and SFR. In the low metallicity dwarf NGC 4449, we observe a slightly decreasing relation, with a slope of $\alpha \, {\sim}\,-0.11$, although the scatter is high with a correlation coefficient $\rho \,{\sim}\, -0.4$.

\begin{figure*}
\centering
\includegraphics[width=0.49\textwidth]{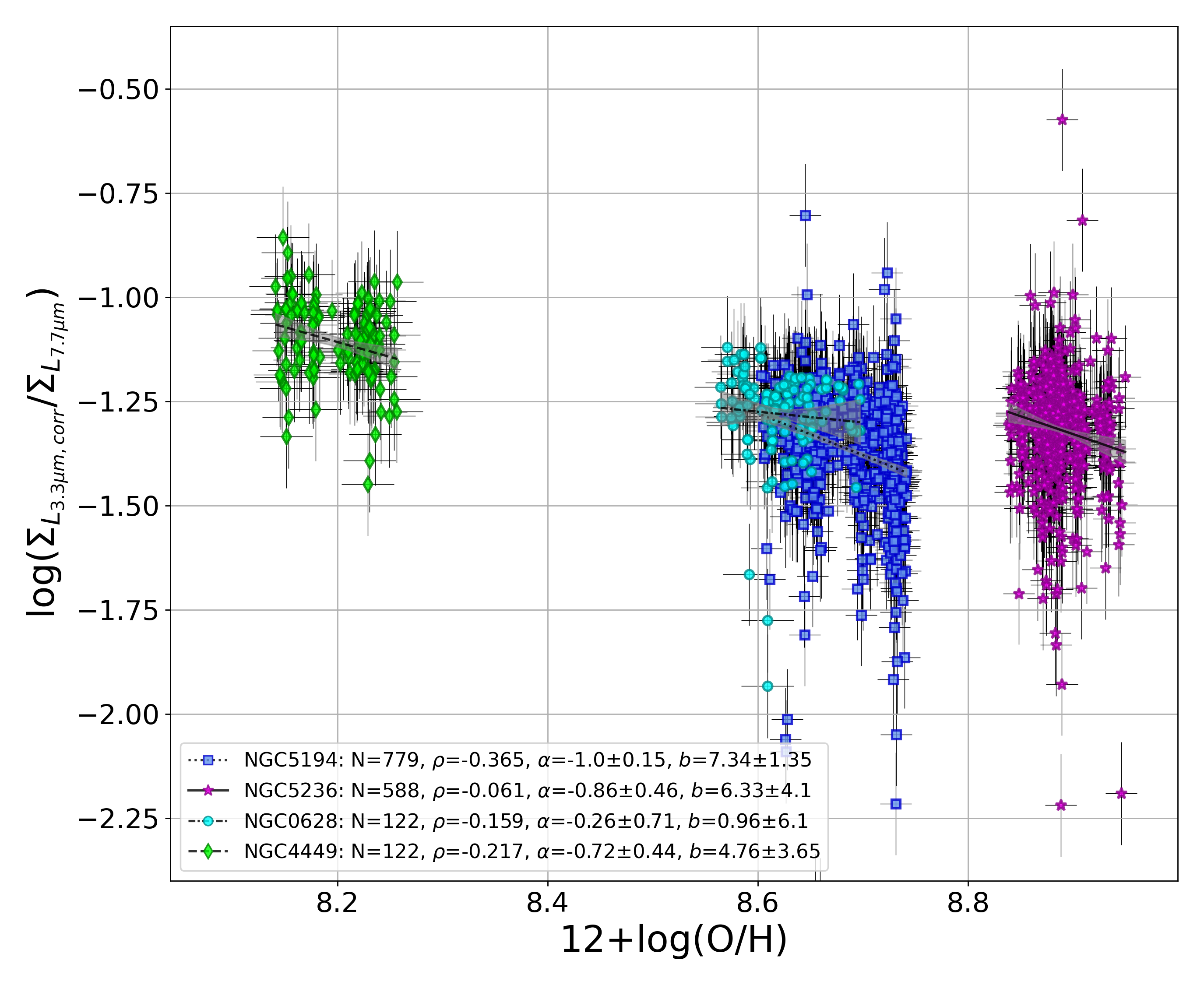}
\includegraphics[width=0.49\textwidth]{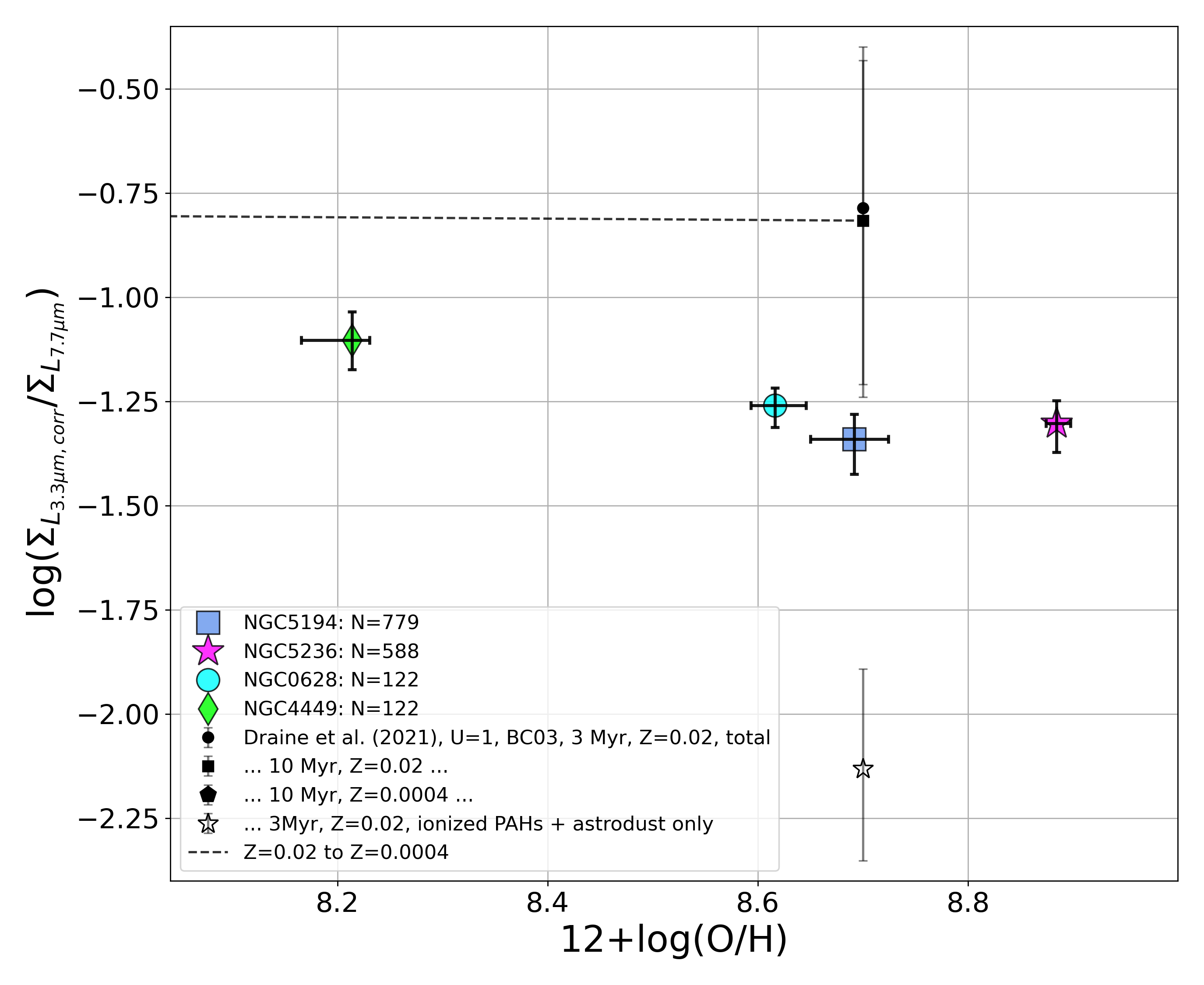}
\caption{The ratio of extinction corrected 3.3~$\mu$m PAH to 7.7~$\mu$m PAH luminosity surface density as a function of nebular oxygen abundance or 12+log(O/H). Left panel: The data points show the measurements around eYSC--I sources color-coded by the galaxy. The best-fit relations for each galaxy are shown by the black lines. Right panel: The colored points show the median values for each galaxy, with error bars corresponding to the middle 50$\%$ of the distribution. The black points show expectations of various models from \cite{2021ApJ...917....3D} for a standard PAH size distribution and ionization, while the error bars show the full range covered by the various sizes/ionizations. The closed points are for the models that consider the total dust emission (neutral PAHs, ionized PAHs, and astrodust), while open is for models that only consider ionized PAHs and astrodust. The black dashed line shows a linear interpolation between the model points at Z=0.02 (12+log(O/H)$\,{\sim}\,$8.7) and Z=0.0004 (12+log(O/H)$\,{\sim}\,$7.0; below the range we show here), assuming the total dust emission and an age of 10 Myr.}
\label{fig:f7}
\end{figure*}

The 3.3/7.7~$\mu$m PAH ratio as a function of the nebular oxygen abundance for each of our measurements is shown in the left panel of Figure \ref{fig:f7}, along with the best-fit relations for each galaxy. We find evidence of a decreasing relation between the 3.3/7.7~$\mu$m PAH ratio and the local ISM metallicity in each of the four galaxies. However, the scatter is quite high with a correlation coefficient $|\rho | \,{\lesssim}\, 0.3$ for all the galaxies. The right panel of Figure \ref{fig:f7} shows the median 3.3/7.7~$\mu$m PAH ratio versus the median oxygen abundance determined for the sources in each galaxy. We observe a significantly higher 3.3/7.7~$\mu$m PAH ratio on average in the low metallicity dwarf NGC 4449, while the three high metallicity spirals exhibit ratios that are relatively similar on average. 

We also show expectations for the 3.3/7.7~$\mu$m PAH ratio for various models from \cite{2021ApJ...917....3D} in the right panel of Figure \ref{fig:f7}. We estimate these model expectations with a method that mimics our observations by convolving the JWST filter throughputs with the model spectra and performing continuum subtractions using the same methods as the observations. We find that the \cite{2021ApJ...917....3D} models that consider the total dust emission (neutral PAHs, ionized PAHs, and astrodust) and a standard PAH size distribution and ionization predict higher values of the 3.3/7.7~$\mu$m PAH ratio than we derive for our observations on average. For NGC 4449, the observed average 3.3/7.7~$\mu$m PAH ratio can be accounted for by the large size distribution and high ionization models, however, the other galaxies are inconsistent with the parameter space covered by these models. Conversely, the models that only consider ionized PAHs and astrodust (no neutral PAHs) predict much lower values for the 3.3/7.7~$\mu$m PAH ratio than we observe. The models predict almost no increase in the 3.3/7.7~$\mu$m PAH ratio if the incident stellar spectrum that heats the dust and PAHs is changed from a near solar metallicity population (Z=0.02) to extremely sub-solar (Z=0.0004), shown by the near flat relation between the model values as a function of metallicity (black dashed line in the right panel of the figure). It is important to note here that the \cite{2021ApJ...917....3D} models do not address the likely changes in the PAH population as a function of the ISM metallicity, but rather, simply investigate the effects of the stellar radiation from stars with varying (high, low) Z. 

\begin{figure*}
\centering
\includegraphics[width=0.49\textwidth]{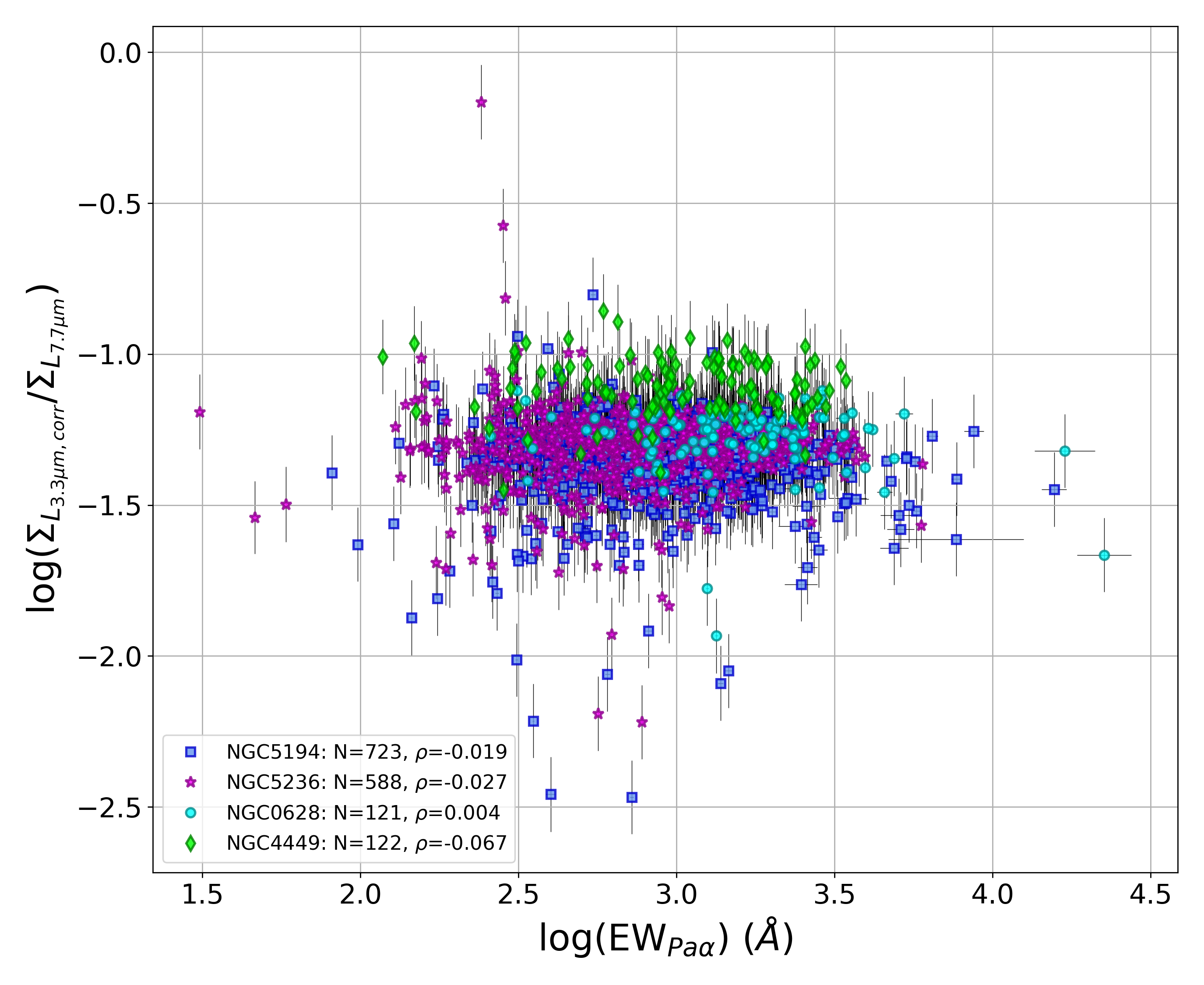}
\includegraphics[width=0.49\textwidth]{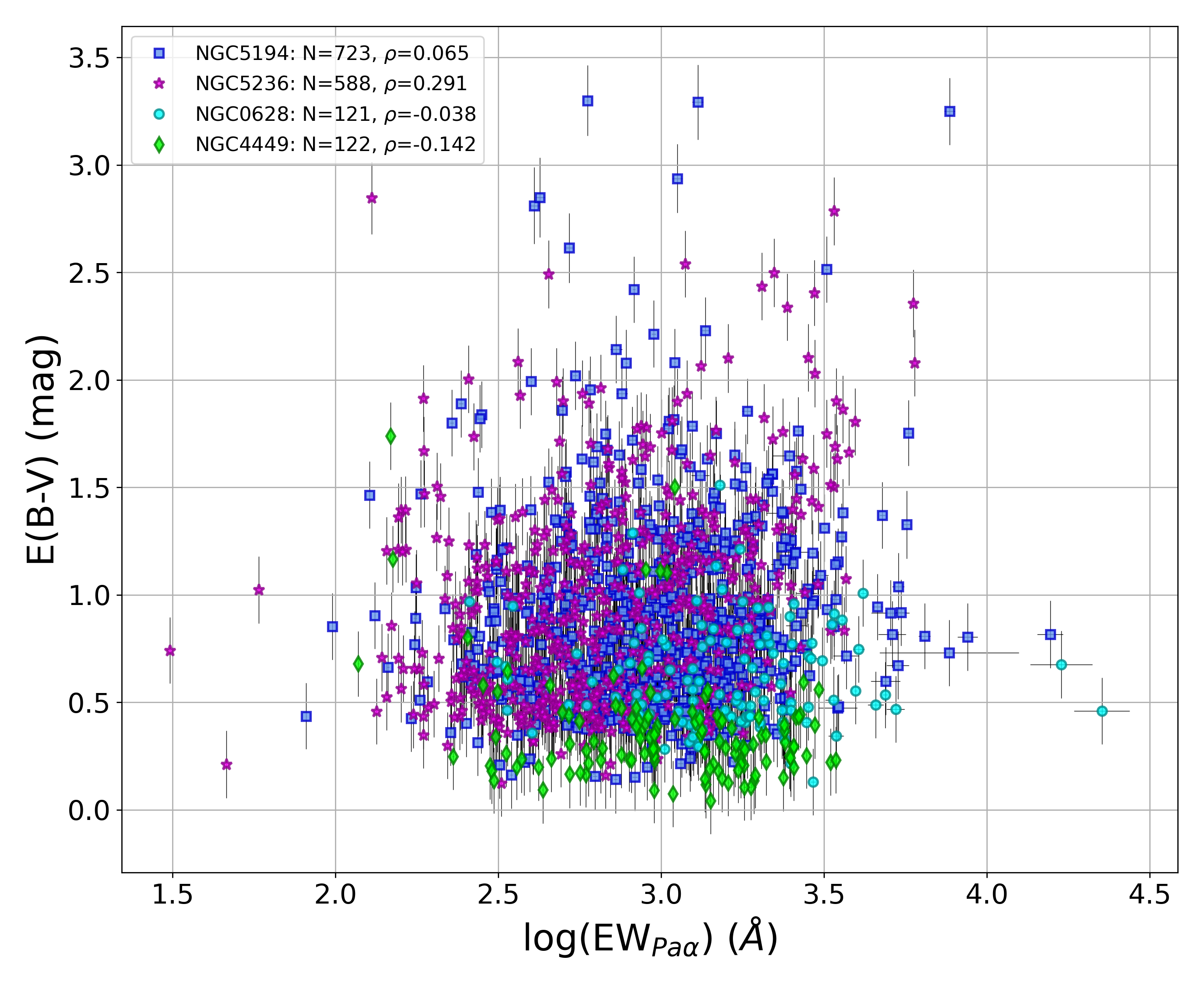}
\includegraphics[width=0.49\textwidth]{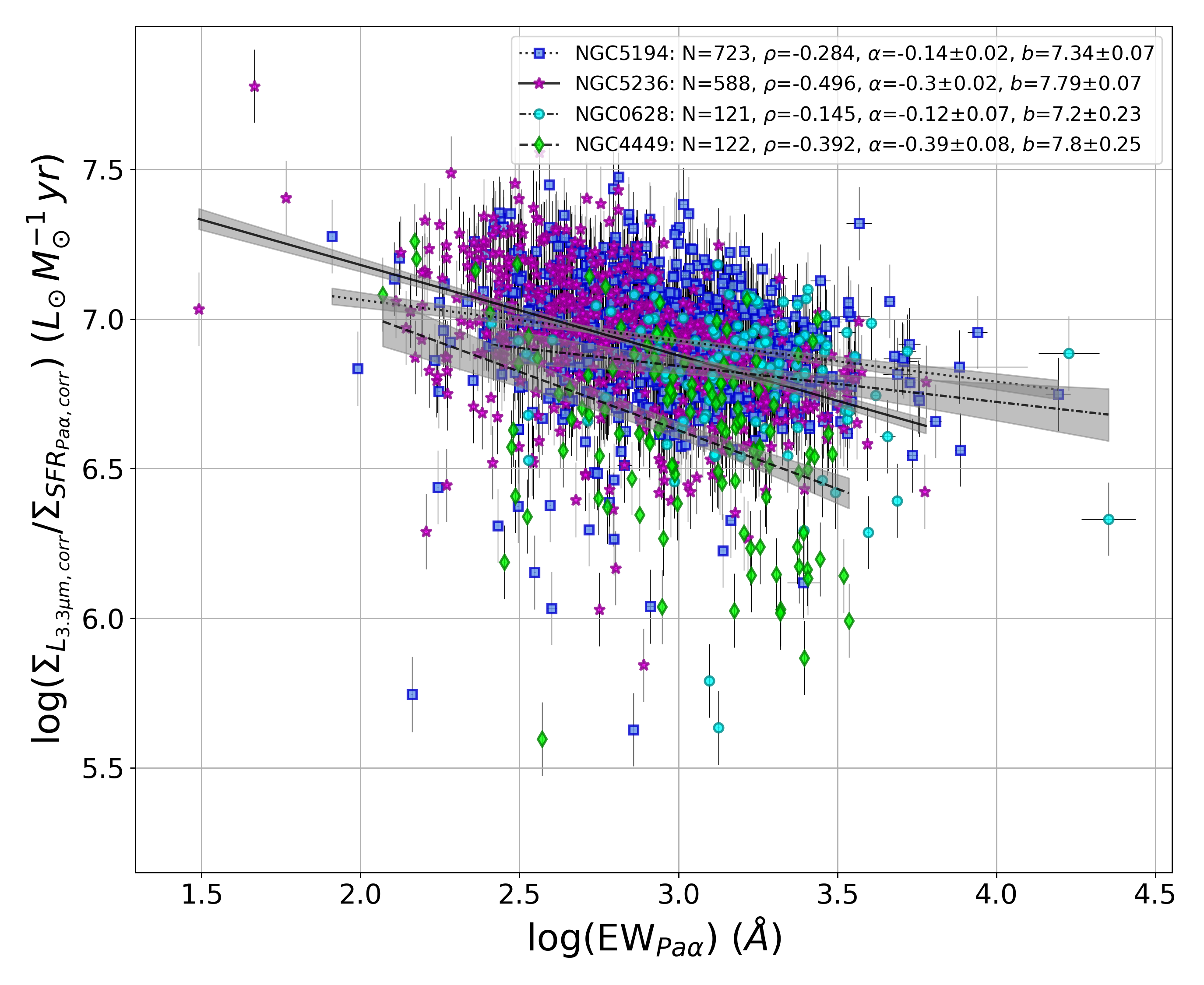}
\includegraphics[width=0.49\textwidth]{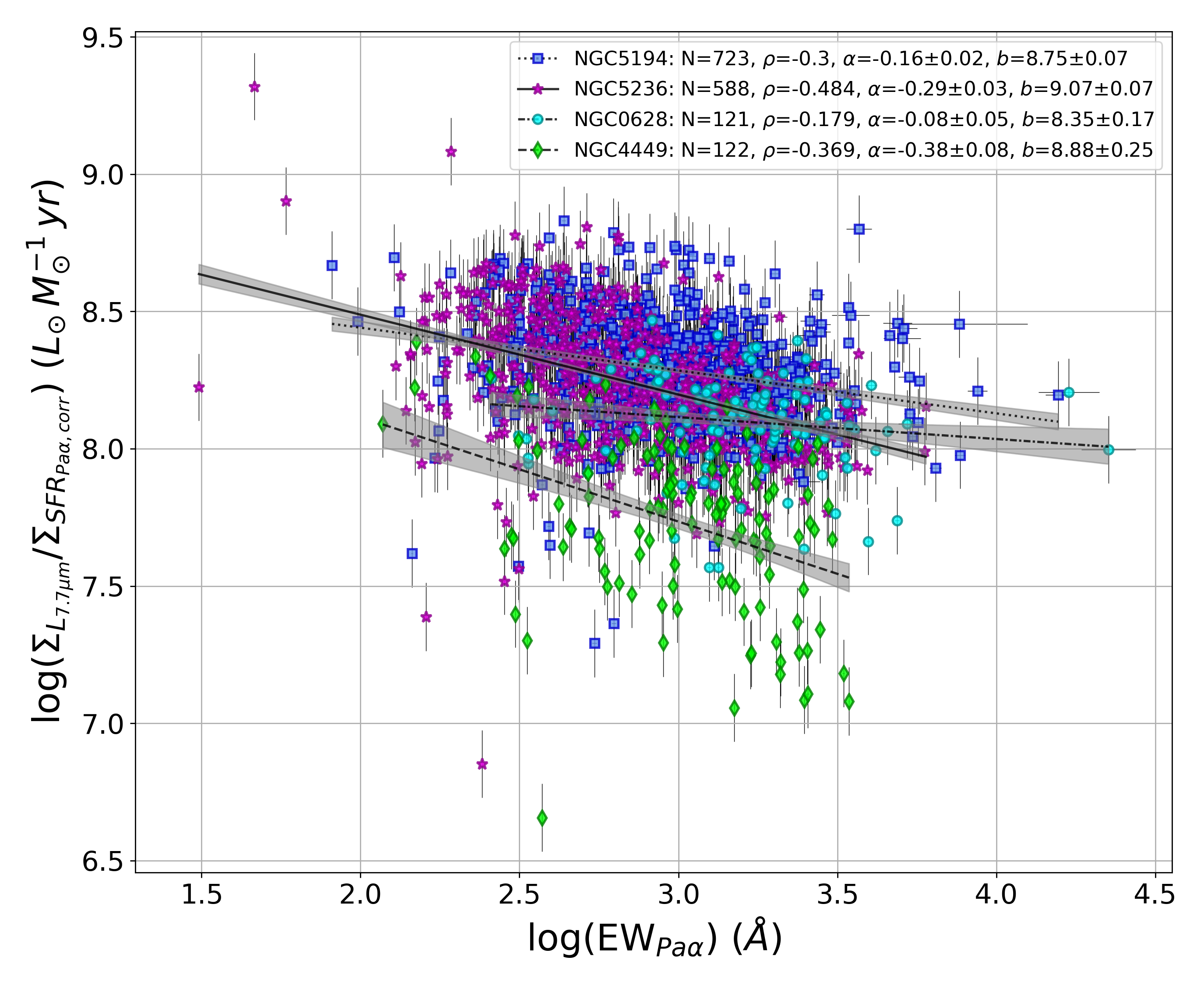}
\caption{Top left: The ratio of extinction corrected 3.3~$\mu$m PAH to 7.7~$\mu$m PAH luminosity surface density as a function of the equivalent width (EW) of Pa$\alpha$. Data points are color-coded by galaxy. Top right: The color excess or E(B$-$V) versus the EW of Pa$\alpha$. Bottom left: The ratio of extinction corrected 3.3~$\mu$m PAH luminosity surface density to SFR surface density versus the EW of Pa$\alpha$. Best-fit relations for each galaxy are shown by the black lines. Bottom right: Same as the bottom left, but instead for the 7.7~$\mu$m PAH luminosity surface density.}
\label{fig:f8}
\end{figure*}

The top left panel of Figure \ref{fig:f8} shows the 3.3/7.7~$\mu$m PAH ratio as a function of the EW of Pa$\alpha$ for each of our measurements. We find that in each of the galaxies, the relation between these parameters is consistent with flat, or no correlation, with a correlation coefficient $|\rho| \, <\, 0.1$ in all cases. The relation between the color excess, or E(B$-$V), and the EW of Pa$\alpha$ is shown in the top right panel of Figure \ref{fig:f8}. Naively, we would expect the youngest star clusters and corresponding HII regions (traced by high Pa$\alpha$ EW) to remain the most embedded in their natal gas and dust, and thus, exhibit higher magnitudes of dust attenuation (traced by high E(B$-$V)). However, we observe that the relation between E(B$-$V) and Pa$\alpha$ EW is fairly flat and exhibits high scatter in each of the galaxies in our sample. NGC 5236 shows some evidence of a slightly positive correlation with a correlation coefficient $\rho \, {\sim}\, 0.3$, although given the high scatter, this remains quite uncertain. This unexpectedly weak correlation between dust attenuation and age has been noted in previous studies \citep[e.g.][]{2021ApJ...909..121M}, even for the eYSCs \citep{2024ApJ...971...32P,2025arXiv250508874K}, and may reflect the complexity of the emergence process. The degeneracy between the mass and age of star clusters with respect to the observed Pa$\alpha$ EW likely plays a role here in scrambling the expected relation; i.e. a higher mass, older cluster can exhibit similar ionized gas properties to that of a lower mass, younger cluster. Controlling for the stellar mass or other important physical properties may help reveal the underlying trends between these parameters, but is beyond the scope of this study. 

In the bottom panels of Figure \ref{fig:f8}, we show the ratio of the extinction corrected 3.3~$\mu$m PAH luminosity (left panel) or 7.7~$\mu$m PAH luminosity (right panel) surface density to the SFR surface density as a function of the EW of Pa$\alpha$ for each of our measurements, along with the best-fit relations for each galaxy. We determine that each galaxy exhibits a decreasing relation between the ratio of PAH luminosity to SFR and the Pa$\alpha$ EW. The scatter is fairly high with a correlation coefficient $\rho$ between about $-$0.1 and $-$0.5, depending on the galaxy and PAH feature. The best-fit slope $\alpha$ is found to be between about $-$0.1 and $-$0.4. 

\begin{center}
\begin{table}
\centering
\caption{\hypertarget{t2}{PAH-SFR relations}}
$\text{log}\Big(\frac{\Sigma_{L_{\text{PAH}}}}{L_{\odot} \, pc^{-2}}\Big)  = \alpha \, \text{log}\Big(\frac{\Sigma_{SFR_{Pa\alpha, corr}}}{M_\odot \, yr^{-1} \, pc^{-2}}\Big) + b$ \\[1mm]
\begin{tabular}{ l c c c c c}
\hline
\hline
\rule{0pt}{4ex} Galaxy/Bin & N & $\rho$ & $\alpha$ & $b$ & d$_{orth}$ \\
\rule{0pt}{3ex} $\,\,\,\,\,\,\,\,\,\,\,$(1) & (2) & (3) & (4) & (5) & (6) \\[2mm]
\hline
\rule{0pt}{3ex}&&&\hspace{-8mm} $\Sigma_{L_{3.3\mu m, corr}}$&&\\[1.5mm]
\hline
\rule{0pt}{3ex} NGC 5194 & 779 & 0.798 & $0.86{\pm}0.02$ & $6.01{\pm}0.14$ & 0.11\\
\rule{0pt}{3ex} NGC 5236 & 588 & 0.806 & $0.77{\pm}0.02$ & $5.48{\pm}0.14$ & 0.12\\
\rule{0pt}{3ex} NGC 628 & 122 & 0.764 & $0.78{\pm}0.08$ & $5.41{\pm}0.49$ & 0.12\\
\rule{0pt}{3ex} NGC 4449 & 122 & 0.563 & $0.66{\pm}0.09$ & $4.35{\pm}0.61$ & 0.19\\
\rule{0pt}{3ex} High$^{\mbox{\textit{a}}}$ & 1489 & 0.801 & $0.82{\pm}0.02$ & $5.78{\pm}0.10$ & 0.11\\
\rule{0pt}{3ex} Low$^{\mbox{\textit{b}}}$ & 122 & 0.563 & $0.66{\pm}0.09$ & $4.34{\pm}0.60$ & 0.19\\[1mm]
\hline
\rule{0pt}{3ex}&&&\hspace{-9mm} $\Sigma_{L_{7.7\mu m}}$&&\\[1.5mm]
\hline
\rule{0pt}{3ex} NGC 5194 & 779 & 0.780 & $0.83{\pm}0.02$ & $7.21{\pm}0.13$ & 0.11\\
\rule{0pt}{3ex} NGC 5236 & 588 & 0.806 & $0.79{\pm}0.02$ & $6.92{\pm}0.14$ & 0.12\\
\rule{0pt}{3ex} NGC 628 & 122 & 0.818 & $0.83{\pm}0.06$ & $7.02{\pm}0.36$ & 0.09\\
\rule{0pt}{3ex} NGC 4449 & 122 & 0.611 & $0.77{\pm}0.09$ & $6.21{\pm}0.63$ & 0.18\\
\rule{0pt}{3ex} High$^{\mbox{\textit{a}}}$ & 1489 & 0.785 & $0.81{\pm}0.01$ & $7.07{\pm}0.09$ & 0.12\\
\rule{0pt}{3ex} Low$^{\mbox{\textit{b}}}$ & 122 & 0.611 & $0.78{\pm}0.09$ & $6.22{\pm}0.64$ & 0.18\\[1mm]
\hline
\end{tabular}
\begin{flushleft} 
\rule{0pt}{3ex}
\currtabletypesize{\sc Note}--- \\
\rule{0pt}{4ex}
\hspace{-2mm} Results for the PAH-SFR relations after removing sources that are likely affected by stochastic stellar IMF sampling. The best-fit parameters ($\alpha$ and $b$) are determined from a Bayesian linear regression (see Figure \ref{fig:f5}). \\[1mm]
Columns: 1) Galaxy or bin name, 2) number of eYSC--I sources, 3) Spearman correlation coefficient, 4) best-fit slope, 5) best-fit y-intercept, and 6) mean orthogonal distance between the data and the best-fit relation (dex).  \\[1mm]
$^{\mbox{\textit{a}}}$ High metallicity bin (NGC 5194, 5236, and 628). NGC 5194 and 5236 dominate the statistics. \\[1mm]
$^{\mbox{\textit{b}}}$ Low metallicity bin (NGC 4449). The small differences ($<<$1$\sigma$) in $\alpha$ and $b$ compared with NGC 4449 are a result of the different instances of the Bayesian linear regression. \\
\end{flushleft}
\end{table}
\end{center}

\hypertarget{5}{\section{Discussion}}

\hypertarget{5.1}{\subsection{The PAH-SFR calibrations}}

\hypertarget{5.1.1}{\subsubsection{Metal-rich environments}}

In this study, we find a fairly tight (correlation coefficient, $\rho \, {\sim}\, 0.8$), sub-linear (power-law exponent, $\alpha \, {\sim}\, 0.8$) relation between the PAH luminosities (3.3 and 7.7~$\mu$m) and SFR traced by ionized gas emission in high metallicity (near solar) environments on small scales (40 pc) around newly formed, emerging star clusters that are bright/massive enough to be mostly unaffected by stochastic sampling of the stellar IMF (Figure \ref{fig:f5}, bottom panels). The power-law exponents are determined to be at least ${\sim}9\sigma$ below a linear relation, given the uncertainties determined from the Bayesian regression. These results are consistent with various previous studies of the PAH emission on resolved scales in nearby galaxies. 

The study by \cite{2024ApJ...971..115G} uses JWST/NIRCam imaging of NGC 628 to study the relation between the 3.3~$\mu$m PAH emission and SFR in similar sources and on similar scales as this study. They derive a sub-linear relation with $\alpha \, {\sim}\, 0.75$. For NGC 628, we determine in this study that $\alpha \, {\sim}\, 0.73 \, \pm \, 0.02$ when we similarly do not remove lower luminosity sources that may be affected by stochastic IMF sampling (Figure \ref{fig:f4}, left panel). Therefore, our results are consistent within about 1$\sigma$ with the study by \cite{2024ApJ...971..115G} and the minor differences can be explained by the slightly different methods (i.e. continuum subtraction techniques, dust attenuation corrections, aperture sizes), and source catalogs. The new 3.3~$\mu$m PAH calibration derived in this study is an improvement over that determined by \cite{2024ApJ...971..115G}, with better methods and statistics and multiple galaxy targets, and thus, should be favored.

Other previous studies similarly find that PAH emission exhibits a sub-linear relation with tracers of ionized gas on resolved scales \citep[e.g.][]{2005ApJ...633..871C,2007ApJ...666..870C}. \cite{2005ApJ...633..871C} determine a power-law exponent $\alpha \, {\sim}\, 0.79$ for the relation between the PAH emission at 7.7~$\mu$m and extinction corrected Pa$\alpha$ at about 500 pc scales using Spitzer/IRAC and HST imaging of NGC 5194. \cite{2007ApJ...666..870C} expands this analysis to a large sample of SINGS galaxies and determines $\alpha \, {\sim}\, 0.94$ on scales between about 30 pc and 1 kpc, depending on the distance. The study by \cite{2023ApJ...944L...9L} utilizes JWST imaging to derive $\alpha \, {\sim}\, 0.6$ for the relation between 7.7~$\mu$m PAH emission and extinction corrected H$\alpha$ at ${\sim}$40 pc scales in NGC 628; however, they do not subtract the local background and may be affected by the heating of PAH grains by older, non-local stellar populations. \cite{2023ApJ...957L..26L} utilize JWST/NIRSpec/IFU spectroscopy to investigate the 3.3~$\mu$m PAH emission in the starburst ring around the AGN of NGC 7469 on ${\sim}$200 pc scales and find a relation between the 3.3~$\mu$m PAH and SFR derived from the [NeII] and [NeIII] emission lines that could be sub-linear, however, it is unclear given the scatter and very narrow range of luminosities probed.  

The key emerging result is that the relation between PAH emission and SFR on resolved scales in galaxies is sub-linear, suggesting a significant impact from secondary processes that complicate the ability of the PAH emission to trace star formation. In addition, we determine that even at high luminosity and high metallicity there remains significant scatter in the calibrations, typically ${\sim}$0.1 dex, which is too large to be accounted for by the measurement errors. There are several effects that can play an important role in driving the observed sub-linear trend, as well as various dependencies that can increase the scatter in the PAH calibrations.

\hypertarget{5.1.1.1}{\paragraph{Stochastic sampling of the stellar IMF}}

One effect is stochastic sampling of the stellar IMF, which can play a role in both increasing the scatter and flattening the relations. Newly formed, low-mass star clusters are comprised of too few stars to fully sample the stellar IMF, particularly at the high-mass end of the IMF, which is the least populated. In the low-mass regime where stochastic IMF sampling is important, young clusters of the same mass may exhibit large differences in the production of ionizing photons as it depends sensitively on how the high-mass stars ($>$15 M$_{\odot}$) have been sampled. This has the effect of increasing the scatter in the ionized gas luminosity (and the derived SFR) towards lower luminosities as stochastic sampling becomes increasingly important. PAHs are heated by non-ionizing UV photons \cite[e.g.][]{2021ApJ...917....3D}, which are relatively less affected by stochastic sampling. \cite{2011ApJ...741L..26F} determine that stochastic IMF sampling produces an asymmetric effect on the FUV/H$\alpha$ luminosity ratio, such that below about 3000 M$_{\odot}$, the ratio is a factor of ${\sim}$3 or more higher relative to higher masses. The FUV/H$\alpha$ ratio is a rough proxy for our ratio of PAH luminosity to SFR derived from Pa$\alpha$, and thus, we expect the ratio to be increasingly higher towards the lowest luminosities/masses, where our sources are the most affected by stochastic IMF sampling. This has the effect of flattening the relationship between the PAH and ionized gas luminosity (or derived SFR) at low luminosity. Assuming the limit for a 3000 M$_{\odot}$, 4 Myr model star cluster presented in Section \hyperlink{4}{4}, we find that about half the sources in our full eYSC--I catalogs are in the regime where stochastic IMF sampling is important. Therefore, the generally lower power-law exponents determined for the full source catalogs (Figure \ref{fig:f4}) compared with the sources above the additional luminosity cut (Figure \ref{fig:f5}, top panels), along with the much higher scatter at the lowest luminosities, are likely a result of the impact of stochastic IMF sampling. However, the additional luminosity cut that we apply should mitigate these effects. We do not expect stochastic sampling of the stellar IMF to play a significant role in driving the observed trends or scatter shown in Figure \ref{fig:f5}, or for the new PAH calibrations given in Equations \ref{eq:1} through \ref{eq:8}. See \cite{2025ApJ...992...96P} for a discussion of additional effects of stochastic IMF sampling on the NIR SED.      

\hypertarget{5.1.1.2}{\paragraph{PAH destruction}}

PAHs are relatively small and fragile dust grains that are well-known to be destroyed in harsh environments, such as in the vicinity of an AGN. This effect has been extensively studied in past works, which conclude that the relative abundance of PAHs decreases in both harder \citep[e.g.][]{2006A&A...446..877M,2014MNRAS.444..757K,2018MNRAS.481.5370M} and more intense radiation fields \citep[e.g.][]{2017ApJ...837..157S,2018ApJ...864..136B}. A more recent study by \cite{2023ApJ...944L..16E} shows evidence for an anti-correlation between the PAH fraction and ionization parameter within HII regions in a sample of nearby galaxies using JWST/MIRI imaging. As a result, we expect that in environments with more intense and harsh radiation fields, i.e. towards high surface densities of ionized gas luminosity or SFR, we may observe a relative decrease in the PAH luminosity as a larger fraction of the grains are disrupted or destroyed. This can have the effect of driving a sub-linear relation between the PAH luminosity and SFR as the highest luminosity sources would exhibit the largest deficit in the PAH emission. The luminosity range where PAH destruction becomes important remains unclear and may depend on the ISM conditions; in high metallicity environments, the shielding provided by an abundance of large dust grains may be more effective at preventing PAH destruction relative to lower metallicity environments. In this study, we find no evidence of a turnover in the PAH calibrations towards high luminosity in high metallicity environments that could indicate the transition as shielding is overcome and PAH destruction becomes efficient (Figure \ref{fig:f5}). Rather, we determine that if PAH destruction plays a dominant role in driving the observed sub-linear trends, then it does so over a broad range of luminosity.  

\hypertarget{5.1.1.3}{\paragraph{Age variations}}

Age variations may also play an important role in the PAH calibrations presented in this work. Standard theoretical stellar population synthesis models show that the ionized gas luminosity in HII regions depends on both the mass and age of the associated young star cluster. For example, \cite{2021ApJ...909..121M} show the expectations for the Yggdrasil \citep{2011ApJ...740...13Z} stellar population models for the relation between the equivalent width of H$\alpha$ and Pa$\beta$ and the age of the associated cluster, along with results combining HST observations with SED fitting. They determine that both the models and observations show a sharp decrease in the equivalent width of ionized gas luminosity with increasing age, reaching near zero by ${\sim}$7 Myr. This has also been studied in previous works by investigating the H$\alpha$ morphology. Typically, more centrally concentrated ionized gas morphologies are found to be associated with younger clusters \citep[e.g.][]{2011ApJ...729...78W,2020ApJ...889..154W,2022MNRAS.512.1294H}. The eYSC--I sources investigated in this study are generally young and in the process of emerging, however, they exhibit a range of ages between about 1-7 Myr, based on SED fitting (see \citealt{2025arXiv250508874K}, \citealt{2025ApJ...992...96P}, A. Adamo et al. in prep., and S. T. Linden et al. in prep.). These variations in the age of our sources likely have the effect of increasing the observed scatter for the PAH calibrations since at fixed cluster mass, the ionized gas luminosity decreases with age. PAH emission is also expected to decrease with age; yet, PAHs are heated by non-ionizing UV photons and may be relatively less affected by age differences of a few Myr. We expect that older sources, residing at generally lower luminosities, may exhibit higher ratios of PAH to ionized gas luminosity. This effect can flatten the relation between the PAH luminosity and the derived SFR. 

The bottom panels of Figure \ref{fig:f8} show the relation between the ratio of PAH luminosity to SFR and the EW of Pa$\alpha$. We find evidence for a decreasing relation with a power-law exponent between about $-$0.1 and $-$0.4, depending on the galaxy and PAH feature, although the scatter is quite significant. Note that the power-law exponents are determined to be significantly higher than the value expected for a simple 1/x versus x relation (e.g. $-$1). These results suggest a typically higher PAH to ionized gas luminosity ratio at low Pa$\alpha$ EW, or for older sources, as the EW of the ionized gas is a direct tracer of the age of the associated star cluster, which is consistent with our expectations. This implies that the observed sub-linear trend may be driven in part by variations in the age of our sources. However, the EW of the ionized gas emission also traces the luminosity, and the trends observed in the bottom panels of Figure \ref{fig:f8} could also be a result of a decreasing PAH to ionized gas luminosity ratio towards high luminosity, i.e. an effect from PAH destruction. We are unable to disentangle these two effects, which requires carefully controlling for the age, and as a result, both may have an important role in driving the observed sub-linear trend. 

\hypertarget{5.1.1.4}{\paragraph{PAH heating}}

Variations in the heating of the PAHs may also be an important consideration, mainly in terms of increasing the scatter. In dense regions within the galaxies, the local PAH grains may be heated by multiple nearby UV-emitting young star clusters. This can contribute to the scatter in the PAH calibrations, especially at high luminosity, as sources in dense regions may exhibit elevated PAH emission relative to the ionized gas. However, the ionized gas may also be excited by multiple local sources of ionizing radiation in these regions, and therefore, the effect will only be important in the case where the non-ionizing and ionizing UV photons leak into neighboring regions at different rates. 

The heating of PAH grains by older, non-local sources of UV photons can have an important impact. Given the close proximity of our sources to a bright eYSC and the fact that our measurements are local background subtracted, we expect that this effect is mostly mitigated in this study. However, the density of PAHs in the ISM is important to consider. The PAH density may be higher in the PDRs than in the more diffuse ISM; for instance, if the PAHs are compressed by the expansion of the HII region. The bright PAH emission in the PDRs could arise from an increased density rather than from a higher intensity of UV photons in these regions. In this case, our sources could be affected by the heating of PAHs by the general UV radiation field of the galaxy, which would have the effect of lowering the slope of the PAH-SFR calibrations. We find that this scenario is unlikely as we would expect our calibrations to be consistent with those derived on larger scales, contrary to our results, which show large offsets (see Section \hyperlink{5.3}{5.3}).

\hypertarget{5.1.1.5}{\paragraph{Variations in the ISM metallicity}}

The local ISM metallicity can also have a large impact on the observed emission from PAHs. However, previous results with the 7.7~$\mu$m PAH feature suggest that the PAH emission depends strongly on metallicity only at relatively low oxygen abundance, 12+log(O/H)$\,{\lesssim}\,$8.3 \citep[e.g.][]{2005ApJ...628L..29E,2007ApJ...666..870C}. Recent studies have found that the PAH mass fraction drops abruptly at 12+log(O/H)$\,{\sim}\,$8.0 \citep{2020ApJ...889..150A}, or alternatively, at 12+log(O/H)$\,{\sim}\,$8.5 and subsequently rapidly decreases with decreasing metallicity \citep{2024ApJ...974...20W}. In this study, we determine that the three high metallicity spiral galaxies (NCC 5194, 5236, and 628) show best-fit relations between the PAH luminosities and SFR that are fairly consistent, although, NGC 628 exhibits a shift down by ${\sim}$0.1 dex (Figure \ref{fig:f5}, top panels). This slight shift down for NGC 628 may be consistent with the slightly lower values of nebular oxygen abundance on average for the sources in this galaxy, corresponding to slightly sub-solar (Figure \ref{fig:f3}). This suggests that variations in the local ISM metallicity may play a minor role in increasing the scatter observed in the PAH calibrations, even at relatively high metallicity. See Section \hyperlink{5.1.2}{5.1.2} for a discussion of the PAH emission in much lower metallicity environments, like NGC 4449.   

\hypertarget{5.1.1.6}{\paragraph{Other considerations}}

The distribution or geometry of the dust and PAHs relative to the ionized gas and stars is another key consideration here. New studies with JWST are beginning to reveal the complexity of the relative distribution and morphology of the stars, gas, and dust in HII regions, like the Orion nebula \cite[e.g.][]{2024A&A...685A..75C,2024A&A...685A..73H,2024A&A...685A..77P,2024A&A...685A..74P, 2024A&A...687A..86V}. Given the selection methods for our eYSC--I sources, we expect at most a minor impact from variations due to differences in the relative geometries, and that viewing orientation should not be a major factor. However, eYSCs tend to be located in the dustier regions of the disk (i.e. the spiral arms), and thus, could be located behind dust lanes that are not associated with the star clusters, but rather are simply along the line of sight. This could affect some of our results, particularly the derivation of E(B-V).

The star formation process is not instantaneous, and as a result, variations in the duration of star formation, i.e. the specific star formation history, for different regions may also play a role, leading to variations in the non-ionizing UV (or PAH) to ionized gas luminosity ratio. Other possible sources of scatter and uncertainty include the continuum subtraction techniques, contamination in the final source catalogs and/or measurements, and overlapping measurements; yet, these are not expected to play a dominant role. In Appendix \hyperlink{B}{B}, we explicitly test the impact of the overlapping measurements and find an insignificant effect on our main results. 

\hypertarget{5.1.2}{\subsubsection{Metal-poor environments}}

In the metal-poor environment of the irregular dwarf galaxy NGC 4449 (12+log(O/H)$\,{\sim}\,$8.2; Figure \ref{fig:f3}), we observe a substantial deficit in the PAH emission features at fixed luminosity, corresponding to about 0.4 and 0.6 dex for the 3.3 and 7.7~$\mu$m features, respectively, at log($\Sigma_{SFR}\,/$ M$_{\odot}$ yr$^{-1}$ pc$^{-2}$)$\,=-6.5$ (Figure \ref{fig:f5}, bottom panels). In addition, we determine that the scatter in the relation between the PAH emission and SFR is significantly higher in these environments, corresponding to about 0.19 and 0.18 dex on average for the 3.3 and 7.7~$\mu$m features, respectively, compared to near solar metallicity environments at about 0.11 and 0.12 dex (Table \hyperlink{t2}{2}). 

These results are consistent with a vast literature of previous studies that investigates the PAH--metallicity relation, finding large deficits in PAH luminosity in low metallicity environments \citep[e.g.][]{2005ApJ...628L..29E,2006A&A...446..877M,2007ApJ...666..870C,2007ApJ...656..770S,2008ApJ...682..336G,2014MNRAS.445..899C,2017ApJ...837..157S,2022ApJ...928..120G,2024A&A...690A..89S}. One common explanation for this deficit is a decreased PAH abundance resulting from the destruction of PAH grains by the harder, or higher average photon energy, radiation in low metallicity environments \citep[e.g.][]{2006A&A...446..877M,2008ApJ...682..336G,2023ApJ...944L..16E}. Metal-poor environments also exhibit generally lower amounts of dust in the ISM \citep{2005ApJ...627..477M} available to shield the PAHs from the harder photons via attenuation. It has also been argued that metals play a key role in the ISM for facilitating the formation and growth of PAH grains, suggesting that the deficit at low metallicity arises from the fact that the grains are not formed in the first place, or from the smaller PAH sizes on average leading to higher rates of destruction \citep{2012ApJ...744...20S}. New observational/theoretical studies suggest that the inhibited PAH grain growth at low metallicity is the dominant driver of the observed PAH deficit, rather than purely a result from PAH destruction \citep[e.g.][]{2024ApJ...974...20W,2025ApJS..280....4Z}. The increased scatter in the PAH calibrations that we observe in low metallicity environments may be a consequence of these processes. At the significantly sub-solar ISM metallicities of our sources in NGC 4449, the emission from the PAH grains depends sensitively on the local ISM conditions due to variations in the stability, survival, growth, and heating of the PAHs, observed as an overall increase in the scatter with tracers of ionized gas.

We find that the magnitude of the PAH deficit in low metallicity environments is substantially larger for the 7.7~$\mu$m feature, compared with the 3.3~$\mu$m. In addition, we find that the best-fit power-law exponent of the relation between the PAH emission and SFR is ${\sim}$2$\sigma$ lower for the metal-poor environments of NGC 4449 compared to near solar metallicities for the 3.3~$\mu$m feature, while the 7.7~$\mu$m feature shows little to no evidence of a lower power-law exponent at low metallicity (Figure \ref{fig:f5}, bottom panels). However, in metal-poor environments like NGC 4449, the scatter in the calibrations is high, making the determination of the power-law exponent inherently uncertain. These results are discussed further and interpreted in the context of the 3.3/7.7~$\mu$m PAH luminosity ratio in the next section. It is important to note that our sample size is much smaller in these low metallicity environments, and the generalization of the results presented here will require expanding the sample beyond NGC 4449 to additional metal-poor galaxies.

\hypertarget{5.2}{\subsection{The PAH ratio}}

We observe an increase in the 3.3/7.7~$\mu$m PAH luminosity ratio towards low ISM metallicity environments both 1) within each galaxy's regions, where the local ISM metallicity is estimated from the observed radial gradients of nebular oxygen abundance, and 2) on average for sources in the low metallicity dwarf NGC 4449 compared to in the near solar metallicity spirals (Figure \ref{fig:f7}). An important caveat here is that the relation determined for each galaxy shows very high scatter with a correlation coefficient $|\rho | \,{\lesssim}\, 0.3$ in all cases (Figure \ref{fig:f7}, left panel). This scatter is likely a result of the substantial intrinsic scatter of the observed radial abundance gradients that we use to derive the local ISM metallicity, which have variations in the oxygen abundance of up to ${\sim}$0.5 dex for fixed galactocentric radius \citep[e.g.][]{2020ApJ...893...96B}. More accurate determinations of the local ISM metallicity for our sources would greatly help reveal the underlying trend here, which may be substantially stronger than we derive in this work. 

Comparing our observed 3.3/7.7~$\mu$m PAH ratios with expectations from the \cite{2021ApJ...917....3D} models, we determine that the models that consider the total dust emission and a standard PAH size distribution and ionization predict higher values of the 3.3/7.7~$\mu$m PAH ratio and are generally inconsistent with our observations (Figure \ref{fig:f7}, right panel). This is consistent with the results of the study by \cite{2024ApJ...971...32P} \citep[also see the erratum,][]{2024ApJ...973...67P}. The overprediction of the 3.3/7.7~$\mu$m PAH ratio for the models is known and discussed in \cite{2021ApJ...917....3D}, attributed mainly to the neglect of important energy loss channels, like photoelectric emission, photodesorption, and fluorescence \citep{1989ApJS...71..733A}, which likely results in an overestimation of the 3.3~$\mu$m feature. The \cite{2021ApJ...917....3D} models assume that the full photon energy absorbed by the grains is converted into vibrational excitation, which is then removed only by infrared emission. Therefore, these results are to be expected, and the consistency between the models and observations may be improved with future and more complete modeling, specifically for the 3.3~$\mu$m feature which remains notably difficult to model accurately.  

The \cite{2021ApJ...917....3D} models predict almost no increase in the 3.3/7.7~$\mu$m PAH ratio when the stellar spectrum that heats the PAHs changes from a near solar metallicity population to extremely sub-solar (Figure \ref{fig:f7}, right panel). Yet, we observe a significantly higher 3.3/7.7~$\mu$m PAH ratio on average for star-forming sources in NGC 4449 compared to the higher metallicity spirals, corresponding to ${\sim}$0.2 dex (Figure \ref{fig:f7}, right panel). These results suggest that the observed increase in the PAH ratio towards low metallicity is not a result of the harder incident radiation field heating the PAHs to higher average temperatures in lower metallicity environments, but rather another physical mechanism.
 
The destruction of PAH grains is expected to depend on the grain size such that the smallest PAHs (traced by the shortest wavelength features) will dissociate at higher rates due to a lower heat capacity \citep[e.g.][]{1989A&A...216..148L,2021ApJ...917....3D}. Naively, this suggests that we may expect a decrease in the 3.3/7.7~$\mu$m PAH ratio at low metallicity as the 3.3~$\mu$m emission traces the smallest PAH grains. However, this is contrary to our observations, which show a strong increase in the ratio towards lower metallicity environments. This increase in the 3.3/7.7~$\mu$m ratio at low metallicity can be a result of either 1) a smaller PAH size distribution, or 2) a larger fraction of neutral PAHs. A number of recent observational studies that do not suffer from the same limitations as our study (i.e. use ratios of PAH features that are sensitive only to the PAH size distribution) find evidence for an increased fractional contribution from smaller PAHs in metal-poor environments, suggesting a shift in the size distribution towards smaller PAHs in these environments \citep[e.g.][]{2012ApJ...744...20S,2020ApJ...905...55L,2024ApJ...974...20W,2025ApJ...991L..56L,2025ApJS..280....4Z}. The study by \cite{2025ApJS..280....4Z} finds evidence for a PAH population with a similar or slightly lower ionization fraction in the low metallicity environment of 30 Doradus in the Large Magellanic Cloud compared to the Orion Bar in the Milky Way. In our case, the increased 3.3/7.7~$\mu$m ratio towards low metallicity is more likely a result of a smaller PAH size distribution, but this remains unclear.

There are several mechanisms that can explain a shift in the size distribution towards smaller PAHs in metal-poor environments. One mechanism, possibly the most likely or dominant, is that the formation and growth of PAH grains is inhibited in low metallicity environments, leading to a smaller size distribution \citep{2012ApJ...744...20S,2024ApJ...974...20W}. Another possible mechanism is that large PAH grains can be shattered into smaller PAHs \citep[e.g.][]{1996A&A...305..602A}, replenishing a fraction of the small grains that are destroyed. There are also mechanisms by which the smallest PAH grains can become relatively more stable against dissociation and destruction in low metallicity environments, namely recurrent fluorescence \citep[see][]{1988PhRvL..60..921L,2017MNRAS.469.4933L,2020Ap&SS.365...58W}, which is a fast radiative cooling process with a relaxation time of the order of milliseconds that is more efficient in smaller grains \citep{2017NIMPB.408...21B,2020ApJ...905...55L}. Alternatively, large PAH grains can emit more like small grains in enhanced radiation fields as the grains reach higher vibrational energy levels, resulting in shorter wavelength emission \citep{2007ApJ...657..810D,2020ApJ...905...55L,2021ApJ...917....3D,2023ApJ...957L..26L}. For this study, we interpret the observed increase in the 3.3/7.7~$\mu$m PAH ratio towards low metallicity as a result of a smaller PAH size distribution in these environments, possibly due to inhibited grain growth. This interpretation can also explain the larger PAH deficit in low metallicity environments for the 7.7~$\mu$m feature compared to the 3.3~$\mu$m (Figure \ref{fig:f5}, bottom panels), as the PAH emission shifts to shorter wavelength features as the metallicity decreases.

Our results also suggest that in low metallicity environments, like for the sources in NGC 4449, the 3.3/7.7~$\mu$m PAH ratio decreases towards higher surface densities of SFR or ionized gas luminosity, while at higher (near solar) metallicities, the ratio remains relatively constant (Figure \ref{fig:f6}). This suggests that in metal-poor (significantly sub-solar) environments, small PAH grains may be preferentially destroyed relative to larger PAHs. This would have the effect of lowering the 3.3/7.7~$\mu$m PAH ratio towards higher intensity ionizing environments traced by high ionized gas luminosity, where destruction has the largest impact. This effect has typically been observed in the past in the harsh regions surrounding AGN \citep[e.g.][]{2023ApJ...957L..26L}, and the hard radiation fields at low metallicity may have a similar effect. Recently, it has also been observed around star clusters by the study of \cite{2025AJ....169..133D}, finding that the smallest PAHs are increasingly destroyed in more intense and harsh environments. Yet, in our case, PAH ionization also has an impact on the 3.3/7.7~$\mu$m PAH ratio, such that more highly ionized PAHs correspond to lower 3.3/7.7~$\mu$m ratios \citep{2021ApJ...917....3D}. This could also explain the trend as PAHs are expected to be more highly ionized (exhibit lower 3.3/7.7~$\mu$m PAH ratios) towards higher intensity ionizing environments. 

In addition to this effect, we see evidence of a slightly lower power-law exponent for the relation between the PAH emission and SFR for the 3.3~$\mu$m feature ($\alpha \, {\sim} \, 0.66$) compared to the 7.7~$\mu$m feature ($\alpha \, {\sim} \, 0.78$) in low metallicity environments (Figure \ref{fig:f5}, bottom panels). This could also indicate the preferential destruction of the smallest PAH grains in these environments, as the shorter wavelength PAH emission would exhibit a larger deficit with tracers of ionized gas at high luminosity. This result is fairly uncertain though given the high scatter in the relations at low metallicity, corresponding to a ${\sim}$1$\sigma$ difference in the power-law exponent. Future studies that include additional PAH features and expand the environments beyond the one metal-poor target investigated in this work will be required to properly understand and interpret these results.  

We determine that in both low and high metallicity environments, the 3.3/7.7~$\mu$m PAH ratio exhibits almost no correlation with the EW of Pa$\alpha$, with a correlation coefficient $|\rho| \, <\, 0.1$ for the sources in each of our galaxies (Figure \ref{fig:f8}, top left panel). A similar result was found by the study of \cite{2024ApJ...971...32P} for regions within NGC 628, where the age of the associated eYSC was determined via SED fitting. These trends, or lack thereof, suggest that the 3.3/7.7~$\mu$m PAH ratio does not vary strongly across the initial stages of the emerging phase of star formation, specifically over the few Myr timescale where the PAH emission remains cospatial with the ionized gas and the local newly formed star cluster. This could suggest that the properties of the PAH grains remain mostly unchanged over this short timescale, however, it is unclear given the high scatter and the various factors/properties that can affect the 3.3/7.7~$\mu$m PAH ratio. A more controlled investigation (e.g. constant metallicity and mass) is likely essential here to accurately determine the underlying effects of aging, along with extra PAH features (e.g. 6.2, 11.3, 17~$\mu$m) from, e.g. spectroscopy, that give a better handle on the PAH properties.

\hypertarget{5.3}{\subsection{The impact of PAH heating by older stellar populations}}

PAHs are heated in ISM of galaxies by non-ionizing UV photons \citep[e.g.][]{2021ApJ...917....3D}, and thus, not only by bright, newly formed stars and star clusters, but also by older ($>$10 Myr), UV-bright stellar populations. The high intensity of UV radiation in the PDRs around massive star clusters and HII regions results in the efficient heating of PAHs in these regions and very bright, localized PAH emission. This results in a close correspondence between maps of ionized gas and PAH emission on highly resolved scales in galaxies, particularly for the brightest peaks of emission (Figure \ref{fig:f2}). However, diffuse PAH emission is also apparent far outside the boundaries of HII regions (Figure \ref{fig:f2}). A fraction of this diffuse emission may arise as a result of PAH heating from non-ionizing UV photons emitted by nearby HII regions, yet, a fraction also likely arises from heating by the general UV radiation field of the galaxy, and therefore, likely does not trace recent ($<$10 Myr) star formation. Previous studies have found that the majority (about 50 to 80$\%$) of the PAH emission at 7.7~$\mu$m is associated with the diffuse interstellar radiation field \citep[e.g.][]{2014ApJ...784..130C,2023ApJ...944L...9L}. The diffuse PAH emission component is effectively removed in this study by subtracting the local background around individual, or at most a few, HII regions/PDRs. We expect that the PAH emission investigated in this study is dominated by the component that is related to star formation, given the close proximity ($\leq$ 20 pc) to the bright, young heating source (i.e. the nearby eYSC) and the removal of the local background outside the PDRs. Similar studies in nearby galaxies with Spitzer \citep[e.g.][]{2007ApJ...666..870C} are relatively more affected by the diffuse PAH contribution and the heating by older stellar populations, as the scales probed are much larger. Studies on the integrated scale of whole galaxies \citep[e.g.][]{2009ApJ...703.1672K} typically do not, or are unable to, properly account for the PAH emission component that is unrelated to star formation. 

\begin{figure}
\centering
\includegraphics[width=0.47\textwidth]{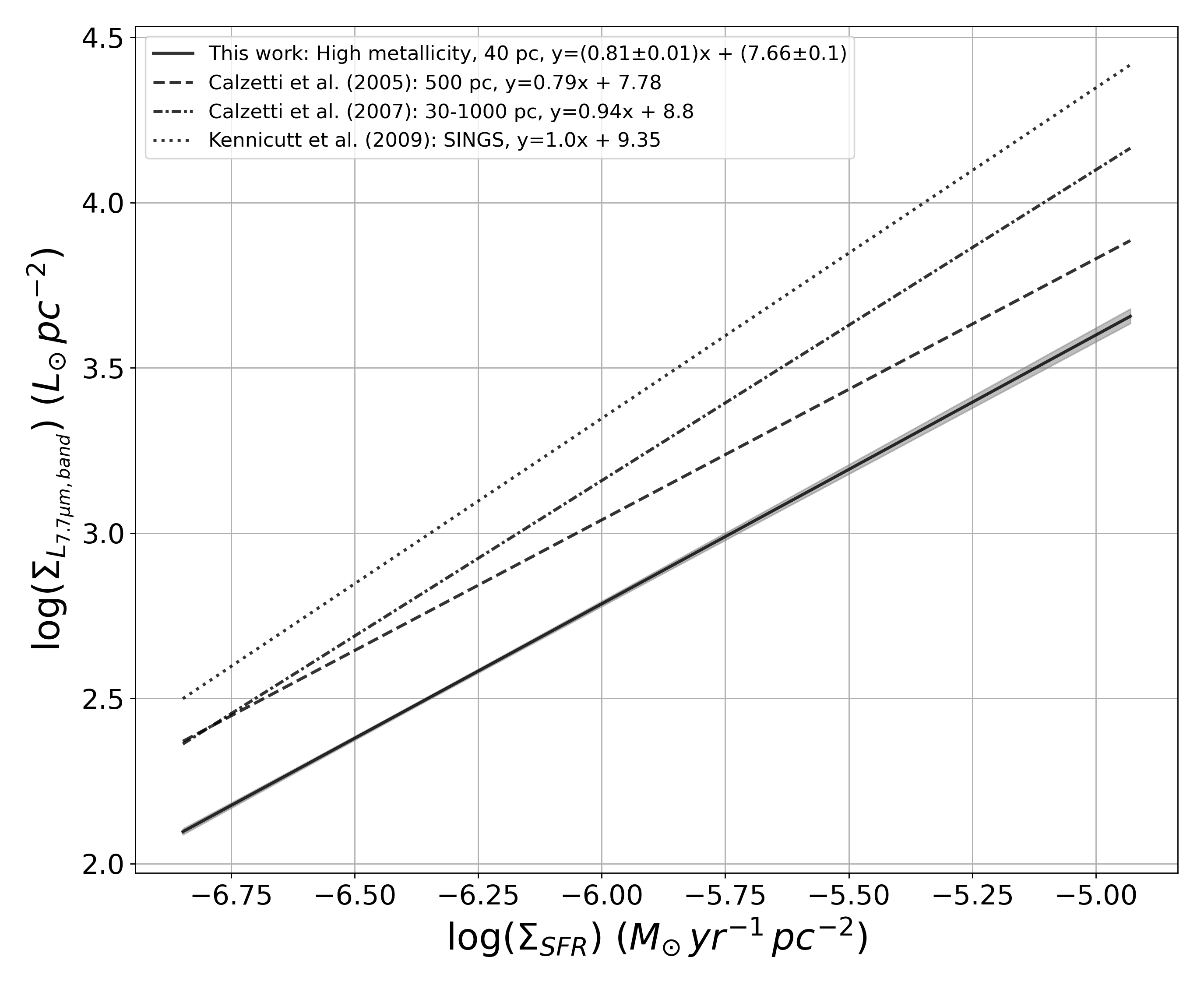}
\caption{The 7.7~$\mu$m PAH luminosity surface density as a function of SFR surface density. The solid black line shows the calibration determined by this study for the high metallicity bin on the 40 pc scale. Here, we use the in-band assumption for calculating the 7.7~$\mu$m luminosity (i.e. multiplying the flux density by the pivot wavelength of the filter rather than the bandwidth), similar to previous studies with Spitzer. The dashed line shows the relation from \cite{2005ApJ...633..871C} on the scale of ${\sim}500$ pc in NGC 5194. The dash-dotted line shows the \cite{2007ApJ...666..870C} relation, corresponding to scales between about 30 pc and 1 kpc in a sample of SINGS galaxies. The dotted line shows the relation determined by \cite{2009ApJ...703.1672K} on the galaxy-integrated scale for the SINGS sample. }
\label{fig:f9}
\end{figure}

Figure \ref{fig:f9} shows a comparison between the 7.7~$\mu$m PAH calibration determined in this study with JWST for the high metallicity bin on the scale of 40 pc and previous calibrations determined in nearby galaxies with Spitzer. Specifically, we compare to relations derived by 1) \cite{2005ApJ...633..871C} on the scale of ${\sim}500$ pc in NGC 5194, 2) \cite{2007ApJ...666..870C} on scales between about 30 pc and 1 kpc in SINGS galaxies, and 3) \cite{2009ApJ...703.1672K} on the galaxy-integrated scale for the SINGS sample, where in each case the calibrations have been converted to the same units used in this study. For this comparison, we use the in-band assumption for deriving the 7.7~$\mu$m luminosity, or in other words, we multiply the flux density by the pivot wavelength of the filter rather than the bandwidth, similar to the previous studies with Spitzer. Note that the previous studies have typically lower surface densities than we observe in this study as they are averaged over larger regions, but here we show them over the range that we observe. We find a clear trend where the calibrations derived on increasingly larger scales exhibit higher surface densities of 7.7~$\mu$m PAH luminosity at fixed SFR surface density (Figure \ref{fig:f9}). At log($\Sigma_{SFR}\,/$ M$_{\odot}$ yr$^{-1}$ pc$^{-2}$)$\,=-6.5$, the \cite{2009ApJ...703.1672K} relation exhibits about 0.5 dex (factor of ${\sim}$3) higher 7.7~$\mu$m PAH luminosity surface density compared to the calibration determined by this study, while the \cite{2005ApJ...633..871C} relation is about 0.3 dex higher (Figure \ref{fig:f9}).

We attribute the significant offsets in the calibrations mainly to the difference in the physical scales probed and the increasing contribution from PAH heating by older stellar populations on larger scales. The difference between the emission probed by the JWST/MIRI/F770W and the Spitzer/IRAC channel 4 (${\sim}$8~$\mu$m) filters could contribute to the offsets given that the bandwidths are slightly different (about 2 and 3~$\mu$m, respectively), however, both are centered on the bright 7.7~$\mu$m feature and the difference is likely no more than a few percent. There are also slight differences in the continuum subtraction of the 7.7~$\mu$m emission as we remove both components of the continuum, stellar and dust, using the F560W, while the studies with Spitzer remove only the stellar continuum via scaling the IRAC 3.6~$\mu$m. This could contribute to the observed offsets, yet in all cases, the 7.7~$\mu$m feature dominates the continuum and thus this is likely a minor effect. Our study is the most affected by the leakage of ionizing photons or by direct absorption of ionizing photons by dust. However, this would only have the effect of increasing the offsets, as in our case, the intrinsic SFRs could be a factor of about 2 higher than shown in Figure \ref{fig:f9}, while studies on larger scales will be less affected. 

These results suggest that on galaxy-integrated scales over half (${\sim}\,$2/3) of the 7.7~$\mu$m PAH emission is unrelated to recent ($<$10 Myr) star formation and rather is excited by older stellar populations (Figure \ref{fig:f9}). On ${\sim}500$ pc scales, up to about half of the PAH emission may be unassociated with newly formed stars (Figure \ref{fig:f9}). This also suggests that the majority of the diffuse PAH emission component in galaxies is related to the heating by older stars rather than from photons emitted by local massive star-forming regions. The study by \cite{2014ApJ...797..129L} determines that in the nearby spiral galaxy M81 about 67$\%$ of the integrated PAH emission at 7.7~$\mu$m is associated with the heating by more evolved stellar populations. This is consistent with the fraction estimated in this study (${\sim}\,$2/3) via the PAH calibrations and suggests that the majority of the integrated PAH emission in local, high metallicity star-forming galaxies arises as a result of processes that are unconnected to new star formation. This casts considerable doubt on the ability of the PAH emission to accurately trace star formation on large scales in galaxies under normal star-forming conditions. 

\hypertarget{6}{\section{Conclusions}}

In this study, we utilize new JWST/NIRCam and MIRI imaging obtained by the FEAST program along with archival HST imaging to map both ionized gas (Pa$\alpha$, Br$\alpha$, and H$\alpha$) and PAH emission (3.3 and 7.7~$\mu$m) across a sample of four nearby galaxies and investigate the ability of the PAH emission to trace recent star formation. We discuss methods of continuum subtraction with the JWST imaging bands, the selection of eYSCs, the measurement of PAH, ionized gas, and physical properties, the calibration of the 3.3 and 7.7~$\mu$m emission as SFR indicators, the 3.3/7.7~$\mu$m PAH ratios, variations with ISM metallicity, and more. Our major findings are the following: 

\begin{itemize}
\item 
We observe a moderately tight (correlation coefficient, $\rho \, {\sim}\, 0.8$), sub-linear (power-law exponent, $\alpha \, {\sim}\, 0.8$) relation between the PAH luminosities (3.3 and 7.7~$\mu$m) and SFR traced by ionized gas emission (extinction corrected Pa$\alpha$) in high metallicity (about solar) environments on small scales (40 pc) around candidate eYSCs (eYSC--I; compact, cospatial peaks in both ionized gas and PAH emission) that have sufficient mass to be mostly unaffected by stochastic stellar IMF sampling (Figure \ref{fig:f5}, bottom panels). We use these results to derive new SFR calibrations from both the 3.3 and 7.7~$\mu$m PAH luminosity surface densities, given in Equations \ref{eq:1} through \ref{eq:8}. The calibration coefficients that we derive may be affected by the leakage of UV photons or direct absorption of UV photons by dust. 

\item 
The sub-linear relation that we observe between the strength of the PAH emission features and SFR is most likely explained by a combination of the destruction of PAHs in more intense ionizing environments and/or variations in the age of our sources. 

\item 
For metal-poor environments, traced by our sources in the dwarf galaxy NGC 4449 (12+log(O/H)$\,{\sim}\,$8.2), we find significant deficits in the PAH emission features at fixed luminosity. The PAH deficits at log($\Sigma_{SFR}\,/$ M$_{\odot}$ yr$^{-1}$ pc$^{-2}$)$\,=-6.5$ correspond to about 0.4 and 0.6 dex for the 3.3 and 7.7~$\mu$m features, respectively (Figure \ref{fig:f5}, bottom panels). The scatter in the relation between the PAH emission and SFR is also substantially higher in these environments, corresponding to about 0.19 and 0.18 dex for the 3.3 and 7.7~$\mu$m features, respectively, compared to about 0.11 and 0.12 dex for near solar metallicity environments (Table \hyperlink{t2}{2}). These results are likely explained by a decreased abundance of PAHs in the ISM at low metallicity, possibly resulting from inhibited PAH grain formation and growth and/or PAH destruction \citep[e.g.][]{2012ApJ...744...20S,2024ApJ...974...20W}.

\item 
The 3.3/7.7~$\mu$m PAH luminosity ratio increases towards low ISM metallicity environments. We observe higher ratios towards low metallicity both 1) within each galaxy's regions (high scatter resulting from the uncertain radial abundance gradients) and 2) on average for sources in NGC 4449 compared to in the higher metallicity spirals (Figure \ref{fig:f7}). We suggest that this is likely due to a shift in the size distribution towards smaller PAHs in low metallicity environments \citep[e.g.][]{2012ApJ...744...20S,2020ApJ...905...55L,2024ApJ...974...20W,2025ApJ...991L..56L,2025ApJS..280....4Z}, possibly due to the inhibited formation and growth of PAH grains. Our observed PAH ratios are generally inconsistent with the higher values expected from standard \cite{2021ApJ...917....3D} models. 

\item 
We find that in the low metallicity environments of NGC 4449, the 3.3/7.7~$\mu$m PAH ratio decreases towards higher surface densities of SFR, while in near solar metallicity environments, the ratio remains relatively constant (Figure \ref{fig:f6}). This in combination with a lower (marginally) power-law exponent for the PAH-SFR relation for the 3.3~$\mu$m feature compared to the 7.7~$\mu$m for sources in NGC 4449 (Figure \ref{fig:f5}, bottom panels) suggests that in significantly sub-solar metallicity environments, small PAH grains may be preferentially destroyed relative to larger PAHs. However, increased PAH ionization at high SFR can also explain the lower PAH ratio.  

\item 
In both low and high metallicity environments, the 3.3/7.7~$\mu$m PAH ratio exhibits little to no correlation with the Pa$\alpha$ equivalent width (Figure \ref{fig:f8}, top left panel). This suggests that over the few Myr timescale where PAH emission is cospatial with ionized gas and the local newly formed star cluster, aging may not be a dominant source of variation in the 3.3/7.7~$\mu$m PAH ratio, compared to the ISM metallicity and the intensity of the ionizing radiation field in harsh environments. 

\item 
Comparing our PAH-SFR calibrations with previous results determined on larger scales in nearby galaxies with Spitzer, we observe a clear trend where calibrations derived on increasingly larger scales show higher surface densities of 7.7~$\mu$m PAH luminosity at fixed SFR surface density (Figure \ref{fig:f9}). On galaxy-integrated scales, the PAH luminosity surface density at log($\Sigma_{SFR}\,/$ M$_{\odot}$ yr$^{-1}$ pc$^{-2}$)$\,=-6.5$ is about 0.5 dex higher than we derive in this study (40 pc scales), while for ${\sim}500$ pc scales, it is about 0.3 dex higher (Figure \ref{fig:f9}). These results suggest that about 2/3 (1/2) of the PAH emission on galaxy-integrated (${\sim}500$ pc) scales is unrelated to recent ($<$10 Myr) star formation, i.e. is excited by older stellar populations, in typical local star-forming environments. 
\end{itemize}

Our study shows the capability of the 3.3 and 7.7~$\mu$m PAH emission features observed by JWST/NIRCam and MIRI to trace recent star formation in nearby galaxy environments on the fundamental scale of star clusters and HII regions. However, there are various complications and uncertainties. The sub-linear ($\alpha{\sim}0.8$) PAH-SFR relation determined on this scale implies a significant impact from secondary processes, i.e. possibly PAH destruction in more intense radiation fields, even in normal, high metallicity star-forming environments. The ISM metallicity is also a key factor and contributes significant variation to the measured PAH emission, particularly towards low metallicities (e.g. NGC 4449), but even some variation in near solar metallicity environments (e.g. NGC 628). In addition, the majority of the PAH emission in local, normal star-forming galaxies may be excited by older stars and unrelated to recent star formation. All of this suggests that the relation between PAH emission and SFR in galaxies is complex and that caution is necessary when using PAHs as indicators of star formation. Yet, PAH emission remains exceptionally useful at high redshift, where the 3.3~$\mu$m feature can be observed with MIRI out to at least z${\sim}$4.5 and can provide a first-order indication of the dust-obscured SFR, assuming the ISM metallicity is near solar (or otherwise accounted for), the photons that heat the dust are dominated by those emitted from newly formed stars (as may be the case in strongly star-forming high redshift systems), and that non-linear calibrations are used. Additional studies are needed to determine if and how the component of PAH emission that is unrelated to the heating by newly formed stars can be removed on galaxy-integrated scales and to establish the connection between local and high redshift estimates, where the very different star-forming environments may greatly affect the results. 

\section*{Acknowledgments}

The authors thank the anonymous referee for many comments that have helped improve this manuscript. 

This work is based in part on observations made with the NASA/ESA/CSA James Webb Space Telescope (JWST). The data were obtained from the Mikulski Archive for Space Telescopes (MAST) at the Space Telescope Science Institute (STScI), which is operated by the Association of Universities for Research in Astronomy, Inc., under NASA contract NAS 5-03127 for JWST. These observations are associated with program $\#$ 1783. Support for program $\#$ 1783 was provided by NASA through a grant from STScI. The specific observations analyzed can be accessed via \dataset[https://doi.org/10.17909/5m3n-6r51]{https://doi.org/10.17909/5m3n-6r51}. Support to MAST for these data is provided by the NASA Office of Space Science via grant NAG 5–7584 and by other grants and contracts.

The authors acknowledge the team of the `JWST-HST-VLT/MUSE-ALMA Treasury of Star Formation in Nearby Galaxies', led by coPIs Lee, Larson, Leroy, Sandstrom, Schinnerer, and Thilker, for developing the JWST observing program $\#$ 2107 with a zero-exclusive-access period.

This work is also based on observations made with the NASA/ESA Hubble Space Telescope, and obtained from the Hubble Legacy Archive, which is a collaboration between STScI/NASA, the Space Telescope European Coordinating Facility (ST-ECF/ESA), and the Canadian Astronomy Data Centre (CADC/NRC/CSA).

This research has made use of the NASA/IPAC Extragalactic Database (NED) which is operated by the Jet Propulsion Laboratory, California Institute of Technology, under contract with NASA.

BG acknowledges support from JWST GO 1783. AA and AP acknowledge support from the Swedish National Space Agency (SNSA) through the grant 2021-00108. KG is supported by the Australian Research Council through the Discovery Early Career Researcher Award (DECRA) Fellowship (project number DE220100766) funded by the Australian Government and by the Australian Research Council Centre of Excellence for All Sky Astrophysics in 3 Dimensions (ASTRO~3D), through project number CE170100013.


\facilities{JWST/NIRCam, JWST/MIRI, HST/ACS, HST/WFC3}

\software{astropy \citep{astropy:2013,astropy:2018}, photutils \citep{larry_bradley_2019_3568287}, SAOImageDS9 \citep{2003ASPC..295..489J}, SEP \citep{1996A&AS..117..393B,2016JOSS....1...58B}, PyNeb \citep{2015A&A...573A..42L}, LINMIX \citep{2007ApJ...665.1489K}}

\appendix
\vspace{-6.8mm}
\hypertarget{A}{\section{The continuum subtraction of MIRI/F770W}}

Over our full sample of galaxies, we currently have access to only the F560W and F770W images from MIRI to use for the continuum subtraction of F770W, i.e. for deriving the strength of the 7.7~$\mu$m PAH emission feature. The obvious choice for the continuum subtraction is simply to subtract the F560W image from the F770W. However, we expect that the dust continuum rapidly increases in strength across the 5-to-8~$\mu$m regime, which implies that this method could lead to an under-subtraction of the F770W continuum. Scaling up the continuum estimated via this method before subtraction could lead to a more accurate estimation of the 7.7~$\mu$m feature strength. In one of our galaxies, NGC 628, an additional filter, the F1000W, has been observed by the PHANGS--JWST program (ID 2107; PI: J. C. Lee), presented in \cite{2023ApJ...944L..17L}, and we use this image here to test a few different continuum subtraction techniques. 

\begin{figure*}
\centering
\includegraphics[width=0.49\textwidth]{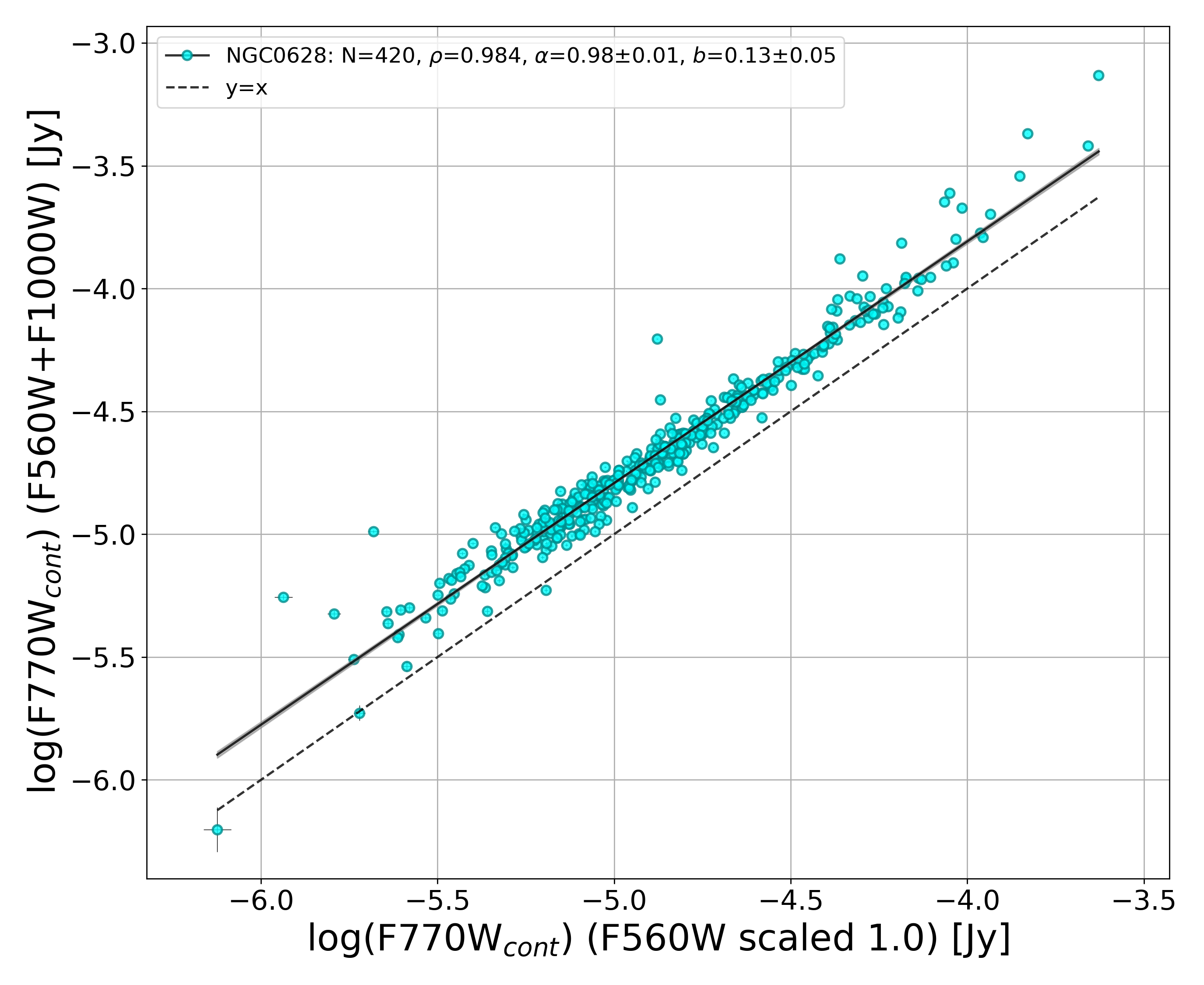}
\includegraphics[width=0.49\textwidth]{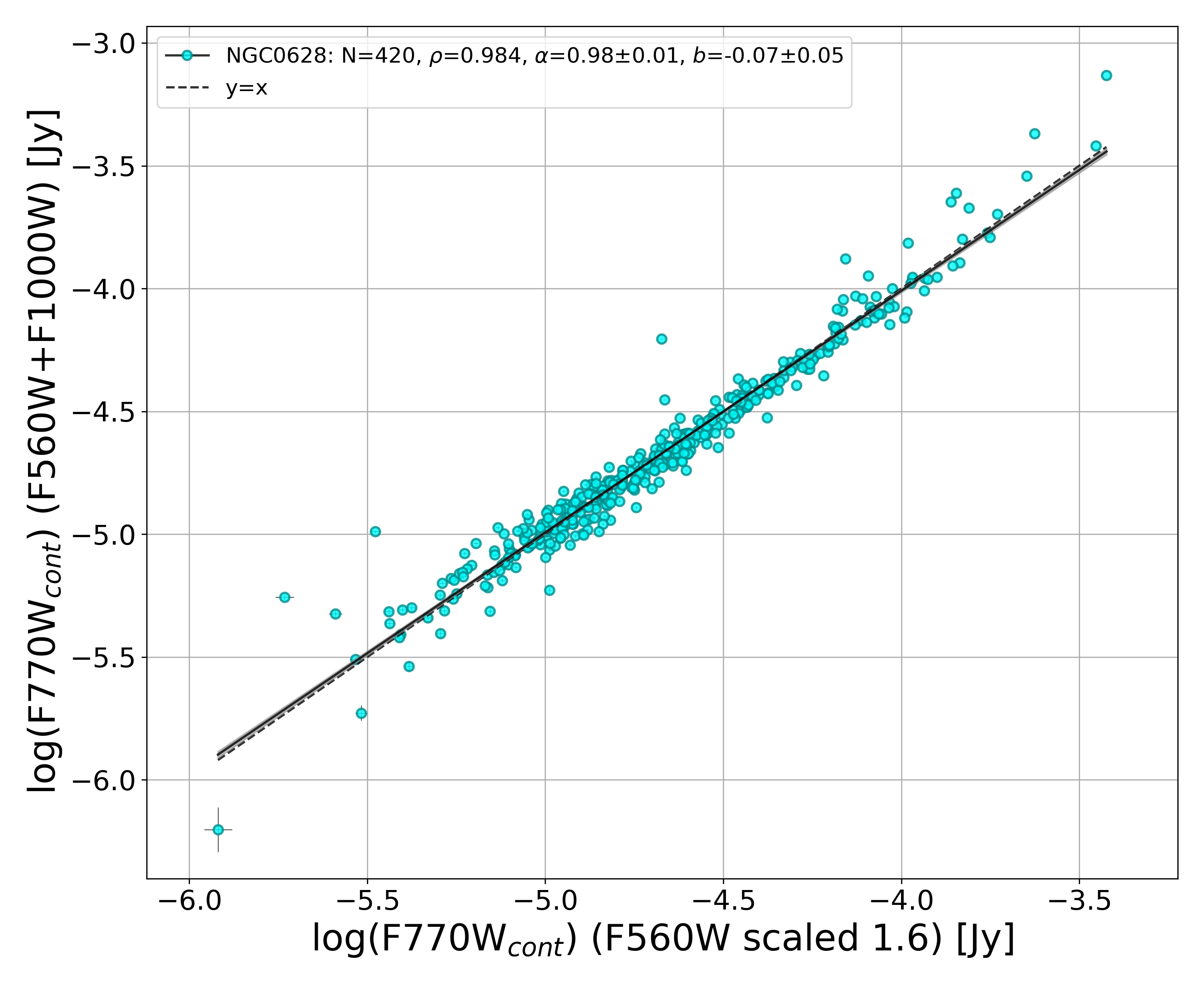}
\caption{Left panel: The continuum in the MIRI/F770W filter determined by linearly interpolating between the F560W and F1000W as a function of the F770W continuum estimated from the unscaled F560W. All images are first matched to the PSF of F1000W. The data points show measurements on the continuum images around eYSC--I sources in NGC 628. The best-fit relation is shown by the solid black line. The black dashed line shows the one-to-one relation or y=x. Right panel: Same as the left, but instead on the x-axis is the F770W continuum estimated from the F560W scaled by a factor of 1.6. In this case, the observed relation is more consistent with the one-to-one relation. }
\label{fig:f10}
\end{figure*}

For the continuum subtraction of F770W in NGC 628, we test 1) subtracting the unscaled F560W image from F770W, 2) subtracting a scaled-up version of the F560W from F770W (for a range of factors between 1.0 and 2.0), and 3) subtracting the continuum determined via a linear interpolation between the F560W and F1000W. Figure \ref{fig:f10} shows a comparison between the estimates of the F770W continuum for the different methods. The left panel (right panel) shows the continuum in F770W determined via the interpolation between F560W and F1000W as a function of the continuum estimated from the unscaled F560W (F560W scaled by a factor of 1.6). The data points correspond to measurements made on the various F770W continuum images of NGC 628 around eYSC--I sources. Each image is matched to the PSF of F1000W before deriving the continuum. We use the same source catalogs and perform the measurements in the same way (20 pc radius apertures, local background subtracted) as presented in Section \hyperlink{3.3}{3.3}.

We find that the relation between the F770W continuum estimated from the interpolation between F560W and F1000W and from only the F560W is strong (correlation coefficient, $\rho \, {\sim}\, 0.98$) and near linear (power-law exponent, $\alpha \, {\sim}\, 0.98$) (Figure \ref{fig:f10}). The F770W continuum derived from the unscaled F560W is found to be systematically lower than determined from the combination of F560W and F1000W  (Figure \ref{fig:f10}, left panel). The F770W continuum from F560W scaled by a factor of 1.6 is consistent on average with the value estimated from both the F560W and F1000W (Figure \ref{fig:f10}, right panel). In this case, the best-fit relation is more consistent with the one-to-one or y=x relation. This suggests that for these sources, the F770W continuum estimated via an interpolation between F560W and F1000W can be accurately approximated with only the F560W when it is scaled up by a factor of 1.6. 

\begin{figure}
\centering
\includegraphics[width=0.47\textwidth]{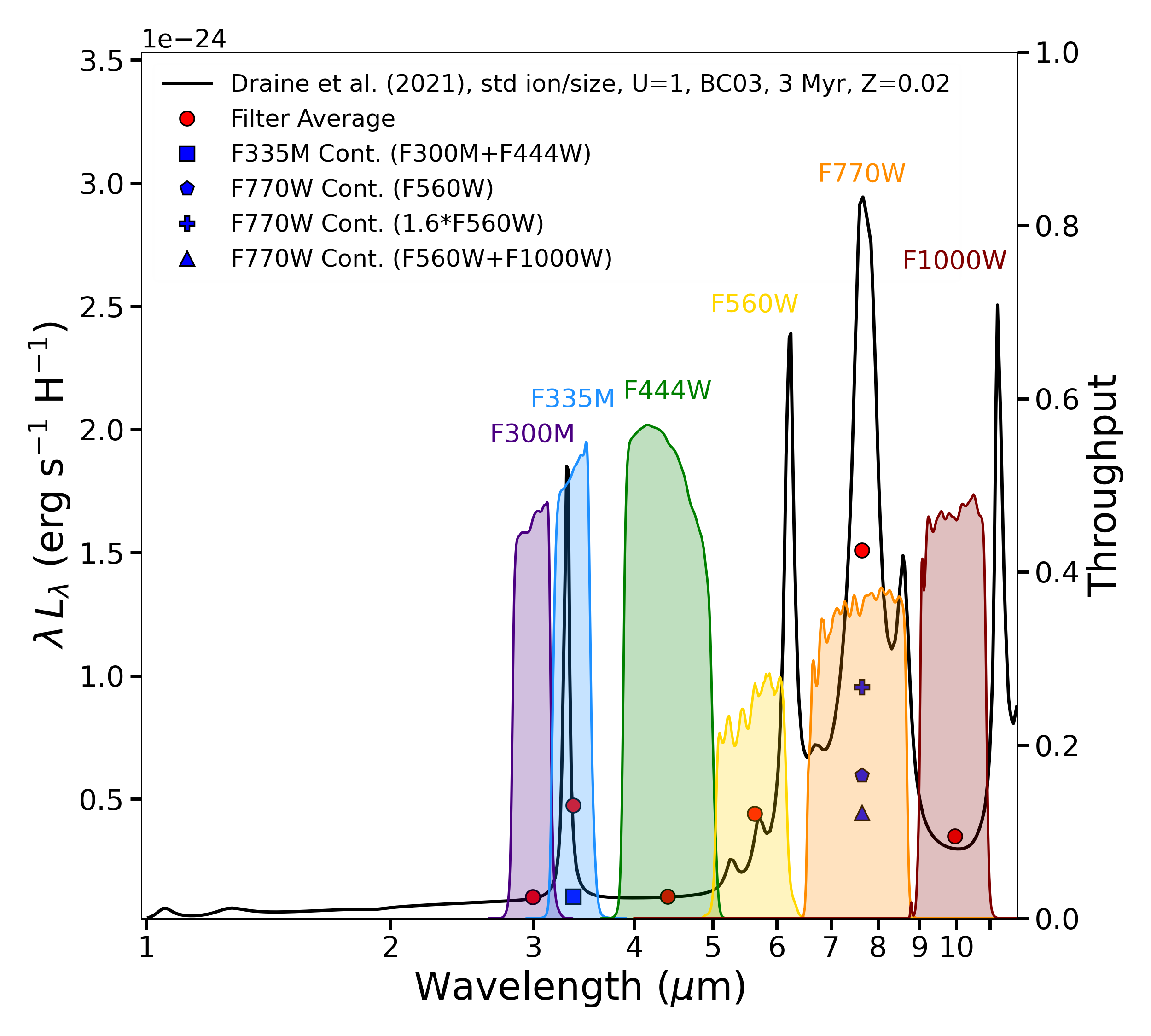}
\caption{The total dust emission spectrum (black line) for a model with standard PAH size distribution and ionization from \cite{2021ApJ...917....3D}. The NIRCam and MIRI filter throughputs (colored curves) that are relevant for estimating the PAH emission features are overlaid on the model. The red data points show the average model luminosity density through each filter at the pivot wavelength, estimated by convolving the filter throughputs with the model. The blue square shows an estimate of the F335M continuum for the model spectrum, derived via an interpolation between F300M and F444W. The blue pentagon shows the F770W continuum for the model from the unscaled F560W, while the blue plus is from the F560W scaled by a factor of 1.6, and the blue triangle is from an interpolation between F560W and F1000W. }
\label{fig:f11}
\end{figure}

In Figure \ref{fig:f11}, we show the total dust emission spectrum for a model from \cite{2021ApJ...917....3D} with standard PAH size distribution and ionization, intensity parameter U$=$1, and assuming the unreddened \cite{2003MNRAS.344.1000B} single-age stellar population model spectra with a metal mass fraction Z$=$0.02 (about solar metallicity) and an age of 3 Myr. Overlaid on the model spectrum are the filter throughput curves for the NIRCam/F300M, F335M, and F444W, and the MIRI/F560W, F770W, and F1000W. We estimate the average model luminosity density through each filter (red data points) by convolving the filter throughputs with the model spectrum. The continuum in F770W for the model spectrum is estimated from the filter luminosity densities via the various methods described above, shown as blue data points in the figure. 

For the model, we determine that the estimates of the F770W continuum from the unscaled F560W and from the interpolation between F560W and F1000W are fairly consistent and that both provide a reasonable approximation of the continuum level (Figure \ref{fig:f11}). On the other hand, the F770W continuum from F560W scaled by a factor of 1.6 appears to slightly over-predict the continuum. However, defining the true continuum level around the 7.7~$\mu$m PAH complex is difficult and requires fitting each emission component. We find that for the model, the F1000W luminosity density is lower than the F560W (Figure \ref{fig:f11}), yet for our observations, we find the opposite trend (see Figure \ref{fig:f10}). It is also apparent in the model spectra shown in Figure \ref{fig:f11} that on top of the dust continuum, the F560W receives a significant contribution from the wing of the bright 6.2~$\mu$m PAH feature, as well as a couple of other lower luminosity dust features, and that these contributions place the F560W above the relevant continuum level. This contamination is evident in our data as the F560W images show clear signatures of PAH emission. The F1000W may receive a contribution from the bright 8.6 and 11.3~$\mu$m features, as well as silicate absorption at ${\sim}$10~$\mu$m. 

Given this contamination in the continuum tracing filters, the more accurate technique for the continuum subtraction of the F770W is unclear. If F560W receives significant contamination, scaling it up further may result in an over-subtraction when it is used to remove the F770W continuum. For our study, we chose to use the unscaled F560W for the subtraction of F770W, rather than the scaled-up version, as it is the simplest method. A detailed study of our eYSCs with MIRI/MRS spectroscopy is required to determine the optimal method of continuum subtraction for the F770W. New studies are beginning to provide insights on how to accurately extract emission line flux from NIRCam and MIRI imaging via spectroscopic calibration \citep[see][]{2025A&A...698A..86C,2025ApJ...983...79D}.

The choice of the continuum subtraction technique for F770W has a relatively insignificant impact on the results presented in this study. Comparing the continuum subtracted F770W images directly, we determine that the flux density is roughly a factor of ${\sim}$1.15 higher on average in bright star-forming regions in NGC 628 when using the unscaled F560W for the subtraction compared to using the F560W scaled by a factor of 1.6. For the relation between the 7.7~$\mu$m PAH luminosity and SFR for sources in NGC 628, we find that the different methods of subtraction result in the same best-fit slope (or power-law exponent of the relation) and only slightly different y-intercepts. The best-fit y-intercept is found to be a value of ${\sim}$0.1 higher when using the unscaled F560W for the subtraction compared to using the scaled-up version. A similar change in the y-intercept (${\sim}$0.1 higher) is observed when using the unsubtracted F770W compared to using the F770W subtracted by the unscaled F560W. 

\hypertarget{B}{\section{The impact of overlapping measurements}}

As discussed in the main text, our photometric measurements exhibit some overlap in dense regions of star formation, shown in Figure \ref{fig:f2}. This is a result of the selection of the eYSC--I sources on the native resolution maps and the large apertures (20 pc radius) used to capture most of the PAH and ionized gas emission excited by the associated eYSC. Here, we test the impact that this overlap has on our results. 

We create catalogs of the eYSC--I sources in each galaxy for which our photometric measurements are completely independent. In the case where sources overlap within the aperture diameter (40 pc), we keep only the brightest source in terms of the measured extinction corrected Pa$\alpha$ luminosity. After the luminosity cut to remove sources that are affected by stochastic sampling of the stellar IMF, this results in 415, 206, 73, and 66 eYSC--I sources remaining for NGC 5194, 5236, 628, and 4449, respectively. Therefore, only about 47$\%$ of the sources remain across the full sample after removing any overlap in the measurements. 

\begin{figure*}
\centering
\includegraphics[width=0.49\textwidth]{calibration_L_3.3_v_SFR_surface_density_all_galaxies_2bins.png}
\includegraphics[width=0.49\textwidth]{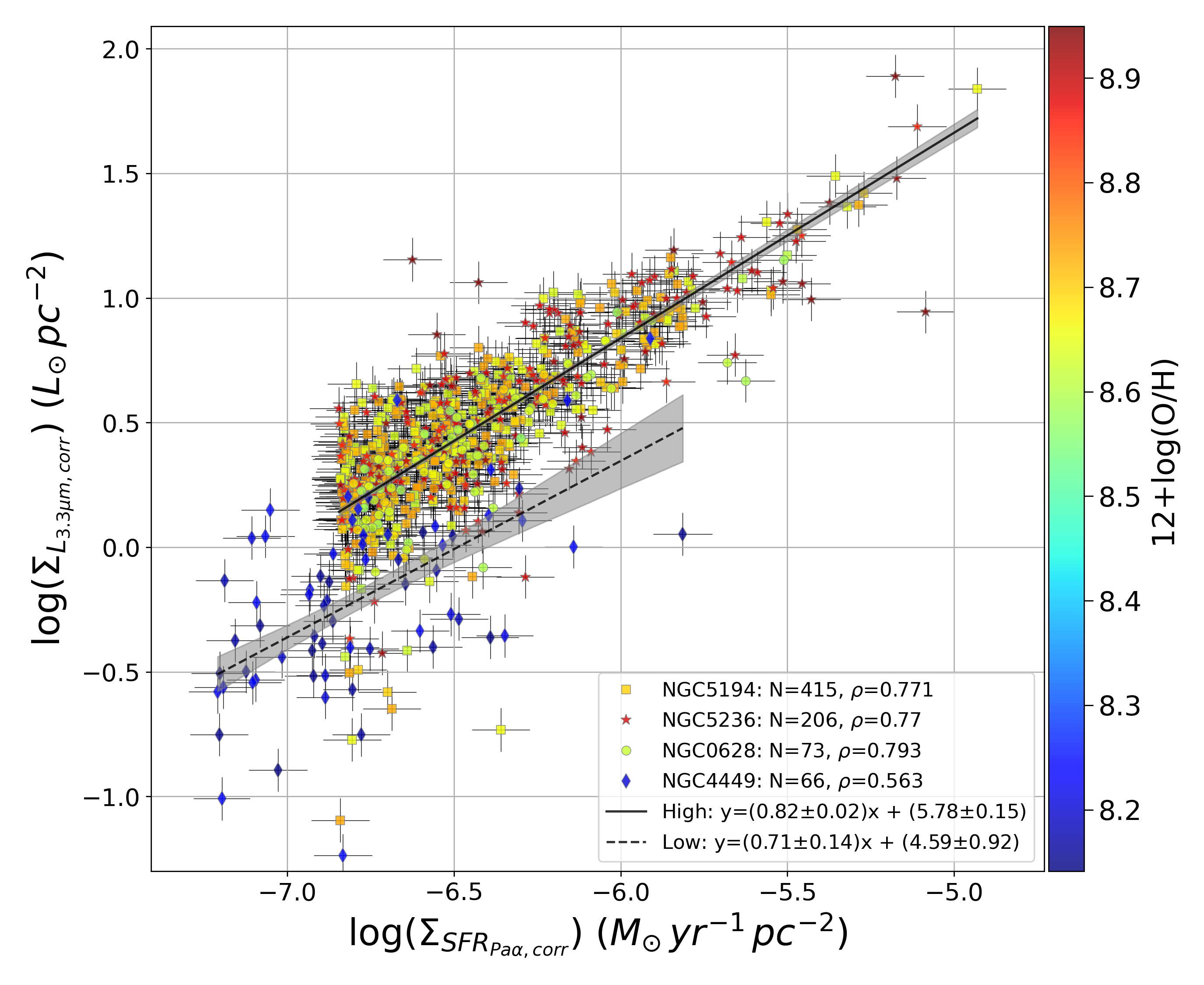}
\caption{Left panel: The same as the bottom left panel of Figure \ref{fig:f5}, shown again for comparison. Right panel: The same as the left panel, however, any overlap in the photometric measurements has been removed. In the case where the eYSC--I sources are separated by less than or equal to the aperture diameters (40 pc), we keep only the brightest source in terms of the measured extincted corrected Pa$\alpha$ luminosity. Note that the best-fit relations do not significantly change when the overlap in the measurements is removed.}
\label{fig:f12}
\end{figure*}

Figure \ref{fig:f12} shows a comparison of the results for the relation between the extinction corrected 3.3~$\mu$m PAH luminosity surface density and the SFR surface density before (left panel) and after (right panel) removing any overlap in the photometric measurements. We find that best-fit relations for either the high or low metallicity bins do not significantly change in terms of the best-fit slope and y-intercept when the overlap in the measurements is removed. Similarly, we also test the impact of the overlapping measurements on our other main results and find insignificant differences (e.g., for the 7.7~$\mu$m feature and the median 3.3/7.7~$\mu$m PAH luminosity ratios). As a result, we determine that the overlap in the photometric measurements has an insignificant impact on the main results presented in this study. 

\clearpage
\bibliographystyle{aasjournalv7}
\bibliography{main.bib}

@software{larry_bradley_2019_3568287,
  author       = {Larry Bradley and
                  Brigitta Sipőcz and
                  Thomas Robitaille and
                  Erik Tollerud and
                  Zé Vinícius and
                  Christoph Deil and
                  Kyle Barbary and
                  Tom J Wilson and
                  Ivo Busko and
                  Hans Moritz Günther and
                  Mihai Cara and
                  Simon Conseil and
                  Michael Droettboom and
                  Azalee Bostroem and
                  E. M. Bray and
                  Lars Andersen Bratholm and
                  P. L. Lim and
                  Matt Craig and
                  Geert Barentsen and
                  Sergio Pascual and
                  Axel Donath and
                  Johnny Greco and
                  Gabriel Perren and
                  Wolfgang Kerzendorf and
                  Miguel de Val-Borro and
                  Nadia Dencheva and
                  Leonardo de Albernaz Ferreira and
                  Harrison Souchereau and
                  Francesco D'Eugenio and
                  Benjamin Alan Weaver},
  title        = {astropy/photutils: v0.7.2},
  month        = dec,
  year         = 2019,
  publisher    = {Zenodo},
  version      = {v0.7.2},
  doi          = {10.5281/zenodo.3568287},
  url          = {https://doi.org/10.5281/zenodo.3568287}
}

@article{astropy:2013,
Adsurl = {http://adsabs.harvard.edu/abs/2013A%26A...558A..33A},
Archiveprefix = {arXiv},
Author = {{Astropy Collaboration} and {Robitaille}, T.~P. and {Tollerud}, E.~J. and {Greenfield}, P. and {Droettboom}, M. and {Bray}, E. and {Aldcroft}, T. and {Davis}, M. and {Ginsburg}, A. and {Price-Whelan}, A.~M. and {Kerzendorf}, W.~E. and {Conley}, A. and {Crighton}, N. and {Barbary}, K. and {Muna}, D. and {Ferguson}, H. and {Grollier}, F. and {Parikh}, M.~M. and {Nair}, P.~H. and {Unther}, H.~M. and {Deil}, C. and {Woillez}, J. and {Conseil}, S. and {Kramer}, R. and {Turner}, J.~E.~H. and {Singer}, L. and {Fox}, R. and {Weaver}, B.~A. and {Zabalza}, V. and {Edwards}, Z.~I. and {Azalee Bostroem}, K. and {Burke}, D.~J. and {Casey}, A.~R. and {Crawford}, S.~M. and {Dencheva}, N. and {Ely}, J. and {Jenness}, T. and {Labrie}, K. and {Lim}, P.~L. and {Pierfederici}, F. and {Pontzen}, A. and {Ptak}, A. and {Refsdal}, B. and {Servillat}, M. and {Streicher}, O.},
Doi = {10.1051/0004-6361/201322068},
Eid = {A33},
Eprint = {1307.6212},
Journal = {\aap},
Month = oct,
Pages = {A33},
Primaryclass = {astro-ph.IM},
Title = {{Astropy: A community Python package for astronomy}},
Volume = 558,
Year = 2013}

@ARTICLE{astropy:2018,
       author = {{Astropy Collaboration} and {Price-Whelan}, A.~M. and
         {Sip{\H{o}}cz}, B.~M. and {G{\"u}nther}, H.~M. and {Lim}, P.~L. and
         {Crawford}, S.~M. and {Conseil}, S. and {Shupe}, D.~L. and
         {Craig}, M.~W. and {Dencheva}, N. and {Ginsburg}, A. and {Vand
        erPlas}, J.~T. and {Bradley}, L.~D. and {P{\'e}rez-Su{\'a}rez}, D. and
         {de Val-Borro}, M. and {Aldcroft}, T.~L. and {Cruz}, K.~L. and
         {Robitaille}, T.~P. and {Tollerud}, E.~J. and {Ardelean}, C. and
         {Babej}, T. and {Bach}, Y.~P. and {Bachetti}, M. and {Bakanov}, A.~V. and
         {Bamford}, S.~P. and {Barentsen}, G. and {Barmby}, P. and
         {Baumbach}, A. and {Berry}, K.~L. and {Biscani}, F. and {Boquien}, M. and
         {Bostroem}, K.~A. and {Bouma}, L.~G. and {Brammer}, G.~B. and
         {Bray}, E.~M. and {Breytenbach}, H. and {Buddelmeijer}, H. and
         {Burke}, D.~J. and {Calderone}, G. and {Cano Rodr{\'\i}guez}, J.~L. and
         {Cara}, M. and {Cardoso}, J.~V.~M. and {Cheedella}, S. and {Copin}, Y. and
         {Corrales}, L. and {Crichton}, D. and {D'Avella}, D. and {Deil}, C. and
         {Depagne}, {\'E}. and {Dietrich}, J.~P. and {Donath}, A. and
         {Droettboom}, M. and {Earl}, N. and {Erben}, T. and {Fabbro}, S. and
         {Ferreira}, L.~A. and {Finethy}, T. and {Fox}, R.~T. and
         {Garrison}, L.~H. and {Gibbons}, S.~L.~J. and {Goldstein}, D.~A. and
         {Gommers}, R. and {Greco}, J.~P. and {Greenfield}, P. and
         {Groener}, A.~M. and {Grollier}, F. and {Hagen}, A. and {Hirst}, P. and
         {Homeier}, D. and {Horton}, A.~J. and {Hosseinzadeh}, G. and {Hu}, L. and
         {Hunkeler}, J.~S. and {Ivezi{\'c}}, {\v{Z}}. and {Jain}, A. and
         {Jenness}, T. and {Kanarek}, G. and {Kendrew}, S. and {Kern}, N.~S. and
         {Kerzendorf}, W.~E. and {Khvalko}, A. and {King}, J. and {Kirkby}, D. and
         {Kulkarni}, A.~M. and {Kumar}, A. and {Lee}, A. and {Lenz}, D. and
         {Littlefair}, S.~P. and {Ma}, Z. and {Macleod}, D.~M. and
         {Mastropietro}, M. and {McCully}, C. and {Montagnac}, S. and
         {Morris}, B.~M. and {Mueller}, M. and {Mumford}, S.~J. and {Muna}, D. and
         {Murphy}, N.~A. and {Nelson}, S. and {Nguyen}, G.~H. and
         {Ninan}, J.~P. and {N{\"o}the}, M. and {Ogaz}, S. and {Oh}, S. and
         {Parejko}, J.~K. and {Parley}, N. and {Pascual}, S. and {Patil}, R. and
         {Patil}, A.~A. and {Plunkett}, A.~L. and {Prochaska}, J.~X. and
         {Rastogi}, T. and {Reddy Janga}, V. and {Sabater}, J. and
         {Sakurikar}, P. and {Seifert}, M. and {Sherbert}, L.~E. and
         {Sherwood-Taylor}, H. and {Shih}, A.~Y. and {Sick}, J. and
         {Silbiger}, M.~T. and {Singanamalla}, S. and {Singer}, L.~P. and
         {Sladen}, P.~H. and {Sooley}, K.~A. and {Sornarajah}, S. and
         {Streicher}, O. and {Teuben}, P. and {Thomas}, S.~W. and
         {Tremblay}, G.~R. and {Turner}, J.~E.~H. and {Terr{\'o}n}, V. and
         {van Kerkwijk}, M.~H. and {de la Vega}, A. and {Watkins}, L.~L. and
         {Weaver}, B.~A. and {Whitmore}, J.~B. and {Woillez}, J. and
         {Zabalza}, V. and {Astropy Contributors}},
        title = "{The Astropy Project: Building an Open-science Project and Status of the v2.0 Core Package}",
      journal = {\aj},
         year = 2018,
        month = sep,
       volume = {156},
       number = {3},
          eid = {123},
        pages = {123},
          doi = {10.3847/1538-3881/aabc4f},
archivePrefix = {arXiv},
       eprint = {1801.02634},
 primaryClass = {astro-ph.IM},
       adsurl = {https://ui.adsabs.harvard.edu/abs/2018AJ....156..123A}
}

@ARTICLE{2018ApJ...862...77C,
       author = {{Casey}, Caitlin M. and {Zavala}, Jorge A. and {Spilker}, Justin and {da Cunha}, Elisabete and {Hodge}, Jacqueline and {Hung}, Chao-Ling and {Staguhn}, Johannes and {Finkelstein}, Steven L. and {Drew}, Patrick},
        title = "{The Brightest Galaxies in the Dark Ages: Galaxies{\textquoteright} Dust Continuum Emission during the Reionization Era}",
      journal = {\apj},
         year = 2018,
        month = jul,
       volume = {862},
       number = {1},
          eid = {77},
        pages = {77},
          doi = {10.3847/1538-4357/aac82d},
archivePrefix = {arXiv},
       eprint = {1805.10301},
 primaryClass = {astro-ph.GA},
       adsurl = {https://ui.adsabs.harvard.edu/abs/2018ApJ...862...77C}
}

@ARTICLE{2020ApJ...902..112B,
       author = {{Bouwens}, Rychard and {Gonz{\'a}lez-L{\'o}pez}, Jorge and {Aravena}, Manuel and {Decarli}, Roberto and {Novak}, Mladen and {Stefanon}, Mauro and {Walter}, Fabian and {Boogaard}, Leindert and {Carilli}, Chris and {Dudzevi{\v{c}}i{\={u}}t{\.{e}}}, Ugn{\.{e}} and {Smail}, Ian and {Daddi}, Emanuele and {da Cunha}, Elisabete and {Ivison}, Rob and {Nanayakkara}, Themiya and {Cortes}, Paulo and {Cox}, Pierre and {Inami}, Hanae and {Oesch}, Pascal and {Popping}, Gerg{\"o} and {Riechers}, Dominik and {van der Werf}, Paul and {Weiss}, Axel and {Fudamoto}, Yoshi and {Wagg}, Jeff},
        title = "{The ALMA Spectroscopic Survey Large Program: The Infrared Excess of z = 1.5-10 UV-selected Galaxies and the Implied High-redshift Star Formation History}",
      journal = {\apj},
         year = 2020,
        month = oct,
       volume = {902},
       number = {2},
          eid = {112},
        pages = {112},
          doi = {10.3847/1538-4357/abb830},
archivePrefix = {arXiv},
       eprint = {2009.10727},
 primaryClass = {astro-ph.GA},
       adsurl = {https://ui.adsabs.harvard.edu/abs/2020ApJ...902..112B}
}

@ARTICLE{2020ApJ...905...55L,
       author = {{Lai}, Thomas S. -Y. and {Smith}, J.~D.~T. and {Baba}, Shunsuke and {Spoon}, Henrik W.~W. and {Imanishi}, Masatoshi},
        title = "{All the PAHs: An AKARI-Spitzer Cross-archival Spectroscopic Survey of Aromatic Emission in Galaxies}",
      journal = {\apj},
         year = 2020,
        month = dec,
       volume = {905},
       number = {1},
          eid = {55},
        pages = {55},
          doi = {10.3847/1538-4357/abc002},
archivePrefix = {arXiv},
       eprint = {2010.05034},
 primaryClass = {astro-ph.GA},
       adsurl = {https://ui.adsabs.harvard.edu/abs/2020ApJ...905...55L}
}

@ARTICLE{2018A&A...617A.130I,
       author = {{Inami}, H. and {Armus}, L. and {Matsuhara}, H. and {Charmandaris}, V. and {D{\'\i}az-Santos}, T. and {Surace}, J. and {Stierwalt}, S. and {Ohyama}, Y. and {Howell}, J. and {Marshall}, J. and {Evans}, A.~S. and {Linden}, S.~T. and {Mazzarella}, J.},
        title = "{The AKARI 2.5-5 micron spectra of luminous infrared galaxies in the local Universe}",
      journal = {\aap},
         year = 2018,
        month = sep,
       volume = {617},
          eid = {A130},
        pages = {A130},
          doi = {10.1051/0004-6361/201833053},
archivePrefix = {arXiv},
       eprint = {1806.05198},
 primaryClass = {astro-ph.GA},
       adsurl = {https://ui.adsabs.harvard.edu/abs/2018A&A...617A.130I}
}

@ARTICLE{1984A&A...137L...5L,
       author = {{Leger}, A. and {Puget}, J.~L.},
        title = "{Identification of the Unidentified Infrared Emission Features of Interstellar Dust}",
      journal = {\aap},
         year = 1984,
        month = aug,
       volume = {137},
        pages = {L5-L8},
       adsurl = {https://ui.adsabs.harvard.edu/abs/1984A&A...137L...5L}
}

@ARTICLE{1985ApJ...290L..25A,
       author = {{Allamandola}, L.~J. and {Tielens}, A.~G.~G.~M. and {Barker}, J.~R.},
        title = "{Polycyclic aromatic hydrocarbons and the unidentified infrared emission bands: auto exhaust along the milky way.}",
      journal = {\apjl},
         year = 1985,
        month = mar,
       volume = {290},
        pages = {L25-L28},
          doi = {10.1086/184435},
       adsurl = {https://ui.adsabs.harvard.edu/abs/1985ApJ...290L..25A}
}

@ARTICLE{1989ApJS...71..733A,
       author = {{Allamandola}, L.~J. and {Tielens}, A.~G.~G.~M. and {Barker}, J.~R.},
        title = "{Interstellar Polycyclic Aromatic Hydrocarbons: The Infrared Emission Bands, the Excitation/Emission Mechanism, and the Astrophysical Implications}",
      journal = {\apjs},
         year = 1989,
        month = dec,
       volume = {71},
        pages = {733},
          doi = {10.1086/191396},
       adsurl = {https://ui.adsabs.harvard.edu/abs/1989ApJS...71..733A}
}

@ARTICLE{2008ARA&A..46..289T,
       author = {{Tielens}, A.~G.~G.~M.},
        title = "{Interstellar polycyclic aromatic hydrocarbon molecules.}",
      journal = {\araa},
         year = 2008,
        month = sep,
       volume = {46},
        pages = {289-337},
          doi = {10.1146/annurev.astro.46.060407.145211},
       adsurl = {https://ui.adsabs.harvard.edu/abs/2008ARA&A..46..289T}
}

@ARTICLE{2000ApJ...532L..21H,
       author = {{Helou}, G. and {Lu}, Nanyao Y. and {Werner}, M.~W. and {Malhotra}, S. and {Silbermann}, N.},
        title = "{The Mid-Infrared Spectra of Normal Galaxies}",
      journal = {\apjl},
         year = 2000,
        month = mar,
       volume = {532},
       number = {1},
        pages = {L21-L24},
          doi = {10.1086/312549},
archivePrefix = {arXiv},
       eprint = {astro-ph/0001461},
 primaryClass = {astro-ph},
       adsurl = {https://ui.adsabs.harvard.edu/abs/2000ApJ...532L..21H}
}

@ARTICLE{2007ApJ...656..770S,
       author = {{Smith}, J.~D.~T. and {Draine}, B.~T. and {Dale}, D.~A. and {Moustakas}, J. and {Kennicutt}, R.~C., Jr. and {Helou}, G. and {Armus}, L. and {Roussel}, H. and {Sheth}, K. and {Bendo}, G.~J. and {Buckalew}, B.~A. and {Calzetti}, D. and {Engelbracht}, C.~W. and {Gordon}, K.~D. and {Hollenbach}, D.~J. and {Li}, A. and {Malhotra}, S. and {Murphy}, E.~J. and {Walter}, F.},
        title = "{The Mid-Infrared Spectrum of Star-forming Galaxies: Global Properties of Polycyclic Aromatic Hydrocarbon Emission}",
      journal = {\apj},
         year = 2007,
        month = feb,
       volume = {656},
       number = {2},
        pages = {770-791},
          doi = {10.1086/510549},
archivePrefix = {arXiv},
       eprint = {astro-ph/0610913},
 primaryClass = {astro-ph},
       adsurl = {https://ui.adsabs.harvard.edu/abs/2007ApJ...656..770S}
}

@ARTICLE{2004ApJS..154..253H,
       author = {{Helou}, G. and {Roussel}, H. and {Appleton}, P. and {Frayer}, D. and {Stolovy}, S. and {Storrie-Lombardi}, L. and {Hurt}, R. and {Lowrance}, P. and {Makovoz}, D. and {Masci}, F. and {Surace}, J. and {Gordon}, K.~D. and {Alonso-Herrero}, A. and {Engelbracht}, C.~W. and {Misselt}, K. and {Rieke}, G. and {Rieke}, M. and {Willner}, S.~P. and {Pahre}, M. and {Ashby}, M.~L.~N. and {Fazio}, G.~G. and {Smith}, H.~A.},
        title = "{The Anatomy of Star Formation in NGC 300}",
      journal = {\apjs},
         year = 2004,
        month = sep,
       volume = {154},
       number = {1},
        pages = {253-258},
          doi = {10.1086/422640},
archivePrefix = {arXiv},
       eprint = {astro-ph/0408248},
 primaryClass = {astro-ph},
       adsurl = {https://ui.adsabs.harvard.edu/abs/2004ApJS..154..253H}
}

@ARTICLE{2006ApJ...652..283B,
       author = {{Bendo}, George J. and {Dale}, Daniel A. and {Draine}, Bruce T. and {Engelbracht}, Charles W. and {Kennicutt}, Robert C., Jr. and {Calzetti}, Daniela and {Gordon}, Karl D. and {Helou}, George and {Hollenbach}, David and {Li}, Aigen and {Murphy}, Eric J. and {Prescott}, Moire K.~M. and {Smith}, John-David T.},
        title = "{The Spectral Energy Distribution of Dust Emission in the Edge-on Spiral Galaxy NGC 4631 as Seen with Spitzer and the James Clerk Maxwell Telescope}",
      journal = {\apj},
         year = 2006,
        month = nov,
       volume = {652},
       number = {1},
        pages = {283-305},
          doi = {10.1086/508057},
archivePrefix = {arXiv},
       eprint = {astro-ph/0607669},
 primaryClass = {astro-ph},
       adsurl = {https://ui.adsabs.harvard.edu/abs/2006ApJ...652..283B}
}

@ARTICLE{2007ApJ...666..870C,
       author = {{Calzetti}, D. and {Kennicutt}, R.~C. and {Engelbracht}, C.~W. and {Leitherer}, C. and {Draine}, B.~T. and {Kewley}, L. and {Moustakas}, J. and {Sosey}, M. and {Dale}, D.~A. and {Gordon}, K.~D. and {Helou}, G.~X. and {Hollenbach}, D.~J. and {Armus}, L. and {Bendo}, G. and {Bot}, C. and {Buckalew}, B. and {Jarrett}, T. and {Li}, A. and {Meyer}, M. and {Murphy}, E.~J. and {Prescott}, M. and {Regan}, M.~W. and {Rieke}, G.~H. and {Roussel}, H. and {Sheth}, K. and {Smith}, J.~D.~T. and {Thornley}, M.~D. and {Walter}, F.},
        title = "{The Calibration of Mid-Infrared Star Formation Rate Indicators}",
      journal = {\apj},
         year = 2007,
        month = sep,
       volume = {666},
       number = {2},
        pages = {870-895},
          doi = {10.1086/520082},
archivePrefix = {arXiv},
       eprint = {0705.3377},
 primaryClass = {astro-ph},
       adsurl = {https://ui.adsabs.harvard.edu/abs/2007ApJ...666..870C}
}

@ARTICLE{2007ApJ...657..810D,
       author = {{Draine}, B.~T. and {Li}, Aigen},
        title = "{Infrared Emission from Interstellar Dust. IV. The Silicate-Graphite-PAH Model in the Post-Spitzer Era}",
      journal = {\apj},
         year = 2007,
        month = mar,
       volume = {657},
       number = {2},
        pages = {810-837},
          doi = {10.1086/511055},
archivePrefix = {arXiv},
       eprint = {astro-ph/0608003},
 primaryClass = {astro-ph},
       adsurl = {https://ui.adsabs.harvard.edu/abs/2007ApJ...657..810D}
}

@ARTICLE{2009ApJ...703.1672K,
       author = {{Kennicutt}, Jr., Robert C. and {Hao}, Cai-Na and {Calzetti}, Daniela and {Moustakas}, John and {Dale}, Daniel A. and {Bendo}, George and {Engelbracht}, Charles W. and {Johnson}, Benjamin D. and {Lee}, Janice C.},
        title = "{Dust-corrected Star Formation Rates of Galaxies. I. Combinations of H{\ensuremath{\alpha}} and Infrared Tracers}",
      journal = {\apj},
         year = 2009,
        month = oct,
       volume = {703},
       number = {2},
        pages = {1672-1695},
          doi = {10.1088/0004-637X/703/2/1672},
archivePrefix = {arXiv},
       eprint = {0908.0203},
 primaryClass = {astro-ph.CO},
       adsurl = {https://ui.adsabs.harvard.edu/abs/2009ApJ...703.1672K}
}

@ARTICLE{2011A&A...533A.119E,
       author = {{Elbaz}, D. and {Dickinson}, M. and {Hwang}, H.~S. and {D{\'\i}az-Santos}, T. and {Magdis}, G. and {Magnelli}, B. and {Le Borgne}, D. and {Galliano}, F. and {Pannella}, M. and {Chanial}, P. and {Armus}, L. and {Charmandaris}, V. and {Daddi}, E. and {Aussel}, H. and {Popesso}, P. and {Kartaltepe}, J. and {Altieri}, B. and {Valtchanov}, I. and {Coia}, D. and {Dannerbauer}, H. and {Dasyra}, K. and {Leiton}, R. and {Mazzarella}, J. and {Alexander}, D.~M. and {Buat}, V. and {Burgarella}, D. and {Chary}, R. -R. and {Gilli}, R. and {Ivison}, R.~J. and {Juneau}, S. and {Le Floc'h}, E. and {Lutz}, D. and {Morrison}, G.~E. and {Mullaney}, J.~R. and {Murphy}, E. and {Pope}, A. and {Scott}, D. and {Brodwin}, M. and {Calzetti}, D. and {Cesarsky}, C. and {Charlot}, S. and {Dole}, H. and {Eisenhardt}, P. and {Ferguson}, H.~C. and {F{\"o}rster Schreiber}, N. and {Frayer}, D. and {Giavalisco}, M. and {Huynh}, M. and {Koekemoer}, A.~M. and {Papovich}, C. and {Reddy}, N. and {Surace}, C. and {Teplitz}, H. and {Yun}, M.~S. and {Wilson}, G.},
        title = "{GOODS-Herschel: an infrared main sequence for star-forming galaxies}",
      journal = {\aap},
         year = 2011,
        month = sep,
       volume = {533},
          eid = {A119},
        pages = {A119},
          doi = {10.1051/0004-6361/201117239},
archivePrefix = {arXiv},
       eprint = {1105.2537},
 primaryClass = {astro-ph.CO},
       adsurl = {https://ui.adsabs.harvard.edu/abs/2011A&A...533A.119E}
}

@ARTICLE{2007ApJ...660..346P,
       author = {{Povich}, Matthew S. and {Stone}, Jennifer M. and {Churchwell}, Ed and {Zweibel}, Ellen G. and {Wolfire}, Mark G. and {Babler}, Brian L. and {Indebetouw}, R{\'e}my and {Meade}, Marilyn R. and {Whitney}, Barbara A.},
        title = "{A Multiwavelength Study of M17: The Spectral Energy Distribution and PAH Emission Morphology of a Massive Star Formation Region}",
      journal = {\apj},
         year = 2007,
        month = may,
       volume = {660},
       number = {1},
        pages = {346-362},
          doi = {10.1086/513073},
       adsurl = {https://ui.adsabs.harvard.edu/abs/2007ApJ...660..346P}
}

@ARTICLE{2008MNRAS.389..629B,
       author = {{Bendo}, G.~J. and {Draine}, B.~T. and {Engelbracht}, C.~W. and {Helou}, G. and {Thornley}, M.~D. and {Bot}, C. and {Buckalew}, B.~A. and {Calzetti}, D. and {Dale}, D.~A. and {Hollenbach}, D.~J. and {Li}, A. and {Moustakas}, J.},
        title = "{The relations among 8, 24 and 160 {\ensuremath{\mu}}m dust emission within nearby spiral galaxies}",
      journal = {\mnras},
         year = 2008,
        month = sep,
       volume = {389},
       number = {2},
        pages = {629-650},
          doi = {10.1111/j.1365-2966.2008.13567.x},
archivePrefix = {arXiv},
       eprint = {0806.2758},
 primaryClass = {astro-ph},
       adsurl = {https://ui.adsabs.harvard.edu/abs/2008MNRAS.389..629B}
}

@ARTICLE{2009ApJ...699.1125R,
       author = {{Rela{\~n}o}, M{\'o}nica and {Kennicutt}, Jr., Robert C.},
        title = "{Star Formation in Luminous H II Regions in M33}",
      journal = {\apj},
         year = 2009,
        month = jul,
       volume = {699},
       number = {2},
        pages = {1125-1143},
          doi = {10.1088/0004-637X/699/2/1125},
archivePrefix = {arXiv},
       eprint = {0905.1158},
 primaryClass = {astro-ph.GA},
       adsurl = {https://ui.adsabs.harvard.edu/abs/2009ApJ...699.1125R}
}

@ARTICLE{2005ApJ...628L..29E,
       author = {{Engelbracht}, C.~W. and {Gordon}, K.~D. and {Rieke}, G.~H. and {Werner}, M.~W. and {Dale}, D.~A. and {Latter}, W.~B.},
        title = "{Metallicity Effects on Mid-Infrared Colors and the 8 {\ensuremath{\mu}}m PAH Emission in Galaxies}",
      journal = {\apjl},
         year = 2005,
        month = jul,
       volume = {628},
       number = {1},
        pages = {L29-L32},
          doi = {10.1086/432613},
archivePrefix = {arXiv},
       eprint = {astro-ph/0506214},
 primaryClass = {astro-ph},
       adsurl = {https://ui.adsabs.harvard.edu/abs/2005ApJ...628L..29E}
}

@ARTICLE{2014MNRAS.445..899C,
       author = {{Cook}, David O. and {Dale}, Daniel A. and {Johnson}, Benjamin D. and {Van Zee}, Liese and {Lee}, Janice C. and {Kennicutt}, Robert C. and {Calzetti}, Daniela and {Staudaher}, Shawn M. and {Engelbracht}, Charles W.},
        title = "{Spitzer Local Volume Legacy (LVL) SEDs and physical properties}",
      journal = {MNRAS},
         year = 2014,
        month = nov,
       volume = {445},
       number = {1},
        pages = {899-912},
          doi = {10.1093/mnras/stu1787},
archivePrefix = {arXiv},
       eprint = {1409.0847},
 primaryClass = {astro-ph.GA},
       adsurl = {https://ui.adsabs.harvard.edu/abs/2014MNRAS.445..899C}
}

@ARTICLE{2017ApJ...837..157S,
       author = {{Shivaei}, Irene and {Reddy}, Naveen A. and {Shapley}, Alice E. and {Siana}, Brian and {Kriek}, Mariska and {Mobasher}, Bahram and {Coil}, Alison L. and {Freeman}, William R. and {Sanders}, Ryan L. and {Price}, Sedona H. and {Azadi}, Mojegan and {Zick}, Tom},
        title = "{The MOSDEF Survey: Metallicity Dependence of PAH Emission at High Redshift and Implications for 24 {\ensuremath{\mu}}m Inferred IR Luminosities and Star Formation Rates at z {\ensuremath{\sim}} 2}",
      journal = {\apj},
         year = 2017,
        month = mar,
       volume = {837},
       number = {2},
          eid = {157},
        pages = {157},
          doi = {10.3847/1538-4357/aa619c},
archivePrefix = {arXiv},
       eprint = {1609.04814},
 primaryClass = {astro-ph.GA},
       adsurl = {https://ui.adsabs.harvard.edu/abs/2017ApJ...837..157S}
}

@ARTICLE{2022ApJ...928..120G,
       author = {{Gregg}, Benjamin and {Calzetti}, Daniela and {Heyer}, Mark},
        title = "{Mid- and Far-infrared Color-Color Relations within Local Galaxies}",
      journal = {\apj},
         year = 2022,
        month = apr,
       volume = {928},
       number = {2},
          eid = {120},
        pages = {120},
          doi = {10.3847/1538-4357/ac558a},
archivePrefix = {arXiv},
       eprint = {2202.09386},
 primaryClass = {astro-ph.GA},
       adsurl = {https://ui.adsabs.harvard.edu/abs/2022ApJ...928..120G}
}

@ARTICLE{2012ApJ...744...20S,
       author = {{Sandstrom}, Karin M. and {Bolatto}, Alberto D. and {Bot}, Caroline and {Draine}, B.~T. and {Ingalls}, James G. and {Israel}, Frank P. and {Jackson}, James M. and {Leroy}, Adam K. and {Li}, Aigen and {Rubio}, M{\'o}nica and {Simon}, Joshua D. and {Smith}, J.~D.~T. and {Stanimirovi{\'c}}, Sne{\v{z}}ana and {Tielens}, A.~G.~G.~M. and {van Loon}, Jacco Th.},
        title = "{The Spitzer Spectroscopic Survey of the Small Magellanic Cloud (S$^{4}$MC): Probing the Physical State of Polycyclic Aromatic Hydrocarbons in a Low-metallicity Environment}",
      journal = {\apj},
         year = 2012,
        month = jan,
       volume = {744},
       number = {1},
          eid = {20},
        pages = {20},
          doi = {10.1088/0004-637X/744/1/20},
archivePrefix = {arXiv},
       eprint = {1109.0999},
 primaryClass = {astro-ph.CO},
       adsurl = {https://ui.adsabs.harvard.edu/abs/2012ApJ...744...20S}
}

@ARTICLE{2006A&A...446..877M,
       author = {{Madden}, S.~C. and {Galliano}, F. and {Jones}, A.~P. and {Sauvage}, M.},
        title = "{ISM properties in low-metallicity environments}",
      journal = {\aap},
         year = 2006,
        month = feb,
       volume = {446},
       number = {3},
        pages = {877-896},
          doi = {10.1051/0004-6361:20053890},
archivePrefix = {arXiv},
       eprint = {astro-ph/0510086},
 primaryClass = {astro-ph},
       adsurl = {https://ui.adsabs.harvard.edu/abs/2006A&A...446..877M}
}

@ARTICLE{2008ApJ...682..336G,
       author = {{Gordon}, Karl D. and {Engelbracht}, Charles W. and {Rieke}, George H. and {Misselt}, K.~A. and {Smith}, J. -D.~T. and {Kennicutt}, Robert C., Jr.},
        title = "{The Behavior of the Aromatic Features in M101 H II Regions: Evidence for Dust Processing}",
      journal = {\apj},
         year = 2008,
        month = jul,
       volume = {682},
       number = {1},
        pages = {336-354},
          doi = {10.1086/589567},
archivePrefix = {arXiv},
       eprint = {0804.3223},
 primaryClass = {astro-ph},
       adsurl = {https://ui.adsabs.harvard.edu/abs/2008ApJ...682..336G}
}

@ARTICLE{2018ApJ...864..136B,
       author = {{Binder}, Breanna A. and {Povich}, Matthew S.},
        title = "{A Multiwavelength Look at Galactic Massive Star-forming Regions}",
      journal = {\apj},
         year = 2018,
        month = sep,
       volume = {864},
       number = {2},
          eid = {136},
        pages = {136},
          doi = {10.3847/1538-4357/aad7b2},
archivePrefix = {arXiv},
       eprint = {1808.00454},
 primaryClass = {astro-ph.GA},
       adsurl = {https://ui.adsabs.harvard.edu/abs/2018ApJ...864..136B}
}

@ARTICLE{2014ApJ...784..130C,
       author = {{Calapa}, Marie D. and {Calzetti}, Daniela and {Draine}, Bruce T. and {Boquien}, M{\'e}d{\'e}ric and {Kramer}, Carsten and {Xilouris}, Manolis and {Verley}, Simon and {Braine}, Jonathan and {Rela{\~n}o}, Monica and {van der Werf}, Paul and {Israel}, Frank and {Hermelo}, Israel and {Albrecht}, Marcus},
        title = "{The Heating of Mid-infrared Dust in the Nearby Galaxy M33: A Testbed for Tracing Galaxy Evolution}",
      journal = {\apj},
         year = 2014,
        month = apr,
       volume = {784},
       number = {2},
          eid = {130},
        pages = {130},
          doi = {10.1088/0004-637X/784/2/130},
archivePrefix = {arXiv},
       eprint = {1402.4668},
 primaryClass = {astro-ph.GA},
       adsurl = {https://ui.adsabs.harvard.edu/abs/2014ApJ...784..130C}
}

@ARTICLE{2014ApJ...797..129L,
       author = {{Lu}, N. and {Bendo}, G.~J. and {Boselli}, A. and {Baes}, M. and {Wu}, H. and {Madden}, S.~C. and {De Looze}, I. and {R{\'e}my-Ruyer}, A. and {Boquien}, M. and {Wilson}, C.~D. and {Galametz}, M. and {Lam}, M.~I. and {Cooray}, A. and {Spinoglio}, L. and {Zhao}, Y.},
        title = "{Quantifying the Heating Sources for Mid-infrared Dust Emissions in Galaxies: The Case of M 81}",
      journal = {\apj},
         year = 2014,
        month = dec,
       volume = {797},
       number = {2},
          eid = {129},
        pages = {129},
          doi = {10.1088/0004-637X/797/2/129},
archivePrefix = {arXiv},
       eprint = {1410.3874},
 primaryClass = {astro-ph.GA},
       adsurl = {https://ui.adsabs.harvard.edu/abs/2014ApJ...797..129L}
}

@ARTICLE{2020NatAs...4..339L,
       author = {{Li}, Aigen},
        title = "{Spitzer's perspective of polycyclic aromatic hydrocarbons in galaxies}",
      journal = {Nature Astronomy},
         year = 2020,
        month = mar,
       volume = {4},
        pages = {339-351},
          doi = {10.1038/s41550-020-1051-1},
archivePrefix = {arXiv},
       eprint = {2003.10489},
 primaryClass = {astro-ph.GA},
       adsurl = {https://ui.adsabs.harvard.edu/abs/2020NatAs...4..339L}
}

@ARTICLE{1998ARA&A..36..189K,
       author = {{Kennicutt}, Jr., Robert C.},
        title = "{Star Formation in Galaxies Along the Hubble Sequence}",
      journal = {\araa},
         year = 1998,
        month = jan,
       volume = {36},
        pages = {189-232},
          doi = {10.1146/annurev.astro.36.1.189},
archivePrefix = {arXiv},
       eprint = {astro-ph/9807187},
 primaryClass = {astro-ph},
       adsurl = {https://ui.adsabs.harvard.edu/abs/1998ARA&A..36..189K}
}

@INCOLLECTION{2013seg..book..419C,
       author = {{Calzetti}, Daniela},
        title = "{Star Formation Rate Indicators}",
    booktitle = {Secular Evolution of Galaxies},
         year = 2013,
       editor = {{Falc{\'o}n-Barroso}, Jes{\'u}s and {Knapen}, Johan H.},
        pages = {419},
       adsurl = {https://ui.adsabs.harvard.edu/abs/2013seg..book..419C}
}

@ARTICLE{2018ApJS..235...23S,
       author = {{Sabbi}, E. and {Calzetti}, D. and {Ubeda}, L. and {Adamo}, A. and {Cignoni}, M. and {Thilker}, D. and {Aloisi}, A. and {Elmegreen}, B.~G. and {Elmegreen}, D.~M. and {Gouliermis}, D.~A. and {Grebel}, E.~K. and {Messa}, M. and {Smith}, L.~J. and {Tosi}, M. and {Dolphin}, A. and {Andrews}, J.~E. and {Ashworth}, G. and {Bright}, S.~N. and {Brown}, T.~M. and {Chandar}, R. and {Christian}, C. and {Clayton}, G.~C. and {Cook}, D.~O. and {Dale}, D.~A. and {de Mink}, S.~E. and {Dobbs}, C. and {Evans}, A.~S. and {Fumagalli}, M. and {Gallagher}, J.~S., III and {Grasha}, K. and {Herrero}, A. and {Hunter}, D.~A. and {Johnson}, K.~E. and {Kahre}, L. and {Kennicutt}, R.~C. and {Kim}, H. and {Krumholz}, M.~R. and {Lee}, J.~C. and {Lennon}, D. and {Martin}, C. and {Nair}, P. and {Nota}, A. and {{\"O}stlin}, G. and {Pellerin}, A. and {Prieto}, J. and {Regan}, M.~W. and {Ryon}, J.~E. and {Sacchi}, E. and {Schaerer}, D. and {Schiminovich}, D. and {Shabani}, F. and {Van Dyk}, S.~D. and {Walterbos}, R. and {Whitmore}, B.~C. and {Wofford}, A.},
        title = "{The Resolved Stellar Populations in the LEGUS Galaxies1}",
      journal = {\apjs},
         year = 2018,
        month = mar,
       volume = {235},
       number = {1},
          eid = {23},
        pages = {23},
          doi = {10.3847/1538-4365/aaa8e510.48550/arXiv.1801.05467},
archivePrefix = {arXiv},
       eprint = {1801.05467},
 primaryClass = {astro-ph.GA},
       adsurl = {https://ui.adsabs.harvard.edu/abs/2018ApJS..235...23S}
}

@ARTICLE{2020ApJ...897..122L,
       author = {{Lang}, Philipp and {Meidt}, Sharon E. and {Rosolowsky}, Erik and {Nofech}, Joseph and {Schinnerer}, Eva and {Leroy}, Adam K. and {Emsellem}, Eric and {Pessa}, Ismael and {Glover}, Simon C.~O. and {Groves}, Brent and {Hughes}, Annie and {Kruijssen}, J.~M. Diederik and {Querejeta}, Miguel and {Schruba}, Andreas and {Bigiel}, Frank and {Blanc}, Guillermo A. and {Chevance}, M{\'e}lanie and {Colombo}, Dario and {Faesi}, Christopher and {Henshaw}, Jonathan D. and {Herrera}, Cinthya N. and {Liu}, Daizhong and {Pety}, J{\'e}r{\^o}me and {Puschnig}, Johannes and {Saito}, Toshiki and {Sun}, Jiayi and {Usero}, Antonio},
        title = "{PHANGS CO Kinematics: Disk Orientations and Rotation Curves at 150 pc Resolution}",
      journal = {\apj},
         year = 2020,
        month = jul,
       volume = {897},
       number = {2},
          eid = {122},
        pages = {122},
          doi = {10.3847/1538-4357/ab995310.48550/arXiv.2005.11709},
archivePrefix = {arXiv},
       eprint = {2005.11709},
 primaryClass = {astro-ph.GA},
       adsurl = {https://ui.adsabs.harvard.edu/abs/2020ApJ...897..122L}
}

@ARTICLE{2015ApJ...806...16B,
       author = {{Berg}, Danielle A. and {Skillman}, Evan D. and {Croxall}, Kevin V. and {Pogge}, Richard W. and {Moustakas}, John and {Johnson-Groh}, Mara},
        title = "{CHAOS I. Direct Chemical Abundances for H II Regions in NGC 628}",
      journal = {\apj},
         year = 2015,
        month = jun,
       volume = {806},
       number = {1},
          eid = {16},
        pages = {16},
          doi = {10.1088/0004-637X/806/1/1610.48550/arXiv.1501.02270},
archivePrefix = {arXiv},
       eprint = {1501.02270},
 primaryClass = {astro-ph.GA},
       adsurl = {https://ui.adsabs.harvard.edu/abs/2015ApJ...806...16B}
}

@ARTICLE{2020ApJ...893...96B,
       author = {{Berg}, Danielle A. and {Pogge}, Richard W. and {Skillman}, Evan D. and {Croxall}, Kevin V. and {Moustakas}, John and {Rogers}, Noah S.~J. and {Sun}, Jiayi},
        title = "{CHAOS IV: Gas-phase Abundance Trends from the First Four CHAOS Galaxies}",
      journal = {\apj},
         year = 2020,
        month = apr,
       volume = {893},
       number = {2},
          eid = {96},
        pages = {96},
          doi = {10.3847/1538-4357/ab7eab},
archivePrefix = {arXiv},
       eprint = {2001.05002},
 primaryClass = {astro-ph.GA},
       adsurl = {https://ui.adsabs.harvard.edu/abs/2020ApJ...893...96B}
}

@ARTICLE{2012A&A...541A..10Y,
       author = {{Yamagishi}, M. and {Kaneda}, H. and {Ishihara}, D. and {Kondo}, T. and {Onaka}, T. and {Suzuki}, T. and {Minh}, Y.~C.},
        title = "{AKARI near-infrared spectroscopy of the aromatic and aliphatic hydrocarbon emission features in the galactic superwind of M 82}",
      journal = {\aap},
         year = 2012,
        month = may,
       volume = {541},
          eid = {A10},
        pages = {A10},
          doi = {10.1051/0004-6361/201218904},
archivePrefix = {arXiv},
       eprint = {1203.2794},
 primaryClass = {astro-ph.GA},
       adsurl = {https://ui.adsabs.harvard.edu/abs/2012A&A...541A..10Y}
}

@ARTICLE{2021ApJ...917....3D,
       author = {{Draine}, B.~T. and {Li}, Aigen and {Hensley}, Brandon S. and {Hunt}, L.~K. and {Sandstrom}, K. and {Smith}, J. -D.~T.},
        title = "{Excitation of Polycyclic Aromatic Hydrocarbon Emission: Dependence on Size Distribution, Ionization, and Starlight Spectrum and Intensity}",
      journal = {\apj},
         year = 2021,
        month = aug,
       volume = {917},
       number = {1},
          eid = {3},
        pages = {3},
          doi = {10.3847/1538-4357/abff51},
archivePrefix = {arXiv},
       eprint = {2011.07046},
 primaryClass = {astro-ph.GA},
       adsurl = {https://ui.adsabs.harvard.edu/abs/2021ApJ...917....3D}
}

@ARTICLE{2023ApJ...944L..17L,
       author = {{Lee}, Janice C. and {Sandstrom}, Karin M. and {Leroy}, Adam K. and {Thilker}, David A. and {Schinnerer}, Eva and {Rosolowsky}, Erik and {Larson}, Kirsten L. and {Egorov}, Oleg V. and {Williams}, Thomas G. and {Schmidt}, Judy and {Emsellem}, Eric and {Anand}, Gagandeep S. and {Barnes}, Ashley T. and {Belfiore}, Francesco and {Be{\v{s}}li{\'c}}, Ivana and {Bigiel}, Frank and {Blanc}, Guillermo A. and {Bolatto}, Alberto D. and {Boquien}, M{\'e}d{\'e}ric and {den Brok}, Jakob and {Cao}, Yixian and {Chandar}, Rupali and {Chastenet}, J{\'e}r{\'e}my and {Chevance}, M{\'e}lanie and {Chiang}, I-Da and {Congiu}, Enrico and {Dale}, Daniel A. and {Deger}, Sinan and {Eibensteiner}, Cosima and {Faesi}, Christopher M. and {Glover}, Simon C.~O. and {Grasha}, Kathryn and {Groves}, Brent and {Hassani}, Hamid and {Henny}, Kiana F. and {Henshaw}, Jonathan D. and {Hoyer}, Nils and {Hughes}, Annie and {Jeffreson}, Sarah and {Jim{\'e}nez-Donaire}, Mar{\'\i}a J. and {Kim}, Jaeyeon and {Kim}, Hwihyun and {Klessen}, Ralf S. and {Koch}, Eric W. and {Kreckel}, Kathryn and {Kruijssen}, J.~M. Diederik and {Li}, Jing and {Liu}, Daizhong and {Lopez}, Laura A. and {Maschmann}, Daniel and {Chen}, Ness Mayker and {Meidt}, Sharon E. and {Murphy}, Eric J. and {Neumann}, Justus and {Neumayer}, Nadine and {Pan}, Hsi-An and {Pessa}, Ismael and {Pety}, J{\'e}r{\^o}me and {Querejeta}, Miguel and {Pinna}, Francesca and {Rodr{\'\i}guez}, M. Jimena and {Saito}, Toshiki and {S{\'a}nchez-Bl{\'a}zquez}, Patricia and {Santoro}, Francesco and {Sardone}, Amy and {Smith}, Rowan J. and {Sormani}, Mattia C. and {Scheuermann}, Fabian and {Stuber}, Sophia K. and {Sutter}, Jessica and {Sun}, Jiayi and {Teng}, Yu-Hsuan and {Tre{\ss}}, Robin G. and {Usero}, Antonio and {Watkins}, Elizabeth J. and {Whitmore}, Bradley C. and {Razza}, Alessandro},
        title = "{The PHANGS-JWST Treasury Survey: Star Formation, Feedback, and Dust Physics at High Angular Resolution in Nearby GalaxieS}",
      journal = {\apjl},
         year = 2023,
        month = feb,
       volume = {944},
       number = {2},
          eid = {L17},
        pages = {L17},
          doi = {10.3847/2041-8213/acaaae},
archivePrefix = {arXiv},
       eprint = {2212.02667},
 primaryClass = {astro-ph.GA},
       adsurl = {https://ui.adsabs.harvard.edu/abs/2023ApJ...944L..17L}
}

@ARTICLE{2021ApJ...916...33G,
       author = {{Gordon}, Karl D. and {Misselt}, Karl A. and {Bouwman}, Jeroen and {Clayton}, Geoffrey C. and {Decleir}, Marjorie and {Hines}, Dean C. and {Pendleton}, Yvonne and {Rieke}, George and {Smith}, J.~D.~T. and {Whittet}, D.~C.~B.},
        title = "{Milky Way Mid-Infrared Spitzer Spectroscopic Extinction Curves: Continuum and Silicate Features}",
      journal = {\apj},
         year = 2021,
        month = jul,
       volume = {916},
       number = {1},
          eid = {33},
        pages = {33},
          doi = {10.3847/1538-4357/ac00b7},
archivePrefix = {arXiv},
       eprint = {2105.05087},
 primaryClass = {astro-ph.GA},
       adsurl = {https://ui.adsabs.harvard.edu/abs/2021ApJ...916...33G}
}

@INPROCEEDINGS{2014SPIE.9143E..3XP,
       author = {{Perrin}, Marshall D. and {Sivaramakrishnan}, Anand and {Lajoie}, Charles-Philippe and {Elliott}, Erin and {Pueyo}, Laurent and {Ravindranath}, Swara and {Albert}, Lo{\"\i}c.},
        title = "{Updated point spread function simulations for JWST with WebbPSF}",
    booktitle = {Space Telescopes and Instrumentation 2014: Optical, Infrared, and Millimeter Wave},
         year = 2014,
       editor = {{Oschmann}, Jacobus M., Jr. and {Clampin}, Mark and {Fazio}, Giovanni G. and {MacEwen}, Howard A.},
       series = {Society of Photo-Optical Instrumentation Engineers (SPIE) Conference Series},
       volume = {9143},
        month = aug,
          eid = {91433X},
        pages = {91433X},
          doi = {10.1117/12.2056689},
       adsurl = {https://ui.adsabs.harvard.edu/abs/2014SPIE.9143E..3XP}
}

@ARTICLE{2011PASP..123.1218A,
       author = {{Aniano}, G. and {Draine}, B.~T. and {Gordon}, K.~D. and {Sandstrom}, K.},
        title = "{Common-Resolution Convolution Kernels for Space- and Ground-Based Telescopes}",
      journal = {\pasp},
         year = 2011,
        month = oct,
       volume = {123},
       number = {908},
        pages = {1218},
          doi = {10.1086/662219},
archivePrefix = {arXiv},
       eprint = {1106.5065},
 primaryClass = {astro-ph.IM},
       adsurl = {https://ui.adsabs.harvard.edu/abs/2011PASP..123.1218A}
}

@ARTICLE{2008ApJS..178..247K,
       author = {{Kennicutt}, Jr., Robert C. and {Lee}, Janice C. and {Funes}, Jos{\'e} G. and {J.}, S. and {Sakai}, Shoko and {Akiyama}, Sanae},
        title = "{An H{\ensuremath{\alpha}} Imaging Survey of Galaxies in the Local 11 Mpc Volume}",
      journal = {\apjs},
         year = 2008,
        month = oct,
       volume = {178},
       number = {2},
        pages = {247-279},
          doi = {10.1086/590058},
archivePrefix = {arXiv},
       eprint = {0807.2035},
 primaryClass = {astro-ph},
       adsurl = {https://ui.adsabs.harvard.edu/abs/2008ApJS..178..247K}
}

@ARTICLE{2007ApJ...665.1489K,
       author = {{Kelly}, Brandon C.},
        title = "{Some Aspects of Measurement Error in Linear Regression of Astronomical Data}",
      journal = {\apj},
         year = 2007,
        month = aug,
       volume = {665},
       number = {2},
        pages = {1489-1506},
          doi = {10.1086/519947},
archivePrefix = {arXiv},
       eprint = {0705.2774},
 primaryClass = {astro-ph},
       adsurl = {https://ui.adsabs.harvard.edu/abs/2007ApJ...665.1489K}
}

@ARTICLE{2015A&A...573A..42L,
       author = {{Luridiana}, V. and {Morisset}, C. and {Shaw}, R.~A.},
        title = "{PyNeb: a new tool for analyzing emission lines. I. Code description and validation of results}",
      journal = {\aap},
         year = 2015,
        month = jan,
       volume = {573},
          eid = {A42},
        pages = {A42},
          doi = {10.1051/0004-6361/201323152},
archivePrefix = {arXiv},
       eprint = {1410.6662},
 primaryClass = {astro-ph.IM},
       adsurl = {https://ui.adsabs.harvard.edu/abs/2015A&A...573A..42L}
}

@ARTICLE{2021ApJ...913...37C,
       author = {{Calzetti}, Daniela and {Battisti}, Andrew J. and {Shivaei}, Irene and {Messa}, Matteo and {Cignoni}, Michele and {Adamo}, Angela and {Dale}, Daniel A. and {Gallagher}, John S. and {Grasha}, Kathryn and {Grebel}, Eva K. and {Kennicutt}, Robert C. and {Linden}, Sean T. and {{\"O}stlin}, G{\"o}ran and {Sabbi}, Elena and {Smith}, Linda J. and {Tosi}, Monica and {Wofford}, Aida},
        title = "{Revisiting Attenuation Curves: The Case of NGC 3351}",
      journal = {\apj},
         year = 2021,
        month = may,
       volume = {913},
       number = {1},
          eid = {37},
        pages = {37},
          doi = {10.3847/1538-4357/abf118},
archivePrefix = {arXiv},
       eprint = {2103.12117},
 primaryClass = {astro-ph.GA},
       adsurl = {https://ui.adsabs.harvard.edu/abs/2021ApJ...913...37C}
}

@ARTICLE{2018MNRAS.481.5370M,
       author = {{Maragkoudakis}, A. and {Ivkovich}, N. and {Peeters}, E. and {Stock}, D.~J. and {Hemachandra}, D. and {Tielens}, A.~G.~G.~M.},
        title = "{PAHs and star formation in the H II regions of nearby galaxies M83 and M33}",
      journal = {\mnras},
         year = 2018,
        month = dec,
       volume = {481},
       number = {4},
        pages = {5370-5393},
          doi = {10.1093/mnras/sty2658},
archivePrefix = {arXiv},
       eprint = {1809.10136},
 primaryClass = {astro-ph.GA},
       adsurl = {https://ui.adsabs.harvard.edu/abs/2018MNRAS.481.5370M}
}

@ARTICLE{2014MNRAS.444..757K,
       author = {{Khramtsova}, M.~S. and {Wiebe}, D.~S. and {Lozinskaya}, T.~A. and {Egorov}, O.~V.},
        title = "{Optical and infrared emission of H II complexes as a clue to the PAH life cycle}",
      journal = {\mnras},
         year = 2014,
        month = oct,
       volume = {444},
       number = {1},
        pages = {757-775},
          doi = {10.1093/mnras/stu1482},
archivePrefix = {arXiv},
       eprint = {1407.8307},
 primaryClass = {astro-ph.SR},
       adsurl = {https://ui.adsabs.harvard.edu/abs/2014MNRAS.444..757K}
}

@ARTICLE{2023ApJ...944L..16E,
       author = {{Egorov}, Oleg V. and {Kreckel}, Kathryn and {Sandstrom}, Karin M. and {Leroy}, Adam K. and {Glover}, Simon C.~O. and {Groves}, Brent and {Kruijssen}, J.~M. Diederik and {Barnes}, Ashley. T. and {Belfiore}, Francesco and {Bigiel}, F. and {Blanc}, Guillermo A. and {Boquien}, M{\'e}d{\'e}ric and {Cao}, Yixian and {Chastenet}, J{\'e}r{\'e}my and {Chevance}, M{\'e}lanie and {Congiu}, Enrico and {Dale}, Daniel A. and {Emsellem}, Eric and {Grasha}, Kathryn and {Klessen}, Ralf S. and {Larson}, Kirsten L. and {Liu}, Daizhong and {Murphy}, Eric J. and {Pan}, Hsi-An and {Pessa}, Ismael and {Pety}, J{\'e}r{\^o}me and {Rosolowsky}, Erik and {Scheuermann}, Fabian and {Schinnerer}, Eva and {Sutter}, Jessica and {Thilker}, David A. and {Watkins}, Elizabeth J. and {Williams}, Thomas G.},
        title = "{PHANGS-JWST First Results: Destruction of the PAH Molecules in H II Regions Probed by JWST and MUSE}",
      journal = {\apjl},
         year = 2023,
        month = feb,
       volume = {944},
       number = {2},
          eid = {L16},
        pages = {L16},
          doi = {10.3847/2041-8213/acac92},
archivePrefix = {arXiv},
       eprint = {2212.09159},
 primaryClass = {astro-ph.GA},
       adsurl = {https://ui.adsabs.harvard.edu/abs/2023ApJ...944L..16E}
}

@ARTICLE{2005ApJ...633..871C,
       author = {{Calzetti}, D. and {Kennicutt}, R.~C., Jr. and {Bianchi}, L. and {Thilker}, D.~A. and {Dale}, D.~A. and {Engelbracht}, C.~W. and {Leitherer}, C. and {Meyer}, M.~J. and {Sosey}, M.~L. and {Mutchler}, M. and {Regan}, M.~W. and {Thornley}, M.~D. and {Armus}, L. and {Bendo}, G.~J. and {Boissier}, S. and {Boselli}, A. and {Draine}, B.~T. and {Gordon}, K.~D. and {Helou}, G. and {Hollenbach}, D.~J. and {Kewley}, L. and {Madore}, B.~F. and {Martin}, D.~C. and {Murphy}, E.~J. and {Rieke}, G.~H. and {Rieke}, M.~J. and {Roussel}, H. and {Sheth}, K. and {Smith}, J.~D. and {Walter}, F. and {White}, B.~A. and {Yi}, S. and {Scoville}, N.~Z. and {Polletta}, M. and {Lindler}, D.},
        title = "{Star Formation in NGC 5194 (M51a): The Panchromatic View from GALEX to Spitzer}",
      journal = {\apj},
         year = 2005,
        month = nov,
       volume = {633},
       number = {2},
        pages = {871-893},
          doi = {10.1086/466518},
archivePrefix = {arXiv},
       eprint = {astro-ph/0507427},
 primaryClass = {astro-ph},
       adsurl = {https://ui.adsabs.harvard.edu/abs/2005ApJ...633..871C}
}

@ARTICLE{2023ApJ...944L...7S,
       author = {{Sandstrom}, Karin M. and {Chastenet}, J{\'e}r{\'e}my and {Sutter}, Jessica and {Leroy}, Adam K. and {Egorov}, Oleg V. and {Williams}, Thomas G. and {Bolatto}, Alberto D. and {Boquien}, M{\'e}d{\'e}ric and {Cao}, Yixian and {Dale}, Daniel A. and {Lee}, Janice C. and {Rosolowsky}, Erik and {Schinnerer}, Eva and {Barnes}, Ashley. T. and {Belfiore}, Francesco and {Bigiel}, F. and {Chevance}, M{\'e}lanie and {Grasha}, Kathryn and {Groves}, Brent and {Hassani}, Hamid and {Hughes}, Annie and {Klessen}, Ralf S. and {Kruijssen}, J.~M. Diederik and {Larson}, Kirsten L. and {Liu}, Daizhong and {Lopez}, Laura A. and {Meidt}, Sharon E. and {Murphy}, Eric J. and {Sormani}, Mattia C. and {Thilker}, David A. and {Watkins}, Elizabeth J.},
        title = "{PHANGS-JWST First Results: Mapping the 3.3 {\ensuremath{\mu}}m Polycyclic Aromatic Hydrocarbon Vibrational Band in Nearby Galaxies with NIRCam Medium Bands}",
      journal = {\apjl},
         year = 2023,
        month = feb,
       volume = {944},
       number = {2},
          eid = {L7},
        pages = {L7},
          doi = {10.3847/2041-8213/acb0cf},
archivePrefix = {arXiv},
       eprint = {2301.00854},
 primaryClass = {astro-ph.GA},
       adsurl = {https://ui.adsabs.harvard.edu/abs/2023ApJ...944L...7S}
}

@ARTICLE{2017ApJ...841...92R,
       author = {{Ryon}, J.~E. and {Gallagher}, J.~S. and {Smith}, L.~J. and {Adamo}, A. and {Calzetti}, D. and {Bright}, S.~N. and {Cignoni}, M. and {Cook}, D.~O. and {Dale}, D.~A. and {Elmegreen}, B.~E. and {Fumagalli}, M. and {Gouliermis}, D.~A. and {Grasha}, K. and {Grebel}, E.~K. and {Kim}, H. and {Messa}, M. and {Thilker}, D. and {Ubeda}, L.},
        title = "{Effective Radii of Young, Massive Star Clusters in Two LEGUS Galaxies}",
      journal = {\apj},
         year = 2017,
        month = jun,
       volume = {841},
       number = {2},
          eid = {92},
        pages = {92},
          doi = {10.3847/1538-4357/aa719e},
archivePrefix = {arXiv},
       eprint = {1705.02692},
 primaryClass = {astro-ph.GA},
       adsurl = {https://ui.adsabs.harvard.edu/abs/2017ApJ...841...92R}
}

@ARTICLE{2016A&A...595A...1G,
       author = {{Gaia Collaboration} and {Prusti}, T. and {de Bruijne}, J.~H.~J. and {Brown}, A.~G.~A. and {Vallenari}, A. and {Babusiaux}, C. and {Bailer-Jones}, C.~A.~L. and {Bastian}, U. and {Biermann}, M. and {Evans}, D.~W. and {Eyer}, L. and {Jansen}, F. and {Jordi}, C. and {Klioner}, S.~A. and {Lammers}, U. and {Lindegren}, L. and {Luri}, X. and {Mignard}, F. and {Milligan}, D.~J. and {Panem}, C. and {Poinsignon}, V. and {Pourbaix}, D. and {Randich}, S. and {Sarri}, G. and {Sartoretti}, P. and {Siddiqui}, H.~I. and {Soubiran}, C. and {Valette}, V. and {van Leeuwen}, F. and {Walton}, N.~A. and {Aerts}, C. and {Arenou}, F. and {Cropper}, M. and {Drimmel}, R. and {H{\o}g}, E. and {Katz}, D. and {Lattanzi}, M.~G. and {O'Mullane}, W. and {Grebel}, E.~K. and {Holland}, A.~D. and {Huc}, C. and {Passot}, X. and {Bramante}, L. and {Cacciari}, C. and {Casta{\~n}eda}, J. and {Chaoul}, L. and {Cheek}, N. and {De Angeli}, F. and {Fabricius}, C. and {Guerra}, R. and {Hern{\'a}ndez}, J. and {Jean-Antoine-Piccolo}, A. and {Masana}, E. and {Messineo}, R. and {Mowlavi}, N. and {Nienartowicz}, K. and {Ord{\'o}{\~n}ez-Blanco}, D. and {Panuzzo}, P. and {Portell}, J. and {Richards}, P.~J. and {Riello}, M. and {Seabroke}, G.~M. and {Tanga}, P. and {Th{\'e}venin}, F. and {Torra}, J. and {Els}, S.~G. and {Gracia-Abril}, G. and {Comoretto}, G. and {Garcia-Reinaldos}, M. and {Lock}, T. and {Mercier}, E. and {Altmann}, M. and {Andrae}, R. and {Astraatmadja}, T.~L. and {Bellas-Velidis}, I. and {Benson}, K. and {Berthier}, J. and {Blomme}, R. and {Busso}, G. and {Carry}, B. and {Cellino}, A. and {Clementini}, G. and {Cowell}, S. and {Creevey}, O. and {Cuypers}, J. and {Davidson}, M. and {De Ridder}, J. and {de Torres}, A. and {Delchambre}, L. and {Dell'Oro}, A. and {Ducourant}, C. and {Fr{\'e}mat}, Y. and {Garc{\'\i}a-Torres}, M. and {Gosset}, E. and {Halbwachs}, J. -L. and {Hambly}, N.~C. and {Harrison}, D.~L. and {Hauser}, M. and {Hestroffer}, D. and {Hodgkin}, S.~T. and {Huckle}, H.~E. and {Hutton}, A. and {Jasniewicz}, G. and {Jordan}, S. and {Kontizas}, M. and {Korn}, A.~J. and {Lanzafame}, A.~C. and {Manteiga}, M. and {Moitinho}, A. and {Muinonen}, K. and {Osinde}, J. and {Pancino}, E. and {Pauwels}, T. and {Petit}, J. -M. and {Recio-Blanco}, A. and {Robin}, A.~C. and {Sarro}, L.~M. and {Siopis}, C. and {Smith}, M. and {Smith}, K.~W. and {Sozzetti}, A. and {Thuillot}, W. and {van Reeven}, W. and {Viala}, Y. and {Abbas}, U. and {Abreu Aramburu}, A. and {Accart}, S. and {Aguado}, J.~J. and {Allan}, P.~M. and {Allasia}, W. and {Altavilla}, G. and {{\'A}lvarez}, M.~A. and {Alves}, J. and {Anderson}, R.~I. and {Andrei}, A.~H. and {Anglada Varela}, E. and {Antiche}, E. and {Antoja}, T. and {Ant{\'o}n}, S. and {Arcay}, B. and {Atzei}, A. and {Ayache}, L. and {Bach}, N. and {Baker}, S.~G. and {Balaguer-N{\'u}{\~n}ez}, L. and {Barache}, C. and {Barata}, C. and {Barbier}, A. and {Barblan}, F. and {Baroni}, M. and {Barrado y Navascu{\'e}s}, D. and {Barros}, M. and {Barstow}, M.~A. and {Becciani}, U. and {Bellazzini}, M. and {Bellei}, G. and {Bello Garc{\'\i}a}, A. and {Belokurov}, V. and {Bendjoya}, P. and {Berihuete}, A. and {Bianchi}, L. and {Bienaym{\'e}}, O. and {Billebaud}, F. and {Blagorodnova}, N. and {Blanco-Cuaresma}, S. and {Boch}, T. and {Bombrun}, A. and {Borrachero}, R. and {Bouquillon}, S. and {Bourda}, G. and {Bouy}, H. and {Bragaglia}, A. and {Breddels}, M.~A. and {Brouillet}, N. and {Br{\"u}semeister}, T. and {Bucciarelli}, B. and {Budnik}, F. and {Burgess}, P. and {Burgon}, R. and {Burlacu}, A. and {Busonero}, D. and {Buzzi}, R. and {Caffau}, E. and {Cambras}, J. and {Campbell}, H. and {Cancelliere}, R. and {Cantat-Gaudin}, T. and {Carlucci}, T. and {Carrasco}, J.~M. and {Castellani}, M. and {Charlot}, P. and {Charnas}, J. and {Charvet}, P. and {Chassat}, F. and {Chiavassa}, A. and {Clotet}, M. and {Cocozza}, G. and {Collins}, R.~S. and {Collins}, P. and {Costigan}, G. and {Crifo}, F. and {Cross}, N.~J.~G. and {Crosta}, M. and {Crowley}, C. and {Dafonte}, C. and {Damerdji}, Y. and {Dapergolas}, A. and {David}, P. and {David}, M. and {De Cat}, P. and {de Felice}, F. and {de Laverny}, P. and {De Luise}, F. and {De March}, R. and {de Martino}, D. and {de Souza}, R. and {Debosscher}, J. and {del Pozo}, E. and {Delbo}, M. and {Delgado}, A. and {Delgado}, H.~E. and {di Marco}, F. and {Di Matteo}, P. and {Diakite}, S. and {Distefano}, E. and {Dolding}, C. and {Dos Anjos}, S. and {Drazinos}, P. and {Dur{\'a}n}, J. and {Dzigan}, Y. and {Ecale}, E. and {Edvardsson}, B. and {Enke}, H. and {Erdmann}, M. and {Escolar}, D. and {Espina}, M. and {Evans}, N.~W. and {Eynard Bontemps}, G. and {Fabre}, C. and {Fabrizio}, M. and {Faigler}, S. and {Falc{\~a}o}, A.~J. and {Farr{\`a}s Casas}, M. and {Faye}, F. and {Federici}, L. and {Fedorets}, G. and {Fern{\'a}ndez-Hern{\'a}ndez}, J. and {Fernique}, P. and {Fienga}, A. and {Figueras}, F. and {Filippi}, F. and {Findeisen}, K. and {Fonti}, A. and {Fouesneau}, M. and {Fraile}, E. and {Fraser}, M. and {Fuchs}, J. and {Furnell}, R. and {Gai}, M. and {Galleti}, S. and {Galluccio}, L. and {Garabato}, D. and {Garc{\'\i}a-Sedano}, F. and {Gar{\'e}}, P. and {Garofalo}, A. and {Garralda}, N. and {Gavras}, P. and {Gerssen}, J. and {Geyer}, R. and {Gilmore}, G. and {Girona}, S. and {Giuffrida}, G. and {Gomes}, M. and {Gonz{\'a}lez-Marcos}, A. and {Gonz{\'a}lez-N{\'u}{\~n}ez}, J. and {Gonz{\'a}lez-Vidal}, J.~J. and {Granvik}, M. and {Guerrier}, A. and {Guillout}, P. and {Guiraud}, J. and {G{\'u}rpide}, A. and {Guti{\'e}rrez-S{\'a}nchez}, R. and {Guy}, L.~P. and {Haigron}, R. and {Hatzidimitriou}, D. and {Haywood}, M. and {Heiter}, U. and {Helmi}, A. and {Hobbs}, D. and {Hofmann}, W. and {Holl}, B. and {Holland}, G. and {Hunt}, J.~A.~S. and {Hypki}, A. and {Icardi}, V. and {Irwin}, M. and {Jevardat de Fombelle}, G. and {Jofr{\'e}}, P. and {Jonker}, P.~G. and {Jorissen}, A. and {Julbe}, F. and {Karampelas}, A. and {Kochoska}, A. and {Kohley}, R. and {Kolenberg}, K. and {Kontizas}, E. and {Koposov}, S.~E. and {Kordopatis}, G. and {Koubsky}, P. and {Kowalczyk}, A. and {Krone-Martins}, A. and {Kudryashova}, M. and {Kull}, I. and {Bachchan}, R.~K. and {Lacoste-Seris}, F. and {Lanza}, A.~F. and {Lavigne}, J. -B. and {Le Poncin-Lafitte}, C. and {Lebreton}, Y. and {Lebzelter}, T. and {Leccia}, S. and {Leclerc}, N. and {Lecoeur-Taibi}, I. and {Lemaitre}, V. and {Lenhardt}, H. and {Leroux}, F. and {Liao}, S. and {Licata}, E. and {Lindstr{\o}m}, H.~E.~P. and {Lister}, T.~A. and {Livanou}, E. and {Lobel}, A. and {L{\"o}ffler}, W. and {L{\'o}pez}, M. and {Lopez-Lozano}, A. and {Lorenz}, D. and {Loureiro}, T. and {MacDonald}, I. and {Magalh{\~a}es Fernandes}, T. and {Managau}, S. and {Mann}, R.~G. and {Mantelet}, G. and {Marchal}, O. and {Marchant}, J.~M. and {Marconi}, M. and {Marie}, J. and {Marinoni}, S. and {Marrese}, P.~M. and {Marschalk{\'o}}, G. and {Marshall}, D.~J. and {Mart{\'\i}n-Fleitas}, J.~M. and {Martino}, M. and {Mary}, N. and {Matijevi{\v{c}}}, G. and {Mazeh}, T. and {McMillan}, P.~J. and {Messina}, S. and {Mestre}, A. and {Michalik}, D. and {Millar}, N.~R. and {Miranda}, B.~M.~H. and {Molina}, D. and {Molinaro}, R. and {Molinaro}, M. and {Moln{\'a}r}, L. and {Moniez}, M. and {Montegriffo}, P. and {Monteiro}, D. and {Mor}, R. and {Mora}, A. and {Morbidelli}, R. and {Morel}, T. and {Morgenthaler}, S. and {Morley}, T. and {Morris}, D. and {Mulone}, A.~F. and {Muraveva}, T. and {Musella}, I. and {Narbonne}, J. and {Nelemans}, G. and {Nicastro}, L. and {Noval}, L. and {Ord{\'e}novic}, C. and {Ordieres-Mer{\'e}}, J. and {Osborne}, P. and {Pagani}, C. and {Pagano}, I. and {Pailler}, F. and {Palacin}, H. and {Palaversa}, L. and {Parsons}, P. and {Paulsen}, T. and {Pecoraro}, M. and {Pedrosa}, R. and {Pentik{\"a}inen}, H. and {Pereira}, J. and {Pichon}, B. and {Piersimoni}, A.~M. and {Pineau}, F. -X. and {Plachy}, E. and {Plum}, G. and {Poujoulet}, E. and {Pr{\v{s}}a}, A. and {Pulone}, L. and {Ragaini}, S. and {Rago}, S. and {Rambaux}, N. and {Ramos-Lerate}, M. and {Ranalli}, P. and {Rauw}, G. and {Read}, A. and {Regibo}, S. and {Renk}, F. and {Reyl{\'e}}, C. and {Ribeiro}, R.~A. and {Rimoldini}, L. and {Ripepi}, V. and {Riva}, A. and {Rixon}, G. and {Roelens}, M. and {Romero-G{\'o}mez}, M. and {Rowell}, N. and {Royer}, F. and {Rudolph}, A. and {Ruiz-Dern}, L. and {Sadowski}, G. and {Sagrist{\`a} Sell{\'e}s}, T. and {Sahlmann}, J. and {Salgado}, J. and {Salguero}, E. and {Sarasso}, M. and {Savietto}, H. and {Schnorhk}, A. and {Schultheis}, M. and {Sciacca}, E. and {Segol}, M. and {Segovia}, J.~C. and {Segransan}, D. and {Serpell}, E. and {Shih}, I. -C. and {Smareglia}, R. and {Smart}, R.~L. and {Smith}, C. and {Solano}, E. and {Solitro}, F. and {Sordo}, R. and {Soria Nieto}, S. and {Souchay}, J. and {Spagna}, A. and {Spoto}, F. and {Stampa}, U. and {Steele}, I.~A. and {Steidelm{\"u}ller}, H. and {Stephenson}, C.~A. and {Stoev}, H. and {Suess}, F.~F. and {S{\"u}veges}, M. and {Surdej}, J. and {Szabados}, L. and {Szegedi-Elek}, E. and {Tapiador}, D. and {Taris}, F. and {Tauran}, G. and {Taylor}, M.~B. and {Teixeira}, R. and {Terrett}, D. and {Tingley}, B. and {Trager}, S.~C. and {Turon}, C. and {Ulla}, A. and {Utrilla}, E. and {Valentini}, G. and {van Elteren}, A. and {Van Hemelryck}, E. and {van Leeuwen}, M. and {Varadi}, M. and {Vecchiato}, A. and {Veljanoski}, J. and {Via}, T. and {Vicente}, D. and {Vogt}, S. and {Voss}, H. and {Votruba}, V. and {Voutsinas}, S. and {Walmsley}, G. and {Weiler}, M. and {Weingrill}, K. and {Werner}, D. and {Wevers}, T. and {Whitehead}, G. and {Wyrzykowski}, {\L}. and {Yoldas}, A. and {{\v{Z}}erjal}, M. and {Zucker}, S. and {Zurbach}, C. and {Zwitter}, T. and {Alecu}, A. and {Allen}, M. and {Allende Prieto}, C. and {Amorim}, A. and {Anglada-Escud{\'e}}, G. and {Arsenijevic}, V. and {Azaz}, S. and {Balm}, P. and {Beck}, M. and {Bernstein}, H. -H. and {Bigot}, L. and {Bijaoui}, A. and {Blasco}, C. and {Bonfigli}, M. and {Bono}, G. and {Boudreault}, S. and {Bressan}, A. and {Brown}, S. and {Brunet}, P. -M. and {Bunclark}, P. and {Buonanno}, R. and {Butkevich}, A.~G. and {Carret}, C. and {Carrion}, C. and {Chemin}, L. and {Ch{\'e}reau}, F. and {Corcione}, L. and {Darmigny}, E. and {de Boer}, K.~S. and {de Teodoro}, P. and {de Zeeuw}, P.~T. and {Delle Luche}, C. and {Domingues}, C.~D. and {Dubath}, P. and {Fodor}, F. and {Fr{\'e}zouls}, B. and {Fries}, A. and {Fustes}, D. and {Fyfe}, D. and {Gallardo}, E. and {Gallegos}, J. and {Gardiol}, D. and {Gebran}, M. and {Gomboc}, A. and {G{\'o}mez}, A. and {Grux}, E. and {Gueguen}, A. and {Heyrovsky}, A. and {Hoar}, J. and {Iannicola}, G. and {Isasi Parache}, Y. and {Janotto}, A. -M. and {Joliet}, E. and {Jonckheere}, A. and {Keil}, R. and {Kim}, D. -W. and {Klagyivik}, P. and {Klar}, J. and {Knude}, J. and {Kochukhov}, O. and {Kolka}, I. and {Kos}, J. and {Kutka}, A. and {Lainey}, V. and {LeBouquin}, D. and {Liu}, C. and {Loreggia}, D. and {Makarov}, V.~V. and {Marseille}, M.~G. and {Martayan}, C. and {Martinez-Rubi}, O. and {Massart}, B. and {Meynadier}, F. and {Mignot}, S. and {Munari}, U. and {Nguyen}, A. -T. and {Nordlander}, T. and {Ocvirk}, P. and {O'Flaherty}, K.~S. and {Olias Sanz}, A. and {Ortiz}, P. and {Osorio}, J. and {Oszkiewicz}, D. and {Ouzounis}, A. and {Palmer}, M. and {Park}, P. and {Pasquato}, E. and {Peltzer}, C. and {Peralta}, J. and {P{\'e}turaud}, F. and {Pieniluoma}, T. and {Pigozzi}, E. and {Poels}, J. and {Prat}, G. and {Prod'homme}, T. and {Raison}, F. and {Rebordao}, J.~M. and {Risquez}, D. and {Rocca-Volmerange}, B. and {Rosen}, S. and {Ruiz-Fuertes}, M.~I. and {Russo}, F. and {Sembay}, S. and {Serraller Vizcaino}, I. and {Short}, A. and {Siebert}, A. and {Silva}, H. and {Sinachopoulos}, D. and {Slezak}, E. and {Soffel}, M. and {Sosnowska}, D. and {Strai{\v{z}}ys}, V. and {ter Linden}, M. and {Terrell}, D. and {Theil}, S. and {Tiede}, C. and {Troisi}, L. and {Tsalmantza}, P. and {Tur}, D. and {Vaccari}, M. and {Vachier}, F. and {Valles}, P. and {Van Hamme}, W. and {Veltz}, L. and {Virtanen}, J. and {Wallut}, J. -M. and {Wichmann}, R. and {Wilkinson}, M.~I. and {Ziaeepour}, H. and {Zschocke}, S.},
        title = "{The Gaia mission}",
      journal = {\aap},
         year = 2016,
        month = nov,
       volume = {595},
          eid = {A1},
        pages = {A1},
          doi = {10.1051/0004-6361/201629272},
archivePrefix = {arXiv},
       eprint = {1609.04153},
 primaryClass = {astro-ph.IM},
       adsurl = {https://ui.adsabs.harvard.edu/abs/2016A&A...595A...1G}
}

@ARTICLE{2023A&A...674A...1G,
       author = {{Gaia Collaboration} and {Vallenari}, A. and {Brown}, A.~G.~A. and {Prusti}, T. and {de Bruijne}, J.~H.~J. and {Arenou}, F. and {Babusiaux}, C. and {Biermann}, M. and {Creevey}, O.~L. and {Ducourant}, C. and {Evans}, D.~W. and {Eyer}, L. and {Guerra}, R. and {Hutton}, A. and {Jordi}, C. and {Klioner}, S.~A. and {Lammers}, U.~L. and {Lindegren}, L. and {Luri}, X. and {Mignard}, F. and {Panem}, C. and {Pourbaix}, D. and {Randich}, S. and {Sartoretti}, P. and {Soubiran}, C. and {Tanga}, P. and {Walton}, N.~A. and {Bailer-Jones}, C.~A.~L. and {Bastian}, U. and {Drimmel}, R. and {Jansen}, F. and {Katz}, D. and {Lattanzi}, M.~G. and {van Leeuwen}, F. and {Bakker}, J. and {Cacciari}, C. and {Casta{\~n}eda}, J. and {De Angeli}, F. and {Fabricius}, C. and {Fouesneau}, M. and {Fr{\'e}mat}, Y. and {Galluccio}, L. and {Guerrier}, A. and {Heiter}, U. and {Masana}, E. and {Messineo}, R. and {Mowlavi}, N. and {Nicolas}, C. and {Nienartowicz}, K. and {Pailler}, F. and {Panuzzo}, P. and {Riclet}, F. and {Roux}, W. and {Seabroke}, G.~M. and {Sordo}, R. and {Th{\'e}venin}, F. and {Gracia-Abril}, G. and {Portell}, J. and {Teyssier}, D. and {Altmann}, M. and {Andrae}, R. and {Audard}, M. and {Bellas-Velidis}, I. and {Benson}, K. and {Berthier}, J. and {Blomme}, R. and {Burgess}, P.~W. and {Busonero}, D. and {Busso}, G. and {C{\'a}novas}, H. and {Carry}, B. and {Cellino}, A. and {Cheek}, N. and {Clementini}, G. and {Damerdji}, Y. and {Davidson}, M. and {de Teodoro}, P. and {Nu{\~n}ez Campos}, M. and {Delchambre}, L. and {Dell'Oro}, A. and {Esquej}, P. and {Fern{\'a}ndez-Hern{\'a}ndez}, J. and {Fraile}, E. and {Garabato}, D. and {Garc{\'\i}a-Lario}, P. and {Gosset}, E. and {Haigron}, R. and {Halbwachs}, J. -L. and {Hambly}, N.~C. and {Harrison}, D.~L. and {Hern{\'a}ndez}, J. and {Hestroffer}, D. and {Hodgkin}, S.~T. and {Holl}, B. and {Jan{\ss}en}, K. and {Jevardat de Fombelle}, G. and {Jordan}, S. and {Krone-Martins}, A. and {Lanzafame}, A.~C. and {L{\"o}ffler}, W. and {Marchal}, O. and {Marrese}, P.~M. and {Moitinho}, A. and {Muinonen}, K. and {Osborne}, P. and {Pancino}, E. and {Pauwels}, T. and {Recio-Blanco}, A. and {Reyl{\'e}}, C. and {Riello}, M. and {Rimoldini}, L. and {Roegiers}, T. and {Rybizki}, J. and {Sarro}, L.~M. and {Siopis}, C. and {Smith}, M. and {Sozzetti}, A. and {Utrilla}, E. and {van Leeuwen}, M. and {Abbas}, U. and {{\'A}brah{\'a}m}, P. and {Abreu Aramburu}, A. and {Aerts}, C. and {Aguado}, J.~J. and {Ajaj}, M. and {Aldea-Montero}, F. and {Altavilla}, G. and {{\'A}lvarez}, M.~A. and {Alves}, J. and {Anders}, F. and {Anderson}, R.~I. and {Anglada Varela}, E. and {Antoja}, T. and {Baines}, D. and {Baker}, S.~G. and {Balaguer-N{\'u}{\~n}ez}, L. and {Balbinot}, E. and {Balog}, Z. and {Barache}, C. and {Barbato}, D. and {Barros}, M. and {Barstow}, M.~A. and {Bartolom{\'e}}, S. and {Bassilana}, J. -L. and {Bauchet}, N. and {Becciani}, U. and {Bellazzini}, M. and {Berihuete}, A. and {Bernet}, M. and {Bertone}, S. and {Bianchi}, L. and {Binnenfeld}, A. and {Blanco-Cuaresma}, S. and {Blazere}, A. and {Boch}, T. and {Bombrun}, A. and {Bossini}, D. and {Bouquillon}, S. and {Bragaglia}, A. and {Bramante}, L. and {Breedt}, E. and {Bressan}, A. and {Brouillet}, N. and {Brugaletta}, E. and {Bucciarelli}, B. and {Burlacu}, A. and {Butkevich}, A.~G. and {Buzzi}, R. and {Caffau}, E. and {Cancelliere}, R. and {Cantat-Gaudin}, T. and {Carballo}, R. and {Carlucci}, T. and {Carnerero}, M.~I. and {Carrasco}, J.~M. and {Casamiquela}, L. and {Castellani}, M. and {Castro-Ginard}, A. and {Chaoul}, L. and {Charlot}, P. and {Chemin}, L. and {Chiaramida}, V. and {Chiavassa}, A. and {Chornay}, N. and {Comoretto}, G. and {Contursi}, G. and {Cooper}, W.~J. and {Cornez}, T. and {Cowell}, S. and {Crifo}, F. and {Cropper}, M. and {Crosta}, M. and {Crowley}, C. and {Dafonte}, C. and {Dapergolas}, A. and {David}, M. and {David}, P. and {de Laverny}, P. and {De Luise}, F. and {De March}, R. and {De Ridder}, J. and {de Souza}, R. and {de Torres}, A. and {del Peloso}, E.~F. and {del Pozo}, E. and {Delbo}, M. and {Delgado}, A. and {Delisle}, J. -B. and {Demouchy}, C. and {Dharmawardena}, T.~E. and {Di Matteo}, P. and {Diakite}, S. and {Diener}, C. and {Distefano}, E. and {Dolding}, C. and {Edvardsson}, B. and {Enke}, H. and {Fabre}, C. and {Fabrizio}, M. and {Faigler}, S. and {Fedorets}, G. and {Fernique}, P. and {Fienga}, A. and {Figueras}, F. and {Fournier}, Y. and {Fouron}, C. and {Fragkoudi}, F. and {Gai}, M. and {Garcia-Gutierrez}, A. and {Garcia-Reinaldos}, M. and {Garc{\'\i}a-Torres}, M. and {Garofalo}, A. and {Gavel}, A. and {Gavras}, P. and {Gerlach}, E. and {Geyer}, R. and {Giacobbe}, P. and {Gilmore}, G. and {Girona}, S. and {Giuffrida}, G. and {Gomel}, R. and {Gomez}, A. and {Gonz{\'a}lez-N{\'u}{\~n}ez}, J. and {Gonz{\'a}lez-Santamar{\'\i}a}, I. and {Gonz{\'a}lez-Vidal}, J.~J. and {Granvik}, M. and {Guillout}, P. and {Guiraud}, J. and {Guti{\'e}rrez-S{\'a}nchez}, R. and {Guy}, L.~P. and {Hatzidimitriou}, D. and {Hauser}, M. and {Haywood}, M. and {Helmer}, A. and {Helmi}, A. and {Sarmiento}, M.~H. and {Hidalgo}, S.~L. and {Hilger}, T. and {H{\l}adczuk}, N. and {Hobbs}, D. and {Holland}, G. and {Huckle}, H.~E. and {Jardine}, K. and {Jasniewicz}, G. and {Jean-Antoine Piccolo}, A. and {Jim{\'e}nez-Arranz}, {\'O}. and {Jorissen}, A. and {Juaristi Campillo}, J. and {Julbe}, F. and {Karbevska}, L. and {Kervella}, P. and {Khanna}, S. and {Kontizas}, M. and {Kordopatis}, G. and {Korn}, A.~J. and {K{\'o}sp{\'a}l}, {\'A}. and {Kostrzewa-Rutkowska}, Z. and {Kruszy{\'n}ska}, K. and {Kun}, M. and {Laizeau}, P. and {Lambert}, S. and {Lanza}, A.~F. and {Lasne}, Y. and {Le Campion}, J. -F. and {Lebreton}, Y. and {Lebzelter}, T. and {Leccia}, S. and {Leclerc}, N. and {Lecoeur-Taibi}, I. and {Liao}, S. and {Licata}, E.~L. and {Lindstr{\o}m}, H.~E.~P. and {Lister}, T.~A. and {Livanou}, E. and {Lobel}, A. and {Lorca}, A. and {Loup}, C. and {Madrero Pardo}, P. and {Magdaleno Romeo}, A. and {Managau}, S. and {Mann}, R.~G. and {Manteiga}, M. and {Marchant}, J.~M. and {Marconi}, M. and {Marcos}, J. and {Marcos Santos}, M.~M.~S. and {Mar{\'\i}n Pina}, D. and {Marinoni}, S. and {Marocco}, F. and {Marshall}, D.~J. and {Martin Polo}, L. and {Mart{\'\i}n-Fleitas}, J.~M. and {Marton}, G. and {Mary}, N. and {Masip}, A. and {Massari}, D. and {Mastrobuono-Battisti}, A. and {Mazeh}, T. and {McMillan}, P.~J. and {Messina}, S. and {Michalik}, D. and {Millar}, N.~R. and {Mints}, A. and {Molina}, D. and {Molinaro}, R. and {Moln{\'a}r}, L. and {Monari}, G. and {Mongui{\'o}}, M. and {Montegriffo}, P. and {Montero}, A. and {Mor}, R. and {Mora}, A. and {Morbidelli}, R. and {Morel}, T. and {Morris}, D. and {Muraveva}, T. and {Murphy}, C.~P. and {Musella}, I. and {Nagy}, Z. and {Noval}, L. and {Oca{\~n}a}, F. and {Ogden}, A. and {Ordenovic}, C. and {Osinde}, J.~O. and {Pagani}, C. and {Pagano}, I. and {Palaversa}, L. and {Palicio}, P.~A. and {Pallas-Quintela}, L. and {Panahi}, A. and {Payne-Wardenaar}, S. and {Pe{\~n}alosa Esteller}, X. and {Penttil{\"a}}, A. and {Pichon}, B. and {Piersimoni}, A.~M. and {Pineau}, F. -X. and {Plachy}, E. and {Plum}, G. and {Poggio}, E. and {Pr{\v{s}}a}, A. and {Pulone}, L. and {Racero}, E. and {Ragaini}, S. and {Rainer}, M. and {Raiteri}, C.~M. and {Rambaux}, N. and {Ramos}, P. and {Ramos-Lerate}, M. and {Re Fiorentin}, P. and {Regibo}, S. and {Richards}, P.~J. and {Rios Diaz}, C. and {Ripepi}, V. and {Riva}, A. and {Rix}, H. -W. and {Rixon}, G. and {Robichon}, N. and {Robin}, A.~C. and {Robin}, C. and {Roelens}, M. and {Rogues}, H.~R.~O. and {Rohrbasser}, L. and {Romero-G{\'o}mez}, M. and {Rowell}, N. and {Royer}, F. and {Ruz Mieres}, D. and {Rybicki}, K.~A. and {Sadowski}, G. and {S{\'a}ez N{\'u}{\~n}ez}, A. and {Sagrist{\`a} Sell{\'e}s}, A. and {Sahlmann}, J. and {Salguero}, E. and {Samaras}, N. and {Sanchez Gimenez}, V. and {Sanna}, N. and {Santove{\~n}a}, R. and {Sarasso}, M. and {Schultheis}, M. and {Sciacca}, E. and {Segol}, M. and {Segovia}, J.~C. and {S{\'e}gransan}, D. and {Semeux}, D. and {Shahaf}, S. and {Siddiqui}, H.~I. and {Siebert}, A. and {Siltala}, L. and {Silvelo}, A. and {Slezak}, E. and {Slezak}, I. and {Smart}, R.~L. and {Snaith}, O.~N. and {Solano}, E. and {Solitro}, F. and {Souami}, D. and {Souchay}, J. and {Spagna}, A. and {Spina}, L. and {Spoto}, F. and {Steele}, I.~A. and {Steidelm{\"u}ller}, H. and {Stephenson}, C.~A. and {S{\"u}veges}, M. and {Surdej}, J. and {Szabados}, L. and {Szegedi-Elek}, E. and {Taris}, F. and {Taylor}, M.~B. and {Teixeira}, R. and {Tolomei}, L. and {Tonello}, N. and {Torra}, F. and {Torra}, J. and {Torralba Elipe}, G. and {Trabucchi}, M. and {Tsounis}, A.~T. and {Turon}, C. and {Ulla}, A. and {Unger}, N. and {Vaillant}, M.~V. and {van Dillen}, E. and {van Reeven}, W. and {Vanel}, O. and {Vecchiato}, A. and {Viala}, Y. and {Vicente}, D. and {Voutsinas}, S. and {Weiler}, M. and {Wevers}, T. and {Wyrzykowski}, {\L}. and {Yoldas}, A. and {Yvard}, P. and {Zhao}, H. and {Zorec}, J. and {Zucker}, S. and {Zwitter}, T.},
        title = "{Gaia Data Release 3. Summary of the content and survey properties}",
      journal = {\aap},
         year = 2023,
        month = jun,
       volume = {674},
          eid = {A1},
        pages = {A1},
          doi = {10.1051/0004-6361/202243940},
archivePrefix = {arXiv},
       eprint = {2208.00211},
 primaryClass = {astro-ph.GA},
       adsurl = {https://ui.adsabs.harvard.edu/abs/2023A&A...674A...1G}
}

@ARTICLE{1996A&AS..117..393B,
       author = {{Bertin}, E. and {Arnouts}, S.},
        title = "{SExtractor: Software for source extraction.}",
      journal = {\aaps},
         year = 1996,
        month = jun,
       volume = {117},
        pages = {393-404},
          doi = {10.1051/aas:1996164},
       adsurl = {https://ui.adsabs.harvard.edu/abs/1996A&AS..117..393B}
}

@ARTICLE{2016JOSS....1...58B,
       author = {{Barbary}, Kyle},
        title = "{SEP: Source Extractor as a library}",
      journal = {The Journal of Open Source Software},
         year = 2016,
        month = oct,
       volume = {1},
       number = {6},
          eid = {58},
        pages = {58},
          doi = {10.21105/joss.00058},
       adsurl = {https://ui.adsabs.harvard.edu/abs/2016JOSS....1...58B}
}

@ARTICLE{2021MNRAS.501.3621A,
       author = {{Anand}, Gagandeep S. and {Lee}, Janice C. and {Van Dyk}, Schuyler D. and {Leroy}, Adam K. and {Rosolowsky}, Erik and {Schinnerer}, Eva and {Larson}, Kirsten and {Kourkchi}, Ehsan and {Kreckel}, Kathryn and {Scheuermann}, Fabian and {Rizzi}, Luca and {Thilker}, David and {Tully}, R. Brent and {Bigiel}, Frank and {Blanc}, Guillermo A. and {Boquien}, M{\'e}d{\'e}ric and {Chandar}, Rupali and {Dale}, Daniel and {Emsellem}, Eric and {Deger}, Sinan and {Glover}, Simon C.~O. and {Grasha}, Kathryn and {Groves}, Brent and {S. Klessen}, Ralf and {Kruijssen}, J.~M. Diederik and {Querejeta}, Miguel and {S{\'a}nchez-Bl{\'a}zquez}, Patricia and {Schruba}, Andreas and {Turner}, Jordan and {Ubeda}, Leonardo and {Williams}, Thomas G. and {Whitmore}, Brad},
        title = "{Distances to PHANGS galaxies: New tip of the red giant branch measurements and adopted distances}",
      journal = {\mnras},
         year = 2021,
        month = mar,
       volume = {501},
       number = {3},
        pages = {3621-3639},
          doi = {10.1093/mnras/staa3668},
archivePrefix = {arXiv},
       eprint = {2012.00757},
 primaryClass = {astro-ph.GA},
       adsurl = {https://ui.adsabs.harvard.edu/abs/2021MNRAS.501.3621A}
}

@ARTICLE{2023Natur.618..708S,
       author = {{Spilker}, Justin S. and {Phadke}, Kedar A. and {Aravena}, Manuel and {Archipley}, Melanie and {Bayliss}, Matthew B. and {Birkin}, Jack E. and {B{\'e}thermin}, Matthieu and {Burgoyne}, James and {Cathey}, Jared and {Chapman}, Scott C. and {Dahle}, H{\^a}kon and {Gonzalez}, Anthony H. and {Gururajan}, Gayathri and {Hayward}, Christopher C. and {Hezaveh}, Yashar D. and {Hill}, Ryley and {Hutchison}, Taylor A. and {Kim}, Keunho J. and {Kim}, Seonwoo and {Law}, David and {Legin}, Ronan and {Malkan}, Matthew A. and {Marrone}, Daniel P. and {Murphy}, Eric J. and {Narayanan}, Desika and {Navarre}, Alex and {Olivier}, Grace M. and {Rich}, Jeffrey A. and {Rigby}, Jane R. and {Reuter}, Cassie and {Rhoads}, James E. and {Sharon}, Keren and {Smith}, J.~D.~T. and {Solimano}, Manuel and {Sulzenauer}, Nikolaus and {Vieira}, Joaquin D. and {Vizgan}, David and {Wei{\ss}}, Axel and {Whitaker}, Katherine E.},
        title = "{Spatial variations in aromatic hydrocarbon emission in a dust-rich galaxy}",
      journal = {\nat},
         year = 2023,
        month = jun,
       volume = {618},
       number = {7966},
        pages = {708-711},
          doi = {10.1038/s41586-023-05998-6},
archivePrefix = {arXiv},
       eprint = {2306.03152},
 primaryClass = {astro-ph.GA},
       adsurl = {https://ui.adsabs.harvard.edu/abs/2023Natur.618..708S}
}

@ARTICLE{2023ApJ...957L..26L,
       author = {{Lai}, Thomas S. -Y. and {Armus}, Lee and {Bianchin}, Marina and {D{\'\i}az-Santos}, Tanio and {Linden}, Sean T. and {Privon}, George C. and {Inami}, Hanae and {U}, Vivian and {Bohn}, Thomas and {Evans}, Aaron S. and {Larson}, Kirsten L. and {Hensley}, Brandon S. and {Smith}, J. -D.~T. and {Malkan}, Matthew A. and {Song}, Yiqing and {Stierwalt}, Sabrina and {van der Werf}, Paul P. and {McKinney}, Jed and {Aalto}, Susanne and {Buiten}, Victorine A. and {Rich}, Jeff and {Charmandaris}, Vassilis and {Appleton}, Philip and {Barcos-Mu{\~n}oz}, Loreto and {B{\"o}ker}, Torsten and {Finnerty}, Luke and {Kader}, Justin A. and {Law}, David R. and {Medling}, Anne M. and {Brown}, Michael J.~I. and {Hayward}, Christopher C. and {Howell}, Justin and {Iwasawa}, Kazushi and {Kemper}, Francisca and {Marshall}, Jason and {Mazzarella}, Joseph M. and {M{\"u}ller-S{\'a}nchez}, Francisco and {Murphy}, Eric J. and {Sanders}, David and {Surace}, Jason},
        title = "{GOALS-JWST: Small Neutral Grains and Enhanced 3.3 {\ensuremath{\mu}}m PAH Emission in the Seyfert Galaxy NGC 7469}",
      journal = {\apjl},
         year = 2023,
        month = nov,
       volume = {957},
       number = {2},
          eid = {L26},
        pages = {L26},
          doi = {10.3847/2041-8213/ad0387},
archivePrefix = {arXiv},
       eprint = {2307.15169},
 primaryClass = {astro-ph.GA},
       adsurl = {https://ui.adsabs.harvard.edu/abs/2023ApJ...957L..26L}
}

@ARTICLE{2023ApJ...944L...9L,
       author = {{Leroy}, Adam K. and {Sandstrom}, Karin and {Rosolowsky}, Erik and {Belfiore}, Francesco and {Bolatto}, Alberto D. and {Cao}, Yixian and {Koch}, Eric W. and {Schinnerer}, Eva and {Barnes}, Ashley. T. and {Be{\v{s}}li{\'c}}, Ivana and {Bigiel}, F. and {Blanc}, Guillermo A. and {Chastenet}, J{\'e}r{\'e}my and {Chen}, Ness Mayker and {Chevance}, M{\'e}lanie and {Chown}, Ryan and {Congiu}, Enrico and {Dale}, Daniel A. and {Egorov}, Oleg V. and {Emsellem}, Eric and {Eibensteiner}, Cosima and {Faesi}, Christopher M. and {Glover}, Simon C.~O. and {Grasha}, Kathryn and {Groves}, Brent and {Hassani}, Hamid and {Henshaw}, Jonathan D. and {Hughes}, Annie and {Jim{\'e}nez-Donaire}, Mar{\'\i}a J. and {Kim}, Jaeyeon and {Klessen}, Ralf S. and {Kreckel}, Kathryn and {Kruijssen}, J.~M. Diederik and {Larson}, Kirsten L. and {Lee}, Janice C. and {Levy}, Rebecca C. and {Liu}, Daizhong and {Lopez}, Laura A. and {Meidt}, Sharon E. and {Murphy}, Eric J. and {Neumann}, Justus and {Pessa}, Ismael and {Pety}, J{\'e}r{\^o}me and {Saito}, Toshiki and {Sardone}, Amy and {Sun}, Jiayi and {Thilker}, David A. and {Usero}, Antonio and {Watkins}, Elizabeth J. and {Whitcomb}, Cory M. and {Williams}, Thomas G.},
        title = "{PHANGS-JWST First Results: Mid-infrared Emission Traces Both Gas Column Density and Heating at 100 pc Scales}",
      journal = {\apjl},
         year = 2023,
        month = feb,
       volume = {944},
       number = {2},
          eid = {L9},
        pages = {L9},
          doi = {10.3847/2041-8213/acaf85},
archivePrefix = {arXiv},
       eprint = {2212.10574},
 primaryClass = {astro-ph.GA},
       adsurl = {https://ui.adsabs.harvard.edu/abs/2023ApJ...944L...9L}
}

@ARTICLE{2023PASP..135d8001R,
       author = {{Rigby}, Jane and {Perrin}, Marshall and {McElwain}, Michael and {Kimble}, Randy and {Friedman}, Scott and {Lallo}, Matt and {Doyon}, Ren{\'e} and {Feinberg}, Lee and {Ferruit}, Pierre and {Glasse}, Alistair and {Rieke}, Marcia and {Rieke}, George and {Wright}, Gillian and {Willott}, Chris and {Colon}, Knicole and {Milam}, Stefanie and {Neff}, Susan and {Stark}, Christopher and {Valenti}, Jeff and {Abell}, Jim and {Abney}, Faith and {Abul-Huda}, Yasin and {Acton}, D. Scott and {Adams}, Evan and {Adler}, David and {Aguilar}, Jonathan and {Ahmed}, Nasif and {Albert}, Lo{\"\i}c and {Alberts}, Stacey and {Aldridge}, David and {Allen}, Marsha and {Altenburg}, Martin and {{\'A}lvarez-M{\'a}rquez}, Javier and {Alves de Oliveira}, Catarina and {Andersen}, Greg and {Anderson}, Harry and {Anderson}, Sara and {Argyriou}, Ioannis and {Armstrong}, Amber and {Arribas}, Santiago and {Artigau}, Etienne and {Arvai}, Amanda and {Atkinson}, Charles and {Bacon}, Gregory and {Bair}, Thomas and {Banks}, Kimberly and {Barrientes}, Jaclyn and {Barringer}, Bruce and {Bartosik}, Peter and {Bast}, William and {Baudoz}, Pierre and {Beatty}, Thomas and {Bechtold}, Katie and {Beck}, Tracy and {Bergeron}, Eddie and {Bergkoetter}, Matthew and {Bhatawdekar}, Rachana and {Birkmann}, Stephan and {Blazek}, Ronald and {Blome}, Claire and {Boccaletti}, Anthony and {B{\"o}ker}, Torsten and {Boia}, John and {Bonaventura}, Nina and {Bond}, Nicholas and {Bosley}, Kari and {Boucarut}, Ray and {Bourque}, Matthew and {Bouwman}, Jeroen and {Bower}, Gary and {Bowers}, Charles and {Boyer}, Martha and {Bradley}, Larry and {Brady}, Greg and {Braun}, Hannah and {Breda}, David and {Bresnahan}, Pamela and {Bright}, Stacey and {Britt}, Christopher and {Bromenschenkel}, Asa and {Brooks}, Brian and {Brooks}, Keira and {Brown}, Bob and {Brown}, Matthew and {Brown}, Patricia and {Bunker}, Andy and {Burger}, Matthew and {Bushouse}, Howard and {Cale}, Steven and {Cameron}, Alex and {Cameron}, Peter and {Canipe}, Alicia and {Caplinger}, James and {Caputo}, Francis and {Cara}, Mihai and {Carey}, Larkin and {Carniani}, Stefano and {Carrasquilla}, Maria and {Carruthers}, Margaret and {Case}, Michael and {Catherine}, Riggs and {Chance}, Don and {Chapman}, George and {Charlot}, St{\'e}phane and {Charlow}, Brian and {Chayer}, Pierre and {Chen}, Bin and {Cherinka}, Brian and {Chichester}, Sarah and {Chilton}, Zack and {Chonis}, Taylor and {Clampin}, Mark and {Clark}, Charles and {Clark}, Kerry and {Coe}, Dan and {Coleman}, Benee and {Comber}, Brian and {Comeau}, Tom and {Connolly}, Dennis and {Cooper}, James and {Cooper}, Rachel and {Coppock}, Eric and {Correnti}, Matteo and {Cossou}, Christophe and {Coulais}, Alain and {Coyle}, Laura and {Cracraft}, Misty and {Curti}, Mirko and {Cuturic}, Steven and {Davis}, Katherine and {Davis}, Michael and {Dean}, Bruce and {DeLisa}, Amy and {deMeester}, Wim and {Dencheva}, Nadia and {Dencheva}, Nadezhda and {DePasquale}, Joseph and {Deschenes}, Jeremy and {Hunor Detre}, {\"O}rs and {Diaz}, Rosa and {Dicken}, Dan and {DiFelice}, Audrey and {Dillman}, Matthew and {Dixon}, William and {Doggett}, Jesse and {Donaldson}, Tom and {Douglas}, Rob and {DuPrie}, Kimberly and {Dupuis}, Jean and {Durning}, John and {Easmin}, Nilufar and {Eck}, Weston and {Edeani}, Chinwe and {Egami}, Eiichi and {Ehrenwinkler}, Ralf and {Eisenhamer}, Jonathan and {Eisenhower}, Michael and {Elie}, Michelle and {Elliott}, James and {Elliott}, Kyle and {Ellis}, Tracy and {Engesser}, Michael and {Espinoza}, Nestor and {Etienne}, Odessa and {Etxaluze}, Mireya and {Falini}, Patrick and {Feeney}, Matthew and {Ferry}, Malcolm and {Filippazzo}, Joseph and {Fincham}, Brian and {Fix}, Mees and {Flagey}, Nicolas and {Florian}, Michael and {Flynn}, Jim and {Fontanella}, Erin and {Ford}, Terrance and {Forshay}, Peter and {Fox}, Ori and {Franz}, David and {Fu}, Henry and {Fullerton}, Alexander and {Galkin}, Sergey and {Galyer}, Anthony and {Garc{\'\i}a Mar{\'\i}n}, Macarena and {Gardner}, Jonathan P. and {Gardner}, Lisa and {Garland}, Dennis and {Garrett}, Bruce and {Gasman}, Danny and {Gaspar}, Andras and {Gaudreau}, Daniel and {Gauthier}, Peter and {Geers}, Vincent and {Geithner}, Paul and {Gennaro}, Mario and {Giardino}, Giovanna and {Girard}, Julien and {Giuliano}, Mark and {Glassmire}, Kirk and {Glauser}, Adrian and {Glazer}, Stuart and {Godfrey}, John and {Golimowski}, David and {Gollnitz}, David and {Gong}, Fan and {Gonzaga}, Shireen and {Gordon}, Michael and {Gordon}, Karl and {Goudfrooij}, Paul and {Greene}, Thomas and {Greenhouse}, Matthew and {Grimaldi}, Stefano and {Groebner}, Andrew and {Grundy}, Timothy and {Guillard}, Pierre and {Gutman}, Irvin and {Ha}, Kong Q. and {Haderlein}, Peter and {Hagedorn}, Andria and {Hainline}, Kevin and {Haley}, Craig and {Hami}, Maryam and {Hamilton}, Forrest and {Hammel}, Heidi and {Hansen}, Carl and {Harkins}, Tom and {Harr}, Michael and {Hart}, Jessica and {Hart}, Quyen and {Hartig}, George and {Hashimoto}, Ryan and {Haskins}, Sujee and {Hathaway}, William and {Havey}, Keith and {Hayden}, Brian and {Hecht}, Karen and {Heller-Boyer}, Chris and {Henriques}, Caroline and {Henry}, Alaina and {Hermann}, Karl and {Hernandez}, Scarlin and {Hesman}, Brigette and {Hicks}, Brian and {Hilbert}, Bryan and {Hines}, Dean and {Hoffman}, Melissa and {Holfeltz}, Sherie and {Holler}, Bryan J. and {Hoppa}, Jennifer and {Hott}, Kyle and {Howard}, Joseph M. and {Howard}, Rick and {Hunter}, Alexander and {Hunter}, David and {Hurst}, Brendan and {Husemann}, Bernd and {Hustak}, Leah and {Ilinca Ignat}, Luminita and {Illingworth}, Garth and {Irish}, Sandra and {Jackson}, Wallace and {Jahromi}, Amir and {Jakobsen}, Peter and {James}, LeAndrea and {James}, Bryan and {Januszewski}, William and {Jenkins}, Ann and {Jirdeh}, Hussein and {Johnson}, Phillip and {Johnson}, Timothy and {Jones}, Vicki and {Jones}, Ron and {Jones}, Danny and {Jones}, Olivia and {Jordan}, Ian and {Jordan}, Margaret and {Jurczyk}, Sarah and {Jurling}, Alden and {Kaleida}, Catherine and {Kalmanson}, Phillip and {Kammerer}, Jens and {Kang}, Huijo and {Kao}, Shaw-Hong and {Karakla}, Diane and {Kavanagh}, Patrick and {Kelly}, Doug and {Kendrew}, Sarah and {Kennedy}, Herbert and {Kenny}, Deborah and {Keski-kuha}, Ritva and {Keyes}, Charles and {Kidwell}, Richard and {Kinzel}, Wayne and {Kirk}, Jeff and {Kirkpatrick}, Mark and {Kirshenblat}, Danielle and {Klaassen}, Pamela and {Knapp}, Bryan and {Knight}, J. Scott and {Knollenberg}, Perry and {Koehler}, Robert and {Koekemoer}, Anton and {Kovacs}, Aiden and {Kulp}, Trey and {Kumari}, Nimisha and {Kyprianou}, Mark and {La Massa}, Stephanie and {Labador}, Aurora and {Labiano}, Alvaro and {Lagage}, Pierre-Olivier and {Lajoie}, Charles-Philippe and {Lallo}, Matthew and {Lam}, May and {Lamb}, Tracy and {Lambros}, Scott and {Lampenfield}, Richard and {Langston}, James and {Larson}, Kirsten and {Law}, David and {Lawrence}, Jon and {Lee}, David and {Leisenring}, Jarron and {Lepo}, Kelly and {Leveille}, Michael and {Levenson}, Nancy and {Levine}, Marie and {Levy}, Zena and {Lewis}, Dan and {Lewis}, Hannah and {Libralato}, Mattia and {Lightsey}, Paul and {Link}, Miranda and {Liu}, Lily and {Lo}, Amy and {Lockwood}, Alexandra and {Logue}, Ryan and {Long}, Chris and {Long}, Douglas and {Loomis}, Charles and {Lopez-Caniego}, Marcos and {Lorenzo Alvarez}, Jose and {Love-Pruitt}, Jennifer and {Lucy}, Adrian and {Luetzgendorf}, Nora and {Maghami}, Peiman and {Maiolino}, Roberto and {Major}, Melissa and {Malla}, Sunita and {Malumuth}, Eliot and {Manjavacas}, Elena and {Mannfolk}, Crystal and {Marrione}, Amanda and {Marston}, Anthony and {Martel}, Andr{\'e} and {Maschmann}, Marc and {Masci}, Gregory and {Masciarelli}, Michaela and {Maszkiewicz}, Michael and {Mather}, John and {McKenzie}, Kenny and {McLean}, Brian and {McMaster}, Matthew and {Melbourne}, Katie and {Mel{\'e}ndez}, Marcio and {Menzel}, Michael and {Merz}, Kaiya and {Meyett}, Michele and {Meza}, Luis and {Miskey}, Cherie and {Misselt}, Karl and {Moller}, Christopher and {Morrison}, Jane and {Morse}, Ernie and {Moseley}, Harvey and {Mosier}, Gary and {Mountain}, Matt and {Mueckay}, Julio and {Mueller}, Michael and {Mullally}, Susan and {Murphy}, Jess and {Murray}, Katherine and {Murray}, Claire and {Mustelier}, David and {Muzerolle}, James and {Mycroft}, Matthew and {Myers}, Richard and {Myrick}, Kaila and {Nanavati}, Shashvat and {Nance}, Elizabeth and {Nayak}, Omnarayani and {Naylor}, Bret and {Nelan}, Edmund and {Nickson}, Bryony and {Nielson}, Alethea and {Nieto-Santisteban}, Maria and {Nikolov}, Nikolay and {Noriega-Crespo}, Alberto and {O'Shaughnessy}, Brian and {O'Sullivan}, Brian and {Ochs}, William and {Ogle}, Patrick and {Oleszczuk}, Brenda and {Olmsted}, Joseph and {Osborne}, Shannon and {Ottens}, Richard and {Owens}, Beverly and {Pacifici}, Camilla and {Pagan}, Alyssa and {Page}, James and {Park}, Sang and {Parrish}, Keith and {Patapis}, Polychronis and {Paul}, Lee and {Pauly}, Tyler and {Pavlovsky}, Cheryl and {Pedder}, Andrew and {Peek}, Matthew and {Pena-Guerrero}, Maria and {Penanen}, Konstantin and {Perez}, Yesenia and {Perna}, Michele and {Perriello}, Beth and {Phillips}, Kevin and {Pietraszkiewicz}, Martin and {Pinaud}, Jean-Paul and {Pirzkal}, Norbert and {Pitman}, Joseph and {Piwowar}, Aidan and {Platais}, Vera and {Player}, Danielle and {Plesha}, Rachel and {Pollizi}, Joe and {Polster}, Ethan and {Pontoppidan}, Klaus and {Porterfield}, Blair and {Proffitt}, Charles and {Pueyo}, Laurent and {Pulliam}, Christine and {Quirt}, Brian and {Quispe Neira}, Irma and {Ramos Alarcon}, Rafael and {Ramsay}, Leah and {Rapp}, Greg and {Rapp}, Robert and {Rauscher}, Bernard and {Ravindranath}, Swara and {Rawle}, Timothy and {Regan}, Michael and {Reichard}, Timothy A. and {Reis}, Carl and {Ressler}, Michael E. and {Rest}, Armin and {Reynolds}, Paul and {Rhue}, Timothy and {Richon}, Karen and {Rickman}, Emily and {Ridgaway}, Michael and {Ritchie}, Christine and {Rix}, Hans-Walter and {Robberto}, Massimo and {Robinson}, Gregory and {Robinson}, Michael and {Robinson}, Orion and {Rock}, Frank and {Rodriguez}, David and {Rodriguez Del Pino}, Bruno and {Roellig}, Thomas and {Rohrbach}, Scott and {Roman}, Anthony and {Romelfanger}, Fred and {Rose}, Perry and {Roteliuk}, Anthony and {Roth}, Marc and {Rothwell}, Braden and {Rowlands}, Neil and {Roy}, Arpita and {Royer}, Pierre and {Royle}, Patricia and {Rui}, Chunlei and {Rumler}, Peter and {Runnels}, Joel and {Russ}, Melissa and {Rustamkulov}, Zafar and {Ryden}, Grant and {Ryer}, Holly and {Sabata}, Modhumita and {Sabatke}, Derek and {Sabbi}, Elena and {Samuelson}, Bridget and {Sapp}, Benjamin and {Sappington}, Bradley and {Sargent}, B. and {Sauer}, Arne and {Scheithauer}, Silvia and {Schlawin}, Everett and {Schlitz}, Joseph and {Schmitz}, Tyler and {Schneider}, Analyn and {Schreiber}, J{\"u}rgen and {Schulze}, Vonessa and {Schwab}, Ryan and {Scott}, John and {Sembach}, Kenneth and {Shanahan}, Clare and {Shaughnessy}, Bryan and {Shaw}, Richard and {Shawger}, Nanci and {Shay}, Christopher and {Sheehan}, Evan and {Shen}, Sharon and {Sherman}, Allan and {Shiao}, Bernard and {Shih}, Hsin-Yi and {Shivaei}, Irene and {Sienkiewicz}, Matthew and {Sing}, David and {Sirianni}, Marco and {Sivaramakrishnan}, Anand and {Skipper}, Joy and {Sloan}, G.~C. and {Slocum}, Christine and {Slowinski}, Steven and {Smith}, Erin and {Smith}, Eric and {Smith}, Denise and {Smith}, Corbett and {Snyder}, Gregory and {Soh}, Warren and {Sohn}, Sangmo Tony and {Soto}, Christian and {Spencer}, Richard and {Stallcup}, Scott and {Stansberry}, John and {Starr}, Carl and {Starr}, Elysia and {Stewart}, Alphonso and {Stiavelli}, Massimo and {Straughn}, Amber and {Strickland}, David and {Stys}, Jeff and {Summers}, Francis and {Sun}, Fengwu and {Sunnquist}, Ben and {Swade}, Daryl and {Swam}, Michael and {Swaters}, Robert and {Swoish}, Robby and {Taylor}, Joanna M. and {Taylor}, Rolanda and {Te Plate}, Maurice and {Tea}, Mason and {Teague}, Kelly and {Telfer}, Randal and {Temim}, Tea and {Thatte}, Deepashri and {Thompson}, Christopher and {Thompson}, Linda and {Thomson}, Shaun and {Tikkanen}, Tuomo and {Tippet}, William and {Todd}, Connor and {Toolan}, Sharon and {Tran}, Hien and {Trejo}, Edwin and {Truong}, Justin and {Tsukamoto}, Chris and {Tustain}, Samuel and {Tyra}, Harrison and {Ubeda}, Leonardo and {Underwood}, Kelli and {Uzzo}, Michael and {Van Campen}, Julie and {Vandal}, Thomas and {Vandenbussche}, Bart and {Vila}, Bego{\~n}a and {Volk}, Kevin and {Wahlgren}, Glenn and {Waldman}, Mark and {Walker}, Chanda and {Wander}, Michel and {Warfield}, Christine and {Warner}, Gerald and {Wasiak}, Matthew and {Watkins}, Mitchell and {Weaver}, Andrew and {Weilert}, Mark and {Weiser}, Nick and {Weiss}, Ben and {Weissman}, Sarah and {Welty}, Alan and {West}, Garrett and {Wheate}, Lauren and {Wheatley}, Elizabeth and {Wheeler}, Thomas and {White}, Rick and {Whiteaker}, Kevin and {Whitehouse}, Paul and {Whiteleather}, Jennifer and {Whitman}, William and {Williams}, Christina and {Willmer}, Christopher and {Willoughby}, Scott and {Wilson}, Andrew and {Wirth}, Gregory and {Wislowski}, Emily and {Wolf}, Erin and {Wolfe}, David and {Wolff}, Schuyler and {Workman}, Bill and {Wright}, Ray and {Wu}, Carl and {Wu}, Rai and {Wymer}, Kristen and {Yates}, Kayla and {Yeager}, Christopher and {Yeates}, Jared and {Yerger}, Ethan and {Yoon}, Jinmi and {Young}, Alice and {Yu}, Susan and {Zak}, Dean and {Zeidler}, Peter and {Zhou}, Julia and {Zielinski}, Thomas and {Zincke}, Cristian and {Zonak}, Stephanie},
        title = "{The Science Performance of JWST as Characterized in Commissioning}",
      journal = {\pasp},
         year = 2023,
        month = apr,
       volume = {135},
       number = {1046},
          eid = {048001},
        pages = {048001},
          doi = {10.1088/1538-3873/acb293},
archivePrefix = {arXiv},
       eprint = {2207.05632},
 primaryClass = {astro-ph.IM},
       adsurl = {https://ui.adsabs.harvard.edu/abs/2023PASP..135d8001R}
}

@ARTICLE{2023ApJ...944L..12C,
       author = {{Chastenet}, J{\'e}r{\'e}my and {Sutter}, Jessica and {Sandstrom}, Karin and {Belfiore}, Francesco and {Egorov}, Oleg V. and {Larson}, Kirsten L. and {Leroy}, Adam K. and {Liu}, Daizhong and {Rosolowsky}, Erik and {Thilker}, David A. and {Watkins}, Elizabeth J. and {Williams}, Thomas G. and {Barnes}, Ashley. T. and {Bigiel}, F. and {Boquien}, M{\'e}d{\'e}ric and {Chevance}, M{\'e}lanie and {Dale}, Daniel A. and {Kruijssen}, J.~M. Diederik and {Emsellem}, Eric and {Grasha}, Kathryn and {Groves}, Brent and {Hassani}, Hamid and {Hughes}, Annie and {Kreckel}, Kathryn and {Meidt}, Sharon E. and {Pan}, Hsi-An and {Querejeta}, Miguel and {Schinnerer}, Eva and {Whitcomb}, Cory M.},
        title = "{PHANGS-JWST First Results: Measuring Polycyclic Aromatic Hydrocarbon Properties across the Multiphase Interstellar Medium}",
      journal = {\apjl},
         year = 2023,
        month = feb,
       volume = {944},
       number = {2},
          eid = {L12},
        pages = {L12},
          doi = {10.3847/2041-8213/acac94},
       adsurl = {https://ui.adsabs.harvard.edu/abs/2023ApJ...944L..12C}
}

@ARTICLE{1999ApJS..123....3L,
       author = {{Leitherer}, Claus and {Schaerer}, Daniel and {Goldader}, Jeffrey D. and {Delgado}, Rosa M. Gonz{\'a}lez and {Robert}, Carmelle and {Kune}, Denis Foo and {de Mello}, Du{\'\i}lia F. and {Devost}, Daniel and {Heckman}, Timothy M.},
        title = "{Starburst99: Synthesis Models for Galaxies with Active Star Formation}",
      journal = {\apjs},
         year = 1999,
        month = jul,
       volume = {123},
       number = {1},
        pages = {3-40},
          doi = {10.1086/313233},
archivePrefix = {arXiv},
       eprint = {astro-ph/9902334},
 primaryClass = {astro-ph},
       adsurl = {https://ui.adsabs.harvard.edu/abs/1999ApJS..123....3L}
}

@ARTICLE{2011ApJ...741L..26F,
       author = {{Fumagalli}, Michele and {da Silva}, Robert L. and {Krumholz}, Mark R.},
        title = "{Stochastic Star Formation and a (Nearly) Uniform Stellar Initial Mass Function}",
      journal = {\apjl},
         year = 2011,
        month = nov,
       volume = {741},
       number = {2},
          eid = {L26},
        pages = {L26},
          doi = {10.1088/2041-8205/741/2/L26},
archivePrefix = {arXiv},
       eprint = {1105.6101},
 primaryClass = {astro-ph.CO},
       adsurl = {https://ui.adsabs.harvard.edu/abs/2011ApJ...741L..26F}
}

@ARTICLE{1995MNRAS.272...41S,
       author = {{Storey}, P.~J. and {Hummer}, D.~G.},
        title = "{Recombination line intensities for hydrogenic ions-IV. Total recombination coefficients and machine-readable tables for Z=1 to 8}",
      journal = {\mnras},
         year = 1995,
        month = jan,
       volume = {272},
       number = {1},
        pages = {41-48},
          doi = {10.1093/mnras/272.1.41},
       adsurl = {https://ui.adsabs.harvard.edu/abs/1995MNRAS.272...41S}
}

@ARTICLE{1997MNRAS.291..827O,
       author = {{Oey}, M.~S. and {Kennicutt}, Jr., R.~C.},
        title = "{Comparison of H II region luminosities with observed stellar ionizing sources in the Large Magellanic Cloud}",
      journal = {\mnras},
         year = 1997,
        month = nov,
       volume = {291},
       number = {4},
        pages = {827-832},
          doi = {10.1093/mnras/291.4.827},
archivePrefix = {arXiv},
       eprint = {astro-ph/9708106},
 primaryClass = {astro-ph},
       adsurl = {https://ui.adsabs.harvard.edu/abs/1997MNRAS.291..827O}
}

@MISC{2006acs..rept....1A,
       author = {{Anderson}, Jay and {King}, Ivan R.},
        title = "{PSFs, Photometry, and Astronomy for the ACS/WFC}",
 howpublished = {Instrument Science Report ACS 2006-01, 34 pages},
         year = 2006,
        month = feb,
        pages = {1},
       adsurl = {https://ui.adsabs.harvard.edu/abs/2006acs..rept....1A}
}

@ARTICLE{2021ApJS..257...43L,
       author = {{Leroy}, Adam K. and {Schinnerer}, Eva and {Hughes}, Annie and {Rosolowsky}, Erik and {Pety}, J{\'e}r{\^o}me and {Schruba}, Andreas and {Usero}, Antonio and {Blanc}, Guillermo A. and {Chevance}, M{\'e}lanie and {Emsellem}, Eric and {Faesi}, Christopher M. and {Herrera}, Cinthya N. and {Liu}, Daizhong and {Meidt}, Sharon E. and {Querejeta}, Miguel and {Saito}, Toshiki and {Sandstrom}, Karin M. and {Sun}, Jiayi and {Williams}, Thomas G. and {Anand}, Gagandeep S. and {Barnes}, Ashley T. and {Behrens}, Erica A. and {Belfiore}, Francesco and {Benincasa}, Samantha M. and {Be{\v{s}}li{\'c}}, Ivana and {Bigiel}, Frank and {Bolatto}, Alberto D. and {den Brok}, Jakob S. and {Cao}, Yixian and {Chandar}, Rupali and {Chastenet}, J{\'e}r{\'e}my and {Chiang}, I-Da and {Congiu}, Enrico and {Dale}, Daniel A. and {Deger}, Sinan and {Eibensteiner}, Cosima and {Egorov}, Oleg V. and {Garc{\'\i}a-Rodr{\'\i}guez}, Axel and {Glover}, Simon C.~O. and {Grasha}, Kathryn and {Henshaw}, Jonathan D. and {Ho}, I. -Ting and {Kepley}, Amanda A. and {Kim}, Jaeyeon and {Klessen}, Ralf S. and {Kreckel}, Kathryn and {Koch}, Eric W. and {Kruijssen}, J.~M. Diederik and {Larson}, Kirsten L. and {Lee}, Janice C. and {Lopez}, Laura A. and {Machado}, Josh and {Mayker}, Ness and {McElroy}, Rebecca and {Murphy}, Eric J. and {Ostriker}, Eve C. and {Pan}, Hsi-An and {Pessa}, Ismael and {Puschnig}, Johannes and {Razza}, Alessandro and {S{\'a}nchez-Bl{\'a}zquez}, Patricia and {Santoro}, Francesco and {Sardone}, Amy and {Scheuermann}, Fabian and {Sliwa}, Kazimierz and {Sormani}, Mattia C. and {Stuber}, Sophia K. and {Thilker}, David A. and {Turner}, Jordan A. and {Utomo}, Dyas and {Watkins}, Elizabeth J. and {Whitmore}, Bradley},
        title = "{PHANGS-ALMA: Arcsecond CO(2-1) Imaging of Nearby Star-forming Galaxies}",
      journal = {\apjs},
         year = 2021,
        month = dec,
       volume = {257},
       number = {2},
          eid = {43},
        pages = {43},
          doi = {10.3847/1538-4365/ac17f3},
archivePrefix = {arXiv},
       eprint = {2104.07739},
 primaryClass = {astro-ph.GA},
       adsurl = {https://ui.adsabs.harvard.edu/abs/2021ApJS..257...43L}
}

@ARTICLE{2021ApJ...909..121M,
       author = {{Messa}, Matteo and {Calzetti}, Daniela and {Adamo}, Angela and {Grasha}, Kathryn and {Johnson}, Kelsey E. and {Sabbi}, Elena and {Smith}, Linda J. and {Bajaj}, Varun and {Finn}, Molly K. and {Lin}, Zesen},
        title = "{Looking for Obscured Young Star Clusters in NGC 1313}",
      journal = {\apj},
         year = 2021,
        month = mar,
       volume = {909},
       number = {2},
          eid = {121},
        pages = {121},
          doi = {10.3847/1538-4357/abe0b5},
archivePrefix = {arXiv},
       eprint = {2011.09392},
 primaryClass = {astro-ph.GA},
       adsurl = {https://ui.adsabs.harvard.edu/abs/2021ApJ...909..121M}
}

@ARTICLE{2011ApJ...740...13Z,
       author = {{Zackrisson}, Erik and {Rydberg}, Claes-Erik and {Schaerer}, Daniel and {{\"O}stlin}, G{\"o}ran and {Tuli}, Manan},
        title = "{The Spectral Evolution of the First Galaxies. I. James Webb Space Telescope Detection Limits and Color Criteria for Population III Galaxies}",
      journal = {\apj},
         year = 2011,
        month = oct,
       volume = {740},
       number = {1},
          eid = {13},
        pages = {13},
          doi = {10.1088/0004-637X/740/1/13},
archivePrefix = {arXiv},
       eprint = {1105.0921},
 primaryClass = {astro-ph.CO},
       adsurl = {https://ui.adsabs.harvard.edu/abs/2011ApJ...740...13Z}
}

@ARTICLE{2011ApJ...729...78W,
       author = {{Whitmore}, Bradley C. and {Chandar}, Rupali and {Kim}, Hwihyun and {Kaleida}, Catherine and {Mutchler}, Max and {Stankiewicz}, Matt and {Calzetti}, Daniela and {Saha}, Abhijit and {O'Connell}, Robert and {Balick}, Bruce and {Bond}, Howard E. and {Carollo}, Marcella and {Disney}, Michael J. and {Dopita}, Michael A. and {Frogel}, Jay A. and {Hall}, Donald N.~B. and {Holtzman}, Jon A. and {Kimble}, Randy A. and {McCarthy}, Patrick J. and {Paresce}, Francesco and {Silk}, Joseph I. and {Trauger}, John T. and {Walker}, Alistair R. and {Windhorst}, Rogier A. and {Young}, Erick T.},
        title = "{Using H{\ensuremath{\alpha}} Morphology and Surface Brightness Fluctuations to Age-date Star Clusters in M83}",
      journal = {\apj},
         year = 2011,
        month = mar,
       volume = {729},
       number = {2},
          eid = {78},
        pages = {78},
          doi = {10.1088/0004-637X/729/2/78},
archivePrefix = {arXiv},
       eprint = {1103.4026},
 primaryClass = {astro-ph.GA},
       adsurl = {https://ui.adsabs.harvard.edu/abs/2011ApJ...729...78W}
}

@ARTICLE{2020ApJ...889..154W,
       author = {{Whitmore}, Bradley C. and {Chandar}, Rupali and {Lee}, Janice and {Ubeda}, Leonardo and {Adamo}, Angela and {Aloisi}, Alessandra and {Calzetti}, Daniela and {Cignoni}, Michele and {Cook}, David and {Dale}, Daniel and {Elmegreen}, B.~G. and {Gouliermis}, Dimitrios and {Grebel}, Eva K. and {Grasha}, Kathryn and {Johnson}, Kelsey E. and {Kim}, Hwihyun and {Sacchi}, Elena and {Smith}, Linda J. and {Tosi}, Monica and {Wofford}, Aida},
        title = "{LEGUS and H$_{{\ensuremath{\alpha}}}$-LEGUS Observations of Star Clusters in NGC 4449: Improved Ages and the Fraction of Light in Clusters as a Function of Age}",
      journal = {\apj},
         year = 2020,
        month = feb,
       volume = {889},
       number = {2},
          eid = {154},
        pages = {154},
          doi = {10.3847/1538-4357/ab59e5},
archivePrefix = {arXiv},
       eprint = {2005.02840},
 primaryClass = {astro-ph.GA},
       adsurl = {https://ui.adsabs.harvard.edu/abs/2020ApJ...889..154W}
}

@ARTICLE{2022MNRAS.512.1294H,
       author = {{Hannon}, Stephen and {Lee}, Janice C. and {Whitmore}, B.~C. and {Mobasher}, B. and {Thilker}, D. and {Chandar}, R. and {Adamo}, A. and {Wofford}, A. and {Orozco-Duarte}, R. and {Calzetti}, D. and {Della Bruna}, L. and {Kreckel}, K. and {Groves}, B. and {Barnes}, A.~T. and {Boquien}, M. and {Belfiore}, F. and {Linden}, S.},
        title = "{H {\ensuremath{\alpha}} morphologies of star clusters in 16 LEGUS galaxies: Constraints on H II region evolution time-scales}",
      journal = {\mnras},
         year = 2022,
        month = may,
       volume = {512},
       number = {1},
        pages = {1294-1316},
          doi = {10.1093/mnras/stac550},
archivePrefix = {arXiv},
       eprint = {2203.01339},
 primaryClass = {astro-ph.GA},
       adsurl = {https://ui.adsabs.harvard.edu/abs/2022MNRAS.512.1294H}
}

@ARTICLE{2020MNRAS.494..642M,
       author = {{Maragkoudakis}, A. and {Peeters}, E. and {Ricca}, A.},
        title = "{Probing the size and charge of polycyclic aromatic hydrocarbons}",
      journal = {\mnras},
         year = 2020,
        month = may,
       volume = {494},
       number = {1},
        pages = {642-664},
          doi = {10.1093/mnras/staa681},
archivePrefix = {arXiv},
       eprint = {2003.02823},
 primaryClass = {astro-ph.GA},
       adsurl = {https://ui.adsabs.harvard.edu/abs/2020MNRAS.494..642M}
}

@INPROCEEDINGS{2003ASPC..295..489J,
       author = {{Joye}, W.~A. and {Mandel}, E.},
        title = "{New Features of SAOImage DS9}",
    booktitle = {Astronomical Data Analysis Software and Systems XII},
         year = 2003,
       editor = {{Payne}, H.~E. and {Jedrzejewski}, R.~I. and {Hook}, R.~N.},
       series = {Astronomical Society of the Pacific Conference Series},
       volume = {295},
        month = jan,
        pages = {489},
       adsurl = {https://ui.adsabs.harvard.edu/abs/2003ASPC..295..489J}
}

@MISC{2017wfc..rept...19B,
       author = {{Bajaj}, V.},
        title = "{Aligning HST Images to Gaia: A Faster Mosaicking Workflow}",
 howpublished = {Instrument Science Report WFC3 2017-19, 4 pages},
         year = 2017,
        month = nov,
        pages = {19},
       adsurl = {https://ui.adsabs.harvard.edu/abs/2017wfc..rept...19B}
}

@ARTICLE{2000PASP..112.1360A,
       author = {{Anderson}, Jay and {King}, Ivan R.},
        title = "{Toward High-Precision Astrometry with WFPC2. I. Deriving an Accurate Point-Spread Function}",
      journal = {\pasp},
         year = 2000,
        month = oct,
       volume = {112},
       number = {776},
        pages = {1360-1382},
          doi = {10.1086/316632},
archivePrefix = {arXiv},
       eprint = {astro-ph/0006325},
 primaryClass = {astro-ph},
       adsurl = {https://ui.adsabs.harvard.edu/abs/2000PASP..112.1360A}
}

@ARTICLE{2009AJ....138..332J,
       author = {{Jacobs}, Bradley A. and {Rizzi}, Luca and {Tully}, R. Brent and {Shaya}, Edward J. and {Makarov}, Dmitry I. and {Makarova}, Lidia},
        title = "{The Extragalactic Distance Database: Color-Magnitude Diagrams}",
      journal = {\aj},
         year = 2009,
        month = aug,
       volume = {138},
       number = {2},
        pages = {332-337},
          doi = {10.1088/0004-6256/138/2/332},
archivePrefix = {arXiv},
       eprint = {0902.3675},
 primaryClass = {astro-ph.CO},
       adsurl = {https://ui.adsabs.harvard.edu/abs/2009AJ....138..332J}
}

@ARTICLE{2004ApJ...613..986P,
       author = {{Peeters}, E. and {Spoon}, H.~W.~W. and {Tielens}, A.~G.~G.~M.},
        title = "{Polycyclic Aromatic Hydrocarbons as a Tracer of Star Formation?}",
      journal = {\apj},
         year = 2004,
        month = oct,
       volume = {613},
       number = {2},
        pages = {986-1003},
          doi = {10.1086/423237},
archivePrefix = {arXiv},
       eprint = {astro-ph/0406183},
 primaryClass = {astro-ph},
       adsurl = {https://ui.adsabs.harvard.edu/abs/2004ApJ...613..986P}
}

@ARTICLE{2016ApJ...818...60S,
       author = {{Shipley}, Heath V. and {Papovich}, Casey and {Rieke}, George H. and {Brown}, Michael J.~I. and {Moustakas}, John},
        title = "{A New Star Formation Rate Calibration from Polycyclic Aromatic Hydrocarbon Emission Features and Application to High-redshift Galaxies}",
      journal = {\apj},
         year = 2016,
        month = feb,
       volume = {818},
       number = {1},
          eid = {60},
        pages = {60},
          doi = {10.3847/0004-637X/818/1/60},
archivePrefix = {arXiv},
       eprint = {1601.01698},
 primaryClass = {astro-ph.GA},
       adsurl = {https://ui.adsabs.harvard.edu/abs/2016ApJ...818...60S}
}

@ARTICLE{2009ApJ...703..270S,
       author = {{Sajina}, Anna and {Spoon}, Henrik and {Yan}, Lin and {Imanishi}, Masatoshi and {Fadda}, Dario and {Elitzur}, Moshe},
        title = "{Detections of Water Ice, Hydrocarbons, and 3.3 {\ensuremath{\mu}}m PAH in z \raisebox{-0.5ex}\textasciitilde 2 ULIRGs}",
      journal = {\apj},
         year = 2009,
        month = sep,
       volume = {703},
       number = {1},
        pages = {270-284},
          doi = {10.1088/0004-637X/703/1/270},
archivePrefix = {arXiv},
       eprint = {0907.5377},
 primaryClass = {astro-ph.CO},
       adsurl = {https://ui.adsabs.harvard.edu/abs/2009ApJ...703..270S}
}

@ARTICLE{2009ApJ...698.1273S,
       author = {{Siana}, B. and {Smail}, Ian and {Swinbank}, A.~M. and {Richard}, J. and {Teplitz}, H.~I. and {Coppin}, K.~E.~K. and {Ellis}, R.~S. and {Stark}, D.~P. and {Kneib}, J. -P. and {Edge}, A.~C.},
        title = "{Detection of Far-Infrared and Polycyclic Aromatic Hydrocarbon Emission from the Cosmic Eye: Probing the Dust and Star Formation of Lyman Break Galaxies}",
      journal = {\apj},
         year = 2009,
        month = jun,
       volume = {698},
       number = {2},
        pages = {1273-1281},
          doi = {10.1088/0004-637X/698/2/1273},
archivePrefix = {arXiv},
       eprint = {0904.1742},
 primaryClass = {astro-ph.CO},
       adsurl = {https://ui.adsabs.harvard.edu/abs/2009ApJ...698.1273S}
}

@ARTICLE{1988PhRvL..60..921L,
       author = {{Leger}, A. and {D'Hendecourt}, L. and {Boissel}, P.},
        title = "{Predicted fluorescence mechanism in highly isolated molecules - The Poincare fluorescence}",
      journal = {\prl},
         year = 1988,
        month = mar,
       volume = {60},
        pages = {921-924},
          doi = {10.1103/PhysRevLett.60.921},
       adsurl = {https://ui.adsabs.harvard.edu/abs/1988PhRvL..60..921L}
}

@ARTICLE{2017MNRAS.469.4933L,
       author = {{Lai}, Thomas S. -Y. and {Witt}, Adolf N. and {Crawford}, Ken},
        title = "{Extended red emission in IC59 and IC63}",
      journal = {\mnras},
         year = 2017,
        month = aug,
       volume = {469},
       number = {4},
        pages = {4933-4948},
          doi = {10.1093/mnras/stx1124},
archivePrefix = {arXiv},
       eprint = {1705.04313},
 primaryClass = {astro-ph.GA},
       adsurl = {https://ui.adsabs.harvard.edu/abs/2017MNRAS.469.4933L}
}

@ARTICLE{2020Ap&SS.365...58W,
       author = {{Witt}, Adolf N. and {Lai}, Thomas S. -Y.},
        title = "{Extended red emission: observational constraints for models}",
      journal = {\apss},
         year = 2020,
        month = mar,
       volume = {365},
       number = {3},
          eid = {58},
        pages = {58},
          doi = {10.1007/s10509-020-03766-w},
archivePrefix = {arXiv},
       eprint = {2003.06453},
 primaryClass = {astro-ph.GA},
       adsurl = {https://ui.adsabs.harvard.edu/abs/2020Ap&SS.365...58W}
}

@ARTICLE{2016ApJ...830...64B,
       author = {{Bresolin}, Fabio and {Kudritzki}, Rolf-Peter and {Urbaneja}, Miguel A. and {Gieren}, Wolfgang and {Ho}, I. -Ting and {Pietrzy{\'n}ski}, Grzegorz},
        title = "{Young Stars and Ionized Nebulae in M83: Comparing Chemical Abundances at High Metallicity.}",
      journal = {\apj},
         year = 2016,
        month = oct,
       volume = {830},
       number = {2},
          eid = {64},
        pages = {64},
          doi = {10.3847/0004-637X/830/2/64},
archivePrefix = {arXiv},
       eprint = {1607.06840},
 primaryClass = {astro-ph.GA},
       adsurl = {https://ui.adsabs.harvard.edu/abs/2016ApJ...830...64B}
}

@ARTICLE{2008ApJS..176..438G,
       author = {{Groves}, Brent and {Dopita}, Michael A. and {Sutherland}, Ralph S. and {Kewley}, Lisa J. and {Fischera}, J{\"o}rg and {Leitherer}, Claus and {Brandl}, Bernhard and {van Breugel}, Wil},
        title = "{Modeling the Pan-Spectral Energy Distribution of Starburst Galaxies. IV. The Controlling Parameters of the Starburst SED}",
      journal = {\apjs},
         year = 2008,
        month = jun,
       volume = {176},
       number = {2},
        pages = {438-456},
          doi = {10.1086/528711},
archivePrefix = {arXiv},
       eprint = {0712.1824},
 primaryClass = {astro-ph},
       adsurl = {https://ui.adsabs.harvard.edu/abs/2008ApJS..176..438G}
}

@PHDTHESIS{2004PhDT.......183G,
       author = {{Groves}, Brent Allan},
        title = "{Photoionization of gases}",
       school = {Australian National University, Canberra},
         year = 2004,
        month = jan,
       adsurl = {https://ui.adsabs.harvard.edu/abs/2004PhDT.......183G}
}

@ARTICLE{2009ApJ...703..517D,
       author = {{Dale}, D.~A. and {Cohen}, S.~A. and {Johnson}, L.~C. and {Schuster}, M.~D. and {Calzetti}, D. and {Engelbracht}, C.~W. and {Gil de Paz}, A. and {Kennicutt}, R.~C. and {Lee}, J.~C. and {Begum}, A. and {Block}, M. and {Dalcanton}, J.~J. and {Funes}, J.~G. and {Gordon}, K.~D. and {Johnson}, B.~D. and {Marble}, A.~R. and {Sakai}, S. and {Skillman}, E.~D. and {van Zee}, L. and {Walter}, F. and {Weisz}, D.~R. and {Williams}, B. and {Wu}, S. -Y. and {Wu}, Y.},
        title = "{The Spitzer Local Volume Legacy: Survey Description and Infrared Photometry}",
      journal = {\apj},
         year = 2009,
        month = sep,
       volume = {703},
       number = {1},
        pages = {517-556},
          doi = {10.1088/0004-637X/703/1/517},
archivePrefix = {arXiv},
       eprint = {0907.4722},
 primaryClass = {astro-ph.CO},
       adsurl = {https://ui.adsabs.harvard.edu/abs/2009ApJ...703..517D}
}

@ARTICLE{2010MNRAS.403..625D,
       author = {{Dobbs}, C.~L. and {Theis}, C. and {Pringle}, J.~E. and {Bate}, M.~R.},
        title = "{Simulations of the grand design galaxy M51: a case study for analysing tidally induced spiral structure}",
      journal = {\mnras},
         year = 2010,
        month = apr,
       volume = {403},
       number = {2},
        pages = {625-645},
          doi = {10.1111/j.1365-2966.2009.16161.x},
archivePrefix = {arXiv},
       eprint = {0912.1201},
 primaryClass = {astro-ph.GA},
       adsurl = {https://ui.adsabs.harvard.edu/abs/2010MNRAS.403..625D}
}

@ARTICLE{2013AJ....146...86T,
       author = {{Tully}, R. Brent and {Courtois}, H{\'e}l{\`e}ne M. and {Dolphin}, Andrew E. and {Fisher}, J. Richard and {H{\'e}raudeau}, Philippe and {Jacobs}, Bradley A. and {Karachentsev}, Igor D. and {Makarov}, Dmitry and {Makarova}, Lidia and {Mitronova}, Sofia and {Rizzi}, Luca and {Shaya}, Edward J. and {Sorce}, Jenny G. and {Wu}, Po-Feng},
        title = "{Cosmicflows-2: The Data}",
      journal = {\aj},
         year = 2013,
        month = oct,
       volume = {146},
       number = {4},
          eid = {86},
        pages = {86},
          doi = {10.1088/0004-6256/146/4/86},
archivePrefix = {arXiv},
       eprint = {1307.7213},
 primaryClass = {astro-ph.CO},
       adsurl = {https://ui.adsabs.harvard.edu/abs/2013AJ....146...86T}
}

@ARTICLE{2014ApJ...784....4C,
       author = {{Colombo}, Dario and {Meidt}, Sharon E. and {Schinnerer}, Eva and {Garc{\'\i}a-Burillo}, Santiago and {Hughes}, Annie and {Pety}, J{\'e}r{\^o}me and {Leroy}, Adam K. and {Dobbs}, Clare L. and {Dumas}, Ga{\"e}lle and {Thompson}, Todd A. and {Schuster}, Karl F. and {Kramer}, Carsten},
        title = "{The PdBI Arcsecond Whirlpool Survey (PAWS): Multi-phase Cold Gas Kinematic of M51}",
      journal = {\apj},
         year = 2014,
        month = mar,
       volume = {784},
       number = {1},
          eid = {4},
        pages = {4},
          doi = {10.1088/0004-637X/784/1/4},
archivePrefix = {arXiv},
       eprint = {1401.3759},
 primaryClass = {astro-ph.GA},
       adsurl = {https://ui.adsabs.harvard.edu/abs/2014ApJ...784....4C}
}

@ARTICLE{2005ApJ...634..281H,
       author = {{Hunter}, Deidre A. and {Rubin}, Vera C. and {Swaters}, Rob A. and {Sparke}, Linda S. and {Levine}, Stephen E.},
        title = "{The Stellar Velocity Dispersion in the Inner 1.3 Disk Scale Lengths of the Irregular Galaxy NGC 4449}",
      journal = {\apj},
         year = 2005,
        month = nov,
       volume = {634},
       number = {1},
        pages = {281-286},
          doi = {10.1086/496949},
archivePrefix = {arXiv},
       eprint = {astro-ph/0508303},
 primaryClass = {astro-ph},
       adsurl = {https://ui.adsabs.harvard.edu/abs/2005ApJ...634..281H}
}

@ARTICLE{2015MNRAS.450.3254P,
       author = {{Pilyugin}, L.~S. and {Grebel}, E.~K. and {Zinchenko}, I.~A.},
        title = "{On the radial abundance gradients in discs of irregular galaxies}",
      journal = {\mnras},
         year = 2015,
        month = jul,
       volume = {450},
       number = {3},
        pages = {3254-3263},
          doi = {10.1093/mnras/stv932},
       adsurl = {https://ui.adsabs.harvard.edu/abs/2015MNRAS.450.3254P}
}

@ARTICLE{2004MNRAS.348L..59P,
       author = {{Pettini}, Max and {Pagel}, Bernard E.~J.},
        title = "{[OIII]/[NII] as an abundance indicator at high redshift}",
      journal = {\mnras},
         year = 2004,
        month = mar,
       volume = {348},
       number = {3},
        pages = {L59-L63},
          doi = {10.1111/j.1365-2966.2004.07591.x},
archivePrefix = {arXiv},
       eprint = {astro-ph/0401128},
 primaryClass = {astro-ph},
       adsurl = {https://ui.adsabs.harvard.edu/abs/2004MNRAS.348L..59P}
}

@ARTICLE{2015AJ....149...51C,
       author = {{Calzetti}, D. and {Lee}, J.~C. and {Sabbi}, E. and {Adamo}, A. and {Smith}, L.~J. and {Andrews}, J.~E. and {Ubeda}, L. and {Bright}, S.~N. and {Thilker}, D. and {Aloisi}, A. and {Brown}, T.~M. and {Chandar}, R. and {Christian}, C. and {Cignoni}, M. and {Clayton}, G.~C. and {da Silva}, R. and {de Mink}, S.~E. and {Dobbs}, C. and {Elmegreen}, B.~G. and {Elmegreen}, D.~M. and {Evans}, A.~S. and {Fumagalli}, M. and {Gallagher}, III, J.~S. and {Gouliermis}, D.~A. and {Grebel}, E.~K. and {Herrero}, A. and {Hunter}, D.~A. and {Johnson}, K.~E. and {Kennicutt}, R.~C. and {Kim}, H. and {Krumholz}, M.~R. and {Lennon}, D. and {Levay}, K. and {Martin}, C. and {Nair}, P. and {Nota}, A. and {{\"O}stlin}, G. and {Pellerin}, A. and {Prieto}, J. and {Regan}, M.~W. and {Ryon}, J.~E. and {Schaerer}, D. and {Schiminovich}, D. and {Tosi}, M. and {Van Dyk}, S.~D. and {Walterbos}, R. and {Whitmore}, B.~C. and {Wofford}, A.},
        title = "{Legacy Extragalactic UV Survey (LEGUS) With the Hubble Space Telescope. I. Survey Description}",
      journal = {\aj},
         year = 2015,
        month = feb,
       volume = {149},
       number = {2},
          eid = {51},
        pages = {51},
          doi = {10.1088/0004-6256/149/2/51},
archivePrefix = {arXiv},
       eprint = {1410.7456},
 primaryClass = {astro-ph.GA},
       adsurl = {https://ui.adsabs.harvard.edu/abs/2015AJ....149...51C}
}

@ARTICLE{2024ApJ...971...32P,
       author = {{Pedrini}, Alex and {Adamo}, Angela and {Calzetti}, Daniela and {Bik}, Arjan and {Gregg}, Benjamin and {Linden}, Sean T. and {Bajaj}, Varun and {Ryon}, Jenna E. and {Ali}, Ahmad A. and {Bortolini}, Giacomo and {Correnti}, Matteo and {Elmegreen}, Bruce G. and {Elmegreen}, Debra Meloy and {Gallagher}, John S. and {Grasha}, Kathryn and {Gutermuth}, Robert A. and {Johnson}, Kelsey E. and {Melinder}, Jens and {Messa}, Matteo and {{\"O}stlin}, G{\"o}ran and {Sabbi}, Elena and {Smith}, Linda J. and {Tosi}, Monica and {Vieira}, Helena Faustino},
        title = "{FEAST: Feedback in Emerging extragAlactic Star ClusTers: JWST Spots Polycyclic Aromatic Hydrocarbon Destruction in NGC 628 during the Emerging Phase of Star Formation}",
      journal = {\apj},
         year = 2024,
        month = aug,
       volume = {971},
       number = {1},
          eid = {32},
        pages = {32},
          doi = {10.3847/1538-4357/ad534d},
archivePrefix = {arXiv},
       eprint = {2406.01666},
 primaryClass = {astro-ph.GA},
       adsurl = {https://ui.adsabs.harvard.edu/abs/2024ApJ...971...32P}
}

@MISC{2016wfc..rept...12A,
       author = {{Anderson}, Jay},
        title = "{Empirical Models for the WFC3/IR PSF}",
 howpublished = {Instrument Science Report WFC3 2016-12, 42 pages},
         year = 2016,
        month = mar,
        pages = {12},
       adsurl = {https://ui.adsabs.harvard.edu/abs/2016wfc..rept...12A}
}

@MISC{2022wfc..rept....5A,
       author = {{Anderson}, Jay},
        title = "{One-Pass HST Photometry with hst1pass}",
 howpublished = {Instrument Science Report WFC3 2022-5, 55 pages},
         year = 2022,
        month = jul,
        pages = {5},
       adsurl = {https://ui.adsabs.harvard.edu/abs/2022wfc..rept....5A}
}

@ARTICLE{2024ApJ...971..115G,
       author = {{Gregg}, Benjamin and {Calzetti}, Daniela and {Adamo}, Angela and {Bajaj}, Varun and {Ryon}, Jenna E. and {Linden}, Sean T. and {Correnti}, Matteo and {Cignoni}, Michele and {Messa}, Matteo and {Sabbi}, Elena and {Gallagher}, John S. and {Grasha}, Kathryn and {Pedrini}, Alex and {Gutermuth}, Robert A. and {Melinder}, Jens and {Kotulla}, Ralf and {P{\'e}rez}, Gustavo and {Krumholz}, Mark R. and {Bik}, Arjan and {{\"O}stlin}, G{\"o}ran and {Johnson}, Kelsey E. and {Bortolini}, Giacomo and {Smith}, Linda J. and {Tosi}, Monica and {Maji}, Subhransu and {Faustino Vieira}, Helena},
        title = "{Feedback in Emerging Extragalactic Star Clusters, FEAST: The Relation between 3.3 {\ensuremath{\mu}}m Polycyclic Aromatic Hydrocarbon Emission and Star Formation Rate Traced by Ionized Gas in NGC 628}",
      journal = {\apj},
         year = 2024,
        month = aug,
       volume = {971},
       number = {1},
          eid = {115},
        pages = {115},
          doi = {10.3847/1538-4357/ad54b4},
archivePrefix = {arXiv},
       eprint = {2405.09667},
 primaryClass = {astro-ph.GA},
       adsurl = {https://ui.adsabs.harvard.edu/abs/2024ApJ...971..115G}
}

@ARTICLE{2024ApJ...971..118C,
       author = {{Calzetti}, Daniela and {Adamo}, Angela and {Linden}, Sean T. and {Gregg}, Benjamin and {Krumholz}, Mark R. and {Bajaj}, Varun and {Bik}, Arjan and {Cignoni}, Michele and {Correnti}, Matteo and {Elmegreen}, Bruce and {Faustino Vieira}, Helena and {Gallagher}, John S. and {Grasha}, Kathryn and {Gutermuth}, Robert A. and {Johnson}, Kelsey E. and {Messa}, Matteo and {Melinder}, Jens and {{\"O}stlin}, G{\"o}ran and {Pedrini}, Alex and {Sabbi}, Elena and {Smith}, Linda J. and {Tosi}, Monica},
        title = "{JWST-FEAST: Feedback in Emerging extrAgalactic Star clusTers: Calibration of Star Formation Rates in the Mid-infrared with NGC 628}",
      journal = {\apj},
         year = 2024,
        month = aug,
       volume = {971},
       number = {1},
          eid = {118},
        pages = {118},
          doi = {10.3847/1538-4357/ad53c0},
archivePrefix = {arXiv},
       eprint = {2406.01831},
 primaryClass = {astro-ph.GA},
       adsurl = {https://ui.adsabs.harvard.edu/abs/2024ApJ...971..118C}
}

@ARTICLE{2001MNRAS.322..231K,
       author = {{Kroupa}, Pavel},
        title = "{On the variation of the initial mass function}",
      journal = {\mnras},
         year = 2001,
        month = apr,
       volume = {322},
       number = {2},
        pages = {231-246},
          doi = {10.1046/j.1365-8711.2001.04022.x},
archivePrefix = {arXiv},
       eprint = {astro-ph/0009005},
 primaryClass = {astro-ph},
       adsurl = {https://ui.adsabs.harvard.edu/abs/2001MNRAS.322..231K}
}

@ARTICLE{2023A&A...678A..44C,
       author = {{Cs{\"o}rnyei}, G. and {Anderson}, R.~I. and {Vogl}, C. and {Taubenberger}, S. and {Blondin}, S. and {Leibundgut}, B. and {Hillebrandt}, W.},
        title = "{Reeling in the Whirlpool galaxy: Distance to M 51 clarified through Cepheids and the type IIP supernova 2005cs}",
      journal = {\aap},
         year = 2023,
        month = oct,
       volume = {678},
          eid = {A44},
        pages = {A44},
          doi = {10.1051/0004-6361/202346971},
archivePrefix = {arXiv},
       eprint = {2305.13943},
 primaryClass = {astro-ph.GA},
       adsurl = {https://ui.adsabs.harvard.edu/abs/2023A&A...678A..44C}
}

@ARTICLE{2024A&A...687A..86V,
       author = {{Van De Putte}, Dries and {Meshaka}, Raphael and {Trahin}, Boris and {Habart}, Emilie and {Peeters}, Els and {Bern{\'e}}, Olivier and {Alarc{\'o}n}, Felipe and {Canin}, Am{\'e}lie and {Chown}, Ryan and {Schroetter}, Ilane and {Sidhu}, Ameek and {Boersma}, Christiaan and {Bron}, Emeric and {Dartois}, Emmanuel and {Goicoechea}, Javier R. and {Gordon}, Karl D. and {Onaka}, Takashi and {Tielens}, Alexander G.~G.~M. and {Verstraete}, Laurent and {Wolfire}, Mark G. and {Abergel}, Alain and {Bergin}, Edwin A. and {Bernard-Salas}, Jeronimo and {Cami}, Jan and {Cuadrado}, Sara and {Dicken}, Daniel and {Elyajouri}, Meriem and {Fuente}, Asunci{\'o}n and {Joblin}, Christine and {Khan}, Baria and {Lacinbala}, Ozan and {Languignon}, David and {Le Gal}, Romane and {Maragkoudakis}, Alexandros and {Okada}, Yoko and {Pasquini}, Sofia and {Pound}, Marc W. and {Robberto}, Massimo and {R{\"o}llig}, Markus and {Schefter}, Bethany and {Schirmer}, Thi{\'e}baut and {Tabone}, Benoit and {Vicente}, S{\'\i}lvia and {Zannese}, Marion and {Colgan}, Sean W.~J. and {He}, Jinhua and {Rouill{\'e}}, Ga{\"e}l and {Togi}, Aditya and {Aleman}, Isabel and {Auchettl}, Rebecca and {Baratta}, Giuseppe Antonio and {Bejaoui}, Salma and {Bera}, Partha P. and {Black}, John H. and {Boulanger}, Francois and {Bouwman}, Jordy and {Brandl}, Bernhard and {Brechignac}, Philippe and {Br{\"u}nken}, Sandra and {Buragohain}, Mridusmita and {Burkhardt}, Andrew and {Candian}, Alessandra and {Cazaux}, St{\'e}phanie and {Cernicharo}, Jose and {Chabot}, Marin and {Chakraborty}, Shubhadip and {Champion}, Jason and {Cooke}, Ilsa R. and {Coutens}, Audrey and {Cox}, Nick L.~J. and {Demyk}, Karine and {Meyer}, Jennifer Donovan and {Foschino}, Sacha and {Garc{\'\i}a-Lario}, Pedro and {Gerin}, Maryvonne and {Gottlieb}, Carl A. and {Guillard}, Pierre and {Gusdorf}, Antoine and {Hartigan}, Patrick and {Herbst}, Eric and {Hornekaer}, Liv and {Issa}, Lina and {J{\"a}ger}, Cornelia and {Janot-Pacheco}, Eduardo and {Kannavou}, Olga and {Kaufman}, Michael and {Kemper}, Francisca and {Kendrew}, Sarah and {Kirsanova}, Maria S. and {Klaassen}, Pamela and {Kwok}, Sun and {Labiano}, {\'A}lvaro and {Lai}, Thomas S. -Y. and {Le Floch}, Bertrand and {Le Petit}, Franck and {Li}, Aigen and {Linz}, Hendrik and {Mackie}, Cameron J. and {Madden}, Suzanne C. and {Mascetti}, Jo{\"e}lle and {McGuire}, Brett A. and {Merino}, Pablo and {Micelotta}, Elisabetta R. and {Morse}, Jon A. and {Mulas}, Giacomo and {Neelamkodan}, Naslim and {Ohsawa}, Ryou and {Omont}, Alain and {Paladini}, Roberta and {Palumbo}, Maria Elisabetta and {Pathak}, Amit and {Pendleton}, Yvonne J. and {Petrignani}, Annemieke and {Pino}, Thomas and {Puga}, Elena and {Rangwala}, Naseem and {Rapacioli}, Mathias and {Rho}, Jeonghee and {Ricca}, Alessandra and {Roman-Duval}, Julia and {Roser}, Joseph and {Roueff}, Evelyne and {Salama}, Farid and {Sales}, Dinalva A. and {Sandstrom}, Karin and {Sarre}, Peter and {Sciamma-O'Brien}, Ella and {Sellgren}, Kris and {Shenoy}, Sachindev S. and {Teyssier}, David and {Thomas}, Richard D. and {Witt}, Adolf N. and {Wootten}, Alwyn and {Ysard}, Nathalie and {Zettergren}, Henning and {Zhang}, Yong and {Zhang}, Ziwei E. and {Zhen}, Junfeng},
        title = "{PDRs4All. VIII. Mid-infrared emission line inventory of the Orion Bar}",
      journal = {\aap},
         year = 2024,
        month = jul,
       volume = {687},
          eid = {A86},
        pages = {A86},
          doi = {10.1051/0004-6361/202449295},
archivePrefix = {arXiv},
       eprint = {2404.03111},
 primaryClass = {astro-ph.GA},
       adsurl = {https://ui.adsabs.harvard.edu/abs/2024A&A...687A..86V}
}

@ARTICLE{2024ApJ...974...20W,
       author = {{Whitcomb}, Cory M. and {Smith}, J. -D.~T. and {Sandstrom}, Karin and {Starkey}, Carl A. and {Donnelly}, Grant P. and {Draine}, Bruce T. and {Skillman}, Evan D. and {Dale}, Daniel A. and {Armus}, Lee and {Hensley}, Brandon S. and {Lai}, Thomas S. -Y. and {Kennicutt}, Robert C.},
        title = "{The Metallicity Dependence of PAH Emission in Galaxies. I. Insights from Deep Radial Spitzer Spectroscopy}",
      journal = {\apj},
         year = 2024,
        month = oct,
       volume = {974},
       number = {1},
          eid = {20},
        pages = {20},
          doi = {10.3847/1538-4357/ad66c8},
archivePrefix = {arXiv},
       eprint = {2405.09685},
 primaryClass = {astro-ph.GA},
       adsurl = {https://ui.adsabs.harvard.edu/abs/2024ApJ...974...20W}
}

@ARTICLE{2005ApJ...627..477M,
       author = {{Massey}, Philip and {Puls}, Joachim and {Pauldrach}, A.~W.~A. and {Bresolin}, Fabio and {Kudritzki}, Rolf P. and {Simon}, Theodore},
        title = "{The Physical Properties and Effective Temperature Scale of O-Type Stars as a Function of Metallicity. II. Analysis of 20 More Magellanic Cloud Stars and Results from the Complete Sample}",
      journal = {\apj},
         year = 2005,
        month = jul,
       volume = {627},
       number = {1},
        pages = {477-519},
          doi = {10.1086/430417},
archivePrefix = {arXiv},
       eprint = {astro-ph/0503464},
 primaryClass = {astro-ph},
       adsurl = {https://ui.adsabs.harvard.edu/abs/2005ApJ...627..477M}
}

@ARTICLE{1996A&A...305..602A,
       author = {{Allain}, T. and {Leach}, S. and {Sedlmayr}, E.},
        title = "{Photodestruction of PAHs in the interstellar medium. I. Photodissociation rates for the loss of an acetylenic group.}",
      journal = {\aap},
         year = 1996,
        month = jan,
       volume = {305},
        pages = {602},
       adsurl = {https://ui.adsabs.harvard.edu/abs/1996A&A...305..602A}
}

@ARTICLE{2017NIMPB.408...21B,
       author = {{Bernard}, J{\'e}r{\^o}me and {Chen}, Li and {Br{\'e}dy}, Richard and {Ji}, Mingchao and {Ort{\'e}ga}, C{\'e}line and {Matsumoto}, Jun and {Martin}, Serge},
        title = "{Cooling of PAH cations studied with an electrostatic storage ring}",
      journal = {Nuclear Instruments and Methods in Physics Research B},
         year = 2017,
        month = oct,
       volume = {408},
        pages = {21-26},
          doi = {10.1016/j.nimb.2017.03.142},
       adsurl = {https://ui.adsabs.harvard.edu/abs/2017NIMPB.408...21B}
}

@ARTICLE{2003MNRAS.344.1000B,
       author = {{Bruzual}, G. and {Charlot}, S.},
        title = "{Stellar population synthesis at the resolution of 2003}",
      journal = {\mnras},
         year = 2003,
        month = oct,
       volume = {344},
       number = {4},
        pages = {1000-1028},
          doi = {10.1046/j.1365-8711.2003.06897.x},
archivePrefix = {arXiv},
       eprint = {astro-ph/0309134},
 primaryClass = {astro-ph},
       adsurl = {https://ui.adsabs.harvard.edu/abs/2003MNRAS.344.1000B}
}

@ARTICLE{1989ApJ...345..230G,
       author = {{Guhathakurta}, P. and {Draine}, B.~T.},
        title = "{Temperature Fluctuations in the Interstellar Grains. I. Computational Method and Sublimation of Small Grains}",
      journal = {\apj},
         year = 1989,
        month = oct,
       volume = {345},
        pages = {230},
          doi = {10.1086/167899},
       adsurl = {https://ui.adsabs.harvard.edu/abs/1989ApJ...345..230G}
}

@ARTICLE{2003ApJ...584..316L,
       author = {{Le Page}, Valery and {Snow}, Theodore P. and {Bierbaum}, Veronica M.},
        title = "{Hydrogenation and Charge States of Polycyclic Aromatic Hydrocarbons in Diffuse Clouds. II. Results}",
      journal = {\apj},
         year = 2003,
        month = feb,
       volume = {584},
       number = {1},
        pages = {316-330},
          doi = {10.1086/345595},
       adsurl = {https://ui.adsabs.harvard.edu/abs/2003ApJ...584..316L}
}

@ARTICLE{2022Univ....8..356S,
       author = {{Sajina}, Anna and {Lacy}, Mark and {Pope}, Alexandra},
        title = "{The Past and Future of Mid-Infrared Studies of AGN}",
      journal = {Universe},
         year = 2022,
        month = jun,
       volume = {8},
       number = {7},
          eid = {356},
        pages = {356},
          doi = {10.3390/universe8070356},
archivePrefix = {arXiv},
       eprint = {2210.02307},
 primaryClass = {astro-ph.GA},
       adsurl = {https://ui.adsabs.harvard.edu/abs/2022Univ....8..356S}
}

@ARTICLE{2023ApJ...944L...8S,
       author = {{Sandstrom}, Karin M. and {Koch}, Eric W. and {Leroy}, Adam K. and {Rosolowsky}, Erik and {Emsellem}, Eric and {Smith}, Rowan J. and {Egorov}, Oleg V. and {Williams}, Thomas G. and {Larson}, Kirsten L. and {Lee}, Janice C. and {Schinnerer}, Eva and {Thilker}, David A. and {Barnes}, Ashley T. and {Belfiore}, Francesco and {Bigiel}, F. and {Blanc}, Guillermo A. and {Bolatto}, Alberto D. and {Boquien}, M{\'e}d{\'e}ric and {Cao}, Yixian and {Chastenet}, J{\'e}r{\'e}my and {Chevance}, M{\'e}lanie and {Chiang}, I-Da and {Dale}, Daniel A. and {Faesi}, Christopher M. and {Glover}, Simon C.~O. and {Grasha}, Kathryn and {Groves}, Brent and {Hassani}, Hamid and {Henshaw}, Jonathan D. and {Hughes}, Annie and {Kim}, Jaeyeon and {Klessen}, Ralf S. and {Kreckel}, Kathryn and {Kruijssen}, J.~M. Diederik and {Lopez}, Laura A. and {Liu}, Daizhong and {Meidt}, Sharon E. and {Murphy}, Eric J. and {Pan}, Hsi-An and {Querejeta}, Miguel and {Saito}, Toshiki and {Sardone}, Amy and {Sormani}, Mattia C. and {Sutter}, Jessica and {Usero}, Antonio and {Watkins}, Elizabeth J.},
        title = "{PHANGS-JWST First Results: Tracing the Diffuse Interstellar Medium with JWST Imaging of Polycyclic Aromatic Hydrocarbon Emission in Nearby Galaxies}",
      journal = {\apjl},
         year = 2023,
        month = feb,
       volume = {944},
       number = {2},
          eid = {L8},
        pages = {L8},
          doi = {10.3847/2041-8213/aca972},
archivePrefix = {arXiv},
       eprint = {2212.11177},
 primaryClass = {astro-ph.GA},
       adsurl = {https://ui.adsabs.harvard.edu/abs/2023ApJ...944L...8S}
}

@ARTICLE{2024A&A...685A..78S,
       author = {{Schroetter}, Ilane and {Bern{\'e}}, Olivier and {Joblin}, Christine and {Canin}, Am{\'e}lie and {Chown}, Ryan and {Sidhu}, Ameek and {Habart}, Emilie and {Peeters}, Els and {Lai}, Thomas S. -Y. and {Candian}, Alessandra and {Chakraborty}, Shubhadip and {Petrignani}, Annemieke and {Trahin}, Boris and {Van De Putte}, Dries and {Alarc{\'o}n}, Felipe},
        title = "{PDRs4All. VII. The 3.3 {\ensuremath{\mu}}m aromatic infrared band as a tracer of physical properties of the interstellar medium in galaxies}",
      journal = {\aap},
         year = 2024,
        month = may,
       volume = {685},
          eid = {A78},
        pages = {A78},
          doi = {10.1051/0004-6361/202348974},
archivePrefix = {arXiv},
       eprint = {2402.16535},
 primaryClass = {astro-ph.GA},
       adsurl = {https://ui.adsabs.harvard.edu/abs/2024A&A...685A..78S}
}

@ARTICLE{2013ApJ...762...79C,
       author = {{Crocker}, Alison F. and {Calzetti}, Daniela and {Thilker}, David A. and {Aniano}, Gonzalo and {Draine}, Bruce T. and {Hunt}, Leslie K. and {Kennicutt}, Robert C. and {Sandstrom}, Karin and {Smith}, J.~D.~T.},
        title = "{Quantifying Non-star-formation-associated 8 {\ensuremath{\mu}}m Dust Emission in NGC 628}",
      journal = {\apj},
         year = 2013,
        month = jan,
       volume = {762},
       number = {2},
          eid = {79},
        pages = {79},
          doi = {10.1088/0004-637X/762/2/79},
archivePrefix = {arXiv},
       eprint = {1211.3332},
 primaryClass = {astro-ph.CO},
       adsurl = {https://ui.adsabs.harvard.edu/abs/2013ApJ...762...79C}
}

@ARTICLE{2010ApJ...712..164H,
       author = {{Hunt}, Leslie K. and {Thuan}, Trinh X. and {Izotov}, Yuri I. and {Sauvage}, Marc},
        title = "{The Spitzer View of Low-Metallicity Star Formation. III. Fine-Structure Lines, Aromatic Features, and Molecules}",
      journal = {\apj},
         year = 2010,
        month = mar,
       volume = {712},
       number = {1},
        pages = {164-187},
          doi = {10.1088/0004-637X/712/1/164},
archivePrefix = {arXiv},
       eprint = {1002.0991},
 primaryClass = {astro-ph.CO},
       adsurl = {https://ui.adsabs.harvard.edu/abs/2010ApJ...712..164H}
}

@ARTICLE{2024A&A...690A..89S,
       author = {{Shivaei}, Irene and {Alberts}, Stacey and {Florian}, Michael and {Rieke}, George and {Wuyts}, Stijn and {Bodansky}, Sarah and {Bunker}, Andrew J. and {Cameron}, Alex J. and {Curti}, Mirko and {D'Eugenio}, Francesco and {Dudzevi{\v{c}}i{\={u}}t{\.{e}}}, Ugn{\.{e}} and {Ji}, Zhiyuan and {Johnson}, Benjamin D. and {Kramarenko}, Ivan and {Lyu}, Jianwei and {Matthee}, Jorryt and {Morrison}, Jane and {Naidu}, Rohan and {P{\'e}rez-Gonz{\'a}lez}, Pablo G. and {Reddy}, Naveen and {Robertson}, Brant and {Sun}, Yang and {Tacchella}, Sandro and {Whitaker}, Katherine and {Williams}, Christina C. and {Willmer}, Christopher N.~A. and {Witstok}, Joris and {Xiao}, Mengyuan and {Zhu}, Yongda},
        title = "{A new census of dust and polycyclic aromatic hydrocarbons at z = 0.7{\textendash}2 with JWST MIRI}",
      journal = {\aap},
         year = 2024,
        month = oct,
       volume = {690},
          eid = {A89},
        pages = {A89},
          doi = {10.1051/0004-6361/202449579},
archivePrefix = {arXiv},
       eprint = {2402.07989},
 primaryClass = {astro-ph.GA},
       adsurl = {https://ui.adsabs.harvard.edu/abs/2024A&A...690A..89S}
}

@ARTICLE{2024A&A...685A..73H,
       author = {{Habart}, Emilie and {Peeters}, Els and {Bern{\'e}}, Olivier and {Trahin}, Boris and {Canin}, Am{\'e}lie and {Chown}, Ryan and {Sidhu}, Ameek and {Van De Putte}, Dries and {Alarc{\'o}n}, Felipe and {Schroetter}, Ilane and {Dartois}, Emmanuel and {Vicente}, S{\'\i}lvia and {Abergel}, Alain and {Bergin}, Edwin A. and {Bernard-Salas}, Jeronimo and {Boersma}, Christiaan and {Bron}, Emeric and {Cami}, Jan and {Cuadrado}, Sara and {Dicken}, Daniel and {Elyajouri}, Meriem and {Fuente}, Asunci{\'o}n and {Goicoechea}, Javier R. and {Gordon}, Karl D. and {Issa}, Lina and {Joblin}, Christine and {Kannavou}, Olga and {Khan}, Baria and {Lacinbala}, Ozan and {Languignon}, David and {Le Gal}, Romane and {Maragkoudakis}, Alexandros and {Meshaka}, Raphael and {Okada}, Yoko and {Onaka}, Takashi and {Pasquini}, Sofia and {Pound}, Marc W. and {Robberto}, Massimo and {R{\"o}llig}, Markus and {Schefter}, Bethany and {Schirmer}, Thi{\'e}baut and {Tabone}, Benoit and {Tielens}, Alexander G.~G.~M. and {Wolfire}, Mark G. and {Zannese}, Marion and {Ysard}, Nathalie and {Miville-Deschenes}, Marc-Antoine and {Aleman}, Isabel and {Allamandola}, Louis and {Auchettl}, Rebecca and {Baratta}, Giuseppe Antonio and {Bejaoui}, Salma and {Bera}, Partha P. and {Black}, John H. and {Boulanger}, Francois and {Bouwman}, Jordy and {Brandl}, Bernhard and {Brechignac}, Philippe and {Br{\"u}nken}, Sandra and {Buragohain}, Mridusmita and {Burkhardt}, Andrew and {Candian}, Alessandra and {Cazaux}, St{\'e}phanie and {Cernicharo}, Jose and {Chabot}, Marin and {Chakraborty}, Shubhadip and {Champion}, Jason and {Colgan}, Sean W.~J. and {Cooke}, Ilsa R. and {Coutens}, Audrey and {Cox}, Nick L.~J. and {Demyk}, Karine and {Meyer}, Jennifer Donovan and {Foschino}, Sacha and {Garc{\'\i}a-Lario}, Pedro and {Gavilan}, Lisseth and {Gerin}, Maryvonne and {Gottlieb}, Carl A. and {Guillard}, Pierre and {Gusdorf}, Antoine and {Hartigan}, Patrick and {He}, Jinhua and {Herbst}, Eric and {Hornekaer}, Liv and {J{\"a}ger}, Cornelia and {Janot-Pacheco}, Eduardo and {Kaufman}, Michael and {Kemper}, Francisca and {Kendrew}, Sarah and {Kirsanova}, Maria S. and {Klaassen}, Pamela and {Kwok}, Sun and {Labiano}, {\'A}lvaro and {Lai}, Thomas S. -Y. and {Lee}, Timothy J. and {Lefloch}, Bertrand and {Le Petit}, Franck and {Li}, Aigen and {Linz}, Hendrik and {Mackie}, Cameron J. and {Madden}, Suzanne C. and {Mascetti}, Jo{\"e}lle and {McGuire}, Brett A. and {Merino}, Pablo and {Micelotta}, Elisabetta R. and {Misselt}, Karl and {Morse}, Jon A. and {Mulas}, Giacomo and {Neelamkodan}, Naslim and {Ohsawa}, Ryou and {Omont}, Alain and {Paladini}, Roberta and {Palumbo}, Maria Elisabetta and {Pathak}, Amit and {Pendleton}, Yvonne J. and {Petrignani}, Annemieke and {Pino}, Thomas and {Puga}, Elena and {Rangwala}, Naseem and {Rapacioli}, Mathias and {Ricca}, Alessandra and {Roman-Duval}, Julia and {Roser}, Joseph and {Roueff}, Evelyne and {Rouill{\'e}}, Ga{\"e}l and {Salama}, Farid and {Sales}, Dinalva A. and {Sandstrom}, Karin and {Sarre}, Peter and {Sciamma-O'Brien}, Ella and {Sellgren}, Kris and {Shenoy}, Sachindev S. and {Teyssier}, David and {Thomas}, Richard D. and {Togi}, Aditya and {Verstraete}, Laurent and {Witt}, Adolf N. and {Wootten}, Alwyn and {Zettergren}, Henning and {Zhang}, Yong and {Zhang}, Ziwei E. and {Zhen}, Junfeng},
        title = "{PDRs4All. II. JWST's NIR and MIR imaging view of the Orion Nebula}",
      journal = {\aap},
         year = 2024,
        month = may,
       volume = {685},
          eid = {A73},
        pages = {A73},
          doi = {10.1051/0004-6361/202346747},
archivePrefix = {arXiv},
       eprint = {2308.16732},
 primaryClass = {astro-ph.GA},
       adsurl = {https://ui.adsabs.harvard.edu/abs/2024A&A...685A..73H}
}

@ARTICLE{2024A&A...685A..75C,
       author = {{Chown}, Ryan and {Sidhu}, Ameek and {Peeters}, Els and {Tielens}, Alexander G.~G.~M. and {Cami}, Jan and {Bern{\'e}}, Olivier and {Habart}, Emilie and {Alarc{\'o}n}, Felipe and {Canin}, Am{\'e}lie and {Schroetter}, Ilane and {Trahin}, Boris and {Van De Putte}, Dries and {Abergel}, Alain and {Bergin}, Edwin A. and {Bernard-Salas}, Jeronimo and {Boersma}, Christiaan and {Bron}, Emeric and {Cuadrado}, Sara and {Dartois}, Emmanuel and {Dicken}, Daniel and {El-Yajouri}, Meriem and {Fuente}, Asunci{\'o}n and {Goicoechea}, Javier R. and {Gordon}, Karl D. and {Issa}, Lina and {Joblin}, Christine and {Kannavou}, Olga and {Khan}, Baria and {Lacinbala}, Ozan and {Languignon}, David and {Le Gal}, Romane and {Maragkoudakis}, Alexandros and {Meshaka}, Raphael and {Okada}, Yoko and {Onaka}, Takashi and {Pasquini}, Sofia and {Pound}, Marc W. and {Robberto}, Massimo and {R{\"o}llig}, Markus and {Schefter}, Bethany and {Schirmer}, Thi{\'e}baut and {Vicente}, S{\'\i}lvia and {Wolfire}, Mark G. and {Zannese}, Marion and {Aleman}, Isabel and {Allamandola}, Louis and {Auchettl}, Rebecca and {Baratta}, Giuseppe Antonio and {Bejaoui}, Salma and {Bera}, Partha P. and {Black}, John H. and {Boulanger}, Fran{\c{c}}ois and {Bouwman}, Jordy and {Brandl}, Bernhard and {Brechignac}, Philippe and {Br{\"u}nken}, Sandra and {Buragohain}, Mridusmita and {Burkhardt}, Andrew and {Candian}, Alessandra and {Cazaux}, St{\'e}phanie and {Cernicharo}, Jose and {Chabot}, Marin and {Chakraborty}, Shubhadip and {Champion}, Jason and {Colgan}, Sean W.~J. and {Cooke}, Ilsa R. and {Coutens}, Audrey and {Cox}, Nick L.~J. and {Demyk}, Karine and {Meyer}, Jennifer Donovan and {Foschino}, Sacha and {Garc{\'\i}a-Lario}, Pedro and {Gavilan}, Lisseth and {Gerin}, Maryvonne and {Gottlieb}, Carl A. and {Guillard}, Pierre and {Gusdorf}, Antoine and {Hartigan}, Patrick and {He}, Jinhua and {Herbst}, Eric and {Hornekaer}, Liv and {J{\"a}ger}, Cornelia and {Janot-Pacheco}, Eduardo and {Kaufman}, Michael and {Kemper}, Francisca and {Kendrew}, Sarah and {Kirsanova}, Maria S. and {Klaassen}, Pamela and {Kwok}, Sun and {Labiano}, {\'A}lvaro and {Lai}, Thomas S. -Y. and {Lee}, Timothy J. and {Lefloch}, Bertrand and {Le Petit}, Franck and {Li}, Aigen and {Linz}, Hendrik and {Mackie}, Cameron J. and {Madden}, Suzanne C. and {Mascetti}, Jo{\"e}lle and {McGuire}, Brett A. and {Merino}, Pablo and {Micelotta}, Elisabetta R. and {Misselt}, Karl and {Morse}, Jon A. and {Mulas}, Giacomo and {Neelamkodan}, Naslim and {Ohsawa}, Ryou and {Omont}, Alain and {Paladini}, Roberta and {Palumbo}, Maria Elisabetta and {Pathak}, Amit and {Pendleton}, Yvonne J. and {Petrignani}, Annemieke and {Pino}, Thomas and {Puga}, Elena and {Rangwala}, Naseem and {Rapacioli}, Mathias and {Ricca}, Alessandra and {Roman-Duval}, Julia and {Roser}, Joseph and {Roueff}, Evelyne and {Rouill{\'e}}, Ga{\"e}l and {Salama}, Farid and {Sales}, Dinalva A. and {Sandstrom}, Karin and {Sarre}, Peter and {Sciamma-O'Brien}, Ella and {Sellgren}, Kris and {Shenoy}, Sachindev S. and {Teyssier}, David and {Thomas}, Richard D. and {Togi}, Aditya and {Verstraete}, Laurent and {Witt}, Adolf N. and {Wootten}, Alwyn and {Zettergren}, Henning and {Zhang}, Yong and {Zhang}, Ziwei E. and {Zhen}, Junfeng},
        title = "{PDRs4All. IV. An embarrassment of riches: Aromatic infrared bands in the Orion Bar}",
      journal = {\aap},
         year = 2024,
        month = may,
       volume = {685},
          eid = {A75},
        pages = {A75},
          doi = {10.1051/0004-6361/202346662},
archivePrefix = {arXiv},
       eprint = {2308.16733},
 primaryClass = {astro-ph.GA},
       adsurl = {https://ui.adsabs.harvard.edu/abs/2024A&A...685A..75C}
}

@ARTICLE{2024A&A...685A..77P,
       author = {{Pasquini}, Sofia and {Peeters}, Els and {Schefter}, Bethany and {Khan}, Baria and {Sidhu}, Ameek and {Chown}, Ryan and {Cami}, Jan and {Tielens}, Alexander and {Alarc{\'o}n}, Felipe and {Canin}, Am{\'e}lie and {Schroetter}, Ilane and {Trahin}, Boris and {Van De Putte}, Dries and {Boersma}, Christiaan and {Dartois}, Emmanuel and {Onaka}, Takashi and {Candian}, Alessandra and {Hartigan}, Patrick and {Lai}, Thomas S. -Y. and {Rouill{\'e}}, Ga{\"e}l and {Sales}, Dinalva A. and {Zhang}, Yong and {Bernard-Salas}, Jeronimo and {Habart}, Emilie and {Bern{\'e}}, Olivier},
        title = "{PDRs4All. VI. Probing the photochemical evolution of PAHs in the Orion Bar using machine learning techniques}",
      journal = {\aap},
         year = 2024,
        month = may,
       volume = {685},
          eid = {A77},
        pages = {A77},
          doi = {10.1051/0004-6361/202348465},
archivePrefix = {arXiv},
       eprint = {2311.01163},
 primaryClass = {astro-ph.GA},
       adsurl = {https://ui.adsabs.harvard.edu/abs/2024A&A...685A..77P}
}

@ARTICLE{2024A&A...685A..74P,
       author = {{Peeters}, Els and {Habart}, Emilie and {Bern{\'e}}, Olivier and {Sidhu}, Ameek and {Chown}, Ryan and {Van De Putte}, Dries and {Trahin}, Boris and {Schroetter}, Ilane and {Canin}, Am{\'e}lie and {Alarc{\'o}n}, Felipe and {Schefter}, Bethany and {Khan}, Baria and {Pasquini}, Sofia and {Tielens}, Alexander G.~G.~M. and {Wolfire}, Mark G. and {Dartois}, Emmanuel and {Goicoechea}, Javier R. and {Maragkoudakis}, Alexandros and {Onaka}, Takashi and {Pound}, Marc W. and {Vicente}, S{\'\i}lvia and {Abergel}, Alain and {Bergin}, Edwin A. and {Bernard-Salas}, Jeronimo and {Boersma}, Christiaan and {Bron}, Emeric and {Cami}, Jan and {Cuadrado}, Sara and {Dicken}, Daniel and {Elyajouri}, Meriem and {Fuente}, Asunci{\'o}n and {Gordon}, Karl D. and {Issa}, Lina and {Joblin}, Christine and {Kannavou}, Olga and {Lacinbala}, Ozan and {Languignon}, David and {Le Gal}, Romane and {Meshaka}, Raphael and {Okada}, Yoko and {Robberto}, Massimo and {R{\"o}llig}, Markus and {Schirmer}, Thi{\'e}baut and {Tabone}, Benoit and {Zannese}, Marion and {Aleman}, Isabel and {Allamandola}, Louis and {Auchettl}, Rebecca and {Baratta}, Giuseppe Antonio and {Bejaoui}, Salma and {Bera}, Partha P. and {Black}, John H. and {Boulanger}, Francois and {Bouwman}, Jordy and {Brandl}, Bernhard and {Brechignac}, Philippe and {Br{\"u}nken}, Sandra and {Buragohain}, Mridusmita and {Burkhardt}, Andrew and {Candian}, Alessandra and {Cazaux}, St{\'e}phanie and {Cernicharo}, Jose and {Chabot}, Marin and {Chakraborty}, Shubhadip and {Champion}, Jason and {Colgan}, Sean W.~J. and {Cooke}, Ilsa R. and {Coutens}, Audrey and {Cox}, Nick L.~J. and {Demyk}, Karine and {Meyer}, Jennifer Donovan and {Foschino}, Sacha and {Garc{\'\i}a-Lario}, Pedro and {Gerin}, Maryvonne and {Gottlieb}, Carl A. and {Guillard}, Pierre and {Gusdorf}, Antoine and {Hartigan}, Patrick and {He}, Jinhua and {Herbst}, Eric and {Hornekaer}, Liv and {J{\"a}ger}, Cornelia and {Janot-Pacheco}, Eduardo and {Kaufman}, Michael and {Kendrew}, Sarah and {Kirsanova}, Maria S. and {Klaassen}, Pamela and {Kwok}, Sun and {Labiano}, {\'A}lvaro and {Lai}, Thomas S. -Y. and {Lee}, Timothy J. and {Lefloch}, Bertrand and {Le Petit}, Franck and {Li}, Aigen and {Linz}, Hendrik and {Mackie}, Cameron J. and {Madden}, Suzanne C. and {Mascetti}, Jo{\"e}lle and {McGuire}, Brett A. and {Merino}, Pablo and {Micelotta}, Elisabetta R. and {Misselt}, Karl and {Morse}, Jon A. and {Mulas}, Giacomo and {Neelamkodan}, Naslim and {Ohsawa}, Ryou and {Paladini}, Roberta and {Palumbo}, Maria Elisabetta and {Pathak}, Amit and {Pendleton}, Yvonne J. and {Petrignani}, Annemieke and {Pino}, Thomas and {Puga}, Elena and {Rangwala}, Naseem and {Rapacioli}, Mathias and {Ricca}, Alessandra and {Roman-Duval}, Julia and {Roser}, Joseph and {Roueff}, Evelyne and {Rouill{\'e}}, Ga{\"e}l and {Salama}, Farid and {Sales}, Dinalva A. and {Sandstrom}, Karin and {Sarre}, Peter and {Sciamma-O'Brien}, Ella and {Sellgren}, Kris and {Shenoy}, Sachindev S. and {Teyssier}, David and {Thomas}, Richard D. and {Togi}, Aditya and {Verstraete}, Laurent and {Witt}, Adolf N. and {Wootten}, Alwyn and {Ysard}, Nathalie and {Zettergren}, Henning and {Zhang}, Yong and {Zhang}, Ziwei E. and {Zhen}, Junfeng},
        title = "{PDRs4All: III. JWST's NIR spectroscopic view of the Orion Bar}",
      journal = {\aap},
         year = 2024,
        month = may,
       volume = {685},
          eid = {A74},
        pages = {A74},
          doi = {10.1051/0004-6361/202348244},
archivePrefix = {arXiv},
       eprint = {2310.08720},
 primaryClass = {astro-ph.GA},
       adsurl = {https://ui.adsabs.harvard.edu/abs/2024A&A...685A..74P}
}

@ARTICLE{2024ApJ...970...61R,
       author = {{Ronayne}, Kaila and {Papovich}, Casey and {Yang}, Guang and {Shen}, Lu and {Dickinson}, Mark and {Kennicutt}, Robert and {Alavi}, Anahita and {Arrabal Haro}, Pablo and {Bagley}, Micaela B. and {Burgarella}, Denis and {Le Bail}, Aur{\'e}lien and {Bell}, Eric F. and {Cleri}, Nikko J. and {Cole}, Justin and {Costantin}, Luca and {de la Vega}, Alexander and {Daddi}, Emanuele and {Elbaz}, David and {Finkelstein}, Steven L. and {Grogin}, Norman A. and {Holwerda}, Benne W. and {Kartaltepe}, Jeyhan S. and {Kirkpatrick}, Allison and {Koekemoer}, Anton M. and {Lucas}, Ray A. and {Magnelli}, Benjamin and {Mobasher}, Bahram and {P{\'e}rez-Gonz{\'a}lez}, Pablo G. and {Prichard}, Laura and {Rafelski}, Marc and {Rodighiero}, Giulia and {Sunnquist}, Ben and {Teplitz}, Harry I. and {Wang}, Xin and {Windhorst}, Rogier A. and {Yung}, L.~Y. Aaron},
        title = "{CEERS: 7.7 {\ensuremath{\mu}}m PAH Star Formation Rate Calibration with JWST MIRI}",
      journal = {\apj},
         year = 2024,
        month = jul,
       volume = {970},
       number = {1},
          eid = {61},
        pages = {61},
          doi = {10.3847/1538-4357/ad5006},
archivePrefix = {arXiv},
       eprint = {2310.07766},
 primaryClass = {astro-ph.GA},
       adsurl = {https://ui.adsabs.harvard.edu/abs/2024ApJ...970...61R}
}

@ARTICLE{2023ApJ...948...55H,
       author = {{Hensley}, Brandon S. and {Draine}, B.~T.},
        title = "{The Astrodust+PAH Model: A Unified Description of the Extinction, Emission, and Polarization from Dust in the Diffuse Interstellar Medium}",
      journal = {\apj},
         year = 2023,
        month = may,
       volume = {948},
       number = {1},
          eid = {55},
        pages = {55},
          doi = {10.3847/1538-4357/acc4c2},
archivePrefix = {arXiv},
       eprint = {2208.12365},
 primaryClass = {astro-ph.GA},
       adsurl = {https://ui.adsabs.harvard.edu/abs/2023ApJ...948...55H}
}

@ARTICLE{1986ApJ...311L..33H,
       author = {{Helou}, George},
        title = "{The IRAS Colors of Normal Galaxies}",
      journal = {\apjl},
         year = 1986,
        month = dec,
       volume = {311},
        pages = {L33},
          doi = {10.1086/184793},
       adsurl = {https://ui.adsabs.harvard.edu/abs/1986ApJ...311L..33H}
}

@ARTICLE{2012ARA&A..50..531K,
       author = {{Kennicutt}, Robert C. and {Evans}, Neal J.},
        title = "{Star Formation in the Milky Way and Nearby Galaxies}",
      journal = {\araa},
         year = 2012,
        month = sep,
       volume = {50},
        pages = {531-608},
          doi = {10.1146/annurev-astro-081811-125610},
archivePrefix = {arXiv},
       eprint = {1204.3552},
 primaryClass = {astro-ph.GA},
       adsurl = {https://ui.adsabs.harvard.edu/abs/2012ARA&A..50..531K}
}

@ARTICLE{2020ApJ...889..150A,
       author = {{Aniano}, G. and {Draine}, B.~T. and {Hunt}, L.~K. and {Sandstrom}, K. and {Calzetti}, D. and {Kennicutt}, R.~C. and {Dale}, D.~A. and {Galametz}, M. and {Gordon}, K.~D. and {Leroy}, A.~K. and {Smith}, J. -D.~T. and {Roussel}, H. and {Sauvage}, M. and {Walter}, F. and {Armus}, L. and {Bolatto}, A.~D. and {Boquien}, M. and {Crocker}, A. and {De Looze}, I. and {Donovan Meyer}, J. and {Helou}, G. and {Hinz}, J. and {Johnson}, B.~D. and {Koda}, J. and {Miller}, A. and {Montiel}, E. and {Murphy}, E.~J. and {Rela{\~n}o}, M. and {Rix}, H. -W. and {Schinnerer}, E. and {Skibba}, R. and {Wolfire}, M.~G. and {Engelbracht}, C.~W.},
        title = "{Modeling Dust and Starlight in Galaxies Observed by Spitzer and Herschel: The KINGFISH Sample}",
      journal = {\apj},
         year = 2020,
        month = feb,
       volume = {889},
       number = {2},
          eid = {150},
        pages = {150},
          doi = {10.3847/1538-4357/ab5fdb},
archivePrefix = {arXiv},
       eprint = {1912.04914},
 primaryClass = {astro-ph.GA},
       adsurl = {https://ui.adsabs.harvard.edu/abs/2020ApJ...889..150A}
}

@ARTICLE{2009ARA&A..47..481A,
       author = {{Asplund}, Martin and {Grevesse}, Nicolas and {Sauval}, A. Jacques and {Scott}, Pat},
        title = "{The Chemical Composition of the Sun}",
      journal = {\araa},
         year = 2009,
        month = sep,
       volume = {47},
       number = {1},
        pages = {481-522},
          doi = {10.1146/annurev.astro.46.060407.145222},
archivePrefix = {arXiv},
       eprint = {0909.0948},
 primaryClass = {astro-ph.SR},
       adsurl = {https://ui.adsabs.harvard.edu/abs/2009ARA&A..47..481A}
}

@ARTICLE{1996ApJS..107..661S,
       author = {{Stasi{\'n}ska}, Grazyna and {Leitherer}, Claus},
        title = "{H II Galaxies versus Photoionization Models for Evolving Starbursts}",
      journal = {\apjs},
         year = 1996,
        month = dec,
       volume = {107},
        pages = {661},
          doi = {10.1086/192377},
       adsurl = {https://ui.adsabs.harvard.edu/abs/1996ApJS..107..661S}
}

\end{document}